\documentclass[a4paper,11pt,twoside]{book}

\RequirePackage[colorlinks,citecolor=blue,urlcolor=blue,linkcolor=blue]{hyperref}

\usepackage{lmodern}

\usepackage{graphicx}
\usepackage{subcaption}

\usepackage{mathtools}
\usepackage{bbold}
\usepackage[normalem]{ulem}
\usepackage{amssymb}
\usepackage{amsmath}
\usepackage[low-sup]{subdepth}

\usepackage{verbatim}

\usepackage{cleveref}

\usepackage[top=1.25in, bottom=1.5in, outer=1.0in, inner=1.0in]{geometry}

\usepackage{enumitem}

\newcommand{\ah}[1]{\left[ #1 \right]_\text{ah}}
\newcommand{\ev}[1]{\left< #1 \right>}
\newcommand{\cm}[2]{\left[ #1, #2 \right]}
\newcommand{\lb}{\left(}
\newcommand{\rb}{\right)}
\newcommand{\nn}{\nonumber \\}

\newcommand{\avg}[1]{{\overline{#1}}}
\DeclarePairedDelimiter\abs{\lvert}{\rvert}%

\newcommand{\LQCD}{\Lambda_{\mathrm{QCD}}}
\newcommand{\dd}{\delta}
\newcommand{\p}{\partial}
\newcommand{\dg}{\dagger}
\newcommand{\tr}{\mathrm{tr}}
\newcommand{\hc}{\mathrm{h.c.}}
\newcommand{\adj}{\mathrm{adj}}
\newcommand{\one}{\mathbb{1}}
\newcommand{\sun}{\mathfrak{su}\mathrm{(N_c)}}
\newcommand{\SUN}{\mathrm{SU(N_c)}}
\renewcommand{\Im}{ \mathrm{Im} }
\DeclareMathOperator\erf{erf}
\newcommand{\e}{\varepsilon}
\newcommand{\fm}{\mathrm{fm}}
\newcommand{\gev}{\mathrm{GeV}}
\newcommand{\LUV}{\Lambda_{\mathrm{UV}}}

\title{Simulations of the Glasma in 3+1D}
\author{David M\"uller}
\date{\today}

\begin{document}
\begin{titlepage}
	\vspace{-3cm}
	\begin{center}
	\includegraphics[scale=0.3]{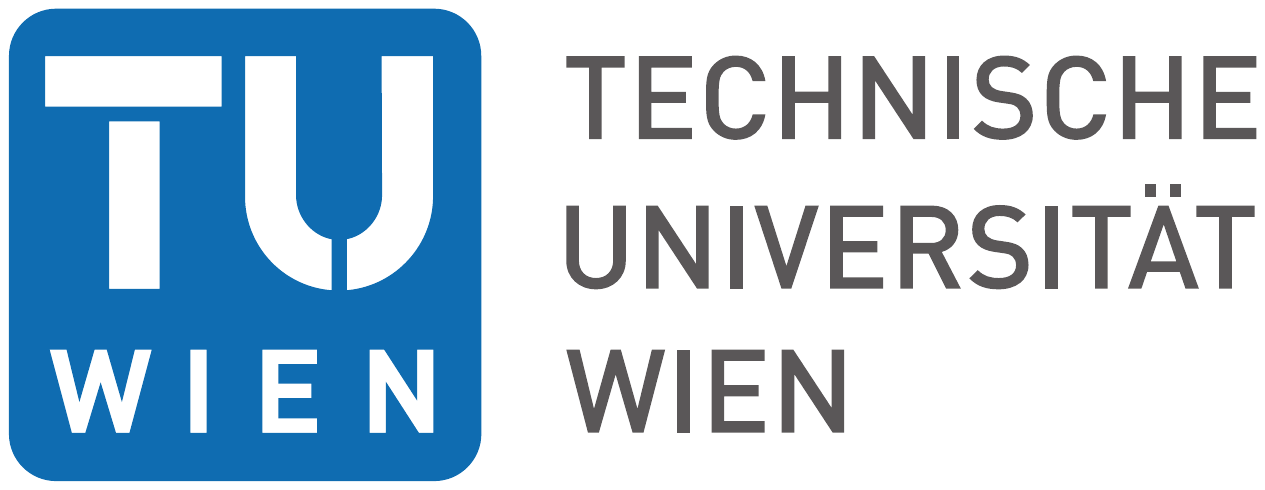}\\[0.5cm]
	\end{center}	
	\vspace{-2cm}
	
	\begin{center}
	{\LARGE DISSERTATION\\[1.0cm]}
	{\LARGE\textbf{Simulations of the Glasma in 3+1D}\\[1.0cm]}
	\end{center}
	
	\begin{center}    
		{\normalsize ausgef\"uhrt zum Zwecke der Erlangung des akademischen Grades eines Doktors der Technischen Wissenschaften unter der Leitung von}\\[1.0cm]
	\end{center}
	\begin{center}    
		{\normalsize 	Priv.Doz.\ Dr.techn.\ Andreas Ipp\\
			Institut f\"ur Theoretische Physik,\\
			Technische Universit\"at Wien, \"Osterreich \\[1.0cm]}
	\end{center}
	\begin{center}    
		{\normalsize eingereicht an der Technischen Universit{\"a}t Wien,\\
			Fakult{\"a}t f{\"u}r Physik} \\[1.0cm]     
	\end{center} 
	\begin{center}    
		{\normalsize von\\
			Dipl.-Ing.\ David M\"uller, BSc\\
			Matrikelnummer 0825271 \\
			Antonie-Alt-Gasse 12/8/11\\
			1100 Wien}\\[1.5cm]
	\end{center} 
	\begin{center} 
	\noindent\begin{tabular}{ll}
		\makebox[2cm]{Wien, am 22.03.2019}\hspace*{4cm} & \makebox[4cm]{\hrulefill}  \\[2.0cm]
	\end{tabular}

	\begin{tabular}{p{5cm}@{}p{5cm}@{}p{5cm}@{}}
		\makebox[4cm]{\hrulefill} & \makebox[4cm]{\hrulefill} & \makebox[4cm]{\hrulefill} \\
		Dr.\ Andreas Ipp    &  Dr.\ Fran\c{c}ois Gelis & Dr.\ Tuomas Lappi    \\
		(Betreuer)                & (Gutachter)               & (Gutachter)                 \\
	\end{tabular}
	\end{center} 
		
\end{titlepage}
\thispagestyle{empty}
\newpage{}


\frontmatter
\chapter*{Abstract}

The Glasma is a gluonic state of matter which can be created in collisions of 
relativistic heavy ions. It only exists for a short period of time before it evolves into the quark-gluon plasma. The existence of the Glasma is a prediction of a first-principles classical effective theory of high energy quantum chromodynamics called the color glass condensate (CGC).
In many analytical and numerical calculations within the CGC framework, the boost invariant approximation is employed. It assumes that the Lorentz-contracted longitudinal extent of the nuclei can be approximated as infinitesimally thin. Consequently, the Glasma produced from such a collision is boost invariant and can be effectively described in 2+1 dimensions. Observables of interest such as energy density, pressure or gluon occupation number of the boost invariant Glasma are by construction independent of rapidity.

The main topic of this thesis is to study how the assumption of infinitesimally thin nuclei can be relaxed.
First, we discuss the properties of the CGC and Glasma by starting with the boost invariant case.
The McLerran-Venugopalan (MV) model is used as a simple model for large heavy nuclei.
The Yang-Mills equations, which govern the dynamics of the Glasma, generally cannot be solved analytically. Numerical solutions to these equations are therefore often the only reliable approach to studying the Glasma. We discuss the methods of real time lattice gauge theory, which are the usual approach to numerically solving the Yang-Mills equations in a gauge-covariant manner. 
Having established the standard tools used to describe the boost invariant Glasma, we focus on developing a numerical method for the non-boost-invariant setting where nuclei are assumed to be thin, but of finite longitudinal extent. 
This small change is in conflict with a number of simplifications and assumptions that are used in the boost invariant case. In particular, one has to describe the collisions in 3+1 dimensions in the laboratory or center-of-mass frame, compared to the co-moving frame of the traditional method. The change of frame forces the explicit inclusion of the color charges of nuclei. In numerical simulations this is achieved using the colored particle-in-cell method.

The new method is tested using a version of the MV model which includes a parameter for longitudinal thickness. It reproduces the boost invariant setting as a limiting case. Studying the pressure components of the Glasma, one finds that the Glasma in 3+1 dimensions does not differ much from the boost invariant case and that the pressure anisotropy remains large. On the other hand, one finds that the energy density of the Glasma depends on rapidity due to the explicit breaking of boost invariance. The width of the observed rapidity profiles is controlled by the thickness of the colliding nuclei. Using only a very simple model for nuclei, the profiles can be shown to agree with experimental data.
If simulation parameters are not carefully chosen, the numerical scheme employed in the 3+1D method suffers from a numerical instability. To eliminate this instability, a completely new numerical scheme for real-time lattice gauge theory is developed. This new scheme is shown to be gauge-covariant and conserves the Gauss constraint even for large time steps. 

\addcontentsline{toc}{section}{Abstract}

\newpage{}
\newpage{}


\chapter*{Zusammenfassung}

\addcontentsline{toc}{section}{Zusammenfassung}

Das Glasma ist ein gluonischer Zustand, welcher in relativistischen Schwerionenkollisionen erzeugt werden kann und nur f\"ur sehr kurze Zeit existiert, bevor er in das Quark-Gluon-Plasma zerf\"allt.
Die Existenz des Glasmas ist eine Vorhersage des Farbglaskondensats (engl.\ ``color glass condensate" (CGC)). Das CGC ist eine klassische effektive Theorie, welche direkt aus der fundamentaleren Theorie der Quantenchromodynamik abgeleitet werden kann.     
In vielen analytischen und numerischen Rechnungen im Rahmen des CGCs kommt die boost-invariante N\"aherung zur Anwendung.
In dieser N\"aherung nimmt man an, dass die d\"unne longitudinale Ausdehnung von Atomkernen (also der Lorentz-kontrahierte Durchmesser entlang der Strahlachse bzw.\ Bewegungsrichtung) infinitesimal ist.
Eine Konsequenz dieser N\"aherung ist, dass das erzeugte Glasma invariant unter Lorentz-Boosts wird und somit effektiv in 2+1 Dimensionen beschrieben werden kann.
Observablen, also im Prinzip beobachtbare Gr\"o\ss en wie die Energiedichte, die Druckkomponenten und die Gluonenbesetzungszahl, sind dadurch per constructionem unabh\"angig von
der Rapidit\"atskoordinate.  

Das Thema dieser Dissertation ist eine neue Methode zu entwickeln, mit der man die Annahme der Boost-Invarianz lockern und umgehen kann. Ich beginne mit einer Diskussion \"uber die physikalischen Eigenschaften 
des Glasmas und des CGCs im boost-invarianten Fall.
Als einfaches Modell f\"ur gro\ss e, schwere Atomkerne kommt das McLerran-Venugopalan-Modell (MV) zum Einsatz.
Die Yang-Mills-Gleichungen, welche die Dynamik des Glasmas bestimmen, k\"onnen im Allgemeinen nicht analytisch gel\"ost werden. Numerische L\"osungsmethoden sind somit oft der einzige verl\"assliche Weg, um das Glasma zu untersuchen. Daher wird Echtzeit-Gittereichtheorie ben\"otigt, welche die Standardmethode zur numerischen L\"osung der Yang-Mills-Gleichungen darstellt.  
Nach dieser Einf\"uhrung in die Standardwerkzeuge, welche verwendet werden, um das boost-invariante Glasma zu beschreiben, liegt der Fokus auf der Entwicklung einer numerischen Methode 
f\"ur den Fall, dass Boost-Invarianz nicht mehr gilt, also wenn man f\"ur relativistische Atomkerne eine kleine, aber endliche Ausdehnung entlang der Bewegungsrichtung annimmt. 
Diese kleine \"Anderung f\"uhrt dazu, dass viele Annahmen und Vereinfachungen, die noch im boost-invarianten Fall verwendet werden konnten, nicht mehr g\"ultig sind. Inbesondere muss die Kollision im Labor- bzw.\ Schwerpunktsystem in drei r\"aumlichen Dimensionen beschrieben werden, anstelle des sich mit dem Glasma mitbewegenden Koordinatensystems im boost-invarianten Szenario. Dieser Koordinatensystemwechsel erfordert unter anderem, dass die Farbladungen der Atomkerne explizit ber\"ucksichtigt werden m\"ussen. In numerischen Simulationen gelingt das mit der Particle-in-Cell-Methode, verallgemeinert auf Farbladungen.

Die neue numerische Methode wird getestet, indem Kollisionen von Kernen mit endlicher Dicke simuliert werden. Als Anfangsbedingung f\"ur diese Simulationen dient ein erweitertes MV-Modell, welches einen neuen Parameter f\"ur longitudinale Ausdehnung besitzt. Es wird gezeigt, dass die neue Methode das boost-invariante Szenario als Grenzfall beschreiben kann. Weiters wird auch die Anisotropie der Druckkomponenten des dreidimensionalen Glasmas untersucht, wobei nur wenige Unterschiede zum boost-invarianten, zweidimensionalen Glasma festgestellt werden k\"onnen. Anders verh\"alt es sich mit der Brechung der Boost-Invarianz: betrachtet man die Energiedichte des Glasmas im lokalen Ruhesystem, kann eine starke Rapidit\"atsabh\"angigkeit festgestellt werden, welche durch die Dicke der kollidierenden Kerne beeinflusst wird. Im Vergleich mit experimentellen Resultaten von echten Kollisionsexperimenten zeigt sich, dass mit diesem sehr einfachen Modell realistische Rapidit\"atsprofile erzeugt werden k\"onnen.
Die numerische Methode, welche f\"ur dreidimensionale Kollisionssimulationen entwickelt wurde, ist auf die Wahl der Simulationsparameter sensibel und kann in gewissen F\"allen instabil werden. Die Ursache dieser numerischen Instabilit\"at wurde identifiziert und eine Erweiterung der urspr\"unglichen Methode entwickelt, welche sich als stabil erweist. Es wird gezeigt, dass diese neue Methode eichkovariant ist und das Gau\ss sche Gesetz w\"ahrend der Simulation auch f\"ur gro\ss e Zeitschritte erf\"ullt bleibt.


\chapter*{Preface}

\addcontentsline{toc}{section}{Preface}

The results and methods presented in this thesis are largely based on the following published articles:

\begin{itemize}
	\item D.~Gelfand, A.~Ipp and D.~M\"uller,
	Phys.\ Rev.\ D {\bf 94}, no.\ 1, 014020 (2016)
	[1605.07184] \cite{Gelfand:2016yho}
	\item   A.~Ipp and D.~M\"uller,
	Phys.\ Lett.\ B {\bf 771}, 74 (2017)
	[1703.00017] \cite{Ipp:2017lho}
	\item   A.~Ipp and D.~M\"uller,
	PoS EPS {\bf HEP2017}, 176 (2017)
	[1710.01732] \cite{Ipp:2017uxo}
	\item   A.~Ipp and D.~M\"uller,
	Eur.\ Phys.\ J.\ C {\bf 78}, no.\ 11, 884 (2018)
	[1804.01995] \cite{Ipp:2018hai}
	
\end{itemize} 
\newpage{}
\newpage{}


\chapter*{Acknowledgements}

\addcontentsline{toc}{section}{Acknowledgements}

There are a lot of people that I owe gratitude toward for their direct or indirect involvement with my PhD studies and this thesis.

First and foremost,
I would like to thank my supervisor Andreas Ipp for giving me the opportunity to work on this project
and providing guidance during my PhD. I thank Daniil Gelfand for collaboration during the early development of the project. I also wish to thank Jean-Paul Blaizot,  Fran\c{c}ois Gelis, Edmond Iancu, Aleksi Kurkela and S\"oren Schlichting for interesting discussions about the CGC, the Glasma and heavy-ion collisions in general.
My colleagues Patrick Kappl and Axel Polaczek deserve special thanks for many interesting discussions about technical details regarding numerical algorithms, programming and lattice gauge theory. I thank Kayran Schmidt for carefully reading this thesis.
I thank Tuomas Lappi for two very productive research stays at the University of Jyv\"askyl\"a.
I also thank Elena Petreska and Carlos A.\ Salgado for inviting me to the University of Santiago de Compostela and enlightening discussions.

Furthermore, the scientific and personal support of my friend and colleague
Alexander Haber has been invaluable throughout the past ten years of studying physics.
The same applies to
Georg Harrer,
Sebastian Sch\"onhuber,
Alexander Soloviev,
Wolfgang Steiger, 
Franz-Stephan Strobl
and many other friends I've made in the past years.
I would also like to thank Gerald Hattensauer for many supportive discussions.
I'd like to give special thanks to Frederic Br\"unner and Christian Ecker, who I've
had the pleasure to share an office with in the last five years.
My research stays in Finland have been very enjoyable due to Kirill Boguslavski, Risto Paatelainen,
Jarkko Peuron and Andrecia Ramnath. Over the past years I've also received valuable 
guidance and advice from my former supervisor Andreas Schmitt.

Finally, I would like to thank my family, in particular my parents Albert and Irene M\"uller, my uncle Paul M\"uller, and my partner Alina G.\ Dragomir.

\newpage{}
\newpage{}

\tableofcontents

\mainmatter

\chapter{Introduction} \label{cha:intro}

Heavy-ion collision experiments such as the ones at the Relativistic Heavy Ion Collider (RHIC) and the Large Hadron Collider (LHC) provide fascinating insights into the properties of strongly interacting nuclear matter under extreme conditions. The main phenomenon of interest in these collisions is the creation and evolution of the quark-gluon plasma (QGP), an exotic state of matter, where the constituents of nuclei, neutrons and protons, come apart and break into their fundamental building blocks, namely quarks and gluons. The study of the QGP is of crucial importance for testing the theory of quantum chromodynamics (QCD) by explaining and predicting experimental data. Moreover, the QGP created in relativistic nuclear collisions can be used as a model for the early universe immediately after the big bang and therefore also has implications for cosmology \cite{Heinz:2013th}.

Although the basic equations of QCD, which describe the interactions between quarks and gluons, are well established, the theoretical description of heavy-ion collisions in terms of the full theory is only partially tractable. This is due to the phenomenon of asymptotic freedom: the strong coupling constant becomes large at low momenta and consequently perturbative techniques are bound to fail. The evolution of the ``fireball" created in heavy-ion collisions is therefore split into various stages with different appropriate models used for each stage which approximate the underlying QCD processes \cite{Brambilla:2014jmp}. Roughly speaking, the three main stages are the pre-equilibrium, the equilibrium and the freeze-out. The pre-equilibrium describes the earliest stage, starting directly after the collision until the evolving matter is in thermal equilibrium, which is a process known as thermalization. The fireball then evolves as a QGP in thermal equilibrium until the freeze-out, where quarks and gluons recombine and form a gas of hadrons. This gas stops interacting after some time and the free streaming hadrons travel towards the detectors. 

One of the most striking results to come out of heavy-ion collision experiments is that the QGP in the equilibrium stage almost behaves like an ideal fluid \cite{Adcox:2004mh, Arsene:2004fa, Adams:2005dq, Romatschke:2007mq}.
The time evolution of the QGP can be described in large parts by relativistic viscous hydrodynamics \cite{Schenke:2010nt, Gale:2013da, Romatschke:2017ejr}, but this successful description only applies to the evolution of the QGP fireball itself and not to the earliest stages of the collision. Relativistic hydrodynamical simulations therefore require initial conditions from the pre-equilibrium stage for which there are different types of models. Popular choices are variants of the phenomenological MC-Glauber model \cite{Miller:2007ri}, but a more sophisticated approach to the initial state of nuclear collisions, developed over the past two and a half decades, is the color glass condensate (CGC) \cite{Gelis:2010nm, Iancu:2003xm, Iancu:2012xa}. It is a classical effective theory for high energy QCD and provides a first-principles model of the early stages of relativistic heavy-ion collisions. 

The main idea behind the CGC is a separation of scales: the hard constituents of a relativistic nucleus, i.e.\ 
partons which carry most of total momentum such as valence quarks, are described as highly Lorentz-contracted, thin sheets of classical color charge. These fast color charges generate a highly occupied color field, which represents the soft partons of the nucleus, namely mostly gluons at lower momenta.
The longitudinal momentum cutoff, at which one performs the separation into soft and hard partons, is entirely arbitrary and by requiring that observables do not depend on this artificial cutoff one can obtain a set of renormalization group equations known as the JIMWLK equations \cite{Iancu:2000hn,Ferreiro:2001qy}.  
The CGC is therefore an effective description of high energy nuclei in terms of classical color fields and color currents, whose dynamics are governed by classical Yang-Mills theory. In the CGC model the result of a collision of two nuclei is a state called Glasma \cite{Lappi:2006fp} (a combination of glass and plasma), which is a precursor to the QGP and can also be treated in terms of classical field theory. The combination of relativistic hydrodynamics with CGC/Glasma initial conditions from Yang-Mills simulations (CGC+Hydro) using models such as MC-KLN \cite{Drescher:2006ca, Drescher:2007ax} or IP-Glasma \cite{Schenke:2012wb, Schenke:2012fw} has been highly successful in explaining observed phenomena in high energy heavy-ion and proton-nucleus collisions: total particle multiplicity, azimuthal anisotropy \cite{Gale:2012rq} and higher flow coefficients \cite{McDonald:2016vlt} observed at the LHC can all be understood in terms of CGC/Glasma and hydrodynamics. Simulations of proton ion collisions even reveal details about the sub-nucleonic structure of the proton \cite{Mantysaari:2017cni, Schlichting:2014ipa}. The CGC also provides an explanation for long-range rapidity correlations (the ridge phenomenon) \cite{Dumitru:2008wn, Dusling:2009ni}.

Despite the obvious success of the CGC+Hydro approach, the description of the initial state is incomplete with regard to rapidity dependence and longitudinal dynamics. 
In the limit of very high collision energies the color field of a nucleus is usually assumed to be an infinitesimally thin shock wave, which makes it possible to derive analytic expressions for the Glasma fields directly after the collision \cite{Kovner:1995ts}. As a result of this approximation, the Glasma behaves in a boost invariant fashion, which means that observables are by construction independent of space-time rapidity and reduce the system from 3+1 dimensions to effectively only 2+1 dimensions. Any non-trivial rapidity dependence of observables is usually obtained from independent boost invariant simulations at different values of rapidity \cite{Lappi:2004sf, Schenke2014, Schenke:2016ksl}. Although 3+1 dimensional Yang-Mills simulations of the Glasma exist, they are mostly performed in the context of studying non-Abelian plasma instabilities \cite{Romatschke:2005ag, Romatschke:2006nk, Romatschke:2006wg, Rebhan:2008uj, Fukushima:2011nq, Attems:2012js, Epelbaum:2013waa, Gelis:2013rba} and not yet for obtaining realistic initial conditions for hydrodynamic simulations with correct rapidity dependence.

In this thesis we develop our new approach for simulations of the pre-equilibrium stage of 3+1 dimensional collisions in the CGC framework, which is able to go beyond the boost-invariant approximation.
This approach is distinguished from the conventional boost invariant scenario by working directly in the laboratory (or center-of-mass) frame of the colliding nuclei using Cartesian coordinates.
Instead of treating the nuclei as infinitesimally thin shock waves, we allow for a finite Lorentz-contracted longitudinal extent along the beam axis, which is proportional to $R/\gamma$, where $R$ is the nuclear radius and $\gamma$ is the Lorentz factor.
The numerical method that is used to solve the classical Yang-Mills equations in this frame takes inspiration from the colored particle-in-cell approach (CPIC) \cite{Moore:1997sn, Dumitru:2005hj} for simulating classical non-Abelian plasmas.
This new approach to simulating collisions not only covers the evolution of the Glasma but also the collision event itself and therefore gives a genuine 3+1 dimensional picture of the early stages of heavy-ion collisions in the CGC/Glasma framework.

This thesis is organized as follows: \cref{chap:bi_glasma} is a short introduction to the CGC and the Glasma in the boost invariant scenario. We discuss the McLerran-Venugopalan model of large nuclei as a simple model for relativistic heavy ions and develop the methods used to perform the classical time evolution of the Glasma numerically on a lattice. The main physical phenomena and some observables in the Glasma are also mentioned. In \cref{chap:glasma3d} the numerical method for simulating 3+1 dimensional collisions in the laboratory frame is developed. In \cref{chap:single_color_sheet} this numerical method is applied to a specific extension of the McLerran-Venugopalan model in 3+1 dimensions and some numerical results are presented. In particular, in \cref{sec:rapidity_profiles} the rapidity profile of the Glasma energy density is computed, which shows significant deviations from the boost-invariant case. Finally, in \cref{cha:semi_implicit} a numerical improvement is developed that allows for more stable and accurate simulations by employing a semi-implicit solving method. Longer, more tedious derivations and calculations are summarized in the appendix. For physical units, sign conventions and other special notation used in this thesis, see \cref{cha:notation}.

\chapter{Aspects of the boost invariant Glasma} \label{chap:bi_glasma}

This chapter is an introduction to the methods and key phenomena associated with the boost invariant Glasma, i.e.\ the Glasma created in an ultrarelativistic heavy-ion collision where the colliding nuclei are assumed to be infinitely thin.
In the following, we will focus on the necessary fundamentals in order to formulate the 3+1 dimensional model of the Glasma presented in the later chapters. For a more general overview see e.g.\ \cite{Gelis:2010nm, Iancu:2003xm, Iancu:2012xa,Lappi:2006fp,Iancu:2005jft}.

First, we introduce the classical field theory description of high energy nuclei using a very simple model, namely the McLerran-Venugopalan (MV) model. Then the boost invariant collision of two nuclei is discussed within this model. We derive the initial state and the equations of motion governing the Glasma after the collision. In order to investigate the time evolution of the Glasma, the equations of motion have to be solved numerically, which can be accomplished using real-time lattice gauge theory. Using these methods it is possible to study quantities like the energy density and the pressure components of the Glasma. Finally, state-of-the-art extensions to the MV model are discussed which allow for correct modeling of realistic nuclei and we discuss how to go beyond the boost invariant approximation.

\section{The McLerran-Venugopalan model} \label{sec:mv_model}

As discussed in \cref{cha:intro}, nuclei at very high energies can be described within the CGC framework, which is a high energy effective theory for QCD and enables an effectively classical description of nuclei in terms of classical color fields and color charges. However, even before the full development of the CGC framework as a genuine effective theory, McLerran and Venugopalan already proposed a classical model for high energy nuclei \cite{MV1, MV2}.

They argued that the valence quarks, which carry most of the total momentum of the nucleus (
referred to as ``hard" degrees of freedom), are to be treated as static, recoilless color charges whose dynamics are ``frozen" due to time dilation. Furthermore they proposed that the color charge density $\rho$ of the valence quarks provides an energy scale much larger than the QCD scale $\LQCD$. This implies that the Yang-Mills coupling constant $g$ can be considered to be weak, which justifies a classical (or eikonal) approximation. A priori the exact positions and charges of the valence quarks are unknown and therefore the classical color charges are considered to be fluctuating random variables. The valence quarks are therefore represented by a random classical color charge density $\rho^a(x)$ (or more generally a classical color current $J^a_\mu(x)$) whose probability distribution is specified by a probability functional $W[\rho]$.

On the other hand, the gluons, which carry only a fraction of the total momentum (the ``soft" degrees of freedom), are considered to be dynamic. McLerran and Venugopalan realized that because of the large number of gluons in high energy nuclei, quantum mechanical effects should be, as a first approximation, neglected\footnote{More precisely, the gluon field is highly occupied and forms a coherent state, which is essentially a classical field state.}. The soft gluons are therefore represented by a classical color field $A^a_\mu(x)$. Since the classical Yang-Mills equations must hold, the color field of the gluons $A^a_\mu$ is fixed by the color charge density $\rho^a$ of the valence quarks, such that the Gauss constraint is fulfilled.

\begin{figure}[t]
	\centering
	\includegraphics{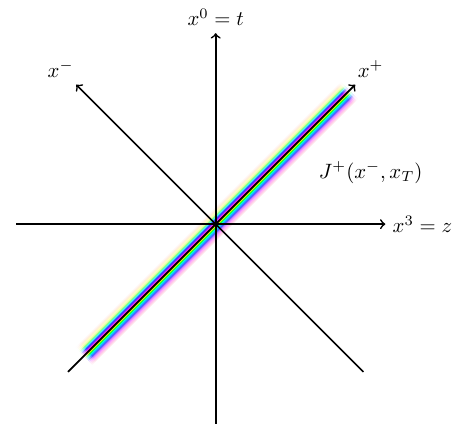}
	\caption{The schematic Minkowski diagram for a single nucleus moving at the speed of light. The transverse coordinate $x_T$ is suppressed. The color structure of the nucleus depends on $x^-$, but is independent of $x^+$ since the current $J^+$ is assumed to be static.
		\label{fig:lightcone1} }
\end{figure}

Using these assumptions we can start solving the classical problem. The equations of motion are most easily solved by employing light cone coordinates $x^{\pm} = \lb x^0 \pm x^3 \rb / \sqrt{2}$, where $t = x^0$ and $z = x^3$ are the laboratory frame coordinates. When using light cone coordinates, Latin indices as in $x^i$ are reserved for transverse coordinates, i.e.\ $i\in \{1,2\}$. We then consider (without loss of generality) the color current $J^\mu_a(x)$ of a nucleus moving in the positive $z$ direction (i.e.\ $x^3$) at the speed of light. The only relevant component is then $J^+_a(x)$ which we associate with the color charge density $\rho_a(x)$. We can drop the $x^+$ dependency of the current because we assume the valence quarks to be static in $x^+$. We are then left with
\begin{equation} \label{eq:mv_color_current1}
J^\mu(x) = \delta^{\mu+} \rho_a(x^-,x_T) t_a,
\end{equation}
where $x_T$ are the coordinates in the transverse plane spanned by $x^1$ and $x^2$, and $t_a$ are the generators of the gauge group in the fundamental representation. The fast moving nucleus that we are describing is highly Lorentz contracted in the $z$ direction which tells us that the support along the $x^-$ direction must be very thin. In the ultrarelativistic limit the longitudinal support becomes infinitesimal and the color current is proportional to $\delta(x^-)$:
\begin{equation} \label{eq:mv_color_current2}
J^\mu(x) = \delta^{\mu+} \delta(x^-) \rho_a(x_T) t_a.
\end{equation}
For the present discussion however, we will keep the more general form \cref{eq:mv_color_current1} with the color current being strongly peaked around $x^- = 0$. A schematic diagram is shown in \cref{fig:lightcone1}.

The next step is to solve the classical Yang-Mills equations (see \cref{sec:ym_conventions} for the conventions used in this thesis)
\begin{equation} \label{eq:ym_equations}
D_\mu F^{\mu\nu} = J^\nu,
\end{equation}
in the presence of the external color current $J^\nu$. In order to make progress we must choose a gauge condition and an appropriate ansatz for the color field. A useful starting point is the covariant (or Lorenz) gauge condition
\begin{equation} \label{eq:cov_gauge}
\p_\mu A^\mu = 0,
\end{equation}
with the ansatz 
\begin{equation} \label{eq:mv_gf_ansatz}
A^\mu = \delta^{+\mu} A^+_a(x^-,x_T) t_a.
\end{equation}
The only remaining field component is $A^+$ and the gauge condition forces us to drop the $x^+$ dependency of the color field due to
\begin{equation}
\p_+ A^+ = 0.
\end{equation}
We also see that the ansatz is compatible with non-Abelian charge conservation, i.e.\
\begin{equation} \label{eq:nonabelian_continuity_eq}
D_\mu J^\mu = \p_+ J^+ + i g \cm{A_+}{J^+} = 0,
\end{equation}
due to $A_+ = A^- = 0$. Inserting the ansatz into the field strength tensor
\begin{equation}
F^{\mu\nu} = \p^\mu A^\nu - \p^\nu A^\mu + i g \cm{A^\mu}{A^\nu},
\end{equation}
we find that the only non-zero components are $F^{i+}$ given simply by
\begin{equation}
F^{i+} = \p^i A^+.
\end{equation}
Plugging this into the Yang-Mills equations yields
\begin{equation}
\p_i \p^i A^+ = J^+,
\end{equation}
which reduces to a Poisson equation in the transverse plane
\begin{equation} \label{eq:mv_poisson}
- \Delta_T A^+_a(x^-,x_T) = \rho_a(x^-,x_T).
\end{equation}
This equation can be readily solved by inverting the two-dimensional Laplace operator ${\Delta_T = \sum_i \p^2_i}$, for instance using the Fourier transform. Defining the partial transform
\begin{align}
\tilde{\rho}_a(x^-,k_T) &= \int d^2 x_T \, \rho_a(x^-,x_T) e^{- i k_T \cdot x_T},
\end{align}
and its inverse
\begin{align}
\rho_a(x^-,x_T) &= \int \frac{d^2 k_T}{\lb 2 \pi \rb^2} \tilde{\rho}_a(x^-,k_T) e^{+ i k_T \cdot x_T},
\end{align}
we find the solution
\begin{equation} \label{eq:poisson_solution_momentum}
A^+_a(x^-, x_T) = \int \frac{d^2 k_T}{\lb 2 \pi \rb^2} \frac{\tilde{\rho}_a(x^-,k_T)}{k_T^2} e^{+ i k_T \cdot x_T}. 
\end{equation}
Note that the color field $A^+$ inherits its longitudinal support and shape from the color charge density $\rho$. It is also instructive to analyze the field strength of the color field: switching back to laboratory frame coordinates $t$ and $z$ we find 
\begin{equation}
A^0(x^-,x_T) = A^3(x^-,x_T) = \frac{1}{\sqrt{2}} A^+(x^-,x_T),
\end{equation}
because $A^-=0$.
The only non-zero field strength components are then $F^{0i}$ and $F^{i3}$ with $i \in \{1,2\}$, which means that the nucleus only has transverse color-electric and color-magnetic fields
\begin{align}
E_i & \equiv F_{0i} = - \p_i A_0, \\
B_i & \equiv - \frac{1}{2} \epsilon_{ijk} F^{jk} = + \epsilon_{ik} \p_k A_0,
\end{align}
which are orthogonal to each other and have the same magnitude. Note that $\epsilon_{ijk}$ refers to the three-dimensional Levi-Civita symbol, while $\epsilon_{ij}$ is the two-dimensional Levi-Civita symbol in the transverse plane. It turns out that the solution for a single propagating nucleus is completely analogous to the electromagnetic field of an electric charge moving at the speed of light, i.e.\ ultrarelativistic Li\'{e}nard-Wiechert potentials. Due to the ansatz and a clever choice of gauge given by \cref{eq:cov_gauge}, all non-linear terms of the Yang-Mills equations can be ignored and the resulting solution ends up being remarkably simple.

The choice of covariant gauge becomes inconvenient when discussing collisions of nuclei, where it turns out that light cone (LC) gauge $A^+ = 0$ is much better suited. Our goal therefore is to find the gauge transformation $V(x)$ acting on the gauge field via
\begin{equation}
A_\mu(x) \rightarrow V(x) \lb A_\mu(x) + \frac{1}{ig} \p_\mu \rb V^\dg(x),
\end{equation}
such that $A^+$ vanishes. This requirement leads us to
\begin{equation}
\p_- V^\dg(x^-,x_T) = - ig A^+(x^-,x_T) V^\dg(x^-,x_T). 
\end{equation}
The solution to this equation is given by the path-ordered exponential
\begin{equation} \label{eq:lc_gauge_wilson_line}
V^\dg(x^-,x_T) = \mathcal{P} \exp \lb - i g \intop_{-\infty}^{x^-} dz^- A^+(z^-,x_T) \rb.
\end{equation}
Note that we use the ``left means later" convention for path ordering, i.e.\  $\mathcal{P} \lb A_\mu(x) A_\nu(y) \rb = A_\mu(x) A_\nu(y)$ if $x > y$, i.e.\ $x$ comes after $y$ along the path.
\Cref{eq:lc_gauge_wilson_line} implies that the Wilson line at the asymptotic boundary $x^- \rightarrow -\infty $ is a unit matrix. 

We identify the gauge transformation $V^\dg(x^-,x_T)$ with the lightlike Wilson line starting at $x^-\rightarrow-\infty$ and ending at $x^-$. Due to $\p_+ V^\dg = 0$ we still have $A^- = 0$ like in covariant gauge. On the other hand, the transverse components of the gauge field are given by
\begin{equation}
A_i(x^-,x_T) = \frac{1}{ig} V(x^-, x_T) \p_i V^\dg(x^-, x_T).
\end{equation}
The color current also has to be transformed accordingly and is given by
\begin{equation}
J^+_{\mathrm{LC}}(x^-, x_T) = \rho_{\mathrm{LC}}(x^-, x_T) = V(x^-, x_T) \rho(x^-, x_T) V^\dg(x^-, x_T), 
\end{equation}
where the subscript ``LC" is used to differentiate the LC gauge current from the covariant gauge current $J^+$. For ultrarelativistic nuclei we can derive a simpler relationship between the transverse gauge fields $A_i$ and the LC gauge current $\rho_{\mathrm{LC}}$: a $\dd$-shaped charge density as in \cref{eq:mv_color_current2} implies that the transverse gauge field has the form of a Heaviside step function given by
\begin{align}
A^i(x^-, x_T) & = \theta(x^-) \alpha^i(x_T), \label{eq:urel_LC_gf}\\
\alpha^i(x_T) & = \frac{1}{ig} V(x_T) \p^i V^\dg(x_T), \label{eq:urel_LC_gf2}
\end{align}
where $V^\dg(x_T)$ is the asymptotic Wilson line
\begin{equation} \label{eq:asym_Wilson_line}
V^\dg(x_T) = \lim_{x^- \rightarrow \infty} V^\dg(x^-,x_T) = \mathcal{P} \exp \lb - i g \intop_{-\infty}^{+\infty} dx^- A^+(x^-,x_T) \rb.
\end{equation}
The ultrarelativistic LC current is given by
\begin{align}
J^+_{\mathrm{LC}}(x^-, x_T)  & = \delta(x^-) V(x_T) \rho(x_T) V^\dg(x_T) \nn
					& = \delta(x^-) \rho_{\mathrm{LC}}(x_T). 
\end{align}  
Inserting the above current and \cref{eq:urel_LC_gf} into the Yang-Mills equations \eqref{eq:ym_equations} yields
\begin{align}
\p_i F^{i+} & = - \p_i \p_- A^i(x^-,x_T) \nn
			& = - \delta(x^-) \p_i \alpha^i(x_T) \nn
			& = \delta(x^-) \rho_{\mathrm{LC}}(x_T),
\end{align}
which results in the relation
\begin{equation}
\p_i \alpha^i(x_T) = - \rho_{\mathrm{LC}}(x_T).
\end{equation}
We see that the space-time picture of the LC gauge solution \cref{eq:urel_LC_gf} is very different compared to the covariant gauge case: the transverse gauge fields are non-zero for $x^- > 0$ and extend to $x^-\rightarrow +\infty$. However, this is merely an artifact of the LC gauge condition as the transverse gauge fields are pure gauge, i.e.~there exists a gauge condition where the field becomes zero (namely the covariant gauge). In contrast, the actual color-electric and color-magnetic field strengths  and the color current are still concentrated around $x^- = 0$. 

Now that the relationship between the color charge density $\rho$ and the color field $A_\mu$ is established, one has to specify what the color charge distribution of a large nucleus looks like. In their original formulation McLerran and Venugopalan assumed that the charge density is $\delta$-shaped as in \cref{eq:mv_color_current2}
and proposed a simple Gaussian probability distribution for the color charge density $\rho^a(x_T)$ of the valence quarks. The distribution is defined by the charge density one- and two-point functions
\begin{align}
\ev{\rho^a (x_T)} & = 0, \label{eq:mv_onep} \\
\ev{\rho^a (x_T) \rho^b (y_T)} & = g^2 \mu^2 \dd^{ab} \delta^{(2)}(x_T - y_T). \label{eq:mv_twop}
\end{align}
The one-point function \cref{eq:mv_onep} guarantees that the nucleus is on average color neutral. The two-point function \cref{eq:mv_twop} fixes the average color charge fluctuation around zero, where $\mu$ is the phenomenological MV parameter given in units of energy or inverse length. For a large nucleus with $A$ nucleons they estimated from the average density of valence quarks that (see \cref{sec:unit_conventions} for the unit conventions used in this thesis)
\begin{equation} \label{eq:mv_mu_phenomenological}
\mu^2 \approx 1.1 A^{1/3} \,\text{fm}^{-2}.
\end{equation}
For a gold nucleus with $A=197$ this estimate gives $\mu \approx 0.5\,\text{GeV}$. The MV model does not assume a finite transverse extent of the nucleus. Instead, it approximates very large nuclei as infinitely thin, but transversely infinite walls of color charge. Within the transverse plane charges at different points are completely uncorrelated due to the $\delta^{(2)}(x_T-y_T)$ term in \cref{eq:mv_twop}. On average, the MV model exhibits translational and rotational invariance in the transverse plane.
The MV model thus gives only a very crude approximation of a realistic nucleus, but due to its simplicity with only one dimensionful parameter $\mu$ and its high symmetry, many otherwise complicated calculations can be performed analytically.  

Using the one- and two-point functions eqs.\ \eqref{eq:mv_onep} and \eqref{eq:mv_twop}, together with the assumption that the random color charges obey a Gaussian distribution, one can define the probability functional $W[\rho]$ as
\begin{equation}
W[\rho] = Z^{-1} \exp{\lb - \int d^2 x_T \frac{\rho_a(x_T) \rho_a(x_T)}{2 g^2 \mu^2} \rb}, 
\end{equation}
where $Z^{-1}$ is a normalization constant. The probability functional is used to define expectation values of arbitrary observables $\mathcal{O}[A_\mu]$ via a functional integral over all charge density configurations
\begin{equation}
\ev{\mathcal{O}[A_\mu]} \equiv \intop \mathcal{D} \rho \mathcal{O}[A_\mu] W[\rho],
\end{equation}
where it is implied that $A_\mu$ is the color field associated with the charge density $\rho$. The probability functional $W[\rho]$, eqs.\ \eqref{eq:mv_onep} and \eqref{eq:mv_twop} are invariant under arbitrary gauge transformations $\Omega(x)$:
\begin{equation}
\rho(x) \rightarrow \Omega(x) \rho(x) \Omega^\dg(x).
\end{equation}

It was later realized \cite{JalilianMarian:1996xn} that for certain derivations the $\delta$-approximation of \cref{eq:mv_color_current2} can be problematic. A more rigorous approach is to first regularize the $\delta$-peak using \cref{eq:mv_color_current1} for intermittent calculation steps and then only performing the ultrarelativistic limit in the final results. A generalization of the original MV model (see eqs.\ \eqref{eq:mv_onep} and \eqref{eq:mv_twop}) with finite longitudinal support is given by \cite{JalilianMarian:1996xn}
\begin{align}
\ev{\rho^a (x^-, x_T)} & = 0, \label{eq:mv2_onep} \\
\ev{\rho^a (x^-,x_T) \rho^b (y^-,y_T)} & = g^2 \mu^2(x^-) \dd^{ab} \dd(x^- - y^-) \delta^{(2)}(x_T - y_T), \label{eq:mv2_twop}
\end{align}
or equivalently
\begin{equation}
W[\rho] = Z^{-1} \exp{\lb - \int d x^- d^2 x_T \frac{\rho_a(x^-, x_T) \rho_a(x^-, x_T)}{2 g^2 \mu^2(x^-)} \rb}.
\end{equation}
Here the function $\mu^2(x^-)$ defines the average color charge fluctuation in the nucleus and the longitudinal shape along the $x^-$ direction. 

\begin{figure}[t]
	\centering
	\begin{subfigure}[b]{0.4\textwidth}
		\centering
		\includegraphics{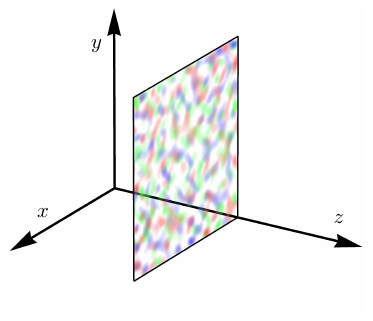}
		\caption{Single color sheet approximation}
	\end{subfigure}
	\qquad
	\begin{subfigure}[b]{0.4\textwidth}
		\centering
		\includegraphics{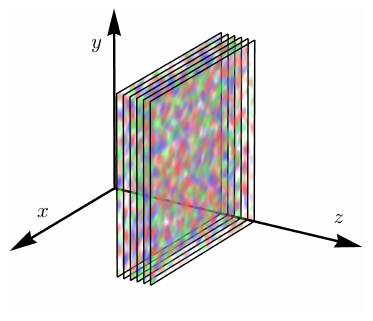}
		\caption{Generalized MV model}
	\end{subfigure}
	\caption{Illustrations of the two versions of the MV model. (a): a nucleus described by the original MV model which can be thought of as an infinitesimal sheet of random color charge. (b): the generalized MV model sketched as a stack of uncorrelated sheets of color charge. In the ultrarelativistic limit the stack collapses to an infinitely thin sheet, but due to path ordering in the infinitesimal extent, the generalized MV model does not reduce to its original formulation.
		\label{fig:color_sheets}}
\end{figure}

The physical picture of the generalized MV model is slightly different compared to the original MV model, see \cref{fig:color_sheets}: for a nucleus with thin but finite longitudinal support, one can think of the nucleus as a stack of uncorrelated, infinitely thin sheets of color charge due to the additional term $\delta(x^- - y^-)$ in the charge density two-point function \cref{eq:mv2_twop}. On the other hand, the original MV model collapses this stack of color sheets into a single, infinitesimal sheet of color charge. Differentiating between the two formulations of the MV model is important because in \cite{Fukushima:2007ki} it was found that the ultrarelativistic limit of the generalized MV model is actually not just simply obtained from the replacement $\mu^2(x^-) \rightarrow \mu^2 \delta(x^-)$ due to subtleties of the path ordering of color charges. Even in the case of infinitesimal width along $x^-$, the asymptotic Wilson line \cref{eq:asym_Wilson_line} ``remembers" the ordering of the uncorrelated sheets of color charge.

In \cite{Fukushima:2007ki} a regularization of \cref{eq:asym_Wilson_line} is presented that can be used for both numerical and analytical calculations: as described above, one imagines the nucleus as a stack of $N_s$ separate, infinitesimally thin, uncorrelated sheets of color charge. The charge density two-point function can then be approximated as
\begin{equation} \label{eq:mv2_twop_reg}
\ev{\rho^a_n (x_T) \rho^b_m (y_T)} = g^2 \mu^2 \frac{1}{N_s} \dd_{nm} \dd^{ab} \delta^{(2)}(x_T - y_T),
\end{equation}
where the indices $n,m \in \{ 1, 2, \dots, N_s \}$ are introduced to denote different sheets. This regularization effectively discretizes the term $\mu^2(x^-) \delta(x^- - y^-)$ in \cref{eq:mv2_twop}.
For each sheet one then solves the Poisson equation
\begin{equation} \label{eq:mv2_poisson_reg}
- \Delta_T A^+_n(x_T) = \rho_n(x_T).
\end{equation}
The lightlike Wilson line \cref{eq:asym_Wilson_line} is given by
\begin{equation} \label{eq:asym_Wilson_line_reg}
V^\dg(x_T) = \prod_{n=1}^{N_s} \exp \lb- i g A^+_{n} (x_T) \rb.
\end{equation}
Taking the limit $N_s \rightarrow \infty$ one obtains the ultrarelativistic limit of the generalized MV model. For $N_s = 1$ one is left with the original formulation of the MV model. We will therefore refer to this limiting case as the ``single color sheet approximation". Obviously, results generally depend on $N_s$ due to the nature of path ordering. This fact is relevant when one tries to relate the MV parameter $\mu$ to another central quantity, the saturation momentum $Q_s$.

One way to define the saturation momentum $Q_s$ within the CGC framework is via the inverse correlation length of the Wilson line two-point function in the fundamental representation
\begin{equation} \label{eq:wilson_corr_fund}
C(x_T - y_T) \equiv \frac{1}{N_c} \ev{\tr \lb V^\dg(x_T) V(y_T) \rb},
\end{equation}
where the fundamental representation Wilson lines $V(x_T)$ are given by \cref{eq:asym_Wilson_line}.
Translational and rotational invariance in the transverse plane implies that the two-point functions only depend on the distance $r = \abs{x_T-y_T}$.
An intuitive physical picture associated with \cref{eq:wilson_corr_fund} is a recoilless quark-antiquark pair ``probing" the nucleus at different transverse coordinates $x_T$ and $y_T$ at the speed of light. Due to the quarks being recoilless (or eikonal), they pass through the nucleus without changing their lightlike trajectories along $x^-$. However, the quarks are not unaffected by the nucleus: as they pass the color field, their color charges rotate in accordance with non-Abelian charge conservation, i.e.\ the continuity equation $D_\mu J^\mu = 0$ has to hold. After passing through, one can compare the rotated color charges of the quarks: if they are very close, meaning that the transverse separation $r = \abs{x_T - y_T}$ is much smaller than any transverse length scale of the nucleus (e.g.\  $\mu^{-1}$), the quarks experience the same color rotation $V(x_T) \simeq V(y_T)$ and thus ${C}(x_T-y_T) \simeq 1$. At very large separation, the color fields and  Wilson lines $V(x_T)$ and $V(y_T)$ are uncorrelated: the color charges of the quarks are completely random and thus ${C}(x_T - y_T) \simeq 0$. The Wilson line two-point functions therefore define a characteristic length scale at which an initially correlated probing recoilless quark pair becomes de-correlated. The inverse of this length is associated with the characteristic transverse momentum scale $Q_s$. Specifically, one defines
\begin{equation} \label{eq:qs_definition}
C \lb r = \frac{\sqrt{2}}{Q_s}   \rb \equiv e^{-\frac{1}{2}}.
\end{equation}
There exist multiple, slightly different definitions of the saturation momentum (see \cite{Lappi:2007ku} for details), involving for instance the adjoint representation Wilson line two-point function instead of \cref{eq:wilson_corr_fund}. Moreover, the above definition involves the two-point function in coordinate space, but it is also possible to define $Q_s$ using the momentum space representation of \cref{eq:wilson_corr_fund}.
On dimensional and parametrical grounds $Q_s$ should be roughly $g^2 \mu$. One factor of $g$ is due to \cref{eq:mv2_twop} and the second one is introduced in the Wilson line \cref{eq:asym_Wilson_line}. In the color sheet regularization the ratio $Q_s / \lb g^2 \mu \rb$ depends on $N_s$ and has to be determined numerically \cite{Lappi:2007ku}. The limiting case $N_s \rightarrow \infty$, however, can be performed analytically (see e.g.\ \cite{JalilianMarian:1996xn} and \cite{Iancu:2005jft} for a detailed derivation). Ignoring all of these complications, it will be sufficient for the purpose of this thesis to assume that the ratio $Q_s / \lb g^2 \mu \rb$ is roughly close to $1$. 

Another interesting two-point function to characterize the MV model is the correlator of the gauge field $A^+$ given by
\begin{equation}
\ev{A^{+}_a(x^-, x_T) A^{+}_b(y^-, y_T)},
\end{equation}
which can be directly related to the charge density correlator \cref{eq:mv2_twop} via \cref{eq:poisson_solution_momentum}. We then find that
\begin{equation}
\ev{A^{+}_a(x^-, x_T) A^{+}_b(y^-, y_T)} = g^2 \mu^2(x^-) \dd(x^- - y^-) \dd_{ab} \int \frac{d^2 k_T}{\lb 2 \pi \rb^2} \frac{1}{k_T^4} e^{+ i k_T \cdot \lb x_T - y_T \rb}.
\end{equation}
The momentum integral in the gauge field two-point function of the MV model turns out to be infrared divergent and has to be regularized. A popular way of curing this divergence is to introduce an infrared regulator $m$ by replacing \cref{eq:poisson_solution_momentum} with
\begin{equation} \label{eq:poisson_solution_momentum_regulated}
A^+_a(x^-, x_T) = \int \frac{d^2 k_T}{\lb 2 \pi \rb^2} \frac{\tilde{\rho}_a(x^-,k_T)}{k_T^2 + m^2} e^{+ i k_T \cdot x_T},
\end{equation}
where $m$ has units of energy. Effectively, $m$ suppresses the long-range behavior of the gauge field $A^+$, which can be illustrated by choosing $\rho$  to be a point charge
\begin{equation}
\rho^a(x^-, x_T) = q^a \dd(x^-) \dd^{(2)}(x_T),
\end{equation}
which yields the gauge field
\begin{equation}
A^+_a(x^-, x_T) = \frac{1}{2 \pi} q_a K_0 \lb m \abs{x_T} \rb \dd(x^-),
\end{equation}
where $K_j(x)$ are modified Bessel functions of the second kind. At large distances $\abs{x_T} \gg m^{-1}$ the gauge field falls of exponentially
\begin{equation}
A^+_a(x^-, x_T) \simeq q_a \frac{e^{-m \abs{x_T}}}{\sqrt{8 \pi m \abs{x_T}}} \dd(x^-).
\end{equation}
The regulator $m$ thus plays the role of a screening mass. A possible physical interpretation one can assign to this infrared regulator is that its inverse $m^{-1}$ mimics the confinement radius (roughly $\LQCD^{-1} \approx 1\,\fm$) by imposing color neutrality at distances larger than the nucleon size. Thus, a more careful way of introducing infrared regulation is to change the color charge density via
\begin{equation} \label{eq:rho_reg_m}
\tilde{\rho}_a(x^-,k_T) \rightarrow \frac{k_T^2}{k_T^2 + m^2} \tilde{\rho}_a(x^-,k_T),
\end{equation}
without modifying \cref{eq:poisson_solution_momentum}. This circumvents the problem of adding a gauge-invariance breaking term in order to obtain \cref{eq:poisson_solution_momentum_regulated}. Moreover, it is now also clear that the color charge density is globally color neutral because the zero mode of $\tilde{\rho}^a(x^-, k_T)$ vanishes.
Finally,  we can compute the (now finite) two-point function of the gauge field
\begin{align} \label{eq:Ap_twopf}
\ev{A^{+}_a(x^-, x_T) A^{+}_b(y^-, y_T)} & = g^2 \mu^2(x^-) \dd(x^- - y^-) \dd_{ab} \int \frac{d^2 k_T}{\lb 2 \pi \rb^2} \frac{1}{\lb k_T^2 + m^2 \rb^2} e^{+ i k_T \cdot \lb x_T - y_T \rb} \nn
& = g^2 \mu^2(x^-) \dd(x^- - y^-) \dd_{ab} \frac{\abs{x_T - y_T} K_1(m \abs{x_T - y_T})}{4 \pi m}.
\end{align}
We will use this two-point function in the introduction of \cref{chap:single_color_sheet}. For now it, is important to know that the MV model contains an infrared divergence that has to be manually regularized. There are other ways to cure the divergence by imposing color neutrality: one can require only global color neutrality by eliminating the zero mode of $\rho$, which is equivalent to subtracting the monopole contribution of the charge density. However, results will then depend on the size of the system. Since the MV model has no notion of finite size in the transverse plane, one is forced to manually choose some kind of system size. In practice the system size is fixed by the transverse size of the lattice in numerical simulations. Optionally one can also choose to not only eliminate the monopole, but also subtract the dipole contribution of $\rho$, which again depends on system size and changes the infrared behavior of $\rho$ in a slightly different manner. For an extended discussion on how to impose color neutrality see \cite{KRASNITZ2003268}. In this thesis we use \cref{eq:rho_reg_m} not only for convenience, but also because this way of regularization has been established in more elaborate models of nuclei such as IP-Glasma \cite{Schenke:2012wb, Schenke:2012fw}.

Even though the results discussed in this section apply to the MV model for nuclei moving in the positive $z$ direction (along $x^+$), completely analogous calculations can be done for the opposite direction along $x^-$. In this case the color current and the gauge field in covariant gauge are given by
\begin{align}
J^\mu(x) & = \delta^{\mu-} \rho_a(x^+,x_T) t_a, \\
A^-_a(x^+, x_T) & = \int \frac{d^2 k_T}{\lb 2 \pi \rb^2} \frac{\tilde{\rho}_a(x^+,k_T)}{k_T^2} e^{+ i k_T \cdot x_T}. 
\end{align}
In LC gauge, now referring to the gauge in which $A^- = 0$, we find
\begin{equation}
A_i(x^+,x_T) = \frac{1}{ig} V(x^+, x_T) \p_i V^\dg(x^+, x_T),
\end{equation}
with the lightlike Wilson line along $x^+$ given by
\begin{equation}
V^\dg(x^+,x_T) = \mathcal{P} \exp \lb - i g \intop_{-\infty}^{x^+} dz^+ A^-(z^+,x_T) \rb.
\end{equation}
The charge density correlator of the MV model reads
\begin{equation}
\ev{\rho^a (x^+,x_T) \rho^b (y^+,y_T)} = g^2 \mu^2(x^+) \dd^{ab} \dd(x^+ - y^+) \delta^{(2)}(x_T - y_T).
\end{equation}

\section{The Glasma initial conditions} \label{sec:glasma_initial}

\begin{figure}[t]
	\centering
	\includegraphics{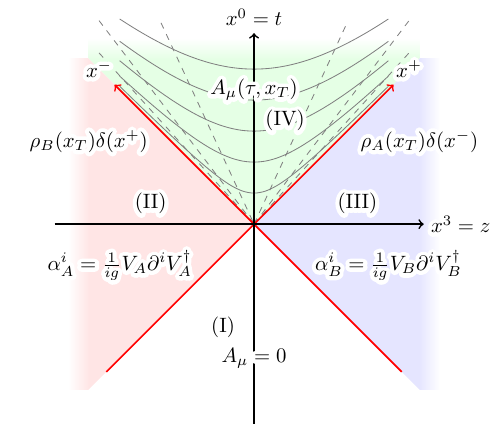}
	\caption{The schematic Minkowski diagram for an ultrarelativistic collision. The red diagonal lines along $x^+$ and $x^-$ are the infinitesimally thin color currents $J^+$ and $J^-$ respectively, which separate space-time into four distinct regions I - IV. The past light cone (region I) is completely unaffected by the collision or the single nuclei and the gauge field can be set to zero. In regions II and III we use the LC gauge solutions of the single nuclei $\alpha^i_A$ and $\alpha^i_B$. In the future light cone (region IV), the $(\tau, \eta)$ coordinate system is shown. The Glasma initial conditions are formulated at $\tau = 0^+$, which forms the boundary of the future light cone. The color field of the Glasma $A_\mu(\tau, x_T)$ for $\tau > 0$ is obtained by solving the boost invariant Yang-Mills equations.
		\label{fig:lightcone2} }
\end{figure}

Equipped with the single nucleus solutions of the classical field equations from the last section, we can start investigating ultrarelativistic collisions and the kinds of color fields produced in these collisions. We denote the two nuclei by ``A" and ``B". The combined color current of both nuclei is given by
\begin{equation} \label{eq:combined_current}
J^\mu(x) = J_A^\mu(x) + J_B^\mu(x) =  \delta^{\mu+} \dd(x^-) \rho_A(x_T) +  \delta^{\mu-} \dd(x^+) \rho_B(x_T),
\end{equation}
where nucleus ``A" moves along the $x^+$ axis and ``B" along $x^-$. 
In the ultrarelativistic limit this color current defines the sharp boundary of the light cone in \cref{fig:lightcone2}, which separates the Minkowski diagram into four distinct regions I - IV. Due to causality the regions I, II and III are unaffected by the collision and therefore we can use the single nuclei solutions from the last section. However, non-linear interaction terms in the Yang-Mills equations, result in a non-trivial color field in region IV, i.e.\ the future light cone. This color field is the Glasma. Our goal is to find this solution to the Yang-Mills equations in the presence of these moving color charges.
Formally, the solution in region IV is a functional of the two charge densities, i.e.\ 
$A_\mu [\rho_A, \rho_B]$.
Using this color field we can compute arbitrary observables via functional integration
\begin{equation}
\ev{\mathcal{O}[A_\mu[\rho_A, \rho_B]]} \equiv \intop \mathcal{D} \rho_B \mathcal{D} \rho_A \mathcal{O}[A_\mu[\rho_A, \rho_B]] W_A[\rho_A] W_B[\rho_B],
\end{equation}
where naturally one has to integrate over both color charge densities.

An important assumption is that the color charges do not suffer any recoil during the collision, i.e.\ they do not lose any longitudinal momentum and stay on their fixed, lightlike trajectories. Consequently, both color currents must be separately conserved:
\begin{align}
D_\mu J^\mu_A &= \p_+ J^+_A + i g \cm{A^-}{J^+_A} = 0, \\
D_\mu J^\mu_B &= \p_- J^-_B + i g \cm{A^+}{J^-_B} = 0.
\end{align}
Even if the trajectories are unaffected, non-Abelian charge conservation still leaves the possibility that the color charges experience color rotation due to the presence of the field, leading to non-trivial $x^+$ ($x^-$) dependence of $J^+_A$ ($J^-_B$). In order to avoid this complication we choose a gauge condition such that $A^-$ ($A^+$) vanishes along the $x^+$ ($x^-$) axis. A suitable choice is the Fock-Schwinger gauge condition
\begin{equation}
x^+ A^- + x^- A^+ = 0,
\end{equation}
which we use in the future light cone $x^+, x^- \geq 0$. In regions I - III we use the LC gauge condition $A^+ = A^- = 0$, which can be smoothly connected to Fock-Schwinger gauge at the boundaries $x^+ = 0$, $x^- \geq 0$ and $x^- = 0$, $x^+ \geq 0$. 

The combined color current \cref{eq:combined_current} features an important symmetry, namely invariance under longitudinal boosts, i.e.\ boost invariance. Specifically, this means that under the Lorentz transformation
\begin{align}
x^\pm &\rightarrow x'^\pm = e^{\pm \beta} x^\pm, \\
J^\pm(x) &\rightarrow J'^\pm(x') =  e^{\pm \beta} J^\pm(x),
\end{align}
the current remains unchanged in its functional form:
\begin{align} 
J^+_A(x^-, x_T) & \rightarrow e^{+\beta}\dd (e^{+\beta} x'^-) \rho_A(x_T) = \dd (x'^-) \rho_A(x_T), \label{eq:boosted_current_A} \\
J^-_B(x^+, x_T) & \rightarrow e^{-\beta}\dd (e^{-\beta} x'^+) \rho_B(x_T) = \dd (x'^+) \rho_B(x_T). \label{eq:boosted_current_B}
\end{align}
Since the color current is the only input that we have for solving the Yang-Mills equations (apart from boundary conditions and choice of gauge), the solution and observables should also reflect this symmetry. One therefore introduces a coordinate system in which boost invariance becomes manifest, namely the $\tau$, $\eta$ coordinate set (see \cref{sec:ym_tau_eta}). We define 
\begin{align}
x^+ = \frac{\tau}{\sqrt{2}} e^{+\eta}, \qquad
x^- = \frac{\tau}{\sqrt{2}} e^{-\eta},
\end{align}
with proper time $\tau \in (0, \infty)$ and space-time rapidity $\eta \in (-\infty, \infty)$ to parametrize the future light cone ($x^+>0$, $x^->0$) with a boundary formed by $\tau \rightarrow 0$. Inverting the above definitions yields
\begin{equation}
\tau = \sqrt{2 x^+ x^-},\qquad
\eta = \frac{1}{2} \ln \lb \frac{x^+}{x^-} \rb.
\end{equation}
Applying the coordinate transformation to the gauge field we find (see \cref{sec:ym_tau_eta})
\begin{align}
A^\tau &= + \frac{1}{\tau} \lb x^+ A^- + x^- A^+ \rb = A_\tau, \label{eq:a_tau_pm}\\
A^\eta &= - \frac{1}{\tau^2} \lb x^+ A^- - x^- A^+ \rb = - \frac{1}{\tau^2} A_\eta.
\end{align}
We can therefore make a manifestly boost invariant ansatz for the color field by ``stiching" together separate functions in the four regions of the Minkowski diagram \cite{Kovner:1995ja}:
\begin{align}
A^i(x) &= \theta(x^+) \theta(x^-) \alpha^i(\tau, x_T) + \theta(-x^+) \theta(x^-) \alpha^i_A(x_T) + \theta(x^+) \theta(-x^-) \alpha^i_B(x_T), \label{eq:sep_ansatz_1} \\
A^\eta(x) &= \theta(x^+) \theta(x^-) \alpha^\eta(\tau, x_T), \label{eq:sep_ansatz_2}
\end{align}
where we suppressed all dependencies on $\eta$, rendering the ansatz boost invariant. In principle it would be possible to perform an $\eta$-dependent gauge transformation without breaking boost invariance in gauge invariant observables, but for the sake of simplicity we also require this symmetry at the level of gauge fields.
The transverse gauge fields of the single nuclei fulfill
\begin{equation}
\p_i \alpha^i_{A,B}(x_T) = - \rho_{A,B}(x_T),
\end{equation}
and are pure gauge in region (II) (in the case of nucleus A) and (III) (nucleus B).
Due to \cref{eq:a_tau_pm} the Fock-Schwinger gauge condition renders the temporal component $A^\tau$ zero. Plugging this ansatz into the Yang-Mills equations one obtains a few worrying terms such as $\dd^2$ and $\dd'$ due to partial derivatives acting on the Heaviside functions. Setting all coefficients of these problematic terms to zero one finds matching conditions at the boundary $\tau \rightarrow 0$ \cite{Kovner:1995ja}:
\begin{align} \label{eq:glasma_initial_1}
\alpha^i(\tau \rightarrow 0, x_T) &= \alpha^i_A(x_T) + \alpha^i_B(x_T), \\
\alpha^\eta(\tau \rightarrow 0, x_T) &= \frac{ig}{2} \cm{\alpha^i_A(x_T)}{\alpha^i_B(x_T)}, \label{eq:glasma_initial_2}
\end{align}
together with $\p_\tau \alpha^i(\tau \rightarrow 0, x_T) = \p_\tau \alpha^\eta(\tau \rightarrow 0, x_T) = 0$. These matching conditions form the starting point for the time evolution of the color field in the future light cone, i.e.\ the Glasma field. 

The longitudinal electric and magnetic field in the laboratory frame is related to the field strength tensor in the co-moving frame via (see \crefrange{eq:lab_to_taueta_1}{eq:lab_to_taueta_4})
\begin{align}
E_3 &= \frac{1}{\tau} F_{\tau \eta}, \\
B_3 &= - F_{12}.
\end{align}
At $\tau \rightarrow 0$ this yields
\begin{align}
E_3(\tau \rightarrow 0, x_T) &= - i g \dd_{ij} \cm{\alpha_{A,i}(x_T)}{\alpha_{B,j}(x_T)}, \label{eq:bi_initial_EL_BL_1}\\
B_3(\tau \rightarrow 0, x_T) &= - i g \epsilon_{ij} \cm{\alpha_{A,i}(x_T)}{\alpha_{B,j}(x_T)}, \label{eq:bi_initial_EL_BL_2}
\end{align}
where $\delta_{ij}$ and $\varepsilon_{ij}$ are the Kronecker delta and the Levi-Civita symbol in two dimensions.
The magnetic contribution of the single nucleus fields $\alpha^i_{A/B}$ vanishes due to them being pure gauge as specified by \cref{eq:urel_LC_gf2}. On the other hand, the transverse electric and magnetic field components vanish for $\tau \rightarrow 0$. The collision of two nuclei therefore produces fields that are initially purely longitudinal. This leads to the following physical picture directly after the collision \cite{Lappi:2006fp}: as the purely transverse color fields of the nuclei move away from the collision region, longitudinal electric and magnetic flux tubes span between the two nuclei. The size of these Glasma flux tubes corresponds to the size of correlated domains in the color fields of the nuclei, i.e.\ roughly $Q^{-1}_s$ \cite{Dumitru:2014nka}. After the initial formation, the flux tubes quickly evolve and start to expand in the transverse directions. In the next section we will introduce methods to study the time evolution of this system numerically.

\section{Numerical time evolution}

\subsection{Boost invariant equations of motion}

The gluonic matter produced in an ultrarelativistic heavy-ion collision is described as a boost invariant classical color field in the forward light cone. In the last section we sketched the derivation of the initial conditions, i.e.\ how this color field is determined by the color fields of the nuclei at the boundary $\tau \rightarrow 0$. For $\tau > 0$ the time evolution is governed by the source-free Yang-Mills equations
\begin{equation}
D_\mu F^{\mu \nu}(x) = 0.
\end{equation}
However, since it is our goal to work in $\lb \tau, \eta \rb$ coordinates, it is reasonable to start from the Yang-Mills action in flat coordinates
\begin{equation}
S = - \frac{1}{2} \int d^4 x \, \tr \left[ F_{\mu\nu} F^{\mu\nu} \right]
\end{equation}
and perform the coordinate transformation there. Furthermore, we implement temporal (or Fock-Schwinger) gauge $A_\tau = 0$ in order to be consistent with the initial conditions at $\tau \rightarrow 0$ and restrict ourselves to manifestly boost invariant gauge fields $A_i(\tau, x_T)$ and $A_\eta(\tau, x_T)$. The boost invariant action then reads (see \cref{sec:ym_tau_eta} for a derivation)
\begin{equation} \label{eq:bi_action}
S = \intop d\tau d^2x_T d\eta \, \tr \left[  \tau \p_\tau A_i  \p_\tau A_i + \frac{1}{\tau} \lb \p_\tau A_\eta \rb^2 - \frac{\tau}{2} F_{ij} F_{ij} - \frac{1}{\tau}  D_i A_\eta D_i A_\eta  \right],
\end{equation}
supplemented by the Gauss constraint
\begin{equation}
D_i P^i + i g \cm{A_\eta}{P^\eta} = 0.
\end{equation} 
The canonical momenta $P^i(\tau, x_T)$ and $P^\eta(\tau, x_T)$ are given by
\begin{align}
P^i &= \tau \p_\tau A_i, \label{eq:bi_momentum_i} \\
P^\eta &= \frac{1}{\tau} \p_\tau A_\eta. \label{eq:bi_momentum_eta}
\end{align}
Varying the action with respect to the gauge fields, one obtains the equations of motion (EOM)
\begin{align}
\p_\tau P^i &=  \tau D_j F_{ji} - \frac{ig}{\tau}  \cm{A_\eta}{D_i A_\eta}, \label{eq:bi_glasma_eom1}\\
\p_\tau P^\eta &= \frac{1}{\tau} D_i \lb D_i A_\eta \rb. \label{eq:bi_glasma_eom2}
\end{align}
In these variables the initial conditions are given by
\begin{align}
P^\eta(\tau \rightarrow 0, x_T) &= - 2 \alpha^\eta(\tau \rightarrow 0, x_T), \label{eq:bi_glasma_ini_1} \\
A_\eta(\tau \rightarrow 0, x_T) &= 0, \label{eq:bi_glasma_ini_2} \\
P^i(\tau \rightarrow 0, x_T) &= 0, \label{eq:bi_glasma_ini_3} \\
A_i(\tau \rightarrow 0, x_T) &= - \alpha^i(\tau \rightarrow 0, x_T). \label{eq:bi_glasma_ini_4}
\end{align}

In general, there exist no closed-form solutions to these equations due to the non-linear self-interaction terms. 
A number of approximate solution approaches have been proposed:
for example, one can solve the boost invariant Yang-Mills equations by expanding in terms of the color fields of the nuclei \cite{Kovner:1995ts}. 
This is only strictly correct if the color fields can be considered weak. Within the color glass condensate framework it is predicted that nuclei have non-perturbatively large fields $A_i \propto 1/g$. Thus, a series expansion is bound to fail at least for nucleus-nucleus collisions, but is valid for proton-nucleus collisions.
Remarkably, one still obtains good agreement with numerical results by approximating the Glasma as weak and non-interacting, but keeping all orders of the color field in the Glasma initial conditions at $\tau \rightarrow 0$ \cite{Fukushima:2007ja, Fujii:2008km}.
A very intriguing semi-analytic approach is to perform a series expansion of the gauge fields in powers of proper time $\tau$ \cite{Fries:2006pv, Chen:2015wia, Li:2017iat, Fries:2017fwk}. A downside is that due to slow convergence of the series, very high orders in $\tau$ are required to obtain reasonable agreement with known numerical results.
A more general approach, which is also easily extendable to models of nuclei more complicated than the MV model and which in principle works up to arbitrary $\tau$, is to solve the equations numerically on a lattice.  

\subsection{Real-time lattice gauge theory} \label{sec:real_time_lattice}

Real-time lattice gauge theory is a numerical method used to solve the classical Yang-Mills equations on the lattice in a gauge covariant manner. It is based on lattice gauge theory (see \cite{Lepage:1998dt, gattringer2009quantum, smit_2002} for introductions to the topic), but instead of solving path integrals in Euclidean space, one formulates a discretized, gauge invariant action in Minkowski space to obtain (discretized) equations of motion. The main strength of this method is that it retains a notion of gauge covariance even for the discretized system. We will  quickly discuss the ideas behind this method.

A naive approach to solving the Yang-Mills equations could be the following: starting from the continuous action
\begin{equation}
S = \int d^4 x \,  \tr  \left[ -\frac{1}{2} F_{\mu\nu} F^{\mu\nu} \right],
\end{equation}
we make a choice on the degrees of freedom to use. An obvious choice are the gauge fields $A_\mu$ and their canonical momenta $P^\mu$. Then, we replace Minkowski space $\mathbb{M}^4$ by a regular hypercubic lattice with lattice spacings $a^\mu$ (see \cref{sec:lattice_notation}):
\begin{equation} \label{eq:minkowski_lattice}
\mathbb{M}^4 \rightarrow \Lambda^4 = \left\{ x \, | \, x = \sum_{\mu=0}^3 n_\mu \hat{a}^\mu, \quad n_\mu \in \mathbb{Z}  \right\},
\end{equation}
where $\hat{a}^\mu = a^\mu \hat{e}_\mu$ with unit vectors $\hat{e}_\mu$. 
In this step one replaces all continuous derivatives (such as in $F_{\mu\nu}$) with finite difference expressions accurate up to some order in $a^\mu$. In principle, this procedure yields a discretized action, and upon variation, discrete equations that have a correct continuum limit. Unfortunately, doing so one destroys the most important property of a gauge theory, namely local gauge invariance. The main issue is that local gauge transformations
\begin{equation}
A_\mu(x) \rightarrow \Omega(x) \lb A_\mu(x) + \frac{1}{i g} \p_\mu \rb \Omega^\dg(x)
\end{equation}
involve derivatives acting on the gauge transformation $\Omega$. Consequently, the discretized action and the equations that follow from variation are only gauge invariant up to some order in $a^\mu$ and gauge invariance too is only approximate. This is particularly troublesome when solving path integrals, which involve integrating over non-smooth field configurations, but breaking gauge invariance is also undesirable when merely solving classical equations.

However, there is a way to remedy this problem. Instead of making the obvious choice of $\lb A_\mu, P^\mu \rb$, we use a different set of degrees of freedom. In \cref{sec:mv_model}, we introduced Wilson lines: given a path $\mathcal{C}(x,y)$ starting at $x$ and ending at $y$, the Wilson line is defined by the path ordered exponential
\begin{equation}
U[\mathcal{C}] = \mathcal{P} \exp \lb - i g \intop_{\mathcal{C}} dx^\mu A_\mu(x) \rb \in \SUN.
\end{equation}
Applying a local gauge transformation to $A_\mu(x)$, one can show that the Wilson line transforms non-locally at the start and end points
\begin{equation}
U[\mathcal{C}] \rightarrow \Omega(y) U[\mathcal{C}]  \Omega^\dg(x).
\end{equation}
In particular, if the path $\mathcal{C}$ is closed (i.e.\ a ``loop"), then one talks about Wilson loops. In this special case $U[\mathcal{C}]$ transforms locally, i.e.
\begin{equation}
U[\mathcal{C}] \rightarrow \Omega(x) U[\mathcal{C}]  \Omega^\dg(x).
\end{equation}
Taking the trace of a Wilson loop yields a gauge invariant quantity
\begin{equation}
\tr \left[ U[\mathcal{C}] \right] \rightarrow \tr \left[ \Omega(x) U[\mathcal{C}]  \Omega^\dg(x) \right] = \tr \left[ U[\mathcal{C}]  \right].
\end{equation}

On a lattice, for instance defined by \cref{eq:minkowski_lattice}, the shortest possible Wilson lines are called gauge links, which connect nearest neighbors on the lattice. For example $U_{x,\mu}$ connects the points $x$ and $x+\hat{a}^\mu$. 
In lattice gauge theory, $U_{x,\mu}$ is actually the anti-path-ordered Wilson line (denoted by $\overline{\mathcal{P}}$)
\begin{equation}
U_{x,\mu} = \overline{\mathcal{P}} \exp \lb + i g \intop_x^{x+\hat{a}^\mu} dx^\mu A_\mu(x) \rb,
\end{equation}
which transforms as
\begin{equation}
U_{x,\mu} \rightarrow \Omega(x) U_{x,\mu} \Omega^\dg(x+\hat{a}^\mu),
\end{equation}
or shorter (see \cref{sec:lattice_notation} for an introduction to this shorthand notation)
\begin{equation} \label{eq:latt_gauge_trans}
U_{x,\mu} \rightarrow \Omega_x U_{x,\mu} \Omega^\dg_{x+\mu}.
\end{equation}
Close to the continuum limit we can use the approximation
\begin{equation}
U_{x,\mu} \simeq \exp \lb i g a^\mu A_\mu(x+\frac{1}{2} \hat{a}^\mu) \rb,
\end{equation}
where the gauge field is placed at the mid-point of the path. 
The smallest possible Wilson loops that we can form on the lattice are the rectangular plaquettes ($1 \times 1$ Wilson loops)
\begin{equation} \label{eq:plaquette_definition}
U_{x,\mu\nu} = U_{x,\mu} U_{x+\mu,\nu} U_{x+\mu+\nu,-\mu} U_{x+\nu,-\nu},
\end{equation}
where we define gauge links with negative directions via $U_{x+\mu,-\mu} = U^\dg_{x,\mu}$. Taking the continuum limit of the plaquette one finds
\begin{equation} \label{eq:plaquette_exponential}
U_{x,\mu\nu} \simeq \exp \lb i \lb g a^\mu a^\nu F_{\mu\nu}(x) + \mathcal{O}(a^3) \rb \rb,
\end{equation}
and consequently
\begin{equation} \label{eq:tr_umunu}
\tr \left[ 2 - U_{x,\mu\nu} -U^\dg_{x,\mu\nu} \right]  \simeq \lb g a^\mu a^\nu\rb^2 \tr \left[ F_{\mu\nu}(x)^2 \right] + \mathcal{O}(a^6),
\end{equation}
which is a gauge invariant expression. Note that the error term is one order higher than one would expect just from the continuum limit of the plaquette. In this expansion the terms proportional to $a^5$ cancel, because the term $\tr \left[ U_{x,\mu\nu} + U^\dg_{x,\mu\nu} \right]$ amounts to taking the real value of the plaquette, which eliminates precisely all $\mathcal{O}(a^5)$ terms (see subsection 2.3.2 of \cite{gattringer2009quantum}, or \cite{Lepage:1998dt, BilsonThompson:2002jk}).
Apart from the correct continuum limit, the most important property of \cref{eq:tr_umunu} is its exact invariance under gauge transformations \cref{eq:latt_gauge_trans}. Therefore, by reducing the space of local gauge transformations $\Omega(x)$ to so-called lattice gauge transformations $\Omega_x$ only defined at lattice points $x \in \Lambda$, one can retain a notion of exact gauge invariance for the discretized system. This allows us to formulate a lattice gauge invariant action in Minkowski space. We start by splitting the action into its electric and magnetic part
\begin{equation}
S = S_E - S_B,
\end{equation} 
with
\begin{align}
S_E &= \intop d^4 x \sum_i \tr \left[ F_{0i}(x)^2 \right], \\
S_B &= \intop d^4 x  \sum_{i,j}\frac{1}{2} \tr \left[  F_{ij}(x)^2 \right],
\end{align}
where $i, j \in \{ 1,2,3 \}$.
We then use the gauge invariant combination of plaquettes \cref{eq:tr_umunu} to approximate
\begin{align}
S_E &\simeq V \sum_x \sum_i \frac{1}{\lb g a^0 a^i \rb^2} \tr \left[ 2- U_{x,0i} - U^\dg_{x,0i} \right], \\
S_B &\simeq V \sum_x \sum_{i, j} \frac{1}{2 \lb g a^i a^j \rb^2} \tr \left[ 2- U_{x,ij} - U^\dg_{x,ij} \right],
\end{align}
where $V = \prod_{\mu} a^\mu$ is the discrete space-time volume. The standard Wilson gauge  action \cite{PhysRevD.10.2445} is therefore given by
\begin{align} \label{eq:wilson_action}
S[U] &=  V \sum_x  \bigg( \sum_i \frac{1}{\lb g a^0 a^i \rb^2} \tr \left[ 2- U_{x,0i} - U^\dg_{x,0i} \right] \nn
& \qquad \qquad -  \sum_{i, j} \frac{1}{2 \lb g a^i a^j \rb^2} \tr \left[ 2- U_{x,ij} - U^\dg_{x,ij} \right] \bigg).
\end{align}
Since we have discretized all of Minkowski space as a lattice, including the time direction $x^0$, the above action is merely a function of gauge links $U_{x,\mu}$ (instead of a functional). Extremizing $S[U]$ with respect to all gauge links yields discretized equations of motion and constraints.

However, coming back to solving the time evolution of the boost invariant Glasma, the Wilson action \cref{eq:wilson_action} in its original formulation is not particularly useful. It is formulated in the $\lb t, z\rb$ frame, instead of the more appropriate $\lb \tau, \eta \rb$ frame in which boost invariance becomes manifest. A rigorous approach to derive the correct action (see e.g.\ \cite{Krasnitz:1998ns, Lappi:2005tt}) is to start from the discretized Wilson action \cref{eq:wilson_action} and take the partial continuum limit $a^0 = a^3 \rightarrow 0$. One then ends up with an action where the transverse plane is discretized as a lattice, while $t = x^0$ and $z = x^3$ are continuous. It is then possible to perform the coordinate transformation to the $\lb \tau, \eta \rb$ system and implement boost invariance by assuming all fields to be independent of $\eta$.

Alternatively, one can also obtain the boost-invariant discretized action in temporal gauge $A_\tau = 0$ by taking a few shortcuts (see \cref{app_bi_latt} for more detailed derivation): the idea is to start directly from the continuous boost invariant case given by \cref{eq:bi_action} and replace the transverse fields $A_i(\tau, x_T)$ with transverse gauge links $U_{x,i}(\tau)$. The $F_{ij}^2$ term can be discretized using \cref{eq:tr_umunu} in a straightforward manner. Introducing the gauge-covariant forward and backward finite differences 
\begin{align}
D^F_i A_{x,\eta}(\tau) & \equiv \lb {U_{x,i}(\tau) A_{x+i,\eta}(\tau) U_{x+i,-i}(\tau) - A_{x,\eta}} \rb / a^i, \label{eq:gc_finite_forward} \\
D^B_i A_{x,\eta}(\tau) & \equiv \lb {A_{x,\eta}(\tau) - U_{x,-i}(\tau) A_{x-i,\eta} U_{x-i,i}(\tau)} \rb / a^i, \label{eq:gc_finite_backward}
\end{align}
we can formulate the discretized action for the boost invariant system:
\begin{align}
S & = \intop d\tau d\eta \sum_x \lb \prod_i a^i \rb \, \tr \bigg[  \frac{1}{\tau} \sum_i \lb P^i_x(\tau) \rb^2 + \tau \lb P^\eta_x(\tau) \rb^2 \nn
& \qquad \qquad - \sum_{i,j} \frac{\tau}{2 \lb g a^i a^j \rb^2} \lb 2 - U_{x,ij}(\tau) - U^\dg_{x,ij} (\tau) \rb - \frac{1}{\tau} \sum_i \lb D^F_i A_{x,\eta}(\tau) \rb^2  \bigg],
\end{align}
with the canonical momenta given by
\begin{align}
P^\eta_x(\tau) &= \frac{1}{\tau} \p_\tau A_{x,\eta}(\tau), \\
P^i_x(\tau) &= - i \frac{\tau}{g a^i} \lb \p_\tau U_{x,i} (\tau) \rb U^\dg_{x,i}(\tau).
\end{align}
The relation for $P^i_x(\tau) \in \sun$ guarantees that the gauge links $U_{x,i}(\tau) \in \SUN$ remain special unitary matrices throughout the time evolution. 

As we have already implemented temporal gauge, it is not possible to find the Gauss constraint through variation of the above action. Nevertheless, an educated guess yields
\begin{equation} \label{eq:bi_latt_gauss_text}
\sum_i D^B_i P^i_x(\tau) + i g \cm{A_{x,\eta}(\tau)}{P^\eta_x(\tau)} = 0.
\end{equation}
If we would have started from an action continuous in $\lb \tau, \eta \rb$ without fixing any gauge condition (as detailed in \cite{Lappi:2005tt}), the above equation would be the result of varying with respect to $A_\tau$.
Performing the variation with respect to $A_{x,\eta}$ is straightforward and yields
\begin{equation}
\p_\tau P^\eta_x  = \frac{1}{\tau} \sum_i D^2_i A_{x,\eta},
\end{equation}
where $D^2_i = D^F_i D^B_i$ (no sum implied) is the second order gauge-covariant finite difference. 
Varying with respect to $U_{x,i}(\tau)$ using (see \cref{app_var})
\begin{equation}
\dd U_{x,i}(\tau) = i g a^i \dd A_{x,i}(\tau) U_{x,i}(\tau)
\end{equation}
yields the equations of motion for the transverse components
\begin{equation}
\p_\tau P^i_x = - \sum_j \frac{\tau}{g a^i \lb a^j \rb^2} \ah{ U_{x,ij} + U_{x,i-j}}
- \frac{i g}{\tau} \cm{A^{(+i)}_{x,\eta}}{D^F_i A_{x,\eta}},
\end{equation}
with the parallel transported field $A^{(+i)}_{x,\eta} \equiv U_{x,i} A_{x+i,\eta} U^\dg_{x,i}$ and the anti-hermitian, traceless part (see \cref{eq:ah_definition_2} of \cref{sec:ym_conventions})
\begin{equation} \label{eq:ah_definition}
\ah{U} \equiv \frac{1}{2i} \lb U - U^\dg \rb - \frac{1}{N_c} \tr \left[ \frac{1}{2i} \lb U - U^\dg \rb \right] \one.
\end{equation}
Even though a bit cumbersome, it is straight-forward to prove that the equations of motion conserve the Gauss constraint \cref{eq:bi_latt_gauss_text}.

The final step is to discretize the proper time coordinate $\tau$ as well. Using a finite time step $\Delta \tau$ one can formulate a leapfrog scheme, where momenta and fields are evaluated at fractional and whole time steps (i.e.\ $\tau_{n+1/2}$ and $\tau_n$ with $\tau_n = n \Delta \tau$) respectively. The complete set of equations then reads
\begin{align}
P^\eta_x(\tau_{n+\frac{1}{2}}) &= P^\eta_x(\tau_{n-\frac{1}{2}}) + \frac{\Delta \tau}{\tau_n} \sum_i D^2_i A_{x,\eta}(\tau_n), \\
P^i_x(\tau_{n+\frac{1}{2}}) &= P^i_x(\tau_{n-\frac{1}{2}}) - \sum_j \frac{\Delta \tau \tau_n}{g a^i \lb a^j \rb^2} \ah{ U_{x,ij}(\tau_n) + U_{x,i-j}(\tau_n)} \nn
& \qquad \qquad - \frac{i g \Delta \tau}{\tau_n} \cm{A^{(+i)}_{x,\eta}(\tau_n)}{D^F_i A_{x,\eta}(\tau_n)}, \\
A_{x,\eta}(\tau_{n+1}) &= A_{x,\eta}(\tau_{n}) + \Delta \tau \tau_{n+\frac{1}{2}} P^\eta_x(\tau_{n+\frac{1}{2}}), \\
U_{x,i}(\tau_{n+1}) &= \exp \lb \frac{i g a^i \Delta \tau}{\tau_{n+\frac{1}{2}}} P^i_x(\tau_{n+\frac{1}{2}}) \rb U_{x,i}(\tau_n).
\end{align}

An advantage of using the leapfrog scheme is its accuracy up to second order in the time step $\Delta \tau$ and its symmetry under time reversal. Furthermore, the use of exponentiation in the evolution equation for $U_{x,i}$ guarantees that gauge links remain special unitary matrices throughout the numerical time evolution, also for finite $\Delta \tau$.
Probably the most important aspect is that the leapfrog equations are gauge covariant, i.e.\ they transform consistently under lattice gauge transformations. Gauge covariance also implies conservation of the discrete Gauss constraint given by
\begin{equation}
\sum_i D^B_i P^i_x(\tau_{n+1/2}) + i g \cm{A_{x,\eta}(\tau_n)}{P^\eta_x(\tau_{n+\frac{1}{2}})} = 0.
\end{equation}
In the expression for the Gauss constraint momenta are evaluated at $\tau_{n+1/2}$ while gauge fields and links are evaluated at $\tau_n$. In this form the Gauss constraint is actually exactly conserved for finite time steps $\Delta \tau$. It is a non-trivial task to find different update equations that are consistent with a discretized Gauss constraint, in particular if one starts from an action that is continuous in $\tau$ instead of discretizing $\tau$ already at the level of the action. This issue will be addressed in more detail in \cref{cha:semi_implicit}.

\subsection{Initial conditions on the lattice} \label{sec:bi_glasma_initial_lattice}

The last missing piece in the numerical description of the boost invariant Glasma is to formulate the initial conditions (see  \cref{eq:bi_glasma_ini_1,eq:bi_glasma_ini_2,eq:bi_glasma_ini_3,eq:bi_glasma_ini_4}) on the lattice in terms of transverse gauge links $U_{x,i}(\tau)$ and the canonical momenta $P^\eta_x(\tau)$. 

We have to find the pure gauge color field of a single nucleus in LC gauge. Starting in covariant gauge we discretize the charge density two-point function \cref{eq:mv2_twop_reg} of the MV model in the transverse plane. The correlator then reads
\begin{equation} \label{eq:mv2_twop_reg_latt}
\ev{\rho^a_{x,n} \rho^b_{y,m}} = \frac{ g^2 \mu^2 }{N_s a^1 a^2} \dd_{nm} \dd^{ab} \delta_{xy},
\end{equation}
where $a^i$ with $i \in \left\{ 1, 2 \right\}$ are transverse lattice spacings and $x,y$ are discretized transverse coordinates.  The indices $n,m$ refer to the set of $N_s$ independent color sheets. The probability functional of the MV model is Gaussian, which makes it easy to generate random numbers  on a computer according to the above two-point function: at each point $x$ in the transverse plane, each color component $a$ (of which there are $N_c^2 - 1$) and each sheet index $n$, the color charge density $\rho^a_{x,n}$ is an independently sampled Gaussian random number with zero mean and variance $g^2 \mu^2 a^1 a^2 / N_s$. Since the MV model describes nuclei with infinite extent in the transverse plane and computers provide only a finite amount of memory, we have to use a finite transverse lattice with periodic boundary conditions. For simplicity, we use a square lattice with $N_T \times N_T$ cells, $a^1 = a^2 = a_T$ lattice spacing and transverse length $L_T = N_T a_T$. 

After generating a random color charge density on the lattice, one has to solve the discretized Poisson equation 
\begin{equation}
- \Delta_T A^{a,+}_{x,n} = - \sum_{i=1,2} \frac{A^{a,+}_{x+i,n} + A^{a,+}_{x-i,n} - 2 A^{a,+}_{x,n}}{\lb a^i \rb^2} = \rho^a_{x,n},
\end{equation}
which is a lattice regularization of \cref{eq:mv2_poisson_reg}. A fast and elegant way of numerically solving the Poisson equation above is to do so in momentum space with the help of the discrete Fourier transformation. We define the Fourier series
\begin{equation}
\rho^a_{x,n} = \frac{1}{\lb 2 \pi \rb^2} \sum_{l=0}^{N_T-1} \, \sum_{m=0}^{N_T-1} \lb \Delta k \rb^2 \tilde{\rho}^a_{k,n} \exp \lb + i \lb l \Delta k x^1 + m \Delta k x^2 \rb \rb 
\end{equation}
where $\Delta k = 2\pi / L_T$. The vector $k = \lb l \Delta k, m \Delta k \rb$ can be interpreted as the discretized momentum on the lattice. The above transformation can be readily calculated on a computer using the fast Fourier transformation (FFT).
Using this definition the Poisson equation simply reads
\begin{equation}
\tilde{k}^2_T \, \tilde{A}^{a,+}_{k,n} =  \tilde{\rho}^a_{k,n},
\end{equation}
with the squared transverse lattice momentum
\begin{equation}
\tilde{k}^2_T = \sum_{i=1,2} \lb \frac{2}{a^i} \rb^2 \sin^2 \lb \frac{k_i a^i}{2} \rb.
\end{equation}
The momentum representation also makes it possible to directly perform infrared and (optional) ultraviolet regulation. The regularized solution is then given by
\begin{equation} \label{eq:poisson_sol_reg_latt}
\tilde{A}^{a,+}_{k,n} = \frac{1}{\tilde{k}^2_T + m^2} \, \tilde{\rho}^a_{k,n} \, \theta \lb \Lambda^2_\mathrm{UV} - \tilde{k}^2_T \rb,
\end{equation}
with the additional condition to eliminate the zero mode of $\rho$, i.e.\  $\tilde{\rho}^a_{0,n} = 0$, in order to guarantee global color neutrality. In addition to the screening mass $m$, we also introduce an ultraviolet regulator $\Lambda_\mathrm{UV}$ to regulate high momentum modes that are poorly resolved on the lattice. Thus, the discretized Poisson equation can be solved in the following way:
\begin{enumerate}
\item Generate a color charge configuration $\rho^a_{x,n}$ according to \cref{eq:mv2_twop_reg_latt} using a Gaussian random number generator. 
\item Perform an FFT to obtain $\tilde{\rho}^a_{k,n}$.
\item Solve the regularized Poisson equation according to \cref{eq:poisson_sol_reg_latt}.
\item Perform an inverse FFT to obtain the solution on the transverse lattice $ A^{a,+}_{x,n}$.
\end{enumerate}
The lightlike Wilson line is then given by (see \cref{eq:asym_Wilson_line_reg})
\begin{equation}
V^\dg_{A,x} = \prod_{n=1}^{N_s} \exp \lb- i g A^{a,+}_{x,n} t^a \rb,
\end{equation}
where the subscript ``A" denotes the right-moving nucleus. The transverse color field of nucleus ``A" is
\begin{equation}
U^A_{x,i} = V_{A,x} V^\dg_{A,x+i},
\end{equation}
which follows from performing the gauge transformation from covariant gauge to LC gauge on the lattice, see \cref{eq:latt_gauge_trans}. 
For the left-moving nucleus ``B" we perform the analogous procedure and find
\begin{equation}
U^B_{x,i} = V_{B,x} V^\dg_{B,x+i}.
\end{equation}
The lattice formulation of eqs.\ \eqref{eq:glasma_initial_1} and \eqref{eq:glasma_initial_2} has been derived in \cite{Krasnitz:1998ns}. The discretization of  \cref{eq:glasma_initial_1} reads
\begin{equation} \label{eq:Ui_initial}
\ah{ \lb U^A_{x,i} + U^B_{x,i} \rb \lb \one + U_{x,i} \lb \tau \rightarrow 0 \rb \rb^\dg} = 0,
\end{equation}
which is an implicit equation for $U_{x,i}$. For $\mathrm{SU(2)}$ there is an explicit solution given by
\begin{equation} \label{eq:Ui_initial_su2_sol}
U_{x,i}\lb \tau \rightarrow 0 \rb = \lb U^A_{x,i} + U^B_{x,i} \rb \lb U^{A,\dg}_{x,i} + U^{B,\dg}_{x,i} \rb^{-1}.
\end{equation}
The inverse can be readily computed, see \cref{sec:su2_inverse}. For $N_c > 2$ \cref{eq:Ui_initial} needs to be solved numerically.
The lattice version of eqs.\ \eqref{eq:glasma_initial_2} and \eqref{eq:bi_glasma_ini_1}, i.e.\ the discretized momentum $P^\eta_x\lb \tau \rightarrow 0 \rb$, is given by
\begin{equation} \label{eq:bi_initial_Peta_lattice}
P^\eta_x\lb \tau \rightarrow 0 \rb = \frac{1}{2g \lb a^i \rb^2} \sum_{i=1,2} \ah{\lb U_{x,i} - 1\rb \lb U^{B,\dg}_{x,i} - U^{A,\dg}_{x,i} \rb + \lb U^\dg_{x-i,i} - 1\rb \lb U^{B}_{x-i,i} - U^{A}_{x-i,i} \rb},
\end{equation}
where $U_{x,i} = U_{x,i} \lb \tau \rightarrow 0 \rb$. The rest of the components are zero:
\begin{equation}
P^i_x \lb \tau \rightarrow 0 \rb = 0, \qquad A_{x,\eta} \lb \tau \rightarrow 0\rb = 0.
\end{equation}
To initialize the leapfrog integrator one has to define the gauge fields at $\tau_0 = 0$, while the momenta are defined at $\tau_{1/2} = \Delta \tau /2$:
\begin{align}
P^\eta_x(\Delta \tau /2) &= P^\eta_x\lb \tau \rightarrow 0 \rb + \mathcal{O}\lb \Delta \tau^2 \rb,  \\
P^i_x(\Delta \tau/2) &= 0 + \mathcal{O}\lb \Delta \tau^2 \rb.
\end{align} 

\section{Glasma observables} \label{sec:bi_observables}
In the following section we discuss a few standard results and phenomena using the numerical procedure outlined in the previous sections. We focus on the energy density and the pressure components of the Glasma, i.e.\ components of the energy momentum tensor $T^{\mu \nu}$. 

Having already worked out the action \cref{eq:bi_action} and the canonical momenta \cref{eq:bi_momentum_i}, \cref{eq:bi_momentum_eta} it is a simple task to perform the Legendre transformation and obtain the Hamiltonian density
\begin{equation} \label{eq:taueta_hamiltonian}
\mathcal{H} = \tr \left[ \frac{1}{\tau} P^i P^i  + \tau \lb P^\eta\rb^2 + \frac{\tau}{2} F_{ij} F_{ij} + \frac{1}{\tau}  D_i A_\eta  D_i A_\eta \right],
\end{equation}
which can be related to the energy density in the $(t, z)$ frame at mid-rapidity $\eta = 0$ via $\varepsilon = \ev{\frac{1}{\tau} \mathcal{H}}$, where the expectation value $\ev{\dots}$ refers to averaging over the color charge densities of both nuclei. In order to identify the various terms in the energy density, we use  the field strength components of the laboratory frame at mid-rapidity
\begin{align}
E_i &= \p_\tau A_i = \frac{1}{\tau} P^i, \\
E_3 &= E_L =  \frac{1}{\tau} \p_\tau A_\eta = P^\eta, \\
B_i &= - \epsilon_{ij} \frac{1}{\tau} D_j A_\eta, \\
B_3 &= B_L =  - F_{12},
\end{align}
where $E_L$ and $B_L$ refer to the longitudinal field components. We also introduce the transverse components $E_T = \sqrt{E_1^2 + E_2^2}$ and analogously $B_T = \sqrt{B_1^2 + B_2^2}$. Using these definitions we can split the energy density $\varepsilon = \varepsilon_{E,L} + \varepsilon_{E,T} + \varepsilon_{B,L} + \varepsilon_{B,T}$ into four contributions:
\begin{align}
\varepsilon_{E,L} &= \ev{\tr \left[ E_L^2 \right]} = \ev{\tr \left[ \lb P^\eta\rb^2 \right]}, \\
\varepsilon_{E,T} &=  \ev{\tr \left[ E_T^2 \right]} = \ev{\frac{1}{\tau^2} \tr \left[ P^i P^i \right]}, \\
\varepsilon_{B,L} &= \ev{\tr \left[ B_L^2 \right]} = \ev{\tr \left[ \frac{1}{2} F_{ij} F_{ij} \right]}, \\
\varepsilon_{B,T} &=  \ev{\tr \left[ B_T^2 \right]} = \ev{\frac{1}{\tau^2} \tr \left[  D_i A_\eta  D_i A_\eta \right]}.
\end{align}
From the expression of the energy momentum tensor
\begin{equation} \label{eq:ym_em_tensor}
T^{\mu\nu} = \tr \left[- F^{\mu \rho} F^\nu_{\,\,\,\rho}  + \frac{1}{4} g^{\mu\nu} F_{\rho\sigma} F^{\rho\sigma} \right],
\end{equation}
we can read off the other diagonal components: the longitudinal pressure $p_L = \ev{T^{33}}$ and the transverse pressure $p_T = \ev{T^{11}} = \ev{T^{22}}$ given by
\begin{align}
p_T &= \varepsilon_{E,L} + \varepsilon_{B,L}, \\
p_L &= \varepsilon_{E,T} + \varepsilon_{B,T} - \varepsilon_{E,L} - \varepsilon_{B,L}.
\end{align}
Due to $T^{\mu\nu}$ being traceless we have $\varepsilon = 2 p_T + p_L$. The four energy density contributions, the energy density and the pressure components are the main observables that we are interested in.

In this section we consider head-on (i.e.\ central) collisions of gold nuclei at ``RHIC-like" collision energies of roughly $\sqrt{s} \approx 200 \, \gev$.
Due to the limitations of the MV model and the ultrarelativistic approximation, the only energy dependence comes from the saturation momentum $Q_s$. However, the MV model by itself does not predict any dependence $Q_s(\sqrt{s})$.
A from-first-principles determination of collision energy dependence of the initial conditions, or rather dependence on the momentum rapidity of the colliding nuclei, is possible through the JIMWLK renormalization group equations.
A proper discussion and application of the JIMWLK equations can be found in \cite{Iancu:2000hn,Ferreiro:2001qy,Weigert:2005us}.
For the purposes of this thesis, 
we simply choose a value for $Q_s$ by hand, such that it more or less corresponds to the case of Gold nuclei colliding at $\sqrt{s} \approx 200 \, \gev$ (see \cite{Lappi:2003bi, Lappi:2006hq}).

Ideally, fixing the saturation momentum would be done indirectly by choosing a value for the MV model parameter $\mu$ and the coupling constant $g$, since naively one expects $Q_s \propto g^2 \mu$ (see discussion below \cref{eq:qs_definition}). However, the number of color sheets $N_s$ and the screening mass $m$, introduced as regularizations of the model, also play a considerable role in the relation between $Q_s$ and $\mu$ \cite{Lappi:2007ku}. Even though the MV model is comparatively simple, its parameter uncertainties are not insignificant. For the purposes of this thesis it will therefore be enough to choose parameters that are phenomenologically somewhat realistic, but it should be stressed that comparisons to experimental results should not be taken too seriously.

With this disclaimer in mind, we proceed as follows: at collision energies of $\sqrt{s} \approx 200 \, \gev$ one expects a saturation momentum $Q_s \approx 2\, \gev$ (see e.g.\ \cite{Lappi:2003bi,Lappi:2006hq}). Since $Q_s$ sets the relevant QCD energy scale, one can use the one-loop beta function to find a coupling constant $\alpha_s(Q_s) = g^2 / 4\pi \approx 0.3$ or $g \approx 2$, which is the popular choice for $g$ in CGC literature. Assuming $Q_s \approx g^2 \mu$ and ignoring all other dependencies, these considerations fix the MV model parameter to be roughly $\mu \approx 0.5 \, \gev$. Alternatively, one can use the phenomenological estimate \cref{eq:mv_mu_phenomenological} and $A=197$ which approximately yields a similar value $\mu \approx 0.5\,\gev$. 

As we are interested in central collisions, the size of the circular overlap region of the two gold nuclei is $\pi R^2_{A}$, where $R_A = r_0 A^{1/3} \approx 7.27\, \fm$ is the nuclear radius with $r_0 \approx 1.25 \, \fm$.
Approximating the overlap region as a square of identical area, we set the transverse length of the simulation box to $L_T = \sqrt{\pi R_A^2} \approx 12.9 \, \fm$.
Due to the use of periodic boundary conditions, boundary effects are completely neglected and the system size in the transverse directions is effectively infinite.
The transverse size $L_T$ has to be large enough in order to accommodate the largest wavelengths in the color field of the nucleus. If color neutrality is realized at the size of the nucleus, the area of the transverse plane must be sufficiently large, which is the reason for choosing $L_T = \sqrt{\pi R_A^2}$.

In order to regulate the infrared modes of the color fields of the single nuclei we use two different prescriptions: either (I) $m = 0$ and global color neutrality by eliminating the zero mode of the charge density, or (II) $m = 0.2 \, \gev \approx 1 \, \fm^{-1}$ which implements color neutrality at the size of nucleons. For prescription (II), the zero mode of the color charge density is also removed. Ultraviolet modes are cut off at $\Lambda_{UV} = 20 \, \gev$, which should only insignificantly affect the results, but eliminates high momentum modes for which the lattice treatment of the system might be less than optimal.

For the number of color sheets we either choose $N_s = 1$, i.e. the single color sheet approximation, or up to $N_s = 100$ in order to approach the continuum limit of the MV model.

The simulations are performed using $\mathrm{SU(2)}$ as the color gauge group. We choose two instead of three colors to make use of a simple and elegant parametrization for $\mathrm{SU(2)}$ (see \cref{app_su2}) which makes the implementation of a real-time lattice Yang-Mills code comparatively easy. Furthermore, since there are only three color components in $\mathrm{SU(2)}$ (compared to eight in $\mathrm{SU(3)}$), performance and memory requirements of the code are much better, which will be particularly important in the three-dimensional setup. Most importantly however, the qualitative physics of the Glasma is essentially the same for any number of colors: phenomena such as the pressure anisotropy and the decay of Glasma flux tubes are similar for all groups $\SUN$. It is also possible to rescale results from $\mathrm{SU(2)}$ simulations with $N_c$ dependent factors to give reasonable agreement with simulations of $\mathrm{SU(3)}$ gauge fields \cite{Krasnitz:1999wc}. Analytical calculations (see e.g.\ \cite{Albacete:2018bbv}) show that the initial energy density scales with the Casimir of the fundamental representation $C_F \lb N_c \rb= \lb N_c^2 - 1\rb / \lb 2 N_c \rb$. This yields a rescaling factor of $C_F(3) / C_F(2) = 16 / 9$, which we will use to extrapolate all energy density contributions.

The code used to solve the lattice equations of motion and produce the results in this section was developed in Python and Cython \cite{behnel2010cython} and is hosted on GitLab\footnote{The code is freely available at \href{https://gitlab.com/dmueller/curraun_cy}{https://gitlab.com/dmueller/curraun\_cy}.}.

\begin{figure}
	\centering
	\begin{subfigure}[b]{0.28\textwidth}
		\centering
		\includegraphics{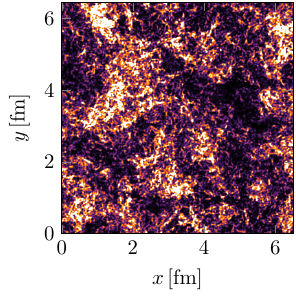}
		\caption{$\tau = 0.01\, \mathrm{fm/c}$}
	\end{subfigure}
	\quad
	\begin{subfigure}[b]{0.28\textwidth}
		\centering
		\includegraphics{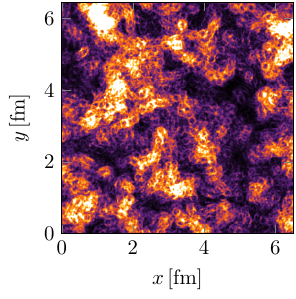}
		\caption{$\tau = 0.1\, \mathrm{fm/c}$}
	\end{subfigure}
	\quad
	\begin{subfigure}[b]{0.28\textwidth}
		\centering
		\includegraphics{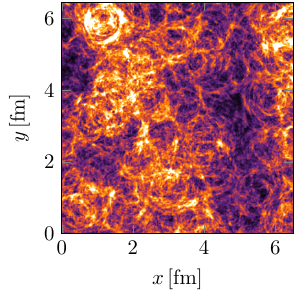}
		\caption{$\tau = 0.5\, \mathrm{fm/c}$}
	\end{subfigure}
	\quad
	\caption{Plots of the energy density $\varepsilon(\tau, x_T)$ as a function of transverse coordinates $x_T=(x,y)$ for various values of proper time $\tau$ in a single collision event. The simulation parameters are chosen as described in the text with $m = 0.2 \, \gev$, $N_s = 1$, a lattice size of $N_T \times N_T = 1024^2$ and a time step set to $\Delta \tau = a_T / 8$. The plots show a quarter of the full simulation box.
	\textit{Left:} the initial distribution of energy density close to $\tau = 0^+$. At the beginning, the system is comprised of purely longitudinal chromo-electric and -magnetic fields in the form of randomly distributed lumps of size $Q_s^{-1} \approx 0.1\, \fm$. These structures are interpreted as Glasma flux tubes.
	\textit{Center:}  immediately after their formation, the flux tubes expand in the transverse plane at the speed of light. After $\tau \approx Q_s^{-1} \approx 0.1 \, \mathrm{fm/c}$ one can observe characteristic circular patterns stemming from the expansion. At this point the longitudinal energy density components have decreased and transverse fields have been generated.
	\textit{Right:} as the expansion continues, the system becomes more homogeneous and dilute. Longitudinal fields have decreased further until all energy is distributed among the four contributions, i.e.\ $\varepsilon_{E,L} \approx \varepsilon_{B,L} \approx \varepsilon_{E,T} \approx \varepsilon_{B,T}$.\label{fig:density2d_plots}
	}
\end{figure}

\begin{figure}
	\centering
	\begin{subfigure}[b]{0.45\textwidth}
		\includegraphics{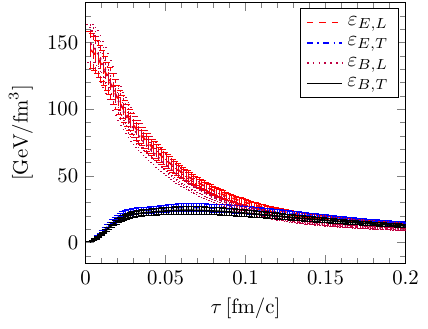}
		\caption{The four energy density contributions averaged across the transverse plane as a function of $\tau$.  }
	\end{subfigure}
	\qquad
	\begin{subfigure}[b]{0.45\textwidth}
		\includegraphics{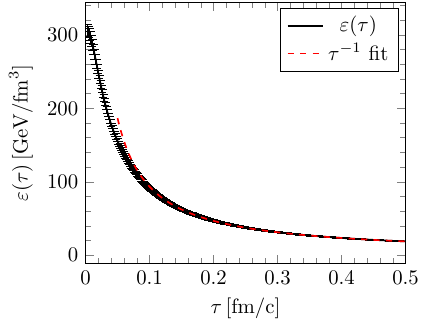}
		\caption{Energy density of the Glasma as a function of $\tau$ compared to characteristic $\tau^{-1}$ behavior and late times. }
	\end{subfigure}
	\caption{(a): the four energy density contributions from the various field components. At first, all energy is in the longitudinal field components and the chromo-electric and -magnetic longitudinal components (red and purple) are roughly the same. (b): energy density of the Glasma and a fit to $\varepsilon_0 / \tau$ for $\tau \in (0.1, 1.0)$. Statistical errors stem from averaging over 30 independent collision events. At late times the system becomes free streaming. The simulation parameters are the same as in \cref{fig:density2d_plots}.
		\label{fig:components_and_density}}
\end{figure}

\Cref{fig:density2d_plots} shows the energy density $\varepsilon(\tau, x_T)$ of a single collision event as a function of proper time and reveals the longitudinal flux tube structure of the Glasma, which, at least in the ultrarelativistic limit, is formed instantaneously at $\tau = 0^+$.
Even though the MV model is homogeneous on average, single events sampled from the probability functional are highly random: the statistical nature of the color charge densities leads to random hotspots of energy. The expansion of flux tubes after $\tau \sim Q_s^{-1}$, which generates transverse chromo-electric and -magnetic fields, leads to characteristic circular patterns in the transverse plane seen in the center plot. 
As the evolution continues, the energy density distributes across the transverse plane and the system becomes more and more dilute. At later times $\tau > Q_s^{-1}$ the four energy density contributions begin to converge towards the same value, as can be seen in \cref{fig:components_and_density} on the left. Afterwards, the system shows free streaming behavior signaled by $\varepsilon(\tau) \propto  \tau^{-1}$ shown in \cref{fig:components_and_density} on the right. The reason for this decrease of energy density is simple: as the colliding nuclei recede, the Glasma expands not only in the transverse plane, but also in the longitudinal direction and energy is transported away from the mid-rapidity region. This expansion occurs in a boost invariant manner and therefore the energy per unit rapidity, i.e.\ $dE/d\eta \propto \tau \, \varepsilon(\tau)$, becomes constant. Since the energy density and the field amplitudes decrease at later times, non-linear interactions become less important and eventually the system can be treated as effectively Abelian.

Let us now focus on some quantitative results for the Glasma energy density. In \cite{Lappi:2006hq} it was shown that the initial energy density at $\tau = 0^+$ of the MV model is not a particularly useful quantity to look at as it suffers from a UV divergence. However, this divergence dies down quickly for $\tau > 0$ and at $\tau_0 = Q_s^{-1} = 0.1 \, \mathrm{fm/c}$ results become independent of UV regularization and the lattice spacing $a_T$ \cite{Fukushima:2007ja, Fujii:2008km, Lappi:2006hq}. At $\tau_0 = Q_s^{-1} = 0.1 \, \mathrm{fm/c}$ we find an energy density of $\varepsilon(\tau_0) \approx 92 \, \mathrm{GeV / fm^3}$ (including the color scaling factor $16/9$) for $m = 0.2 \, \gev$ and $\varepsilon(\tau_0) \approx 101 \, \mathrm{GeV / fm^3}$ for $m=0$. Similar simulations \cite{Lappi:2006hq} with $\mathrm{SU(3)}$ as the gauge group and $m=0$ yield a value of $130 \, \mathrm{GeV/fm^3}$. Performing the numerical evolution with $\mathrm{SU(2)}$ and using the $C_F(3) / C_F(2)$ color scaling factors seems to underestimate the Glasma energy density. In \cite{Krasnitz:2001qu} a different color scaling factor of $D_A(3) / D_A(2) = 8 / 3$ was used, where $D_A(N_c) = N_c^2 - 1$ is the dimension of the adjoint representation for $\SUN$. Using this other factor we find $\varepsilon(\tau_0) \approx 138 \, \mathrm{GeV / fm^3}$ for $m = 0.2 \, \gev$ and $\varepsilon(\tau_0) \approx 152 \, \mathrm{GeV / fm^3}$ for $m=0$. We see that this time the energy density is overestimated compared to $\mathrm{SU(3)}$.
Obviously, playing such extrapolation games can never be exact, but when considering all other parameter uncertainties of the MV model, it is clear that $\mathrm{SU(2)}$ still provides phenomenologically relevant ``ballpark" results. 

\begin{figure}[t]
	\centering
	\includegraphics{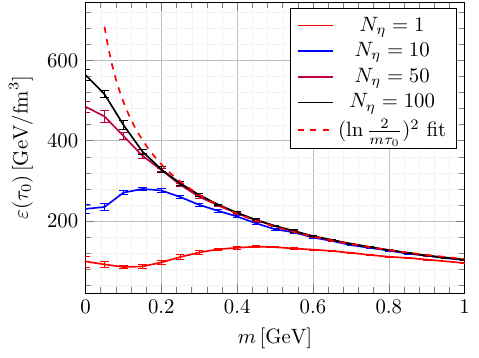}
	\caption{Energy density of the Glasma at $\tau_0 = Q_s^{-1} = 0.1 \, \mathrm{fm/c}$ for various values of $m$ and $N_s$. The simulations for this plot have been performed on a $512^2$ lattice with $\Delta \tau = a_T / 2$ and an ensemble average over 10 collisions.  All other parameters are chosen as described in the main text. The red dashed line corresponds to a fit $\varepsilon(m) = a + b \lb \ln \frac{2}{m\tau_0}\rb^2$ in the range $m \in (0.25, 1.0)$ with fitting parameters $a$ and $b$.
		\label{fig:energydensity_scan}}
\end{figure}

For fixed values of $\mu$, $g$ and $m$, the energy density also strongly depends on the number of color sheets. Plotting $\varepsilon(\tau_0)$ as a function of $m$ for various values of $N_s$ we find the results presented in \cref{fig:energydensity_scan}. In general, $\varepsilon(\tau_0)$ increases with $N_s$, but this growth stops around $N_s \sim 100$ with convergence being faster for larger values of $m$. In the continuum limit $N_s \rightarrow \infty$ one finds an approximate logarithmic dependency $\varepsilon(\tau_0) \propto \ln \lb 2 / ( m \tau_0 ) \rb ^2$ \cite{Fukushima:2007ja, Fujii:2008km}, which is well reproduced in \cref{fig:energydensity_scan}. On the other hand, the single color sheet approximation $N_s = 1$ shows more complicated behavior with a peculiar minimum around $m \approx 0.1 \, \gev$. In \cite{Fukushima:2007ki} it was shown that the $N_s = 1$ case inadequately describes the generalized MV model of \cref{eq:mv2_twop} as the path ordering of the lightlike Wilson lines is completely neglected. In light of these results we can expect that the form of this curve depends on the number of colors $N_c$ as well and the minimum should probably not be taken too seriously. 

The plot shows that for a given $\mu$ the energy density $\varepsilon(\tau_0)$ is significantly higher in the continuum limit compared to the naive single color sheet approximation. This is something to keep in mind and shows that for phenomenological applications one has to be careful when choosing $\mu$. It is always possible to simply view $\mu$ as a phenomenological parameter that can be fitted to known results \cite{Fukushima:2007ki}. Nevertheless, it is still interesting to study the interplay of parameters and their consequences on observables.

\begin{figure}
\centering
\includegraphics{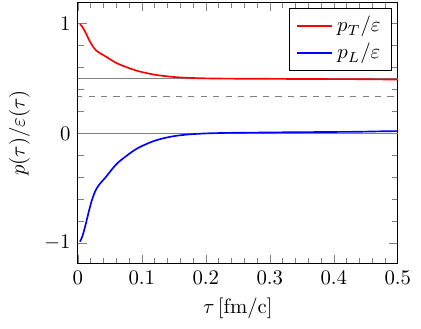}
\caption{Pressure to energy density ratios as a function of $\tau$. The simulation parameters are the same as in \cref{fig:components_and_density}. Initially the Glasma flux tube structure is maximally anisotropic with negative longitudinal pressure. After $\tau \sim 0.1 \, \mathrm{fm/c}$ the system settles into an anisotropic, free streaming state.
\label{fig:pressures}}
\end{figure}

The formation of purely longitudinal Glasma flux tubes has a peculiar feature: when computing the two pressure components $p_L$ and $p_T$, one finds that, because transverse fields only build up later in the evolution, the longitudinal pressure $p_L$ is initially negative. In fact, due to $\varepsilon_{B,T}(\tau=0^+) = \varepsilon_{E,T}(\tau=0^+) = 0$ we have $\varepsilon(\tau=0^+) = \varepsilon_{E,L} + \varepsilon_{B,L} = p_T = -p_L$ and consequently the Glasma is maximally anisotropic directly after the collision with $p_T / \varepsilon = 1$ and $p_L / \varepsilon = -1$. The time evolution of the pressure components can be seen in \cref{fig:pressures}. For $\tau > 0 $ the flux tubes expand and generate $\varepsilon_{E,T}$ and $\varepsilon_{B,T}$, which decreases the anisotropy in the system until $p_L$ is slightly positive. In the purely boost-invariant Glasma one never encounters isotropization in the sense that $p_L \approx p_T$ or $p_L / \varepsilon \approx p_T / \varepsilon \approx 1/3$. After roughly $\tau \sim Q_s^{-1}$ (around the same time when $\varepsilon(\tau) \propto \tau^{-1}$) the system settles into a free streaming expansion characterized by $p_L \approx 0$ and $p_T \approx \varepsilon / 2$. This happens because the expansion of flux tubes leads to equilibration of the four energy density components. It would be necessary to generate more transverse fields than possible through such an expansion to reach an isotropized state. It turns out that the free streaming limit is a generic feature of the boost invariant expansion of longitudinal flux tubes even in the Abelian limit: in \cite{Fujii:2008dd} it was shown that the Abelian evolution of electric or magnetic flux tubes of size $Q_s^{-1}$ qualitatively agrees with fully non-Abelian Glasma simulations with free streaming setting in after $\tau \sim Q_{s}^{-1}$.

The pressure anisotropy of the Glasma poses a serious problem for heavy-ion collision modeling, since a transition from the Glasma state into the QGP somewhere around $\tau \approx 1 \, \mathrm{fm/c}$ is necessary. Famously, the QGP is well described by relativistic viscous hydrodynamics \cite{Romatschke:2017ejr}, which is able to support some pressure anisotropy $1 - p_L / p_T$, but not as large as predicted by the classical Yang-Mills evolution of the Glasma where $p_L / p_T \approx 0$. An ad-hoc solution to this problem is to simply assume some kind of mechanism that cures the anisotropy when switching from the Glasma to hydrodynamics. Even though this approach is successful (see e.g.\ \cite{Gale:2012rq}), it is unsatisfying from a theoretical viewpoint. A more recent and elaborate approach is to introduce an intermediate kinetic theory stage between the Glasma and the QGP \cite{Kurkela:2018vqr, Kurkela:2018wud}, which drives the system towards isotropization within $0.2 \, \mathrm{fm/c} \lesssim \tau \lesssim 1.0 \, \mathrm{fm/c}$. On the other hand, there exist possible mechanisms for reaching an isotropized system from the Glasma within classical Yang-Mills theory, which will be discussed in the next section.  

Up until now we have only discussed observables derived from the energy-momentum tensor $T^{\mu\nu}$, but it is also possible to analyze the gluon spectrum of the Glasma by computing gluon occupation numbers. Although this quantity is not studied in this section, it is still important to mention the procedure, as it allows one to make more direct connections to experiments.
We saw earlier that the expansion of the Glasma at later times is characterized by $\varepsilon(\tau) \propto \tau^{-1}$. As the field amplitudes die down, non-linear interactions are less relevant and consequently one can treat the system as an ensemble of weakly interacting gluons \cite{Lappi:2003bi, Krasnitz:2000gz}. Furthermore, using the assumption that the electric and magnetic contribution is equal at late times (at least on average), the Hamiltonian of the system \eqref{eq:taueta_hamiltonian} can be approximated by
\begin{equation}
H = \intop d^2 x_T \mathcal{H} \simeq 2 \intop d^2 x_T \tr \left[ \frac{1}{\tau} P^i (\tau, x_T) P^i (\tau, x_T) + \tau \lb P^\eta (\tau, x_T) \rb^2 \right].
\end{equation}
Performing a Fourier transformation one finds
\begin{equation}
H \simeq 2 \intop \frac{d^2k_T}{\lb 2\pi \rb^2} \tr \left[ \frac{1}{\tau}  P^i (\tau, k_T) P^i (\tau, -k_T) + \tau P^\eta (\tau, k_T) P^\eta (\tau,- k_T) \right].
\end{equation}
Identifying this with the Hamiltonian describing free particles
\begin{equation}
H_\mathrm{free} = \intop \frac{d^2k_T}{\lb 2\pi \rb^2} \omega(k_T) n(\tau, k_T),
\end{equation}
where $\omega(k_T) = \abs{k_T}$ is the free dispersion relation and $n(\tau, k_T)$ is the gluon occupation number in momentum space, one finds the relation
\begin{equation}
n(\tau, k_T) = \frac{2}{\abs{k_T}} \tr \left[ \frac{1}{\tau}  P^i (\tau, k_T) P^i (\tau, -k_T) + \tau P^\eta (\tau, k_T) P^\eta (\tau,- k_T) \right].
\end{equation}
However, due to the system being described by gauge theory, this identification is not unique. The residual gauge freedom, i.e. time-independent gauge transformations, has to be fixed. The standard practice \cite{Lappi:2003bi, Krasnitz:2000gz, Lappi:2016ato, Lappi:2017ckt, Boguslavski:2018beu} is to choose transverse Coulomb gauge
\begin{equation}
\p_i A_c^i(\tau, x_T) = 0,
\end{equation}
which minimizes the functional $\int d^2 x_T \tr \left[ A^i(\tau, x_T)^2 \right]$.  Computing the gluon occupation number $n(\tau, k_T)$, one can determine the total number of gluons per unit rapidity produced in a collision by integrating over all momenta. Assuming that all gluons decay into hadrons, it is possible to make direct comparisons to the number of charged particles per rapidity measured in heavy-ion collision experiments \cite{Schenke:2012fw, Schenke:2016ksl, Lappi:2003bi, Schenke:2015aqa}. This can be used to fix the MV model parameter $\mu$ directly using only classical Yang-Mills simulations of the Glasma. Other quantities such as the energy density at $\tau_0 = Q_s^{-1}$ can not be measured directly, making the gluon spectrum a valuable tool for phenomenological applications.

\section{Beyond the MV model} \label{sec:beyond_mv}

The goal of this chapter was to give a short presentation on the standard techniques and main phenomena of the boost invariant Glasma using the MV model as a simple approximation of nuclei. The MV model is highly restrictive due to its simplicity: we approximate the colliding nuclei to be infinitesimally thin in order to justify boost invariance and thus neglect any rapidity dependence of observables. We can only describe a crude approximation to central, head-on collisions, because the MV model is infinitely extended in the transverse plane. As such there is no notion of impact parameter dependence. Finally, the model only depends on one dimensionful parameter $\mu$ (apart from regularization parameters). Consequently, the transverse structure of the model is very simple: even though color charge densities are random on an event-by-event basis, the average density is completely homogeneous. A realistic model of a nucleus should also include its nucleonic and possibly even sub-nucleonic structure. 

In this section we discuss a few extensions to this simple picture, which will naturally lead us to the main topic of this thesis.

\subsection{Glasma instability and isotropization} \label{sec:glasma_instability}

As we saw in the previous section, the boost invariant expansion of the Glasma leads to a highly anisotropic system with vanishing longitudinal pressure. Even at very large proper times the system will remain in this state.
This asymptotic behavior of the boost-invariant system stands in contrast to experimental indications of a nearly isotropized quark-gluon
plasma, where $p_{T}\simeq p_{L}$ after a few $\fm/c$ \cite{ Romatschke:2007mq, Kolb:1999it, Romatschke:2009im}. This scenario of fast isotropization is necessary for traditional relativistic viscous
hydrodynamics to be applicable at early times. 

Since boost invariance was put in by hand, it is a reasonable question to ask whether a system like this is actually stable against small variations which break this symmetry. This question led to a large number of publications (see e.g.\ \cite{Romatschke:2005ag, Romatschke:2006nk, Fukushima:2011nq,  Epelbaum:2013waa, Gelis:2013rba, Fukushima:2006ax,  Dusling:2010rm, Berges:2012cj}) studying a similar setup: an initially boost invariant Glasma at $\tau = 0^+$ is perturbed with rapidity dependent fluctuations and then evolved using 3+1 dimensional Yang-Mills equations in the $\tau, \eta$ frame. These studies found that the initially boost invariant Glasma  is indeed unstable and that the pressure anisotropy becomes less severe at late times $\tau \gg Q_s^{-1}$. Rapidity-dependent fluctuations (whose origin is thought to be quantum) therefore provide a possible mechanism to explain the transition from the pre-equilibrium Glasma to the equilibrated QGP. However, the general consensus is that this mechanism is not efficient enough to fully isotropize the system within the short time frame before the system becomes hydrodynamic at around $\tau \approx 1 \, \mathrm{fm/c}$. Without an intermediate stage described by kinetic theory \cite{Kurkela:2018vqr, Kurkela:2018wud}, this either means that the Glasma description (even including fluctuations) is insufficient or that the system might not be isotropic at early times.
It should be noted that the scenario of fast isotropization ($\sim 0.5$ -- $2 \, \fm/c$) is not undisputed. 
The development of anisotropic relativistic hydrodynamics (aHydro) \cite{Martinez:2010sc, Florkowski:2010cf} over the past few years has shown that
the QGP can evolve hydrodynamically while still maintaining a large pressure anisotropy throughout its 
lifetime (see \cite{Strickland:2014pga} and \cite{Alqahtani:2017mhy} for recent reviews and pedagogical introductions to this topic).  
In any case, it is clear that the boost invariant approximation neglects more than just simple rapidity dependence of observables and that there are interesting dynamics to be found in the 3+1 dimensional Glasma.

\subsection{Realistic models of nuclei}

In the beginning of this section we have outlined the shortcomings of the MV model and how it is only a very crude approximation of high energy nuclei. Fortunately, the MV model (see eqs.\ \eqref{eq:mv2_onep} and \eqref{eq:mv2_twop}) can be easily generalized by promoting the parameter $\mu$ to a function of the transverse coordinate $x_T$ via
\begin{align}
\ev{\rho^a(x^-,x_T)} &= 0, \\
\ev{\rho^a(x^-,x_T) \rho^b(y^-,y_T)} &= g^2 \mu^2(x^-, x_T) \dd^{ab}  \dd(x^- - y^-)  \dd^2(x_T - y_T).
\end{align}
The color charge densities are still considered to be Gaussian, but now the variance of the fluctuation varies locally. This extension allows us to consider nuclei with finite radius $R_A$ by choosing $\mu^2$ to have e.g.\ the form of a Wood-Saxon distribution with appropriate parameters. In turn, finite nuclei allow us to describe off-central collisions. This is in fact how the earliest numerical studies of the Glasma with impact parameter dependence have been performed \cite{Lappi:2003bi, Krasnitz:2002ng}.

The above model assumes nuclei to be homogeneous disks of diameter $\sim 2 R_A$ with possibly some smooth boundary, thus neglecting any structure inside the nucleus. This can be improved upon, by also considering the nucleonic structure of nuclei: since nuclei are comprised of nucleons it is reasonable to assume that the color charge density inside is not homogeneous (not even on average) and only becomes large in the vicinity of nucleons \cite{Krasnitz:2002ng}.
One can formulate even more accurate models for nuclei from its building blocks (i.e.\  nucleons) if one has an accurate description for single nucleons. The IP-Glasma \cite{Schenke:2012wb, Schenke:2012fw} is a state-of-the-art model for high energy nuclei that is constructed in the above described manner. In the IP-Glasma, the saturation momentum of a single proton $Q_{s,p}$ is considered as a function of $x_T$ relative to the center of the proton, which can be related to the color charge density fluctuation $g^2 \mu(x_T)$. The question then arises how one should choose $Q_{s,p}(x_T)$ in order to arrive at an accurate representation of a full nucleus. This leads to the most important feature of the IP-Glasma model: the exact form of $Q_{s,p}$ is extracted from experimental deep inelastic scattering (DIS) data \cite{Schenke:2012wb, Schenke:2012fw}. The color charge density of neutrons is determined in the same manner. Using this procedure the IP-Glasma has become the most successful model when applying CGC/Glasma to phenomenology. For instance, it successfully describes the charged particle multiplicity \cite{Schenke2014} computed from the gluon occupation numbers at both RHIC and LHC not just for the nucleus-nucleus case but also for proton-nucleus and proton-proton collisions. When coupled to relativistic viscous hydrodynamics after the Yang-Mills evolution, it correctly describes anisotropic hydrodynamic flow at RHIC and LHC \cite{Gale:2012rq}.

\subsection{Rapidity dependence and finite longitudinal extent of nuclei}

Even though the Glasma initial conditions were derived under the assumption of boost invariance, it is still possible to simulate collisions with rapidity dependence, at least in an approximate manner. The main idea, as outlined in \cite{Lappi:2004sf}, is to introduce a momentum rapidity $y$ dependent MV model parameter $\mu$ for both nuclei ``A" and ``B" via
\begin{equation}
\mu^2_{(A,B)} = \mu^2 \exp \lb \pm \lambda y \rb,
\end{equation}
with a parameter $\lambda \approx 0.25 - 0.3$ (see also \cite{McLerran:1997fk} for the rapidity dependence of $\mu^2$). This relation can be justified by recalling that the saturation momentum $Q_s$ depends on the longitudinal momentum fraction $x$ of the gluons and that the relevant $x$ can be related to the momentum rapidity $y$ of the colliding nuclei \cite{Iancu:2012xa}. The IP-Glasma uses a very similar procedure to obtain rapidity dependent initial conditions \cite{Schenke2014}.

The rapidity $y$ is then viewed as an external parameter and one performs independent, boost invariant simulations for a range of values of $y$. The computed observables for each $y$, such as the energy density or gluon numbers, are identified with the observables at that rapidity.

The same simulation procedure is used in \cite{Schenke:2016ksl}, but the rapidity dependent initial conditions are not obtained as outlined above. Instead, the initial conditions are computed from a JIMWLK evolution, which takes into account quantum effects that lead to the saturation momentum growing with rapidity. 

Treating each rapidity direction independently obviously neglects any interactions between fields at different rapidities, which is a reasonable approximation at later times in the evolution, but can not be completely accurate at the earliest stages of the Glasma. Furthermore, the procedure relies on using the boost invariant initial conditions eqs.\ \eqref{eq:glasma_initial_1} and \eqref{eq:glasma_initial_2}, as there are no analytical results for rapidity-dependent Glasma initial conditions. One could interpret this procedure as using a different effective (boost invariant) description of the nucleus for each rapidity slice and simply ``stitching" together the results for each value of $y$. On the other hand, if one assumes that there should be a single classical field configuration valid for a single nucleus associated with rapidity dependent fields, then it is clear that such a field configuration can not have a $\dd$-shape as assumed in the ultrarelativistic limit. This leads us to the main goal of this thesis. In the next chapters we develop a numerical method to solve the classical Yang-Mills equations for collisions of nuclei with finite longitudinal extent. This method forces us to work in the $(t, z)$ frame instead of using comoving coordinates $(\tau, \eta)$ in order to correctly incorporate initial conditions for nuclei with finite thickness along $x^-$ or $x^+$. By virtue of breaking boost invariance explicitly, one finds rapidity dependent observables such as the rapidity profile of the energy density of the Glasma (see \cref{sec:rapidity_profiles}). This rapidity dependence is a consequence of the finite longitudinal extent and the longitudinal structure of the color fields of the nucleus before the collision. It should be interpreted as a purely classical effect as there is \textit{a priori} no notion of growing $Q_s$ with rapidity. Since the problem will be formulated in a different frame and previously important assumptions like boost invariance will be loosened, we will have to construct the method from the ground up. This will be the main task in the next chapter. 

\chapter{The 3+1D Glasma on the lattice} \label{chap:glasma3d}

In this chapter we develop the numerical method we use to simulate heavy ion collisions in the CGC framework without relying on the ultrarelativistic limit and the symmetry under longitudinal boosts. In \cref{sec:coll_with_finite} we argue that the laboratory (or center of mass) frame is a more suitable choice of frame when considering nuclei with finite extent along the longitudinal axis. We also mention some assumptions that render the numerical problem more tractable. In \cref{sec:cpic} we introduce the colored particle-in-cell (CPIC) method that we use to solve the problem numerically on a lattice and derive the discretized equations of motion for the color fields and color currents used in a simulation. As the numerical description differs from the boost invariant case, the initial conditions for single nuclei on the lattice have to be formulated in laboratory frame temporal gauge (see \cref{sec:initial_lattice}). In \cref{sec:em_tensor_latt} we explain how we compute the energy-momentum tensor on the lattice, which will be the most important observable that we study. Finally, in \cref{sec:implementation}, we discuss some details regarding the actual implementation of the code such as optimizations, some subtleties regarding the initial and boundary conditions, and how to parallelize the simulation.

\section{Collisions of nuclei with finite longitudinal extent} \label{sec:coll_with_finite}

\begin{figure}
	\centering
	\includegraphics{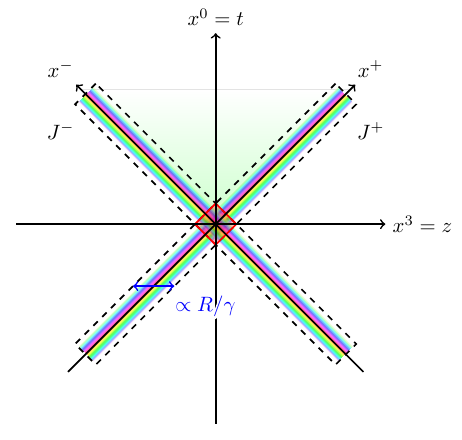}
	\caption{The schematic Minkowski diagram for a collision of two nuclei with finite longitudinal extent, which is proportional to the Lorentz-contracted radius $R / \gamma$. The transverse coordinates $x_T$ are suppressed. In contrast to the boost invariant case (\cref{fig:lightcone2}), the collision region is not a single point but a finite area (red square) proportional to $\lb R / \gamma \rb^2$, which makes the separation into distinct regions ambiguous. Furthermore, the Glasma does not form instantaneously but is produced in finite time. 
		\label{fig:lightcone3} }
\end{figure}

Let us quickly recapitulate how to derive the color fields of a single nucleus (for details see \cref{sec:mv_model}). Starting from a color charge density $\rho_{A,B}(x^\mp,x_T)$ moving along the $x^\pm$ axis and fixing the covariant gauge $\p_\mu A^\mu=0$, the only non-vanishing component of the color field $A^\pm$ is found by solving the Poisson equation
\begin{equation}
- \Delta_T A^\pm_{A,B}(x^\mp,x_T) = \rho_{A,B}(x^\mp,x_T).
\end{equation}
The subscripts ``A" and ``B" are used to differentiate the two nuclei.
This solution is exact for arbitrary color charges if the charge densities only depend on one of the light cone coordinates $x^\pm$. In the last chapter we considered the longitudinal support along the longitudinal axes $x^\mp$ finite, but only in order to account for path ordering along $x^\mp$ when switching to LC gauge. Here, we would like to associate finite longitudinal support with the actual physical Lorentz contracted extent of relativistic nuclei proportional to $R / \gamma$, where $R$ is the nuclear radius and $\gamma$ is the Lorentz factor. Therefore, at no step in the calculation can we take this longitudinal extent to be infinitesimal as we did in the previous chapter. This immediately poses a problem when considering collisions of two such color fields. A schematic picture of this is shown in \cref{fig:lightcone3}. The charge densities (which are basically the only input for the calculation) are not invariant under longitudinal boosts. The Glasma fields produced in a collision will depend on space-time rapidity $\eta$ and the system cannot be reduced to 2+1 dimensions. With finite longitudinal support the collision and creation of the Glasma is not instantaneous, although the formation still happens very fast with the formation time being proportional to $R / \gamma$. Without employing any approximations (such as expanding in $R / \gamma$) one has to solve the fully non-perturbative Yang-Mills equations in order to compute the rapidity-dependent color field after the collision. Furthermore, without a point-like origin, the Minkowski diagram cannot be separated into distinct, causally independent regions in an unambiguous way. The separation ansatz, eqs.\ \eqref{eq:sep_ansatz_1} and \eqref{eq:sep_ansatz_2}, is not appropriate in this more general case. In conclusion, it seems that many assumptions that apply to the boost invariant Glasma simply cannot be used for the collisions we are considering and a different approach is necessary.

This brings us to the main novel feature of our 3+1 dimensional setup. Since the separation of space-time into distinct regions is ambiguous and a future light cone cannot be identified, we choose a ``brute force" approach and simply stay in the center of mass frame. We therefore use the laboratory time $t$ and the beam axis coordinate $z$ as our coordinate set. This means that we also have to consider the dynamics of the color current. Recall that in the boost invariant case, the color charges of the nuclei are only relevant for finding the initial Glasma fields. The infinitesimally thin color currents only form the boundary of the future light cone and do not affect the time evolution of the Glasma after its formation.  On the other hand, with finite longitudinal extent the color currents, in addition to the color fields of the nuclei, have to be included in the description and treated as dynamic degrees of freedom. 

There is one restriction we can put on the color currents: we can assume them to be eikonal, i.e.\ the color charges do not recoil or lose any momentum. Thus, the color charges are only affected by parallel transport, meaning that they can rotate in color space as they undergo the collision, but their lightlike trajectories remain fixed. This color rotation is a consequence of non-Abelian charge conservation via the continuity equation (see \cref{eq:nonabelian_continuity_eq}). Writing this conservation law down for the nucleus ``A" we find
\begin{equation}
D_\mu J_A^\mu(x) = \p_+ \rho_A(x) + i g \cm{A^-(x)}{\rho_A(x)} = 0. 
\end{equation}
In the asymptotic past of the nucleus $x^+ \rightarrow -\infty$ the density $\rho_A(x^-,x_T)$ is indeed independent of $x^+$, but as soon as the two nuclei collide the color current is parallel transported due to the color field $A^-(x) \neq 0$ which is comprised of the color field of the second nucleus ``B" and the ``$-$" component of the Glasma fields. Since the color charges are only rotated, we can use the ansatz
\begin{equation}
\rho_A(x) = W(x) \tilde{\rho}_A(x^-,x_T) W^\dg(x),
\end{equation}
where $\tilde{\rho}_A(x^-,x_T)$ is the unaffected current at $x^+ \rightarrow -\infty$.
The unitary matrix $W(x)$ fulfills the relation
\begin{equation}
\p_+ W(x) = - i g A^-(x) W(x),
\end{equation}
with the general solution given by the Wilson line
\begin{equation}
W(x^+, x^-, x_T) = \mathcal{P} \exp \lb - i g \intop_{-\infty}^{x^+} dy^+ A^-(y^+, x^-, x_T) \rb.
\end{equation}
An analogous solution can be found for the current of the left-moving nucleus ``B". In the previous chapter the currents are unaffected due to the choice of Fock-Schwinger gauge which eliminates the $A^-$ ($A^+$ resp.) component at the $x^+$ ($x^-$ resp.) boundary of the future light cone, but in collisions with finite extent we cannot exploit gauge freedom to set both light cone components $A^+$ and $A^-$ to zero simultaneously everywhere. Parallel transport therefore needs to be accounted for in the 3+1 dimensional setting.

Now that we have understood the scope of the problem at hand, let us formulate it in detail. First we have to pick an appropriate gauge. The obvious choice  for numerical time evolution is the laboratory temporal gauge $A^0 = 0$. 
Transforming from the covariant gauge to the temporal gauge we have to solve
\begin{equation}
A^0 = \Omega \lb A^0_\mathrm{cov} + \frac{1}{ig} \p^0 \rb \Omega^\dg = 0,
\end{equation}
where the subscript ``cov" is used to explicitly denote covariant gauge fields. The general solution of this equation is given by the time-like Wilson line
\begin{equation} \label{eq:cpic_temporal_wilson_line}
\Omega^\dg(t,\vec{x}) = \mathcal{P} \exp \lb - ig \intop_{-\infty}^t dt' A^0_{\mathrm{cov}}(t',\vec{x}) \rb,
\end{equation}
where the spatial coordinate vector is given by $\vec{x} = \lb x_T, z \rb$.
If we apply temporal gauge to the color field of nucleus ``A", the Wilson line reads
\begin{equation}
\Omega_A^\dg(t,\vec{x}) = \mathcal{P} \exp \lb - \frac{ig}{\sqrt{2}} \intop_{-\infty}^t dt' A^+_{\mathrm{cov}}(x'^{-}, x_T) \rb,
\end{equation}
where we used $A^0_\mathrm{cov} = A^3_\mathrm{cov} = A^+_\mathrm{cov} / \sqrt{2}$ and $x'^{-}= \lb t' - z \rb / \sqrt{2}$. Using the fact that $A^+_\mathrm{cov}$ only depends on $t$ and $z$ via $x^-$ we can reparametrize the integral and find that $\Omega^\dg_A$ is in fact the same transformation matrix associated with switching from covariant gauge to LC gauge \cref{eq:lc_gauge_wilson_line}. Writing the path-ordered exponential first as a product integral
\begin{equation}
\Omega_A^\dg(t,\vec{x}) = \mathcal{P} \left[ \prod_{-\infty}^t \lb \one - \frac{ig}{\sqrt{2}} dt' A_\mathrm{cov}^+ (\frac{t'-z}{\sqrt{2}},x_T) \rb \right],
\end{equation}
and performing the coordinate transformation from $t'$ to $x'^{-}$, while keeping $z$ fixed, yields
\begin{align}
\Omega_A^\dg(t,\vec{x}) &= \mathcal{P} \left[ \prod_{-\infty}^{\frac{t-z}{\sqrt{2}}} \lb \one - ig dx'^{-} A_\mathrm{cov}^+ (x'^{-},x_T) \rb \right] \nn
&= \mathcal{P} \exp \lb - ig \intop_{-\infty}^{x^-} dx'^{-} A^+_\mathrm{cov}(x'^{-},x_T) \rb. 
\end{align}
\begin{figure}[t]
	\centering
	\includegraphics{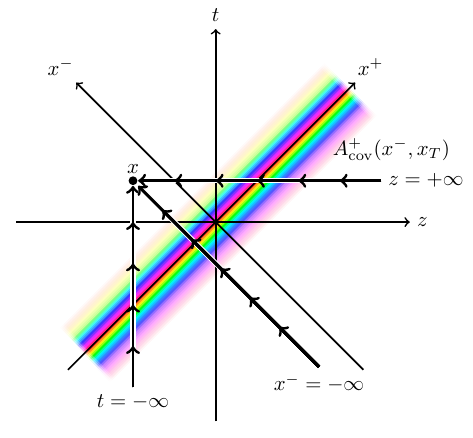}
	\caption{An illustration of the color field of a right-moving nucleus in covariant gauge and three identical Wilson lines ending at $x$. Since the color field is independent of $x^+$ and $A^0_\mathrm{cov} = A^3_\mathrm{cov}$, the time-ordered Wilson line (vertical thick line) passes through the same ``layers" of the nucleus in the same order as the light-like Wilson line (parallel to $x^-$) and the spatial Wilson line (horizontal).
		\label{fig:lightcone4} }
\end{figure}\\
This fact can also be intuitively understood in a diagram, see \cref{fig:lightcone4}. Using an analogous reparametrization one finds that $\Omega^\dg_A$ can be identified with the spatial Wilson line starting at $z \rightarrow \infty$:
\begin{align}
\Omega_A^\dg(t,\vec{x}) &= \mathcal{P} \exp \lb - \frac{ig}{\sqrt{2}} \intop_{+\infty}^z dz' A^+_{\mathrm{cov}}(x''^{-}, x_T) \rb \nn
&= \mathcal{P} \exp \lb - ig \intop_{+\infty}^z dz' A^z_{\mathrm{cov}}(t, x_T, z') \rb,
\end{align}
where $x''^{-} = \lb t - z' \rb / \sqrt{2}$. As a consequence, it is easy to show that the longitudinal component $A^3$ vanishes in temporal gauge. Since the spatial Wilson line fulfills
\begin{equation}
\p_3 \Omega^\dg_A = + i g A^3_\mathrm{cov} \Omega^\dg_A,
\end{equation}
we find
\begin{align} \label{eq:cpic_longitudinal_zero}
A^3 &= \Omega_A \lb A^3_\mathrm{cov} + \frac{1}{ig} \p^3 \rb \Omega^\dg_A = 0.
\end{align}
The only remaining components are the transverse gauge fields given by
\begin{equation}
A^i_A(t,\vec{x}) = \frac{1}{ig} \Omega_A(t, \vec{x}) \p^i \Omega_A^\dg(t,\vec{x}).
\end{equation}
Analogous arguments can be made for the left-moving nucleus. The color fields of single nuclei in temporal gauge are therefore completely equivalent to the LC gauge solutions.

Before the collision, at some initial time $t_0$, when the nuclei are sufficiently well separated, it is possible to superimpose the transverse gauge fields 
\begin{equation}
A^i(t_0, \vec{x}) = A^i_A(t_0, \vec{x}) + A^i_B(t_0, \vec{x}).
\end{equation}
The total color current is also a superposition
\begin{equation}
J^i(t_0, \vec{x}) = J^i_A(t_0, \vec{x}) + J^i_B(t_0, \vec{x}),
\end{equation}
where only the longitudinal component is non-zero:
\begin{equation}
J^i_{A,B}(t_0, \vec{x}) = \pm \dd^{i3} \frac{1}{\sqrt{2}} \Omega_{A,B}(t_0, \vec{x}) \tilde{\rho}_{A,B}(x^-,x_T) \Omega^\dg_{A,B}(t_0, \vec{x}).
\end{equation}
This initial condition is then evolved to $t > t_0$ by solving the Yang-Mills equations (see \cref{eq:ym_equations}), which read
\begin{equation}
\p^2_0 A^i(t,\vec{x}) = - D_j F^{ji}(t,\vec{x}) + J^i_{A}(t,\vec{x}) + J^i_{B}(t,\vec{x}), 
\end{equation}
where the color currents are given by
\begin{align}
J^i_{A,B}(t,\vec{x}) &= W_{A,B}(t, \vec{x}) J^i_{A,B}(t_0,\vec{x}) W^\dg_{A,B}(t, \vec{x}).
\end{align}
Parallel transport is accomplished by
\begin{align}
W_{A}(t, \vec{x}) &= \mathcal{P} \exp \lb + ig \intop_{t_0}^t dt' A^3(t',x_T,z+t'-t) \rb, \label{eq:para_tr_A}\\
W_{B}(t, \vec{x}) &= \mathcal{P} \exp \lb - ig \intop_{t_0}^t dt' A^3(t',x_T,z+t-t') \rb. \label{eq:para_tr_B}
\end{align}

The parallel transport of the color currents (and the signs used in the exponential) can be better understood if one thinks of the color charge density of the nuclei to be comprised of an ensemble of point-like particles.
Consider a single point-like particle in 1+1 in a background color field $A^3(t,z)$ in temporal gauge.  Assuming that it follows a lightlike trajectory defined by ${z(t) = t - \lb t_0 - z_0 \rb}$ (such that $z(t_0) = z_0$), the color charge density and longitudinal current is given by
\begin{align}
\rho(t,z) &= Q(t) \dd(z-z(t)), \\
J(t, z) &= + \rho(t, z).
\end{align}
Due to non-Abelian charge conservation the particle's color charge $Q(t)$ is time dependent. Requiring that the continuity equation (see \cref{eq:nonabelian_continuity_eq})
\begin{equation}
D_\mu J^\mu(t,z) = \p_0 \rho(t,z) + \p_3 \rho(t,z) - ig \cm{A^3(t,z)}{\rho(t,z)} = 0,
\end{equation}
is fulfilled, yields
\begin{align}
Q(t) &= W(t;z_0) Q(t_0) W^\dg(t;z_0), \\
W(t;z_0) &= \mathcal{P} \exp \lb + i g \intop_{t_0}^t dt' A^3(t', z(t')) \rb.
\end{align}
We can generalize this to a smooth distribution of color charge by letting $Q(t_0) \rightarrow Q(t_0; z_0)$ depend on $z_0$. The color charge density is then given by
\begin{align}
\rho(t,z) &= \intop dz_0 Q(t;z_0) \delta(z - z(t)) \nn
&= Q(t;t-z+t_0) \nn
&= W(t;t-z+t_0) Q(t_0;z_0) W^\dg(t;t-z+t_0),
\end{align}
with
\begin{equation}
W(t; z+t_0-t) = \mathcal{P} \exp \lb + i g \intop_{t_0}^t dt' A^3(t', z+ t'-t)) \rb,
\end{equation}
which is exactly \cref{eq:para_tr_A}. Repeating the same steps for a particle moving in the opposite direction defined by $z(t) = - t + (t_0 + z_0)$ we find
\begin{equation}
W(t;z+t-t_0) = \mathcal{P} \exp \lb - ig \intop_{t_0}^t dt' A^3(t',x_T,z+t-t') \rb,
\end{equation}
which corresponds to \cref{eq:para_tr_B}. The negative sign in the exponential can be traced back to the opposite sign occurring in the spatial color current, i.e.\ $J(t, z) = - \rho(t, z)$.

We see that the equations for the 3+1 dimensional setup are quite complicated. Not only do we need to include the additional longitudinal axis $z$, but due to parallel transport the color fields couple to the color currents in a highly non-linear way. Fortunately, there is  a numerical method that can handle dynamic color charges in a gauge covariant manner. This will be the topic of the following sections.

\section{The colored particle-in-cell method} \label{sec:cpic}

The colored particle-in-cell (CPIC) method is a numerical approach to simulating an ensemble of colored point-like particles coupled to non-Abelian gauge fields. It combines the real-time lattice description of gauge fields with classical particles obeying Wong's equations \cite{Wong}, which is the non-Abelian equivalent of the Lorentz force. To name a few applications, CPIC has been successfully used in hard thermal loop simulations \cite{Moore:1997sn,Hu:1996sf}, where colored particles act as hard (i.e.\ ultraviolet) degrees of freedom which interact with soft (infrared) classical color fields. It can also be used to study non-Abelian plasma instabilities \cite{Dumitru:2005hj, Strickland:2007} in the QGP. When including collisions among the colored particles, it is possible to simulate energy loss and momentum broadening of jets in the QGP \cite{Schenke2009}.

The CPIC method originates from the popular particle-in-cell (PIC) method used in (Abelian) plasma physics \cite{Verboncoeur2005}. Pedagogical introductions can be found in \cite{BirdsallLangdon, Hockney:1988:CSU:62815, LAPENTA2012795}. PIC was developed to accurately simulate electromagnetic plasmas with typically very large numbers of charged particles coupled to the electromagnetic field. In a PIC simulation fields are usually sampled in cells on a numerical grid and a finite difference approximation is used to numerically solve Maxwell's equations. On top of this lattice, an ensemble of $N_p$ charged particles is able to move through the simulation volume. The particle trajectories are only affected by the fields within their closest (and possibly also neighboring) cells. This approach solves a number of problems encountered in the simulation of plasmas: instead of having to compute the Lorentz force of each particle pair in the plasma individually, which is an operation that scales quadratically with the number of particles $N_p$, one only has to consider the field in the immediate vicinity of each particle (in the simplest approximations just a single cell), which only involves $N_p$ operations. The PIC method therefore makes it possible to perform large scale simulations of systems with high numbers of particles (usually $N_p > 10^6$). Furthermore, compared to early electrostatic simulations, it allows for a fully dynamic treatment of the electromagnetic field. Consequently, the fields exhibit a finite speed of propagation and can be consistently coupled to relativistic particles, enabling the simulation of fully relativistic systems such as the acceleration of electrons in laser wakefield experiments \cite{martins2010exploring}.

An important aspect of PIC is how to relate the field degrees of freedom to the particle degrees of freedom. Fields are defined at lattice sites, whereas particles have continuous positions and velocities and move freely across the lattice. This is accomplished through a procedure called interpolation or weighting. Interpolation is necessary in two directions: interpolating from the particles to the grid in order to compute charge densities and currents on the lattice and interpolation from the grid to particles to obtain the force acting on them. Each particle is given a certain shape characterized by a continuous shape factor or weight $S$. Given the continuous position of a particle, the shape factor determines how its point-like charge is distributed among the cells on the discrete lattice and, conversely, how the fields act on the particle.

\begin{figure}[t]
	\centering
	\begin{subfigure}[b]{0.45\textwidth}
		\includegraphics{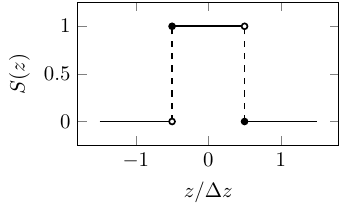}
		\caption{Nearest grid point (NGP) shape factor.}
	\end{subfigure}
	\qquad
	\begin{subfigure}[b]{0.45\textwidth}
		\includegraphics{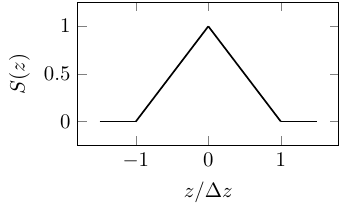}
		\caption{Cloud-in-cell (CIC) shape factor. }
	\end{subfigure}
	\caption{The two lowest order shape factors used in PIC simulations.
		\label{fig:shape_factors}}
\end{figure}

The concept of interpolation is best explained by giving a specific example: consider a simple 1+1 dimensional setting with a single particle with charge $Q$ following a trajectory $z(t)$. How should one relate this charge to the discrete charge density $\rho_i$ defined at the lattice points $z_i = i \, \Delta z$, where $\Delta z$ is the lattice spacing? The solution is to specify the shape of the particle via the shape factor $S(z)$ and perform the mapping
\begin{equation}
\rho_i(t) = \frac{Q}{\Delta z} S(z_i - z(t)).
\end{equation}
Depending on the exact form of $S(z)$ the charge $Q$ is distributed among multiple neighboring cells or just a single cell. Charge conservation is guaranteed by requiring that the total charge on the lattice is always the same at any time $t$ independent of the particle position, i.e.
\begin{equation}
\Delta z \sum_i \rho_i(t) = Q = \mathrm{const}.
\end{equation}
This leads to the normalization condition for charge conserving shape factors
\begin{equation}
\sum_i S(z_i - z) = 1,
\end{equation}
where $z$ is an arbitrary position and the sum runs over all lattice sites $z_i$. In principle, shape factors can be arbitrary functions as long as the above constraint holds, but for simplicity they are usually constructed from piecewise polynomials and categorized by the order of the polynomials. In \cref{fig:shape_factors} two simple shape factors are presented: the nearest-grid-point (NGP) or zero-order weight, and the cloud-in-cell (CIC) or first-order weight. The NGP scheme is defined by
\begin{equation}
S(z) = 
\begin{cases}
1,& \qquad  -\frac{\Delta z}{2} \leq z < \frac{\Delta z}{2} \\
0,& \qquad  \mathrm{otherwise},
\end{cases}
\end{equation}
while the CIC scheme is given by
\begin{equation}
S(z) = 
\begin{cases}
	1 - \frac{\abs z}{\Delta z},& \qquad  \abs z \leq \Delta z \\
	0,& \qquad  \mathrm{otherwise}.
\end{cases}
\end{equation}
In this one-dimensional example the NGP scheme interpolates the charge to the (as the name implies) nearest neighboring lattice site, whereas the CIC scheme distributes the charge among two neighboring cells (except if the particle is exactly at a lattice site $z_i$).
Higher order shape factors can be systematically constructed using quadratic, cubic or higher order polynomials \cite{Esirkepov2001}.

The interpolation step can easily be generalized to multiple particles $N_p$ by summing over each individual particle
\begin{equation}
\rho_i(t) = \sum_{\alpha=1}^{N_p}\, \frac{Q_\alpha}{\Delta z} \, S(z_i - z_\alpha(t)).
\end{equation}
In more than one spatial dimension the shape factor is defined as the product of the one-dimensional shape factors, e.g.\ for three spatial dimensions and a trajectory $\vec{r}(t) = \lb x(t), y(t), z(t) \rb$ we have
\begin{equation}
S(\vec{r}_i - \vec{r}(t)) \equiv S(x_i- x(t)) S(y_i- y(t)) S(z_i - z(t)),
\end{equation}
where $\vec{r}_i$ are the coordinates of lattice site $i$.
Shape factors of particles are also relevant when performing the inverse interpolation step, i.e.\ interpolation from fields to particles. In this simple 1+1 dimensional example using the temporal gauge condition, there is only one gauge field component $A_{i,z}$ defined at the lattice sites $z_i$. The local gauge field $A_z(z(t))$ acting on a particle at position $z(t)$ is computed using the mapping
\begin{equation}
A_z(z(t)) = \sum_i\, A_{i,z} \, S(z_i - z(t)).
\end{equation}
The use of the same shape factors for both interpolation directions is crucial, as this eliminates numerical instabilities caused by an unphysical self-force, i.e.\ a spurious force due to particles generating fields that act on themselves (see \cite{birdsall_langdon_1970} and section 5.3.3 of \cite{Hockney:1988:CSU:62815}). This elimination of numerical instabilities only works in the electrostatic case, where one assumes that particle velocities are much smaller than the propagation speed of electric and magnetic fields, i.e.\ the speed of light. In the more general electromagnetic case, new numerical instabilities due to self-interaction can occur. A prominent example is the numerical Cherenkov instability, which will be discussed in \cref{cha:semi_implicit}. 

As particles move across the lattice, they generate a charge current that affects the time evolution of the fields. In order to preserve the Gauss constraint, the current must be locally charge conserving and satisfy the continuity equation.
In the 1+1 dimensional example, the continuity equation reads
\begin{equation}
\p_\mu J^\mu = \p_t \rho(t, z) + \p_z j(t, z) = 0.
\end{equation}
A straightforward discretization of the above equation is given by
\begin{equation} \label{eq:discrete_cont_eq}
\frac{1}{\Delta t} \lb \rho_i(t+\Delta t) - \rho_i(t) \rb + \frac{1}{\Delta z} \lb j_{i+\frac{1}{2}}(t+\frac{\Delta t}{2}) - j_{i-\frac{1}{2}}(t+\frac{\Delta t}{2}) \rb = 0.
\end{equation}
Here $\Delta t$ refers to the time step and $j_{i+\frac{1}{2}}$ to the charge current density on the lattice. The current is evaluated at shifted lattice sites and at fractional time steps. This assignment is known as the Yee scheme \cite{Yee66numericalsolution} which is an example of a staggered grid. The main benefit of using the Yee scheme is that simple forward differences can be interpreted as central differences, which have quadratic accuracy:
\begin{align}
\frac{1}{\Delta t} \lb \rho_i(t+\Delta t) - \rho_i(t) \rb &\simeq \frac{\p}{\p t}\rho(z_i, t+\frac{\Delta t}{2}) + \mathcal{O}(\Delta t^2), \\
\frac{1}{\Delta z} \lb j_{i+\frac{1}{2}}(t+\frac{\Delta t}{2}) - j_{i-\frac{1}{2}}(t+\frac{\Delta t}{2}) \rb &\simeq \frac{\p}{\p z} j(z_{i}, t+\frac{\Delta t}{2}) + \mathcal{O}(\Delta z^2).
\end{align}
We now consider a single particle with charge $Q$ and trajectory $z(t)$ in the NGP scheme. The charge current generated by the moving particle can be computed directly from the discretized continuity equation with a few simple assumptions. We consider a particle moving in the positive $z$ direction. Notice that if the particle moves in such a way that its nearest neighbor cell does not change from one time step to the next, then the discretized charge density $\rho_i$ stays the same. In this case no current is generated. However, once the particle crosses from one cell into the next the time derivative of $\rho_i$ is non-zero and a non-vanishing current must be generated. Assuming that the NGP of $z(t)$ is $z_i$ and the NGP of $z(t+\Delta t)$ is $z_{i+1}$ we have $\rho_{i+1}(t+\Delta t) = Q / \Delta z$ and $\rho_i(t) = Q / \Delta z$. Plugging this into \cref{eq:discrete_cont_eq} evaluated at $z_i$ and $z_{i+1}$ we obtain
\begin{align}
-\frac{Q}{\Delta z \Delta t} &= - \frac{1}{\Delta z} \lb j_{i+\frac{1}{2}}(t+\frac{\Delta t}{2}) - j_{i-\frac{1}{2}}(t+\frac{\Delta t}{2}) \rb, \\
+\frac{Q}{\Delta z \Delta t} &= - \frac{1}{\Delta z} \lb j_{i+\frac{3}{2}}(t+\frac{\Delta t}{2}) - j_{i+\frac{1}{2}}(t+\frac{\Delta t}{2}) \rb.
\end{align}
Using the boundary condition that no spurious current should be generated far away from the particle position, it is intuitive to induce the current at the mid-point $z_{i+\frac{1}{2}}$. The charge current that satisfies the continuity equation is therefore
\begin{equation}
j_{i+\frac{1}{2}} = \frac{Q}{\Delta t}.
\end{equation}
At all other lattice sites, the current is set to zero. Reversing the above example and considering a particle moving from $z_{i+1}$ to $z_i$ yields the same current with opposite sign
\begin{equation}
j_{i+\frac{1}{2}} = -\frac{Q}{\Delta t}.
\end{equation}
The NGP scheme is peculiar because current is only generated when a boundary crossing happens. In the case of a single particle, this leads to sudden and high spikes in the discrete current density, which introduce unwanted numerical noise when coupled to the fields. This artifact can be avoided with the CIC and higher order schemes which smear the particle charge among multiple cells and induce smoother currents.  In \cref{sec:implementation} we will see how this problem can be mitigated even for the NGP scheme.

The focus of this section is how interpolation is performed in PIC simulations. Although we left out details regarding the computation of forces and the time evolution of fields, a summary about PIC is incomplete without mentioning the typical simulation cycle. A PIC simulation proceeds as follows:
  
\begin{enumerate}
	\item \textit{Interpolation from particles to fields}: given all particle positions, velocities and the shape factor, determine the lattice charge and current density.
	\item \textit{Field evolution}: solve Maxwell's equations on the lattice using the current density and previous fields.
	\item \textit{Interpolation from fields to particles}: determine the local field acting on each particle with the same shape factor as in the first step.
	\item \textit{Particle evolution}: use interpolated fields to compute the force acting on each particle and update velocities and positions accordingly.
\end{enumerate}
This cycle is repeated for each time step until the simulation is complete. At the end of each cycle one can compute various observables such the energy momentum tensor of the fields or particle densities. 

\begin{figure}
	\centering
	
	\qquad\qquad\includegraphics[scale=0.14]{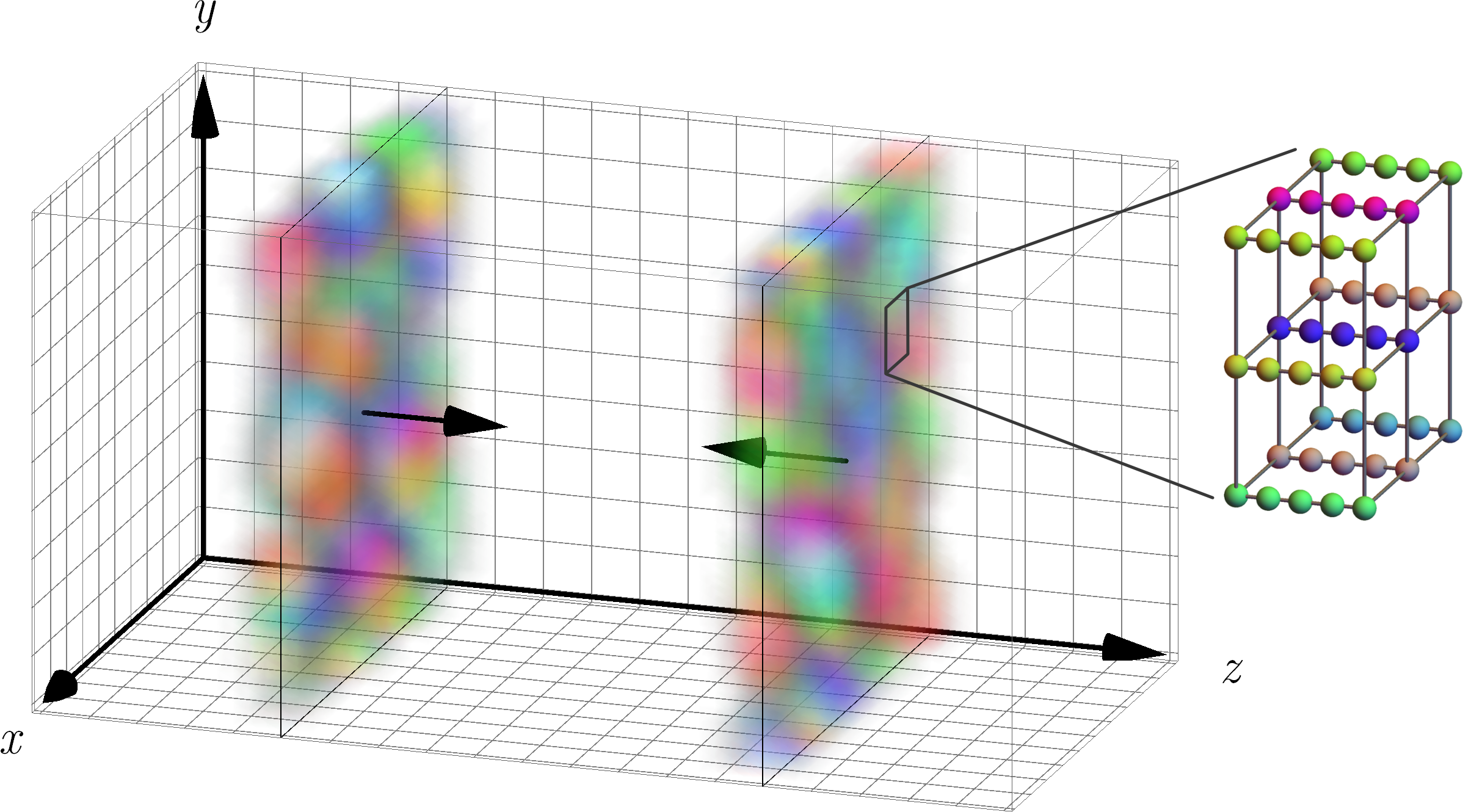}
	\caption{An illustration of the CPIC approach to simulating CGC collisions in 3+1 dimensions from \cite{Gelfand:2016yho}. The smooth color charge densities of the colliding nuclei are replaced by an ensemble of colored point-like particles with appropriate color charges. In the eikonal approximation the color charges are not affected by recoil and thus only move along the longitudinal axis $z$. The movement of colored particles is therefore effectively 1+1 dimensional.
		\label{fig:cpic_overview}}
\end{figure}

The CPIC method generalizes the concept of dynamic point-like particles in cells coupled to discretized fields to non-Abelian gauge theory. The main ingredient in PIC, namely the interpolation step, can be readily generalized to reflect non-Abelian gauge covariance. Color currents induced by the movement of colored particles can be computed from the non-Abelian continuity equation.
We apply CPIC to the problem of colliding nuclei in the sense that we approximate the smooth color charge densities of the nuclei as an ensemble of colored particles. A schematic overview of this approach is shown in \cref{fig:cpic_overview}.  For this purpose, we only use CPIC in a very limited way, because we assume the colliding nuclei to be recoilless. We can therefore neglect any deceleration of color charges and only care about parallel transport. This greatly simplifies the problem since particle movement can be treated one-dimensionally. In the following subsections we derive the equations for gauge fields on the lattice and colored particles moving along lightlike trajectories.

\subsection{Field equations} \label{sec:latt_field__eqs}

The starting point of our derivation is the Wilson action \cref{eq:wilson_action} coupled to an external color current (see \cref{sec:lattice_notation} for some clarifications regarding the shorthand lattice notation)
\begin{align} \label{eq:wilson_action_j}
S[U] &=  V \sum_x  \bigg( \sum_i \frac{1}{\lb g a^0 a^i \rb^2} \tr \left[ 2- U_{x,0i} - U^\dg_{x,0i} \right] \nn
& \qquad \qquad -  \sum_{i, j} \frac{1}{2 \lb g a^i a^j \rb^2} \tr \left[ 2- U_{x,ij} - U^\dg_{x,ij} \right] \nn
& \qquad \qquad + 2 \, \tr \, \bigg[ - \rho_x A_{x,0} + \sum_i j_{x,i} A_{x,i} \bigg] \bigg),
\end{align}
where $\rho_x$ and $j_{x,i}$ are the color charge and current density. The current $j_{x,i}$ is identified with the spatial components of the four-vector $J_\mu$.
This action is slightly inconsistent in the sense that gauge links are used to formulate the pure Yang-Mills part, while the interaction term with the external current $J^\mu$ is written in terms of the gauge fields $A_\mu$. Even worse, due to the presence of the $J_\mu A^\mu$ term, the action is not even strictly gauge invariant under arbitrary gauge transformations. This is not an artifact of the lattice description, but is also true in the continuum case \cite{Arodz:1980gg}. Nevertheless, the action still yields gauge covariant equations of motion (EOM) and constraints upon variation. The constrained variation of gauge links is described in \cref{app_var}. The variation of the coupling term is simply given by
\begin{equation}
\dd S_J[U] = 2 V \sum_x \tr \left[  - \rho_x \dd A_{x,0} + \sum_i j_{x,i} \dd A_{x,i} \right].
\end{equation}
which is invariant under $J_{x,\mu} \rightarrow \Omega_x J_{x,\mu} \Omega_x^\dg$ and $\delta A_{x,\mu} \rightarrow \Omega_x \dd A_{x,\mu} \Omega^\dg_x$.
Performing the variation of the Yang-Mills part (see \cref{app_leapfrog}) with respect to temporal components $U_{x,0}$ and $A_{x,0}$ yields the Gauss constraint
\begin{equation} \label{eq:leapfrog_gauss_cpic}
\sum_{i}\frac{1}{\left(a^{0}a^{i}\right)^{2}}\ah{U_{x,0i}+U_{x,0-i}} = \frac{g}{a^0} \rho_x,
\end{equation}
while variation with respect to spatial components $U_{x,i}$ and $A_{x,i}$ yields the EOM
\begin{equation} \label{eq:leapfrog_eom_cpic}
\frac{1}{\lb a^0 a^i \rb^2} \ah{U_{x,i0} + U_{x,i-0}} =
\sum_j \frac{1}{\lb a^i a^j \rb^2} \ah{U_{x,ij} + U_{x,i-j}} - \frac{g}{a^i} j_{x,i}.
\end{equation}
Here, $\ah{\dots}$ denotes the anti-hermitian, traceless part, see \cref{eq:ah_definition_2}.
Equivalently, we can write the Gauss constraint as
\begin{equation}
\sum_{i}\frac{1}{\left(a^{0}a^{i}\right)^{2}}P^{a}\left(U_{x,0i}+U_{x,0-i}\right)=\frac{g}{a^0} \rho^a_{x},
\end{equation}
and the EOM as
\begin{equation} \label{eq:cpic_lagrange_eom_links}
\frac{1}{\lb a^0 a^i \rb^2} P^a \lb U_{x,i0} + U_{x,i-0}\rb =
\sum_j \frac{1}{\lb a^i a^j \rb^2} P^a \lb U_{x,ij} + U_{x,i-j}\rb - \frac{g}{a^i} j^a_{x,i},
\end{equation}
where we have used
\begin{equation} \label{eq:Pa_definition}
P^a(X) \equiv 2 \, \Im \, \tr \left[ t^a X \right],
\end{equation}
and the identity (see \cref{eq:Pa_and_ah})
\begin{equation}
\sum_a t^{a}P^{a}\left(X\right) = \ah{X}.
\end{equation}
Both the constraint and EOM are gauge covariant under lattice gauge transformations. In order to perform time evolution we use temporal gauge fixing
\begin{equation}
U_{x,0} = \one,\qquad \forall x \in \Lambda^4,
\end{equation}
where $\Lambda^4$ is the regular hypercubic lattice (see \cref{eq:lambda4_def}). The temporal plaquettes then reduce to the product of consecutive spatial gauge links (see \cref{eq:plaquette_definition})
\begin{equation}
U_{x,i0} = U_{x,i} U^\dg_{x+0,i}.
\end{equation}
The goal is now to express all gauge links in the next time slice $U_{x+0,i}$ in terms of previous gauge links. Once the temporal plaquettes $U_{x,i0}$ are known from the EOM, the next gauge links can be computed via
\begin{equation} \label{eq:leapfrog_gauge_link_update}
U_{x+0,i} = U^\dg_{x,i0} U_{x,i}.
\end{equation}

At a first glance, the EOM can only be solved for the anti-hermitian traceless part of the temporal plaquette $\ah{U_{x,i0}}$, but since elements of $\SUN$ in the fundamental representation are completely specified by $N_c^2 - 1$ real parameters, there are enough equations to reconstruct the full matrix $U_{x,i0}$. However, obtaining $U_{x,i0}$ from $\ah{U_{x,i0}}$ depends on the parametrization and on the specific gauge group. For $\mathrm{SU(2)}$ we use the parametrization (see \cref{app_su2}) 
\begin{equation} 
U_{x,i0} = u^0 \, \one + i \sum_a u^a \sigma^a,
\end{equation}
where $\sigma^a$, $a\in \{1,2,3\}$ are the Pauli matrices. The real-valued parameters $u^0$, $u^a$ fulfill the constraint (see \cref{eq:app_su2_constraint}) 
\begin{equation} \label{eq:su2_constraint}
(u^0)^2 + \sum_a \lb u^a \rb^2  = 1.
\end{equation}
Using
\begin{equation}
P^a(U_{x,i0}) = 2 u ^a,
\end{equation}
the EOM can be written as
\begin{equation} \label{eq:su2_evolve1}
u^a = \frac{1}{2} \lb -P^a\lb U_{x,i-0} \rb + \sum_j \lb \frac{a^0}{a^j} \rb^2 P^a \lb U_{x,ij} + U_{x,i-j}\rb - (a^0 )^2 g a^i j^a_{x,i} \rb.
\end{equation}
To obtain $u^0$, we take the positive branch of  \cref{eq:su2_constraint}
\begin{equation} \label{eq:su2_evolve2}
u^0=\sqrt{1 - \sum_a \lb u^a \rb^2},
\end{equation}
assuming that the changes from one time slice to the next are ``small", i.e.\ the temporal plaquette $U_{x,0i}$ will be ``closer" to $\one$ than to $-\one$. Obviously, if $\sum_a (u^a)^2$ exceeds $1$, the time evolution is ill-defined. This puts a limit on the field amplitudes and lattice spacings $a^\mu$ used in the simulation. The equations can be cast into a more familiar form by introducing the chromo-electric field via
\begin{equation} \label{eq:cpic_lagrange_efield}
E_{x,i} = \frac{1}{g a^i a^0} \ah{U_{x,0i}}.
\end{equation}
The lattice EOM then read
\begin{equation} \label{eq:cpic_lagrange_eom}
E_{x,i} = - \sum_j \frac{a^0}{g a^i \lb a^j \rb^2} \ah{U_{x,ij} + U_{x,i-j}} + a^0 j_{x,i} + E_{x-0,i},
\end{equation}
and the temporal plaquette in $\mathrm{SU(2)}$ is (see \cref{eq:su2_plaq_from_ah})
\begin{equation} \label{eq:cpic_lagrangian_temporal_plaq}
U_{x,0i} = \sqrt{1 - \frac{1}{4} \lb g a^0 a^i \rb^2 \sum_a E_{x,i}^a E_{x,i}^a} \, \one + i g a^0 a^i \sum_a t^a E_{x,i}^a,
\end{equation}
which is used to update gauge links via \cref{eq:leapfrog_gauge_link_update}.
The Gauss constraint reads
\begin{equation}
\sum_i \frac{1}{a^i} \lb E_{x,i} - U^\dg_{x-i,i} E_{x-i,i} U_{x-i,i} \rb = \rho_x,
\end{equation}
or simply
\begin{equation}
\sum_i D^B_i E_{x,i} = \rho_x,
\end{equation}
where $D^B_i$ is the backward gauge-covariant finite difference.

At this point a remark about the accuracy of these discrete EOM is necessary. Recall that the Wilson action is accurate up to quadratic terms in the lattice spacing $a^\mu$ and the EOM derived from this action should exhibit the same accuracy.
At first glance the EOM \cref{eq:cpic_lagrange_eom} seem to resemble a forward Euler scheme, due to the electric field $E_{x,i}$ and the gauge links $U_{x,i}$ ``living" in the same time-slice according how they transform under lattice gauge transformations. Forward Euler schemes are only linearly accurate in the time step $a^0$, which seems to contradict the accuracy of the action these equations were derived from.
This apparent contradiction is resolved by interpreting the electric field to be evaluated in-between two time steps. The reasoning for this is as follows: the plaquette $U_{x,\mu\nu}$ best approximates the field strength tensor $F_{\mu\nu}$ at the center of the plaquette $x_0 = x + \frac{1}{2} \hat{\mu} + \frac{1}{2} \hat{\nu}$. Electric fields are defined via \cref{eq:cpic_lagrange_efield}, i.e.\ they are related to temporal plaquettes $U_{x,0i}$, which are naturally centered in-between time slices. The field $E_{x,i}$ therefore corresponds to $E_i ( t + \frac{1}{2} a^0, x + \frac{1}{2} a^i \hat{e_i} )$ (no sum over $i$ is implied) in the continuum. The EOM \cref{eq:cpic_lagrange_eom} are therefore actually equivalent to a leapfrog scheme, which is quadratically accurate in the time step. We refer to this method as the Lagrangian leapfrog approach from now on.

Although the above presented time evolution follows directly from varying the Wilson action, it is less commonly used \cite{Lappi:2016ato, Lappi:2017ckt, Peuron:2018gho}. The more popular method is the Hamiltonian leapfrog approach. It based on the Kogut-Susskind Hamiltonian \cite{PhysRevD.11.395}, which can be derived from the Wilson action by taking the continuous time limit $a^0 \rightarrow 0$ and performing a Legendre transformation. Since we have already derived the discrete equations \cref{eq:leapfrog_eom_cpic}, it easy to take the limit $a^0 \rightarrow 0$ using
\begin{equation}
U_{x\pm0,i} \simeq U_{x,i}(t) \pm a^0 \p_0 U_{x,i}(t) + \frac{1}{2} \lb a^0 \rb^2 \p_0^2 U_{x,i}(t) \pm \frac{1}{6} \lb a^0 \rb^3 \p_0^3 U_{x,i}(t) + \mathcal{O}\big( \big( a^0 \big)^4 \big).
\end{equation}
In this limit the lattice $\Lambda^4$ reduces to the spatial three-dimensional lattice $\Lambda^3 \times \mathbb{R}$, where $\mathbb R$ denotes the continuous time coordinate.
The EOM in temporal gauge then read
\begin{equation}
\ah{U_{x,i}(t) \p_0^2 U^\dg_{x,i}(t)} + \mathcal{O}\big( \big( a^0 \big)^2 \big) = \sum_j \frac{1}{\lb a^j \rb^2} \ah{U_{x,ij}(t) + U_{x,i-j}(t)} - g a^i j_{x,i}(t).
\end{equation}
The quadratic accuracy arises naturally from the time reflection symmetry of the Wilson action. Identifying the time derivative of spatial gauge links with the chromo-electric field
\begin{equation} \label{eq:leapfrog_link_eq}
\p_0 U_{x,i}(t) = i g a^i E_{x,i}(t) U_{x,i}(t),
\end{equation}
we find
\begin{equation} 
\p_0^2 U_{x,i}(t) = i g a^i \p_0 E_{x,i}(t) U_{x,i}(t) - \big( g a^i \big)^2 E^2_{x,i}(t) U_{x,i}(t),
\end{equation}
and therefore
\begin{equation} \label{eq:leapfrog_electric_eq}
\p_0 E_{x,i}(t) = - \sum_j \frac{1}{g a^i \lb a^j \rb^2} \ah{U_{x,ij}(t) + U_{x,i-j}(t)} + j_{x,i}(t).
\end{equation}
In the Hamiltonian approach the chromo-electric field is identified with the conjugate momentum of the gauge links.
Repeating similar steps, the Gauss constraint reads
\begin{equation}
\sum_i \frac{1}{a^i} \lb E_{x,i}(t) - U_{x-i,i}(t) E_{x-i,i}(t) U^\dg_{x-i,i}(t) \rb = \rho_x(t),
\end{equation}
or simply
\begin{equation}
\sum_i D^B_i E_{x,i}(t) = \rho_x(t).
\end{equation}
In order to solve the first-order differential equations eqs.\ \eqref{eq:leapfrog_link_eq} and \eqref{eq:leapfrog_electric_eq} numerically, the time coordinates is (again) discretized and one obtains the leapfrog equations
\begin{align} \label{eq:cpic_leapfrog_eom_orig}
E_{x,i}(t+\frac{a^0}{2}) &= - \sum_j \frac{a^0}{g a^i \lb a^j \rb^2} \ah{U_{x,ij}(t) + U_{x,i-j}(t)} + a^0 j_{x,i}(t) + E_{x,i}(t-\frac{a^0}{2}), \\
U_{x,i}(t+a^0) &= \exp \bigg( i g a^i a^0 E_{x,i}(t+\frac{a^0}{2}) \bigg) U_{x,i}(t),
\end{align}
which are accurate up to second order in $a^0$ like the Lagrangian leapfrog approach. The momenta (electric fields) are shifted by half a time step compared to the gauge links, which is again reminiscent of a leapfrog scheme. Note that the use of the exponential in the gauge link update guarantees that gauge links remain in $\SUN$.

Both methods agree up to quadratic order in $a^0$. The only difference lies in the update of gauge links from one time slice to the next and the definition of the electric field. To see this we first note that in the Hamiltonian leapfrog method, the temporal plaquette is given by
\begin{equation}
U_{x,0i} = U_{x,i}(t+a^0) U^\dg_{x,i}(t) =  \exp \bigg(i g a^i a^0 E_{x,i}(t+\frac{a^0}{2}) \bigg).
\end{equation}
For $\mathrm{SU(2)}$ the exponential can be computed via
\begin{equation}
U_{x,0i} = \cos \lb \frac{1}{2} \sqrt{\mathcal{E}^b \mathcal{E}^b} \rb \one + i \sum_a  \sigma^a e^a \sin \lb \frac{1}{2} \sqrt{\mathcal{E}^b \mathcal{E}^b} \rb,
\end{equation}
where $\mathcal{E} = g a^0 a^i E_{x,i}(t+a^0/2)$ and $e^a = \mathcal{E}^a / \sqrt{\mathcal{E}^b \mathcal{E}^b}$. Expanding up to third order in $a^0$ we find
\begin{equation}
U_{x,0i} \simeq \lb 1 - \frac{1}{8} \mathcal{E}^a \mathcal{E}^a\rb \one + i \mathcal{E}  + \mathcal{O}\lb \mathcal{E}^3 \rb. 
\end{equation}
In the Lagrangian leapfrog method, the chromo-electric field is identified as (see \cref{eq:cpic_lagrange_efield})
\begin{equation}
E_{x,i} = \frac{1}{g a^i a^0} \ah{U_{x,0i}},
\end{equation}
which can be thought of as naturally centered around the mid-point between $x$ and $x+0$.
The temporal gauge EOM look identical to the Hamiltonian leapfrog method. The temporal plaquette in $\mathrm{SU(2)}$ is given by
\begin{equation}
U_{x,0i} = \sqrt{1 - \sum_a u^a u^a} \, \one + i \sum_a \sigma^a u^a,
\end{equation}
where $u^a = P^a(U_{x,0i}) / 2$. With the above definition of the electric field we then find
\begin{align}
U_{x,0i} &= \sqrt{1 - \frac{1}{4} \mathcal{E}^a \mathcal{E}^a } \, \one + i \mathcal{E} \nn
&\simeq \lb 1 - \frac{1}{8} \mathcal{E}^a \mathcal{E}^a \rb \one + i \mathcal{E} + \mathcal{O}(\mathcal{E}^3).
\end{align}
The Lagrangian and Hamiltonian leapfrog methods agree up to a cubic term in $a^0$ and are interchangeable since both are only correct up to quadratic order. The Hamiltonian leapfrog equations were used in the first two publications \cite{Gelfand:2016yho, Ipp:2017lho}, but during the development of the numerical improvements introduced in \cite{Ipp:2018hai} and presented in \cref{cha:semi_implicit} it became apparent that the Lagrangian leapfrog approach is necessary for more complicated actions than the standard Wilson action.

\subsection{Particle equations} \label{sec:latt_particle__eqs}

In this section we derive a locally charge conserving description of colored particles coupled to lattice Yang-Mills fields. As already outlined earlier, the trajectories of particles are completely fixed and we only need to derive how color charges are parallel transported as they move across the lattice.

Local charge conservation is closely related to the Gauss constraint and gauge invariance. Without external color sources, the lattice EOM \cref{eq:leapfrog_eom_cpic} exactly preserve the Gauss constraint \cref{eq:leapfrog_gauss_cpic} (see \cref{app_gauss}) due to lattice gauge invariance. If external currents are included, the constraint is only conserved if and only if the non-Abelian continuity equation holds
\begin{equation}
\frac{1}{a^0} \lb \rho_x -\rho_{x-0} \rb = \sum_i \frac{1}{a^i} \lb j_{x,i} - U^\dg_{x-i,i} j_{x-i,i} U_{x-i,i} \rb.
\end{equation}
As colored particles move across the lattice, their charge density and color current must satisfy the above equation. Similar to the interpolation step derived previously in the Abelian case, we can use the non-Abelian continuity equation to derive the charge conserving current given a certain interpolation scheme. 

Our setup is effectively 1+1 dimensional in terms of particle movement due to the eikonal approximation. It is therefore enough to consider a single point-like particle with color charge $Q = Q^a t^a$ moving along the longitudinal axis $z$ at the speed of light. In general, $Q(t)$ is a function of time because of parallel transport and the color current only has a non-vanishing longitudinal component $j_{x,3}$.

We choose the NGP scheme for the interpolation of charges to the lattice due to its simplicity. If the particle moves in such a way that its NGP does not change, there is no current generated, even if the actual particle position varies. This situation would be slightly more complicated if we had not fixed temporal gauge. Then, even if the particle is at rest, its color charge would be parallel transported along the time axis such that the charge would be gauge-covariantly constant
\begin{equation}
D_0 Q(t) = 0.
\end{equation}  
In our setup, temporal gauge is used to solve the EOM, so we can ignore this complication of the more general case. It is then only necessary to study boundary crossings of particles as in the Abelian case. For this, we consider a particle moving in the positive $z$ direction. Assume that at time $t$ its NGP is $x$ and in the next time slice its NGP is $x+\hat{0}+\hat{3}$. The charge density on the lattice then reads
\begin{align}
\rho_{x} &= \frac{Q(t)}{V}, \\
\rho_{x+3+0} &= \frac{Q(t+a^0)}{V},
\end{align}
where $V = \prod_i a^i$ is the volume of a cell. The charge density is zero everywhere else. We have also accounted for a possible color rotation of the charge. Plugging this into the continuity equation evaluated at $x+\hat{0}$ and $x+\hat{0}+\hat{3}$ we find
\begin{align}
-\frac{1}{a^0} \frac{Q(t)}{V} &= \frac{1}{a^3} \lb j_{x+0,3} - U^\dg_{x+0-3,3} j_{x+0-3,3} U_{x+0-3,3} \rb, \\
+\frac{1}{a^0} \frac{Q(t+a^0)}{V} &= \frac{1}{a^3} \lb j_{x+0+3,3} - U^\dg_{x+0,3} j_{x+0,3} U_{x+0,3} \rb.
\end{align}
Analogous to the PIC derivation, we assume that only $j_{x+0,3}$ is non-zero. We then find from the first equation
\begin{equation}
j_{x+0,3} = -\frac{a^3}{a^0} \frac{Q(t)}{V}.
\end{equation}
Using this in the second equation fixes the parallel transport of the particle charge
\begin{equation} \label{eq:q_pt_rm}
Q(t+a^0) = U^\dg_{x+0,3} Q(t) U_{x+0,3}.
\end{equation}
We can repeat the same steps for a particle moving in the opposite direction. Assuming that at $t$ its NGP is $x+\hat{3}$ and in the next time slice the NGP is $x+\hat{0}$, the charge density is given by
\begin{align}
\rho_{x+3} &= \frac{Q(t)}{V}, \\
\rho_{x+0} &= \frac{Q(t+a^0)}{V}.
\end{align}
The continuity equation yields
\begin{align}
+\frac{1}{a^0} \frac{Q(t+a^0) }{V} &= \frac{1}{a^3} \lb j_{x+0,3} - U^\dg_{x+0-3,3} j_{x+0-3,3} U_{x+0-3,3} \rb, \\
-\frac{1}{a^0} \frac{Q(t)}{V} &= \frac{1}{a^3} \lb j_{x+0+3,3} - U^\dg_{x+0,3} j_{x+0,3} U_{x+0,3} \rb.
\end{align}
In this case, the color current and the color charge are given by
\begin{equation}
j_{x+0,3} = +\frac{a^3}{a^0} \frac{Q(t+a^0)}{V},
\end{equation}
and
\begin{equation} \label{eq:q_pt_lm}
Q(t+a^0) = U_{x+0,3} Q(t) U^\dg_{x+0,3}.
\end{equation}
Note that the slightly unintuitive signs of the color currents (i.e.\ a negative sign for the right-moving particle and vice versa) arise because $j_{x,i}$ are the spatial components of $J_\mu$ with lowered indices. Raising the indices yields $j^i_x = - j_{x,i}$ and one obtains the expected signs.

In principle, one could also consider non-Abelian generalizations of higher order interpolation schemes such as the first-order CIC scheme. The CIC scheme for CPIC simulations has been derived in \cite{Strickland:2007}, but it has a slight flaw: the trace of the color charge squared of each particle should be a conserved quantity, because the charge is only affected by rotations. In the CIC scheme this conservation is violated and only restored in the continuum limit (see section C of \cite{Strickland:2007}). On the other hand, the NGP scheme conserves this quantity exactly, i.e.\ for each individual particle we have
\begin{equation}
\tr \left[ Q(t) ^2 \right] = \tr \left[ Q(t+a^0)^2 \right].
\end{equation}
The main disadvantage of the NGP scheme is that current is only generated at each boundary crossing. This problem can be circumvented by simply using a high number of particles per cell, such that at each time step the color current is non-zero. 

\subsection{Simulation cycle} \label{sec:simulation_cycle}

Having derived all the necessary equations, we can summarize what a single simulation step entails in the Lagrangian leapfrog approach. The Hamiltonian leapfrog approach is completely analogous. Assuming that the ensemble of particles, $U_{x,i}$ and $E_{x-0,i}$ are known initially, the simulation cycle reads:

\begin{enumerate}
\item \textit{Interpolation from particles to fields}: for each particle, determine the NGP from their current position $\vec{x}(t-a^0)$ and compute the charge density on the lattice. Assuming that the NGP is $x$, the charge density receives the contribution
\begin{equation}
\rho_{x-0} = \frac{Q(t-a^0)}{V},
\end{equation}
where $V = \prod_i a^i$.

Optionally, check if the Gauss constraint is satisfied
\begin{equation}
\sum_i D^B_i E_{x-0,i} = \rho_{x-0}
\end{equation}
to make sure that the interpolation procedure works correctly.

Determine if a boundary crossing happens between the current and the next time slice and compute the longitudinal current density $j_{x,3}$ on the lattice. For right-moving particles the contribution is
\begin{equation} \label{eq:cycle_current_rm}
j_{x,3} = -\frac{a^3}{a^0} \frac{Q(t-a^0)}{V},
\end{equation}
and for left-moving particles it is
\begin{equation} \label{eq:cycle_current_lm}
j_{x,3} = +\frac{a^3}{a^0} U_{x,3} \frac{Q(t-a^0)}{V} U^\dg_{x,3}.
\end{equation}

\item \textit{Update fields}: Given $E_{x-0,i}$, $U_{x,i}$ and $j_{x,3}$ from the previous step, compute $E_{x,i}$ via the EOM \cref{eq:cpic_lagrange_eom}
\begin{equation}
E_{x,i} = - \sum_j \frac{a^0}{g a^i \lb a^j \rb^2} \ah{U_{x,ij} + U_{x,i-j}} + a^0 j_{x,i} + E_{x-0,i}.
\end{equation}
Compute $U_{x+0,i}$ from $E_{x,i}$ and $U_{x,i}$ via \cref{eq:cpic_lagrangian_temporal_plaq}
\begin{equation}
U_{x,0i} = \sqrt{1 - \frac{1}{4} \lb g a^0 a^i \rb^2 \sum_a E_{x,i}^a E_{x,i}^a} \, \one + i g a^0 a^i \sum_a t^a E_{x,i}^a,
\end{equation}
and \cref{eq:leapfrog_gauge_link_update}
\begin{equation}
U_{x+0,i} = U^\dg_{x,i0} U_{x,i}.
\end{equation}

\item \textit{Interpolation from fields to particles}: update the color charge of each particle using the gauge links $U_{x,i}$. Right-moving particles are updated via \cref{eq:q_pt_rm}
\begin{equation} \label{eq:cycle_parallel_rm}
Q(t) = U^\dg_{x,3} Q(t-a^0) U_{x,3},
\end{equation}
whereas left-moving particle use \cref{eq:q_pt_lm}
\begin{equation} \label{eq:cycle_parallel_lm}
Q(t) = U_{x-3,3} Q(t-a^0) U^\dg_{x-3,3},
\end{equation}
given that $x$ is their NGP at time $t-a^0$.

\item \textit{Particle evolution}: update the position $\vec{x}(t-a^0)$ of each right-moving particle via
\begin{equation}
\vec{x}(t) = \vec{x}(t-a^0) + a^0 \hat{e}_3,
\end{equation}
and for each left-moving particle
\begin{equation}
\vec{x}(t) = \vec{x}(t-a^0) - a^0 \hat{e}_3,
\end{equation}
where $\hat{e}_3$ is the unit vector in the longitudinal direction.
\end{enumerate}
This form of a typical simulation cycle closely resembles traditional PIC simulations, but in practice some steps can be simplified. For example, the update to color charges and particle positions can be immediately applied after computing charge densities and currents, thus avoiding multiple loops over each particle. Similarly, the update of the electric fields and the gauge links at a given lattice site can be combined as well.

\section{Initial conditions on the lattice} \label{sec:initial_lattice}

In this section we discuss how to put the temporal gauge initial conditions derived in \cref{sec:coll_with_finite} on the lattice. Consider a right-moving nucleus described by the generalized MV model (see eqs.\ \eqref{eq:mv2_onep} and \eqref{eq:mv2_twop})
\begin{equation}
\ev{\rho^a (x^-,x_T) \rho^b (x'^-,x'_T)} = g^2 \mu^2(x^-) \dd^{ab} \dd(x^- - x'^-) \delta^{(2)}(x_T - x'_T).
\end{equation}
At some initial time $t^0$ we can express this correlator in laboratory frame coordinates.
\begin{equation}
\ev{\rho^a (t^0,x_T,z) \rho^b (t^0,x'_T,z')} =  \sqrt{2} g^2 \mu^2(\frac{t^0-z}{\sqrt{2}}) \dd^{ab} \dd(z-z') \delta^{(2)}(x_T - x'_T).
\end{equation}
The additional factor of $\sqrt{2}$ is due to the scaling property of the Dirac distribution $\delta(x/\sqrt{2}) = \sqrt{2} \delta(x)$.
On the three-dimensional lattice this correlator can be readily discretized via
\begin{equation}
\ev{\rho^a_x \rho^b_{x'}} =  \frac{\sqrt{2} g^2}{a_L a_T^2}  \mu^2(\frac{t^0-z}{\sqrt{2}}) \dd^{ab} \dd_{x,x'},
\end{equation}
where $a_L=a^3$ is the longitudinal lattice spacing  and $a_T=a^1=a^2$ is the transverse spacing. In practice such a random configuration can be easily generated, because for $x \neq x'$ the charge density is completely uncorrelated. It is therefore sufficient to generate $N_c^2 - 1$ individual Gaussian random numbers for each color component with $z$ dependent variance  $\sqrt{2} g^2 \mu^2 / \lb a_L a_T^2 \rb$ and zero mean at each point on the lattice.

\subsection{Sampling charge density with particles} \label{sec:replaceing_density}

In principle, it is a simple task to generate ensembles of particles that mimic a given charge density $\rho_x$ on the lattice, but there are some subtleties that are discussed in the following.

Starting with a simple example, one could, for each lattice site $x$, generate a single particle with charge $Q = V \rho_x$, where $V$ is the cell volume. This would exactly replicate the original charge density $\rho_x$ when performing the NGP interpolation step. However, in the derivation of the current $j_{x,i}$ in the NGP interpolation scheme, we saw that particles only generate current when they cross cell boundaries such that their NGP changes from one time step to the next. If the time step is $a^0 = a^3 /2$, this only happens every second time step. While correct in principle, in the sense that the Gauss constraint is satisfied and the EOM are solved correctly, the time evolution will be adversely affected by numerical noise introduced by the current.

A simple solution to this problem is to use more than one particle per cell such that at each time step at least one cell boundary crossing per cell occurs. This leads to smoother currents and suppresses noise. Assuming that the time step is $a^0 = a^3 / n_p$, where $n_p$ is an integer with $n_p \geq 2$, one uses $n_p$ particles per cell and places them equidistantly within each cell. To circumvent rounding errors in the computation of the NGP, we avoid placing particles exactly at the boundaries of cells. The charge within a cell $V \rho_x$ can be distributed equally among the particles $Q_i = V \rho_x / n_p$ with $i \in \{ 1,2,\dots n_p \}$.

This procedure only partially solves the problem of sudden changes in the current. Due to the equal distribution of charge within a cell, the same particle charges cross the boundary for $n_p$ consecutive time steps. As a result, the current generated by the interpolation step stays the same until all $n_p$ particles per cell have moved into the next cell. When all particles have moved, the current jumps a full lattice spacing $a^3$. This sudden change also negatively impacts the time evolution of the fields.

The smoothness of the current can be further improved because the distribution of color charge among multiple particles per cell is underdetermined. Consider all particle (color) charges $q_i$ for a fixed transverse position $x_T$ lined up along the longitudinal axis. All charges $q_i$ with index $n_p j \leq i < n_p (j+1)$ belong to the same NGP cell with index $j$. Within each cell we can modify the particle charges via
\begin{align}
q'_i &= q_i - \Delta q_i, \\
q'_{i+1} &= q_{i+1} + \Delta q_i,
\end{align}
without changing the total charge in the cell. We iteratively apply such small changes to smoothen the sub-lattice distribution of charges among particles. For example, one can pick the new first order finite difference of modified charges $q'_i$ to be the average of adjacent derivatives via
\begin{equation}
\lb q'_{i+1} - q'_{i} \rb = \frac{1}{2} \left[ \lb q_{i+2} - q_{i+1} \rb + \lb q_{i} - q_{i-1} \rb \right].
\end{equation}
The above relation can be solved for $\Delta q_i$:
\begin{equation}
\Delta q_i = \frac{1}{4} \lb q_{i+2} - 3 q_{i+1} + 3 q_{i} - q_{i-1}  \rb.
\end{equation}
In order not to affect the overall charge in an NGP cell, one has to leave out the charges at the right boundary of each cell, i.e.\ $i \neq n_p j - 1$. The above procedure eventually leads to convergence by applying it to each particle (except at boundaries) in each cell multiple times. The result of the above iteration is that the second order finite difference becomes constant in each cell. Since $\Delta q_i$ is directly proportional to the third order finite difference, we can write
\begin{equation}
\Delta q_i = \frac{1}{4} \left[ \lb q_{i+2} - 2 q_{i+1} + q_{i} \rb - \lb q_{i+1} - 2 q_{i} + q_{i-1} \rb \right].
\end{equation}
Requiring that the procedure converges, we have $\Delta q_i = 0$ and thus
\begin{equation}
q_{i+2} - 2 q_{i+1} + q_{i} = q_{i+1} - 2 q_{i} + q_{i-1},
\end{equation}
which shows that the second order difference is constant. We therefore refer to the above iteration procedure as second-order charge refinement.

\begin{figure}
	\centering
	\includegraphics[scale=0.6]{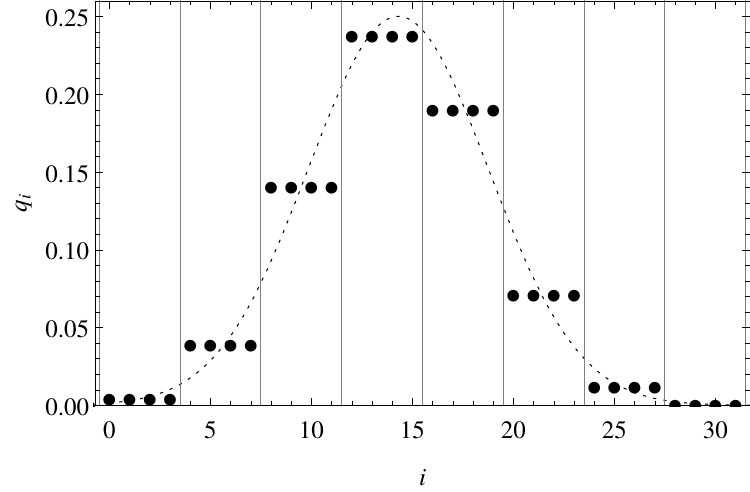}
	\includegraphics[scale=0.6]{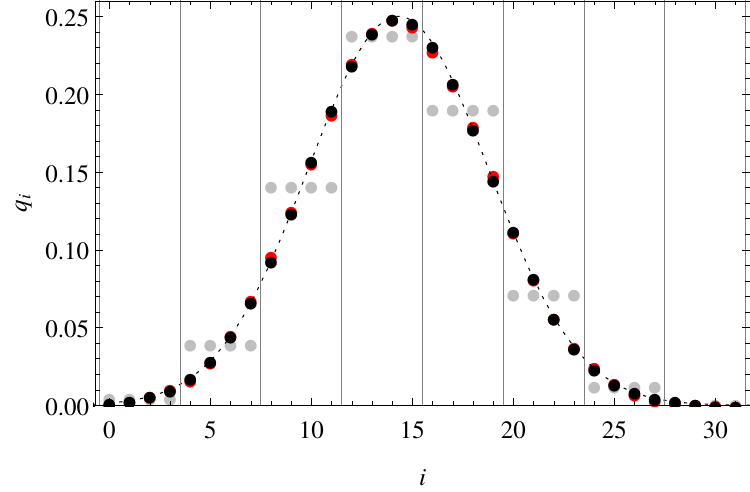}
	\caption{An example of charge refinement from \cite{Gelfand:2016yho}: the particle charges are given by $q_i$ with $i$ being the position along the longitudinal axis. As indicated by the vertical lines, there are $n_p = 4$ particles per cell. The continuous charge density is plotted as a dashed line. The left plot shows the initial distribution of particle charges before refinement. The total charge per cell is equally distributed among the four particles. The right plot shows the distribution after performing the refinement procedure. Red dots are the result of the second-order method and black dots result from the fourth-order method. By construction, the refinement procedure does not change the total charge per cell: the interpolated charge density is the same for the left and right plot. It is evident that the refined sub-lattice distribution of charges approximates the continuous charge density much better than the initial distribution without refinement. 
		\label{fig:cpic_refinement}}
\end{figure}

Similarly, we can derive a fourth-order refinement procedure by smoothing out the third order differences. This leads to
\begin{align}
q_{i+2} - 3 q'_{i+1} + 3 q'_i - q_{i-1} =& \frac{1}{2} \bigg[ \lb q_{i+3} - 3 q_{i+2} + 3 q_{i+1} - q_{i} \rb \nn
& + \lb q_{i+1} - 3 q_{i} + 3 q_{i-1} - q_{i-2} \rb \bigg],
\end{align}
which yields
\begin{equation}
\Delta q_i = \frac{1}{12} \lb -q_{i+3} + 5 q_{i+2} -10 q_{i+1} + 10 q_{i} - 5 q_{i-1} + q_{i-2} \rb.
\end{equation}
Again, since $\Delta q_i$ is proportional to the fifth-order finite difference, we can show that this leads to constant fourth-order finite differences. The above procedures work for both Abelian and non-Abelian charges: for color charges one performs the refinement steps for each color component independently. 

In practice, we find the fastest convergence if we first apply the second-order refinement and afterwards the fourth-order refinement.  The result of using these refinement procedures is shown in \cref{fig:cpic_refinement}.

\subsection{Initial fields} \label{sec:initial_fields_cpic}

The gauge field in covariant gauge is obtained through solving the two-dimensional Poisson equation. In each transverse plane we solve
\begin{equation}
- \Delta_T A^{a,+}_{x} = - \sum_{i=1,2} \frac{A^{a,+}_{x+i} + A^{a,+}_{x-i} - 2 A^{a,+}_{x}}{\lb a^i \rb^2} = \rho^a_{x}.
\end{equation}
This can be done completely analogous to the boost invariant case using the discrete Fourier transformation, see \cref{sec:bi_glasma_initial_lattice}. In momentum representation the equation reads
\begin{equation}
\tilde{k}^2_T \, \tilde{A}^{a,+}_{k} =  \tilde{\rho}^a_{k},
\end{equation}
with the squared transverse lattice momentum
\begin{equation}
\tilde{k}^2_T = \sum_i \lb \frac{2}{a^i} \rb^2 \sin^2 \lb \frac{k_i a^i}{2} \rb.
\end{equation}
Implementing infrared and ultraviolet regularization, the solution is given by
\begin{equation} \label{eq:poisson_sol_reg_latt_2}
\tilde{A}^{a,+}_{k} = \frac{1}{\tilde{k}^2_T + m^2} \, \tilde{\rho}^a_{k} \, \theta \lb \Lambda^2_\mathrm{UV} - \tilde{k}^2_T \rb,
\end{equation}
where $m$ is the infrared regulator (screening mass) and $\Lambda_{UV}$ is the ultraviolet regulator. After performing the inverse discrete Fourier transformation, one is left with the covariant gauge field $A^{a,+}_x$ at each lattice point $x$.

In \cref{sec:coll_with_finite} we showed that the solution to the Yang-Mills equations in temporal gauge involves a temporal Wilson line. Through reparametrization of the path-ordered exponential, it is possible to show that the temporal Wilson line is equivalent to the spatial Wilson line along $z$. The longitudinal component in covariant gauge is simply $A^3 = A^+ / \sqrt{2}$. The spatial Wilson line is computed by constructing the path-ordered exponential starting at $z \rightarrow \infty$ and ending at some finite $z$.
\begin{equation}
\Omega^\dg(t_0,\vec{x}) = \mathcal{P} \exp \lb - ig \intop_{+\infty}^z dz' A^3(t^0, x_T, z') \rb.
\end{equation}
On the lattice, the Wilson line can be computed layer by layer, starting at some large $z_0$ far away from the nucleus where $A^+ \approx 0$. We then have
\begin{equation}
\Omega^\dg(t^0, x_T, z - a^3) = \exp \lb - i g a^3 A^3(t^0, x_T, z) \rb \Omega^\dg_x(t^0, x_T, z),
\end{equation}
with the initial condition
\begin{equation}
\Omega^\dg(t^0, x_T, z_0) = \one.
\end{equation}
The transverse gauge links in temporal gauge are then given by
\begin{equation}
U_{x,i} = \Omega_x \Omega^\dg_{x+i},\qquad i \in \{1,2\},
\end{equation}
which follows from the lattice gauge transformations of gauge links, see \cref{eq:latt_gauge_trans}. 
Longitudinal links $U_{x,3}$ are set to unit matrices in accordance with temporal gauge, see \cref{eq:cpic_longitudinal_zero}. For left-moving nuclei, the procedure is almost identical, except that the spatial Wilson line starts at $z \rightarrow -\infty$. In practice, one simply performs a mirrored computation of the Wilson line starting at some $z_0$ left of the nucleus.

The above procedure allows us to determine the gauge links at some initial time step for arbitrary charge densities given that their longitudinal structure only depends on either $x^-$ or $x^+$. Since we are solving a second order partial differential equation, 
we also need to specify the gauge links at the next time step $U_{x+0,i}$ or alternatively the electric field $E_{x,i}$. From the analytical initial conditions we know that the single right-moving nucleus does not depend on $x^+$. A single time step $t_0 \rightarrow t_0 + a^0$ therefore corresponds to shifting the charge density by $a^0$ to the right along the longitudinal axis. In order to guarantee numerical stability, the time step must be smaller than the smallest lattice spacing, which in our case is usually the longitudinal spacing $a^3$ (see sections \ref{sec:wave_leapfrog} and \ref{sec:abelian_leapfrog} for more details on stability conditions). A shift by $a^0$ to the right therefore corresponds to shifting the lattice charge density by a fraction of a longitudinal spacing, which is slightly ambiguous. One solution to this problem is interpolating through the charge density $\rho_x$ and sampling the interpolated function at the shifted positions. An alternative method, which is more closely related to the CPIC approach, is to sample the charge density with a large number of colored particles as outlined previously. The ensemble of particles (with each particle having a continuous position) can then be easily shifted by $a^0$ to the right. Using the interpolation procedure described in the previous section, we then obtain the charge density $\rho_{x+0}$ in the next time slice. This shifted charge density is used to repeat the computation of the Wilson line $\Omega^\dg_{x+0}$. Finally, we determine the transverse gauge links in the next time slice via
\begin{equation}
U_{x+0,i} = \Omega^\dg_{x+0} \Omega_{x+0+i},\qquad i \in \{ 1, 2\}.
\end{equation}
Using $U_{x,i}$ and $U_{x+0,i}$ the electric field can be immediately computed from 
\begin{equation}
E_{x,i} = \frac{1}{g a^i a^0} \ah{U_{x,0i}}.
\end{equation}

The initialization described above has to be performed for each nucleus separately. In order to arrive at the correct field configuration, the nuclei initially have to be far away from each other such that there is no overlap in their color fields. In the implementation, the simulation box is split in half along the longitudinal axis. Each half box is occupied by a single nucleus. We initialize the color fields in each half-box separately. With compact longitudinal support, one simply has to guarantee that the support of each nucleus is contained in their respective half-boxes. In the case of non-compact support (as used in \cref{chap:single_color_sheet}, where the longitudinal profile is assumed to be Gaussian) slight errors are inevitable, but strongly suppressed when the initial separation of the nuclei is sufficiently large.

\begin{figure}[t]
	\centering
	\includegraphics[scale=0.2]{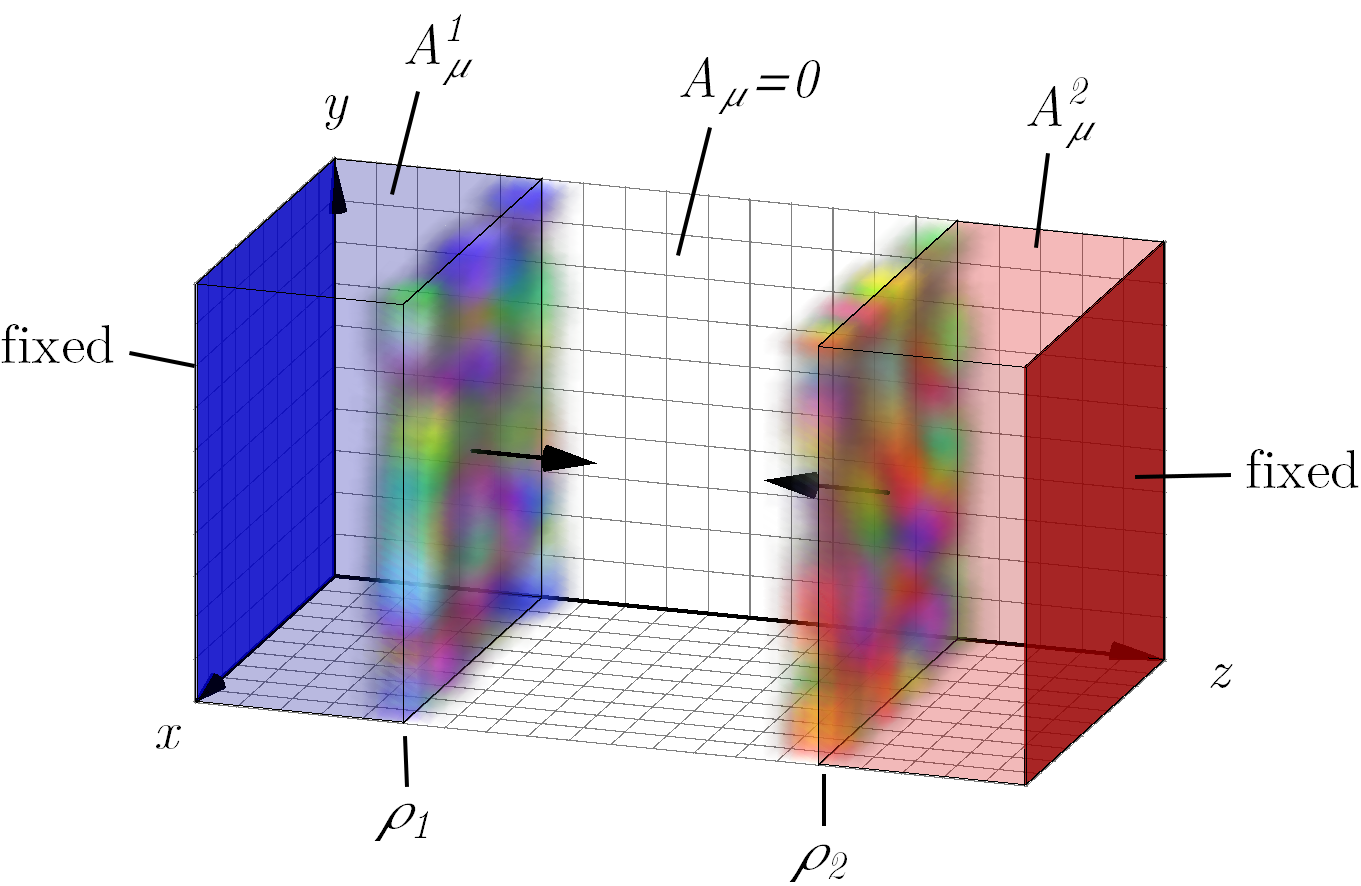}
	\caption{Schematic overview of the initial conditions in temporal gauge before the collision from \cite{Gelfand:2016yho}. The color charge densities $\rho_{(1,2)}$ of the colliding nuclei ($1,2$ referring to A,B respectively) are depicted as colorful clouds. Associated with each color charge density is the temporal gauge field $A^{(1,2)}_\mu$, which is transverse and asymptotically pure gauge behind the nuclei. At the longitudinal boundaries the color field has to be fixed to its asymptotic value. The transverse boundaries can be treated periodically. In the center region the color field is chosen to be zero. 
		\label{fig:cpic_initial}}
\end{figure}

As already mentioned in  \cref{sec:coll_with_finite}, the single nuclei color fields in temporal gauge are essentially the same as in LC gauge. In particular, they are purely transverse and asymptotically pure gauge behind each nucleus. One therefore arrives at the picture presented in \cref{fig:cpic_initial}. The use of temporal gauge forces us to use fixed boundary conditions on the outermost transverse planes, which form the longitudinal boundary.

\subsection{Temporal gauge charge density}

After performing the gauge transformation, the gauge links and color electric fields obey the temporal gauge condition. However, the original charge density used to solve the Poisson equation in each transverse plane is the covariant gauge charge density. To restore gauge covariance one can explicitly perform the lattice gauge transformation of $\rho_x$. Alternatively, an even simpler approach is to use the Gauss constraint
\begin{equation}
\sum_i D^B_i E_{x,i} = \rho_x,
\end{equation}
and re-compute the temporal gauge charge density from the electric field and the gauge links. This trivially guarantees that the Gauss constraint is satisfied for the initial conditions. Moreover, any floating point errors introduced by the initialization procedure are absorbed in the temporal gauge charge density. The particle ensemble mimicking the charge density is obtained as described in \cref{sec:replaceing_density}.

\section{Energy-momentum tensor} \label{sec:em_tensor_latt}

The most important observables that we want to study (at least for the purposes of this thesis) are the energy density and the pressure components of the Glasma, which are components of the energy-momentum tensor. Since we always consider initial conditions of the MV model type, we do not care about the transverse coordinate dependence of most observables. It is therefore sufficient to consider quantities that are averaged over the transverse plane. We also perform the average over the ensemble of charge densities defined by the probability functional $W[\rho]$. Due to homogeneity and isotropy (with respect to the transverse plane) of the MV model, the averaged energy-momentum tensor (see \cref{eq:ym_em_tensor}) in the laboratory frame reads
\begin{equation}
\ev{T^{\mu\nu}}=\left(\begin{array}{cccc}
\ev{\varepsilon} & 0 & 0 & \ev{S_{L}}\\
0 & \ev{p_{T}} & 0 & 0\\
0 & 0 & \ev{p_{T}} & 0\\
\ev{S_{L}} & 0 & 0 & \ev{p_{L}}
\end{array}\right).
\end{equation}
Here $\varepsilon$ is the energy density, $p_T$ and $p_L$ are the two pressure components, and $S_L$ is the Poynting vector (or energy flux) along the longitudinal direction $z$.
The energy-momentum tensor is traceless in the sense of $T^\mu_{\,\,\mu} = 0$, which yields
\begin{equation}
\ev{\varepsilon} = 2 \ev{p_T} + \ev{p_L}.
\end{equation}
Due to the averaging procedure, all components of $\ev{T^{\mu\nu}}$ are functions of $t$ and $z$ only. The specific form of $\ev{T^{\mu\nu}}$ is explained by the symmetries of the MV model. For example, there is no net energy flux in the transverse direction $x$ and $y$. In the MV model, nuclei have infinite extent in the transverse directions and the average fluctuation of the color charge density is homogeneous. Therefore there is no distinguished direction within the plane. Net energy flux in any of the transverse directions would violate these symmetries. This is not true when considering single events: each collision produces an inhomogeneous distribution of color flux tubes, which expand in the transverse plane. Therefore, we have locally non-vanishing $S_x$ and $S_y$, i.e.\ transverse components of the Poynting vector, on an event-by-event basis. However, once we average over all charge densities, these components must vanish.
Similar arguments can be made for the shear stress components. These involve expectation values $\ev{E_i E_j}$ and  $\ev{B_i B_j}$ with $i \neq j$. Again, due to rotational symmetry and homogeneity, these terms must vanish. 

Analogously to the boost invariant case in \cref{sec:bi_observables}, the energy density and the pressure can be written in terms of longitudinal and transverse components of the chromo-electric and -magnetic fields:
\begin{align}
\ev{\varepsilon} & =  \ev{\e_{E,T}} + \ev{\e_{B,T}} + \ev{\e_{E,L}} + \ev{\e_{B,L}},\\
\ev{p_{T}} & =  \ev{\e_{E,L}} + \ev{\e_{B,L}},\\
\ev{p_{L}} & =  \ev{\e_{E,T}} + \ev{\e_{B,T}} - \ev{\e_{E,L}} - \ev{\e_{B,L}}.
\end{align}
In the continuum limit, these four contributions are defined as
\begin{align}
\e_{E,L}(t,z) &= \frac{1}{A_T} \int d^2 x_T \tr \left[ \lb E_3(x) \rb ^2 \right], \\
\e_{B,L}(t,z) &= \frac{1}{A_T} \int d^2 x_T\tr \left[ \lb B_3(x) \rb ^2 \right], \\
\e_{E,T}(t,z) &= \frac{1}{A_T} \int d^2 x_T\sum_{i=1,2} \tr \left[ \lb E_i(x) \rb ^2 \right], \\
\e_{B,T}(t,z) &= \frac{1}{A_T} \int d^2 x_T\sum_{i=1,2} \tr \left[ \lb B_i(x) \rb ^2 \right],
\end{align}
where $A_T$ is the area of the transverse plane. We also use the shorthand $\e_L = \e_{E,L} + \e_{B,L}$ and $\e_T = \e_{E,T} + \e_{B,T}$. The Poynting vector or longitudinal energy flux is given by
\begin{equation} \label{eq:cpic_poynting_def}
S_L(t, z) = \frac{1}{A_T} \int d^2 x_T \, 2 \tr \left[ E_1(x) B_2(x) - E_2(x) B_1(x) \right].
\end{equation}

On the lattice, we can express these quantities in terms of gauge links. Recall that the plaquette (see \cref{eq:plaquette_definition})
\begin{equation}
U_{x,\mu\nu} = U_{x,\mu} U_{x+\mu, \nu} U_{x+\mu+\nu, -\mu} U_{x+\nu, -\nu}
\end{equation}
can be used to approximate the squared field strength tensor (see \cref{eq:tr_umunu})
\begin{equation}
\tr \left[ 2 - U_{x,\mu\nu} -U^\dg_{x,\mu\nu} \right]  \simeq \lb g a^\mu a^\nu\rb^2 \lb  \tr \left[ F_{\mu\nu}(x_0)^2 \right] + \mathcal{O}(a^2) \rb.
\end{equation}
Here, $x_0$ refers to the center of the plaquette, i.e.\ $x_0 = x + a^\mu \hat{e}^\mu / 2 +  a^\nu \hat{e}^\nu/ 2$. The quadratic error term is of the form $\mathcal{O} \lb a^2 \rb = \mathcal{O} \lb \lb a^\mu \rb^2 \rb + \mathcal{O} \lb \lb a^\nu \rb^2 \rb$, i.e.\ there are no mixed terms of the form $a^\mu a^\nu$ (see e.g.\ \cite{Lepage:1998dt} or \cite{BilsonThompson:2002jk}  for a proof). The four energy density components can therefore be written as
\begin{align}
\e_{E,L} &= \frac{1}{N_T^2} \sum_{x'} \frac{1}{\lb g a^0 a^3 \rb^2} \tr \left[2 - U_{x',03} - U^\dg_{x',03} \right], \\
\e_{B,L} &= \frac{1}{N_T^2} \sum_{x'} \frac{1}{\lb g a^1 a^2 \rb^2} \tr \left[2 - U_{x',12} - U^\dg_{x',12} \right], \\
\e_{E,T} &= \frac{1}{N_T^2} \sum_{i=1,2} \sum_{x'} \frac{1}{\lb g a^0 a^i \rb^2} \tr \left[2 - U_{x',0i} - U^\dg_{x',0i} \right], \\
\e_{B,T} &= \frac{1}{N_T^2} \sum_{i=1,2} \sum_{x'} \frac{1}{\lb g a^i a^3 \rb^2} \tr \left[2 - U_{x',3i} - U^\dg_{x',3i} \right].
\end{align}
The sum $\sum_{x'}$ runs over all lattice sites of a given transverse plane. The approximation of field strengths in terms of plaquettes is only accurate up to quadratic order if one considers the field strengths to be evaluated at the centers of plaquettes. For instance, $\e_{E,L}$ is naturally centered around $\lb t+ \Delta t / 2, z + a_L /2 \rb$. If we need to evaluate quantities at positions other than the ones they are naturally defined at, we use averaging. In the case of $\e_{E,L}$ evaluated at $(t,z)$ we use
\begin{align}
\e_{E,L}(t, z) =& \frac{1}{4} \bigg( \e_{E,L}(t+\frac{\Delta t}{2} , z + \frac{a_L}{2}) + \e_{E,L}(t-\frac{\Delta t}{2} , z + \frac{a_L}{2}) \nn
& + \e_{E,L}(t+\frac{\Delta t}{2}, z - \frac{a_L}{2}) + \e_{E,L}(t-\frac{\Delta t}{2}, z - \frac{a_L}{2}) \bigg),
\end{align}
which is both a spatial and temporal average. Similar averages can be defined for all other components. By using symmetric averages we are able to retain quadratic accuracy for all quantities.

In order to compute the Poynting vector, we need to find approximations of the (un-squared) field strength tensor. This is accomplished by taking the anti-hermitian, traceless part of plaquettes:
\begin{align} \label{eq:u_munu_ah}
\ah{U_{x,\mu\nu}} &= \frac{1}{2i} \lb U_{x,\mu\nu} - U^\dg_{x,\mu\nu} \rb - \frac{1}{N_c} \tr \left[ \frac{1}{2i} \lb U_{x,\mu\nu} - U^\dg_{x,\mu\nu} \rb \right] \nn
&\simeq g a^\mu a^\nu \lb F_{\mu\nu}(x_0) + \mathcal{O}(a^2) \rb,
\end{align}
where $x_0$ is the center of the plaquette. This expression follows from inserting \cref{eq:plaquette_exponential} into \cref{eq:ah_definition} and expanding for small lattice spacing $a^\mu$.
Eq.\ \eqref{eq:u_munu_ah} has already been used to define the chromo-electric field in the Lagrangian leapfrog method (see \cref{eq:cpic_lagrange_efield})
\begin{equation} \label{eq:electric_plaq}
E_i(x+\frac{a^i}{2} \hat{e}^i + \frac{a^0}{2} \hat{e}^0 ) \simeq E_{x,i} = \frac{1}{g a^0 a^i} \ah{U_{x,0i}}.
\end{equation}
Although $E_{x,i}$ transforms locally at the lattice site $x$, the above relation shows that it approximates the electric field in the continuum at $x+a^i \hat{e}^i /2 + a^0 \hat{e}^0 /2$ up to quadratic order. Similarly, the magnetic field can be approximated via
\begin{equation} \label{eq:magnetic_plaq}
B_i(x + \frac{a^j}{2} \hat{e}^j + \frac{a^k}{2} \hat{e}^k) \simeq B_{x,i} = - \frac{1}{2} \sum_{j,k} \varepsilon_{ijk} \frac{1}{g a^j a^k} \ah{U_{x,jk}},
\end{equation} 
where $j,k$ on the LHS are orthogonal directions to $i$.
Since $E_i$ and $B_i$ are not defined at the same positions, averaging is also necessary here. In the case of longitudinal energy flux $S_L$ we use
\begin{equation}
E_1(x) B_2(x) \simeq \frac{1}{4} \lb E_{x,1} + E_{x-0,1} \rb \lb B_{x,2} + B_{x-3,2} \rb, 
\end{equation}
and
\begin{equation}
E_2(x) B_1(x) \simeq \frac{1}{4} \lb E_{x,2} + E_{x-0,2} \rb \lb B_{x,1} + B_{x-3,1} \rb.
\end{equation}
Here, a time average is used for electric fields and spatial averaging for magnetic fields. The properly discretized version of the longitudinal Poynting vector component $S_L$ from \cref{eq:cpic_poynting_def} then reads
\begin{equation} \label{eq:poynting_notsogood}
S_L = \frac{1}{N_T^2} \sum_{x'} \frac{1}{2} \tr \left[ \lb E_{x',1} + E_{x'-0,1} \rb \lb B_{x',2} + B_{x'-3,2} \rb -  \lb E_{x',2} + E_{x'-0,2} \rb \lb B_{x',1} + B_{x'-3,1} \rb \right].
\end{equation}
There are a few things to note: first, the two terms $E_1(x) B_2(x)$ and $E_2(x) B_1(x)$ are not actually defined at the same transverse positions $x_T$. This poses no problem if we are only interested in quantities averaged over the transverse plane. Secondly, the spatial average of the magnetic field is slightly problematic because it is not gauge-covariant. A better idea might be to use a gauge-covariant spatial average:
\begin{align}
E_1(x) B_2(x) &\simeq \frac{1}{4} \lb E_{x,1} + E_{x-0,1} \rb \lb B_{x,2} + U^\dg_{x-3,3} B_{x-3,2} U_{x-3,3} \rb, \\
E_2(x) B_1(x) &\simeq \frac{1}{4} \lb E_{x,2} + E_{x-0,2} \rb \lb B_{x,1} + U^\dg_{x-3,3} B_{x-3,1} U_{x-3,3} \rb.
\end{align}
In practice, we found that this does not change results significantly, as long as one does not apply highly discontinuous gauge transformations and the fields can be assumed to vary only slowly from lattice site to lattice site.

A more rigorous expression for the Poynting vector on the lattice can be found the following way:
in the continuum, the Poynting vector $\vec{S}(x)$ and the energy density $\varepsilon(x)$ satisfy the Poynting theorem
\begin{equation} \label{eq:cpic_poynting_theorem}
\frac{\p \varepsilon}{\p t} + \vec{\nabla} \cdot \vec{S} = 0,
\end{equation}
if no external currents are considered. This theorem simply follows from the temporal component of the more general energy momentum conservation law
\begin{equation}
\p_\mu T^{\mu \nu} = 0.
\end{equation}
In a discretized system, one can require that a discrete analogue of \cref{eq:cpic_poynting_theorem} should hold. Given a discretized version of the energy density $\varepsilon(x)$,
one can then try to find a lattice expression of the Poynting vector, which exactly satisfies the discretized Poynting theorem (see \cite{pkappl_master} for such a derivation). It is not guaranteed that such an expression for $\vec{S}$ exists for arbitrary discretizations of the local energy conservation law. Furthermore, discretized systems do not even necessarily satisfy local energy conservation exactly. 
In our numerical studies \cite{Gelfand:2016yho,Ipp:2017lho,Ipp:2017uxo} \cref{eq:poynting_notsogood} proved to work sufficiently well for our applications.  

In some cases, we are interested in components of the energy-momentum tensor without averaging over the transverse plane. For instance, in \cref{sec:approaching_bi} we study the energy density due to the longitudinal color electric field as a function of the transverse coordinate $x_T$. We then simply leave out the summation over transverse coordinates. We have to keep in mind where the various energy density components are naturally defined, i.e.\ in the center of plaquettes. 

\section{Implementation details} \label{sec:implementation}

During the development of my thesis several implementations have been created based on the numerical procedures presented in this chapter. The first version of the code was built on top of an already existing code base for Abelian particle-in-cell simulations called OpenPixi\footnote{The code is publicly available at \href{https://github.com/openpixi/openpixi}{https://github.com/openpixi/openpixi}. The original Abelian PIC simulation can be found at \href{https://github.com/openpixi/openpixi_pic}{https://github.com/openpixi/openpixi\_pic}.} and co-developed with my colleague Daniil Gelfand and my supervisor Andreas Ipp. This code was written in Java in an object-oriented fashion. This first version was designed to be modular such that components  of the simulation (e.g.\ various solvers for the equations of motion, different implementations of interpolation procedures, initial conditions and gauge groups) can be easily swapped out and replaced.
It also features a graphical user interface, which can be used to, for example, monitor observables while the simulation runs or create two- and three-dimensional plots. These features facilitate easier debugging during development and allow users to experiment with the simulation without making changes to the source code.
Unfortunately, the original version suffers from performance issues, as the Java programming language is not the best choice for high-performance computing applications. In particular, due to the slightly unpredictable memory management behavior in Java, large simulations are often at risk of running out of available memory. Nevertheless, all results of the first two publications \cite{Gelfand:2016yho,Ipp:2017lho} were achieved using the Java version running on the Vienna Scientific Cluster 3 (VSC3). This required great computational resources: for example in \cite{Ipp:2017lho} the simulation of one single collision took roughly two days to complete. Due to the stochastic nature of charge densities in the CGC framework, one has to perform such simulations multiple times in order to compute expectation values. These long simulation times severely limit the way one can experiment with different initial conditions and parameters. As a result, I wrote a second version\footnote{The code is publicly available at \href{https://gitlab.com/monolithu/pyglasma3d}{https://gitlab.com/monolithu/pyglasma3d}.} which was implemented in Cython \cite{behnel2010cython} with a focus on a minimal set of features. Using Cython, it was possible to implement non-critical parts of the simulation (such as data in- and output and general program flow) in the Python programming language, while performance critical code (e.g.\ time evolution, initial conditions) is translated to highly optimized C. This led to a substantial decrease in simulation time by a factor of roughly 7 - 8, while also decreasing memory use by a factor of 6 - 7. The main disadvantage of this code is that it is not object-oriented and tends to be less readable compared to the Java version. Furthermore, due to certain design choices early on in the development, it is also restricted to $\mathrm{SU(2)}$ as the color gauge group. Extension to other gauge groups would require large parts of the code to be rewritten. All results of the most recent publication \cite{Ipp:2018hai} were obtained using this faster version. A third version\footnote{The code is publicly available at \href{https://gitlab.com/openpixi/openpixi_c}{https://gitlab.com/openpixi/openpixi\_c}.} written in C++ is currently (at the time of writing) being developed by my colleague Patrick Kappl with comparable performance to the second version, while at the same time maintaining the object-oriented design principle and increased readability of the code \cite{pkappl_master}. It is also readily extendable to other gauge groups. 

In this last section we discuss a few technical details regarding the actual implementation of these codes and some optimizations that were found to improve performance and, in particular memory consumption, of these simulations. 

\paragraph{Grouping particles into charge planes} The first thing to notice is that one does not actually have to individually track each particle in the simulation as would be the case in completely general PIC simulations. Since all the trajectories are fixed due to the eikonal approximation (i.e.\ no recoil) and all particles only move along the longitudinal axis, one can group particles within the same transverse plane into so-called ``charge planes". A single charge plane is defined by its longitudinal position $z$, its velocity or orientation (either left- or right-moving), and the (color) charge density distribution in the transverse plane. Since all particles within a charge plane move collectively, one simply has to check if the NGP associated with the longitudinal position of a charge plane is about to change from one time slice to the next. Computing the current and applying parallel transport is then performed for all charges within the plane. Grouping particles this way avoids iterating over each individual particle and performing unnecessary repeated calculations. Furthermore, as a particle's position can be trivially inferred from its initial position and the number of already performed simulation cycles, it is also not necessary to update the position as described in \cref{sec:simulation_cycle} or actually compute the NGP. It is much easier to keep track of the sub-cell position $s \in \{ 0, 1, 2, \dots, n_p -1 \}$ of each charge plane within each cell, where $n_p$ is the number of particles per cell and its last (integer) lattice position along the longitudinal axis $z$. After each time step the sub-cell position $s$ is increased by one for each right-moving plane and conversely decreased for each left-moving plane. At the beginning of each simulation cycle, all right-moving charge planes with $s = n_p - 1$ and all left-moving planes with $s = 0$ will perform a boundary crossing. It is therefore immediately clear that these particular planes contribute to the color current interpolation step and require parallel transport. Accordingly, the lattice positions $z$ are updated and the sub-cell positions $s$ set to their new values. 

\paragraph{Combined current calculation and parallel transport} A further improvement is to combine the current interpolation and the parallel transport into one single step. Recall that current is only generated when particles move from one cell into the next. Parallel transport is also only applied at such boundary crossings. It is therefore natural to perform both calculations simultaneously. If one attempts to do this, it turns out that one has to discriminate between right and left-moving charges: 
in the computation of the current density of right-moving particles \cref{eq:cycle_current_rm}, one has to use the particle charge $Q(t-a^0)$ of the previous time slice. On the other hand, the current for left-moving particles \cref{eq:cycle_current_lm} involves the parallel transported charge $U_{x,3} Q(t-a^0) U^\dg_{x,3}$, which corresponds to the charge in the next time slice $Q(t)$, see \cref{eq:cycle_parallel_lm}. For right-moving particles the combined step therefore involves first computing the color current \cref{eq:cycle_current_rm} and then applying parallel transport \cref{eq:cycle_parallel_rm} immediately after. For left-moving particles the order of steps is reversed: we first apply parallel transport \cref{eq:cycle_parallel_lm} and then (using the updated charge $Q(t)$) compute the current density \cref{eq:cycle_current_lm}. This optimization avoids performing parallel transport for left-moving particles twice (which would increase computational effort) or alternatively storing an ``old" and ``new" version of each particle charge (which would increase memory consumption). This idea can be applied to whole groups of particles as described in the previous paragraph.

\begin{figure}
	\centering
	\includegraphics{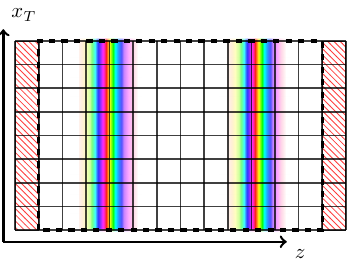}
	\caption{An illustration of fixed boundary conditions on the lattice. The fields of the nuclei are shown as colorful clouds. Only one of the transverse coordinates is shown. We use additional transverse layers at the longitudinal boundaries (hatched cells), which are considered ``inactive" and thus never updated during the time evolution step. Only the inner cells (black dashed rectangle) are considered to be dynamic. The field values in the boundary layers are still taken into account during the time evolution, for instance when computing plaquettes. 
		\label{fig:cpic_boundary}}
\end{figure}

\paragraph{Boundary conditions at longitudinal boundaries} As discussed in the previous section, the color fields are non-zero at the longitudinal boundaries of the simulation box due to the choice of temporal gauge. On the lattice, we can implement these Dirichlet-type boundary conditions by adding additional ``inactive" transverse planes at the left- and right-most boundary, as shown in \cref{fig:cpic_boundary}. During the time evolution, these fixed boundaries are simply not updated and therefore the fields within remain constant throughout the simulation. Obviously, the simulation has to be stopped once the nuclei are about to hit these fixed boundaries.

\paragraph{Parallelization with shared memory} A major speed-up is to perform certain calculations in parallel by making use of multiple cores available in modern CPUs. The cores of a typical multi-core CPU can access the same shared memory. In such a shared memory setup, the parallelization of finite difference solvers -- such as the real-time lattice gauge theory approach to solving the Yang-Mills equations used in this thesis -- are trivially parallelizable. For instance, the computation of the electric field in the next time slice (\cref{eq:cpic_lagrange_eom} for the Lagrangian leapfrog method or \cref{eq:cpic_leapfrog_eom_orig} for the Hamiltonian leapfrog method) only depends on quantities of previous time slices. Therefore, the calculation of electric fields $E_{x,i}$ and $E_{x',i}$ at different lattice sites $x$ and $x'$ can be performed independently. The same applies to the update of the gauge links, the interpolation step, parallel transport of charges, and various parts of the initialization procedure.
The first version of the code (Java) uses the built-in thread functionality to achieve parallelization. The second version (Cython) and the third (C++) both use OpenMP \cite{dagum1998openmp}. All three versions of the code exhibit high degrees of parallelism and scale well with the number of cores used.

\begin{figure}[t]
	\centering
	\includegraphics{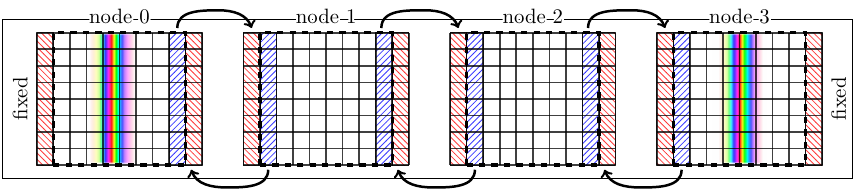}
	\caption{A sketch of how to split up the simulation across four nodes using MPI. The left layer of node 0 and the right layer of node 3 are set to fixed values in order to implement the temporal gauge boundary conditions. The interior longitudinal boundaries act as communication layers. In the synchronization step of the simulation cycle, the field values of the blue hatched layers are sent to their connected red layers as indicated by the black arrows. One also has to check if particles have to be moved from one node to a neighboring node and perform a transfer if required. 
		\label{fig:cpic_mpi}}
\end{figure}

\paragraph{Parallelization with distributed memory} Shared memory parallelization is usually restricted to simulations on single computers with multi-core processors. A further speed-up of the calculation can be achieved by employing distributed memory parallelization techniques through a Message Passing Interface (MPI) \cite{gropp1999using}. In this approach parallelizable tasks are distributed among multiple computers communicating e.g.\ via network interfaces. Since each individual computer (or node) can only access its own memory directly, such approaches are referred to as parallelization with distributed memory. Communication between nodes is achieved by passing and receiving messages. In the 3+1D Glasma setup a straightforward way of splitting up the computation is to divide the simulation box along the longitudinal coordinate. Each node is assigned one of the smaller simulation boxes as seen in \cref{fig:cpic_mpi}. Analogous to how fixed boundary conditions were implemented using additional transverse planes, we use communicating boundaries to synchronize the individual smaller boxes. This introduces an additional synchronization step at the end of each simulation cycle. Additionally, when particles cross into communication layers, they are removed from the originating node and reintroduced in the communication layer of the neighboring node. In practice, whole charge planes (i.e.\ all charges within a transverse plane grouped together) are transfered if such a node crossing happens.\footnote{This simple transfer of particles (or charge planes) only works with the NGP scheme. Using higher order schemes (e.g.\ the CIC scheme), it might be necessary to keep copies of particles in two adjacent simulation boxes simultaneously during an extended transition step.}
At the time of writing, only the Cython version of the code is able to perform parallel calculations in this manner.

\chapter{Single color sheet approximation} \label{chap:single_color_sheet}

Having developed the necessary numerical methods in the previous chapters, we are now finally able to discuss heavy-ion collision simulations in 3+1 dimensions. Much of the material covered in this chapter is based on the first two publications \cite{Gelfand:2016yho, Ipp:2017lho} and conference proceedings \cite{Ipp:2017uxo}. In addition, we include extended discussions and some new results.

Our starting point is a simple regularization of the original MV model. In the boost invariant MV model (see \cref{sec:mv_model}), the (covariant gauge) color current of a right-moving nucleus is given by \cref{eq:mv_color_current2}
\begin{equation} \label{eq:scsa_mv_current}
J^\mu(x^-, x_T) = \delta^{\mu+} \delta(x^-) \rho(x_T),
\end{equation}
with the associated gauge field \eqref{eq:mv_gf_ansatz}
\begin{equation}
A^\mu(x^-, x_T)  =  \delta^{\mu+} A^+(x^-, x_T).
\end{equation}
The gauge fields inherits the longitudinal profile of the color current \eqref{eq:scsa_mv_current}. Explicitly separating the $x^-$ dependence, we can write 
\begin{equation}
A^+(x^-, x_T) = \delta(x^-) \phi(x_T),
\end{equation}
where $\phi(x_T)$ describes the color structure in the transverse plane.
Inserting this into the Yang-Mills equations \eqref{eq:ym_equations}, we find
\begin{equation}
\Delta_T \phi(x_T) = - \rho(x_T),
\end{equation}
see \cref{eq:mv_poisson}.
The transverse charge density $\rho(x_T)$ obeys a Gaussian probability functional defined by the one- and two-point functions (see eqs.\ \eqref{eq:mv_onep} and \eqref{eq:mv_twop})
\begin{align}
\ev{\rho^a(x_T)} &= 0, \\
\ev{\rho^a(x_T) \rho^b(y_T)} &= g^2 \mu^2 \delta^{ab} \delta^{(2)}(x_T - y_T).
\end{align}
As already mentioned in \cref{sec:mv_model}, this model neglects structure along the longitudinal axis. We therefore refer to this as the single color sheet approximation, as the nucleus resembles only a single infinitesimally thin sheet of color charge defined by the transverse color charge density $\rho(x_T)$.

The first step towards three-dimensional collisions using this model is to express everything in terms of laboratory frame coordinates. In this frame we have $A^0 = A^3 = A^+ / \sqrt{2}$. Analogously, for the current in the laboratory frame we find
\begin{equation}
J^0(t,x_T,z) =\frac{1}{\sqrt{2}} J^+(x^-,x_T) = \delta(t-z) \rho(x_T),
\end{equation}
where we have used $\dd(\lambda x) = \lambda^{-1} \delta(x)$ for some constant $\lambda > 0$. Similarly, we have
\begin{equation}
A^0(t, x_t, z) =  \delta(t - z) \phi(x_T).
\end{equation}

In order to explicitly break boost invariance, we choose to regularize the delta distribution and replace it with a smooth longitudinal profile function $f(t-z)$. Our ansatz for the single color sheet model in 3+1 dimensions therefore reads
\begin{align} 
J^0(t, x_t, z) &= f(t - z) \rho(x_T), \label{eq:scsa_J} \\
A^0(t, x_t, z) &= f(t - z) \phi(x_T), \label{eq:scsa_A}
\end{align}
with $J^0 = J^3$ and $A^0 = A^3$. We choose $f(t-z)$ to be a Gaussian given by
\begin{equation}
f(t-z) = \frac{1}{\sqrt{2 \pi } \sigma} \exp \lb - \frac{(t-z)^2}{2 \sigma^2} \rb,
\end{equation}
where $\sigma$ controls the longitudinal width in the laboratory frame. It holds that
\begin{equation}
\intop_{-\infty}^{\infty} dx f(x) = 1.
\end{equation}
In contrast to the singular color fields of the boost invariant model, the regularized color fields are smooth along $z$ and can be resolved on a finite lattice. The boost invariant limit is restored by taking $\sigma \rightarrow 0$. It is obvious that such a limit cannot be taken in practice, due to the finite resolution of the lattice. In \cref{sec:approaching_bi} we will see that numerical accuracy is lost for very thin nuclei at fixed lattice spacing. 

One can relate $\sigma$ to the Lorentz contracted width $2 R_A / \gamma$ of the nucleus, where $R_A$ is the nuclear radius and $\gamma$ is the Lorentz factor. However, since the Gaussian profile does not have compact support, there is some ambiguity in choosing a proportionality factor for $\sigma$. In \cite{Gelfand:2016yho, Ipp:2017lho} we simply set
\begin{equation} \label{eq:scsa_four_sigma}
4 \sigma = \frac{2 R_A}{\gamma},
\end{equation}
such that practically all of the color charge density lies within the longitudinal width $4 \sigma$. The Lorentz factor can be directly computed from the collision energy per nucleon pair $\sqrt{s_\mathrm{NN}}$ via
\begin{equation} \label{eq:scsa_gamma}
\gamma = \frac{\sqrt{s_{NN}}}{2 m_N},
\end{equation}
where $m_N \approx 1\,\gev$ is the nucleon mass. If we consider the MV model parameter $\mu$ and the Yang-Mills coupling $g$ to be constants, then the energy dependence only enters via the thickness parameter $\sigma$.

The regularization of the delta distribution in eqs.\ \eqref{eq:scsa_J} and \eqref{eq:scsa_A} allows us to compute the total field energy of a single nucleus analytically. From \cref{sec:mv_model} we already know that the color-electric and -magnetic fields of single nuclei moving at the speed of light are purely transverse and, after averaging over color charge configurations, give the same contribution to the energy density. We can therefore write
\begin{equation}
\ev{\varepsilon(x)} = \sum_{i=1,2} \sum_a \ev{E^a_i(x) E^a_i(x)}.
\end{equation} 
In covariant gauge we find for the transverse electric field $E_i = \p_i A^0$ and thus
\begin{align}
\ev{\varepsilon(x)} &= \sum_{i=1,2} \sum_a \ev{\lb \p_i A^{a}_0(x) \rb^2} \nn
&= f(t- z)^2 \sum_{i=1,2} \sum_a \ev{\lb \p_i \phi^a(x_T) \rb^2}.
\end{align}
First, we notice that due to the presence of the $f(t- z)^2$ term, the boost invariant limit, where we take $f(t-z) \rightarrow \delta(t-z)$, is ill-defined. Fortunately, this is what one would expect. Integrating the energy density over $x_T$ and $z$ and using the Gaussian profile for $f(t-z)$, we end up with
\begin{equation}
\ev{E} = \int d^3 x \ev{\varepsilon(x)} = \frac{1}{\sqrt{4 \pi} \sigma} \int d^2 x_T \sum_{i=1,2} \sum_a \ev{\lb \p_i \phi^a(x_T) \rb^2}.
\end{equation}
The total energy of a nucleus thus diverges with $\sigma^{-1}$ when taking $\sigma \rightarrow 0$. From our previous considerations we know that $\sigma \propto \gamma^{-1}$. The ill-defined boost invariant limit is therefore related to ultrarelativistic nuclei carrying infinite energy.

The next step is to compute the expectation value $\ev{\lb \p_i \phi^a(x_T) \rb^2}$, which can be easily related to the two-point function $\ev{\phi^a(x_T) \phi^b(y_T)}$. From the infrared regulated two-point function \cref{eq:Ap_twopf} we can read off
\begin{equation}
\ev{\phi^a(x_T) \phi^b(y_T)} = g^2 \mu^2 \delta^{ab} \frac{\abs{x_T - y_T} K_1 \lb m \abs{x_T - y_T} \rb}{4 \pi m}.
\end{equation}
Using integration by parts and the fact that the MV model is invariant under translations in the transverse plane we can use
\begin{equation}
\sum_{i, a} \ev{\lb \p_i \phi^a(x_T) \rb^2} \simeq - \lim_{y_T \rightarrow x_T}\sum_{i, a} \ev{\p^2_i \phi^a(x_T) \phi^a(y_T)}
\end{equation}
When plugging in the result for the two-point function from above however, one quickly realizes that this does not yield a finite result. Even though the two-point function is well defined through infrared regularization, taking the second derivative does not work at the origin $y_T \rightarrow x_T$. For the energy density $\ev{\varepsilon(x)}$ in the laboratory frame, ultraviolet regulation is essential. We therefore compute the required expectation value directly in momentum space:
\begin{align}
\lim_{y_T \rightarrow x_T} \ev{\p^2_i \phi^a(x_T) \phi^a(y_T)} &= - \lim_{y_T \rightarrow x_T} \ev{\rho^a(x_T) \phi^a(y_T)} \nn
&= \intop^{\Lambda_{\mathrm{UV}}}_0 \frac{d^2 p_T}{\lb 2\pi \rb^2}  \intop^{\Lambda_{\mathrm{UV}}}_0 d^2 q_T \frac{p_T^2 \lb g \mu \rb^2 \lb N_c^2 - 1\rb}{\lb p_T^2 + m^2 \rb \lb q_T^2 + m^2 \rb} \delta^2(p_T + q_T)   e^{-i \lb p_T + q_T \rb \cdot x_T} \nn
&= \intop^{\Lambda_{\mathrm{UV}}}_0 \frac{d^2 p_T}{\lb 2\pi \rb^2} \frac{p_T^2 \lb g \mu \rb^2 \lb N_c^2 - 1\rb}{\lb p_T^2 + m^2 \rb^2} \nn
&= \lb g \mu \rb^2 \lb N_c^2 - 1\rb \frac{1}{4\pi} \lb \ln \lb 1 + \lb \frac{\Lambda_\mathrm{UV}}{m} \rb^2 \rb  - \frac{\Lambda_\mathrm{UV}^2}{m^2 + \Lambda_\mathrm{UV}^2}\rb \nn
&\simeq \frac{\lb g \mu \rb^2 \lb N_c^2 - 1 \rb}{2\pi}  \ln \lb \frac{\Lambda_\mathrm{UV}}{m} \rb.
\end{align}
In the last line we used $m \ll \Lambda_{\mathrm{UV}}$. The total field energy then reads
\begin{equation}
\ev{E} \simeq \lb N_c^2 - 1 \rb \frac{\lb g \mu \rb^2}{\sqrt{16 \pi^3} \sigma} A_T  \ln \lb \frac{\Lambda_\mathrm{UV}}{m} \rb,
\end{equation}
where the integral over $x_T$ is replaced with the transverse area $A_T$.

In the single color sheet approximation it is easy to compute the temporal Wilson line for transforming from covariant gauge to temporal gauge as required for the numerical method. The temporal Wilson line (see \cref{eq:cpic_temporal_wilson_line}) is given by
\begin{equation} 
\Omega^\dg(t,\vec{x}) = \mathcal{P} \exp \lb -i g \intop_{-\infty}^t dt' A^0(t', \vec{x})\rb,
\end{equation}
where $A^0$ is given in covariant gauge. Using $A^0(t,x_T,z) = f(t-z) \phi(x_T)$ we see that the color structure along the longitudinal direction is trivial, because the profile function simply smears out the color field in $z$. As a consequence, we can drop the path ordering and find
\begin{equation} \label{eq:temp_wilson_line_single_color}
\Omega^\dg(t,\vec{x}) = \exp \lb -i g \phi(x_T) F(t - z)\rb,
\end{equation}
where the integrated profile function is given by
\begin{equation}
F(t-z) = \intop_{-\infty}^t dt' f(t'-z) = \frac{1}{2} \lb 1 + \erf \lb \frac{t-z}{\sqrt{2} \sigma }\rb \rb.
\end{equation}
For $t-z \gg 1$, i.e.\ far behind the right-moving nucleus, we have $F(t,z) \simeq 1$ and recover the asymptotic pure gauge solution known from the original MV model with a single color sheet (see \cref{eq:asym_Wilson_line_reg} for $N_s = 1$)
\begin{equation}
\Omega^\dg(t,\vec{x}) = \exp \lb -i g \phi(x_T)\rb.
\end{equation}
The fact that the Wilson line can be computed analytically greatly simplifies the initialization procedure of the simulation. Instead of actually constructing the path ordered Wilson line on the lattice as detailed in \cref{sec:initial_fields_cpic}, in the single color sheet approximation we can simply evaluate \cref{eq:temp_wilson_line_single_color} at each point of the lattice to compute the temporal gauge initial conditions. 

It is noteworthy that the Gaussian profile function $f(t-z)$ does not have compact support along the longitudinal axis as often assumed in the previous chapter. In practice this does not pose a problem, because the Gaussian profile tends to zero exponentially fast. It is sufficient if the two nuclei are well separated at the beginning of the simulation. Any errors can be suppressed below the floating point accuracy used in the code and are therefore negligible. 

\section{Simulation parameters}

Before presenting numerical results, we summarize all parameters that define the initial conditions and the simulation itself. First, there are a number of physical parameters related to the initial conditions:

\begin{itemize}
\item The coupling constant $g$ fixes the strength of interactions. The canonical value in CGC simulations is $g = 2$, which can be obtained by evaluating the strong coupling constant at the relevant energy scale $Q_s \approx 2 \, \gev$. 

\item The MV model parameter $\mu$, which controls the average color charge density of the nucleus: we use the estimate provided by McLerran and Venugopalan \cite{MV1, MV2} which yields approximately $\mu \approx 0.5 \, \gev$ in the case of gold nuclei (see also sections \ref{sec:mv_model} and \ref{sec:bi_observables}).

\item The longitudinal thickness $\sigma$: it can be related to the collision energy per nucleon pair $\sqrt{s_{NN}}$ via $\sigma = m_N R_A / \sqrt{s_{NN}}$ (see eqs.\ \eqref{eq:scsa_four_sigma} and \eqref{eq:scsa_gamma}) where $m_N \approx 1 \, \gev$ is the nucleon mass and $R_A$ is the nuclear radius. For collisions at RHIC with $\sqrt{s_\mathrm{NN}} = 200 \, \gev$ we have $\sigma \approx 0.036\,\fm$. 

\item Infrared modes are regulated by $m$. Usually this is set to $m \approx 0.2 \, \gev$, such that we can mimic the effects of confinement. In order to study the dependence of this parameter on collisions, we vary this up to a few $\gev$.

\item Ultraviolet modes are cutoff by $\Lambda_{\mathrm{UV}}$, which is set to values between $10$ and $20 \, \gev$. For some applications we leave out manual UV regulation and let the modes be regulated by the transverse lattice spacing $a_T$.
\end{itemize}

On the other hand, there are parameters related to the lattice and size of the simulation.

\begin{itemize}
\item The longitudinal and transverse system size $L_L$ and $L_T$ fix the finite simulation volume. The transverse area (spanned by $x$ and $y$) is assumed to be square, i.e.\ $L_T = L_x = L_y$. Periodic boundary conditions are used in the transverse directions, which leads to a system size dependent infrared cutoff proportional to the inverse of $L_T$. One therefore is effectively dealing with two separate infrared regulators. In order to not have any interference between these two regulators, it is necessary to set the screening mass to larger values $m > L^{-1}_T$. The finite length along the longitudinal axis $L_L = L_z$ limits the maximum time the simulation can run before the colliding nuclei approach the fixed longitudinal boundaries of the box. If the two nuclei are initially placed at $z = -L_L / 4$ and $z = + L_L / 4$, they will overlap completely at $t_{c} = L_L / 4$ and $z = 0$ and have switched positions at $t_{f} = L_L / 2$, which is the typical time where we stop the simulation. If the nuclei are very thin compared to $L_L$, one can keep the simulation running longer while making sure that there is no interference from the boundary conditions.

\item The lattice size is defined by the number of cells along the longitudinal and transverse directions $N_L$ and $N_T$. The lattice spacings are given by $a_L = L_L / N_L$ and $a_T = L_T / N_T$. In practice, we work with dimensionless quantities in the simulation. This is done by choosing an energy scale defined by the transverse lattice spacing $E_0 = a_T^{-1}$. All dimensionful quantities are then expressed in terms of $E_0$. In particular, the transverse lattice spacing can be set to $1$ and the longitudinal spacing $a_L$ becomes the length ratio of $a_L$ and $a_T$. There are two transverse scales that have to be properly resolved on the lattice. First, there is the screening range induced by the infrared regulator $m^{-1}$. It roughly translates to the size of color neutral regions in the transverse plane. In order to get close to the continuum limit with our lattice calculation, the transverse spacing $a_T$ must be much smaller than $m^{-1}$.
However, there is a second (usually) even smaller length scale provided by $Q^{-1}_s \sim \lb g^2 \mu \rb^{-1}$ which corresponds to the transverse size of Glasma flux tubes produced in the collision. Obviously, the simulation must be able to resolve this scale as well and therefore $a_T$ must be chosen such that it is smaller than the average Glasma flux tube.
This leads to a hierarchy of scales on the lattice: $a_T \ll Q_s^{-1} < m^{-1} \ll L_T$. 
The longitudinal spacing $a_L$ must be small enough in order to resolve the full longitudinal width $\sigma$ of the nucleus and possibly also a detailed structure within the longitudinal extent. It also acts as an effective momentum cutoff proportional to $a^{-1}_L$. The single color sheet approximation, which has a very smooth color structure along $z$, does not require a highly resolved longitudinal direction.

\item The time step $\Delta t$ or $a^0$ must be strictly smaller than the smallest lattice spacing, which is usually $a_L$. For the methods developed in \cref{chap:glasma3d}, numerical stability is guaranteed for $\Delta t = a_L / 2 \leq a_T / 2$ (see sections \ref{sec:wave_leapfrog} and \ref{sec:abelian_leapfrog} for a derivation of the stability criterion). The NGP scheme for particles works best if we set $\Delta t$ to a fraction of $a_L$ such that the number of particles per cell $n_p = a_L / \Delta t$ yields a whole number. This way color current is generated at each time step and no additional numerical noise is introduced into the simulation.
In \cref{cha:semi_implicit} we develop a numerical integrator that allows us to set $\Delta t = a_L$, but for the results presented in this chapter we use the methods of \cref{chap:glasma3d}.
\end{itemize}

\begin{figure}
	\centering
	\includegraphics[scale=1.1]{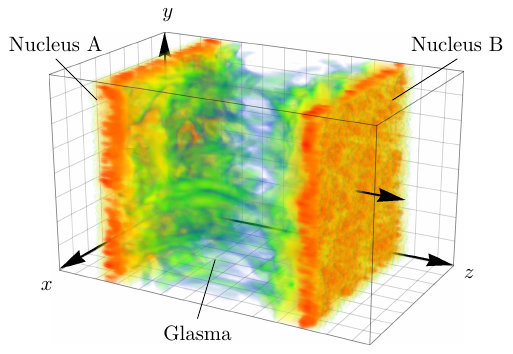}
	\caption{A three-dimensional density plot of the field energy density in a typical collision from \cite{Ipp:2017lho}. This plot is the result of an actual simulation. An animated version can be found online\protect\footnotemark. The Glasma is created after the two nuclei collide and exhibits characteristic flux tube structure.
	\label{fig:glasma3d}}
\end{figure}

\section{Approaching the boost invariant limit} \label{sec:approaching_bi}

\footnotetext{See the supplementary material of the published version of \cite{Ipp:2017lho}.}

The result of a typical collision simulated with our code is shown in \cref{fig:glasma3d}, where we plot the energy density of the color fields. As is evident from the figure, it is possible to study the fully three-dimensional structure of the Glasma. 

The first step we take is to validate that the simulation produces meaningful results when compared to the boost invariant case. As discussed earlier, the ultrarelativistic limit should be reached by setting $\sigma \rightarrow 0$. It is clear that for any fixed lattice resolution this limit cannot be taken directly. If $\sigma$ is comparable to $a_L$, then the nuclei are not properly resolved on the lattice and the numerical solution will suffer from lattice artifacts. Nevertheless, we can check if the Glasma created in a 3+1 dimensional collision resembles the boost invariant limit by choosing $\sigma$ to be very small and comparing the fields from our simulation to the boost invariant initial conditions described in  \cref{sec:glasma_initial}.

\begin{figure}[t]
	\centering
	\begin{subfigure}[b]{0.31\textwidth}
		\includegraphics{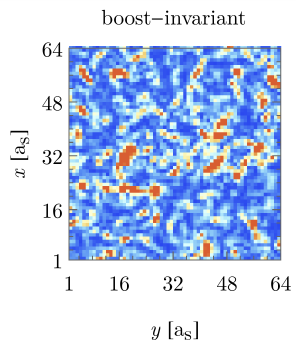}
	\end{subfigure}
	\begin{subfigure}[b]{0.31\textwidth}
		\includegraphics{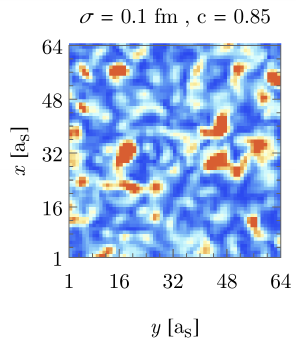}
	\end{subfigure}
	\begin{subfigure}[b]{0.31\textwidth}
		\includegraphics{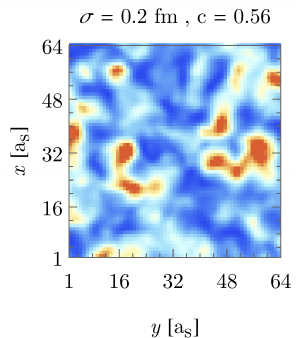}
	\end{subfigure}
	\caption{A density plot of the energy density component $\varepsilon_{E,L}$ as a function of the transverse coordinate $x_T = (x, y)$ in the center region of the collision for a single event from \cite{Gelfand:2016yho}. The left panel shows the boost invariant	(``analytic") result for $\tau = 0^+$ computed from \cref{eq:bi_initial_EL_BL_1}. The middle and right panels are from our simulation for two different values of the thickness parameter $\sigma$. The correlation coefficient $c$ (see \cref{eq:correlation_coeff}) quantifies how similar the energy density distribution from our results is to the boost invariant case. Thinner nuclei (middle) lead to a correlation coefficient of $c = 0.85$ and the two distributions are clearly similar. On the other hand, the energy density distribution for thicker nuclei (right) is smeared out and has a lower value of $c = 0.56$. The lattice parameters are $N_L = 256$, $N_T = 128$ with $a_T = a_L = 0.028 \, \fm$ and $\Delta t = 0.5 a_L$. The physical parameters are $g=2$, $\mu = 0.505 \, \gev$, $m = 2 \, \gev$, and $\Lambda_\mathrm{UV} = 10 \, \gev$. The transverse length $L_T$ is $3.58 \, \fm$, but we only show a quarter of the full transverse plane.
		\label{fig:ssa_comparison}}
\end{figure}

The question then arises what quantities one should actually compare. In the boost invariant case the initial conditions can be formulated as initial values for the longitudinal chromo-electric and magnetic fields as given by eqs.\ \eqref{eq:bi_initial_EL_BL_1} and \eqref{eq:bi_initial_EL_BL_2}. These fields are given in the co-moving temporal gauge $A^\tau = 0$, which is different from the laboratory frame temporal gauge condition $A^0 = 0$. Even though both gauge conditions agree at mid-rapidity $\eta = 0$, it is a better idea to compare gauge invariant quantities. We therefore choose the longitudinal electric part of the energy density
\begin{equation}
\varepsilon_{E,L}(x) = \tr \left[ E^2_L(x) \right]
\end{equation}
as the main observable. Given the lattice charge densities of two colliding nuclei, we can compute $\varepsilon_{E,L}$ at $\tau = 0^+$ directly from lattice initial conditions in the boost invariant case, see \cref{eq:bi_initial_Peta_lattice}. On the other hand, in the 3+1 dimensional setup we use the same charge densities but smeared out along $z$ with some finite width $\sigma$ and record the longitudinal chromo-electric field on the lattice. One has to decide when and where the two results should be compared. For simplicity, we choose the center of the simulation box at $z=0$, which corresponds to the mid-rapidity region $\eta = 0$. This is also the position at which the two nuclei will overlap completely if their initial placement was symmetric around the center $z = 0$. Since the interaction of the nuclei during the collision is not instantaneous anymore, the time $t$ at which the fields should be compared is ambiguous. A natural choice would be $t=t_c$ when the maximum overlap takes place, but for very thick nuclei the collision starts much earlier when the color fields start to overlap. We therefore record the longitudinal electric field in the transverse plane at $z = 0$ as a function of time $t$ to study this effect. The transverse structure is shown in \cref{fig:ssa_comparison}, where we see that for small $\sigma$ the 3+1 dimensional setup generally agrees with the boost invariant fields (at least visually). 

In order to make a more meaningful and quantitative comparison, we compute the correlation coefficient $c$ between the boost invariant, ``analytical" result $\tr \left[ E^2_L(x) \right]_\mathrm{ana}$ (as determined from \cref{eq:bi_initial_Peta_lattice}) and the numerical result $\tr \left[ E^2_L(x) \right]_\mathrm{num}$ from our 3+1 dimensional simulations. The correlation is  given by
\begin{equation} \label{eq:correlation_coeff}
c \lb \tr \left[ E^2_L \right]_\mathrm{ana}, \, \tr \left[ E^2_L \right]_\mathrm{num}\rb = \frac{\mathrm{cov} \lb \left[ E^2_L \right]_\mathrm{ana}, \, \tr \left[ E^2_L \right]_\mathrm{num} \rb}{s_\mathrm{ana} s_\mathrm{num}},
\end{equation}
where the covariance across the transverse plane is defined by
\begin{equation}
\mathrm{cov} \lb \tr \big[ E^2_L \big]_\mathrm{ana}, \, \tr \big[ E^2_L \big]_\mathrm{num}\rb \equiv \frac{1}{N_T^2} \sum_{x_T} \lb \lb \tr \big[ E^2_L \big]_\mathrm{ana} - \overline{\tr \big[ E^2_L \big]_\mathrm{ana}} \rb \lb \tr \big[ E^2_L \big]_\mathrm{num} - \overline{\tr \big[ E^2_L \big]_\mathrm{num}} \rb  \rb,
\end{equation}
and $\overline{\tr \big[ E^2_L \big]_\mathrm{ana}}$,\, $\overline{\tr \big[ E^2_L \big]_\mathrm{ana}}$ are the mean values of the energy density in the transverse plane
\begin{equation}
\overline{\tr \big[ E^2_L \big]_{\mathrm{ana/num}}} = \frac{1}{N_T} \sum_{x_T} \tr \big[ E^2_L \big]_{\mathrm{ana/num}}.
\end{equation}
The standard deviations $s_\mathrm{ana}$, $s_\mathrm{num}$ are given by
\begin{equation}
s_{\mathrm{ana/num}}^2 = \frac{1}{N_T} \sum_{x_T} \lb \tr \big[ E^2_L \big]_{\mathrm{ana/num}}  \rb^2 - \lb \overline{\tr \big[ E^2_L \big]_{\mathrm{ana/num}}} \rb ^2.
\end{equation}
Using $c$ has the advantage that it only measures how similar the two energy densities are without taking their absolute values into account. Lower collision energies produce a less energetic Glasma compared to the ultrarelativistic case, but one can still expect similar distributions of flux tubes in the transverse plane, especially for small $\sigma$.

\begin{figure} [t]
	\centering
	\begin{subfigure}[b]{0.45\textwidth}
		\includegraphics[scale=0.9]{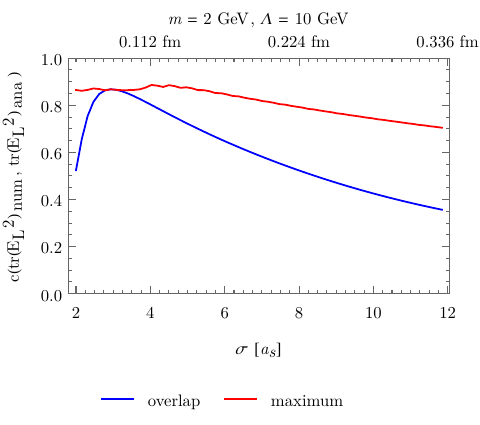}
		\caption{Correlations as a function of the thickness parameter $\sigma$ in units of spatial lattice spacings $a_s = a_L = a_T$.}
		\vspace{9.5pt}
	\end{subfigure}
	\qquad
	\quad
	\begin{subfigure}[b]{0.45\textwidth}
		\includegraphics[scale=0.9]{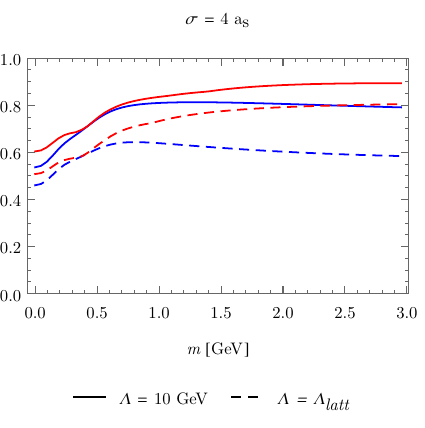}
		\caption{Correlations as a function of the IR regulator $m$. For the thick curves we use $\Lambda_{\mathrm{UV}} = 10\,\gev$ as a UV cutoff. Dashed lines use the UV cutoff $\Lambda_{\mathrm{latt}}$ provided by the lattice spacing.}
	\end{subfigure}
	\caption{Comparison of simulations to boost-invariant initial conditions from \cite{Gelfand:2016yho}. This plot shows the correlation coefficient of $\tr \big[ E^2_L \big]_\mathrm{num}$ in the central region with the boost-invariant result for $\tr \big[ E^2_L \big]_\mathrm{ana}$ (at $\tau = 0^+$) as a function of $\sigma$ and $m$. A correlation coefficient of 1 implies perfect agreement between the numerical and the analytical result (apart from the absolute value). The blue solid line shows the correlation when the nuclei completely overlap and the red line is the maximum	correlation achieved during the evolution. The correlation increases for thinner nuclei. Below a certain threshold, small values of $m$ and $\sigma$ lead to decreased correlations due to numerical instabilities, which appear at high field amplitudes. The simulation parameters are the same as in \cref{fig:ssa_comparison} except that we vary the thickness parameter $\sigma$ and IR regulator $m$.
		\label{fig:correlations_initial}}
\end{figure}

Fixing some initial two-dimensional charge densities $\rho_A$ and $\rho_B$ for the two colliding nuclei, we can run the simulation for various values of $\sigma$ and compute the correlation in each case. Since the correlation depends on when we make the comparison, we record $c$ as a function of $t$.
The results of this are shown in in \cref{fig:correlations_initial} (a):
Here, the blue curve corresponds to $t=t_{c}$ where
the two nuclei overlap completely. Starting from very large values of the thickness parameter, the correlation increases as we decrease $\sigma$. However, at some point the correlation is quickly lost and decreases again. In the plot this happens around $\sigma \approx 3 a_L$. 
As mentioned earlier, the thickness $\sigma$ has to be resolved properly on the lattice in order to obtain results similar to the continuum limit. For $\sigma = 3 a_L$ this is not the case anymore. In the case of these very thin, improperly resolved nuclei we observe a numerical instability. Even before the collision, the color fields of the single nuclei start to disperse and lose their original longitudinal profile. At the same time, unphysical longitudinal fields start to show up. We will come back to this instability and provide a solution in \cref{cha:semi_implicit}. For the time being, we will make sure not to go below a threshold of $\sigma_\mathrm{min} = 4 a_L$ for the thickness parameter. Unfortunately, we will see that $\sigma$ is not the only parameter that affects the development of this instability.

The red curve in \cref{fig:correlations_initial} (a)
is the maximum value of the correlation recorded during the collision.
Even in the case of large $\sigma$, relatively high values can be achieved, although at earlier times than the overlap $t_c$. For very thick nuclei the color fields start overlapping sooner, producing the characteristic longitudinal Glasma fields. These initial fields therefore start evolving much earlier and consequently have changed by the time both nuclei overlap.

Furthermore, we also study how infrared and ultraviolet regulation affects our results while keeping $\sigma$ fixed. This is shown in \cref{fig:correlations_initial} (b), where we plot the correlation for $\sigma = 4 a_L$ as a function of $m$ and two choices of the UV regulator. The solid lines correspond to $\LUV = 10 \, \gev$ which is below the lattice cutoff $\Lambda_{\mathrm{latt}} = \pi / a_T \approx 22 \, \gev$. The dashed lines correspond to $\LUV= \Lambda_{\mathrm{latt}}$.
The same infrared and ultraviolet regulators are used in the boost-invariant scenario.
There are a few things to notice here:  the manually regulated simulations are consistently above the lattice regulated cases. The use of the ultraviolet regulator helps us achieve more accurate results. The infrared regulator affects the correlation only slightly except when going to very small values. Using small $m$ generally leads to large field amplitudes as is evident from the solution of the Poisson equation \eqref{eq:poisson_sol_reg_latt_2}. We observe that large field amplitudes lead to similar unstable field configurations as in the case of small thickness $\sigma$. Our results are therefore generally unreliable when we study thin nuclei with large color fields without appropriate lattice resolution. The effects of the instability can be suppressed or postponed to later times when we use very fine lattices, but only at the cost of larger computational effort.

Studying the correlation between our numerical results and the analytic
expressions for the boost invariant initial conditions shows that
we are able to correctly describe boost invariant collisions in the
limit of thin nuclei. It also reveals that one has to be careful
in choosing simulation parameters, in particular $\sigma \gtrsim 4a_{L}$.

\section{Pressure and energy density} \label{sec:scsa_pressure_energy_density}

In this section we study the energy density and the pressure components of the Glasma in more detail. We are particularly interested in deviations from the 2+1 dimensional Glasma and the evolution of the system in the laboratory frame.

\subsection{Pressure anisotropy} \label{sec:scsa_pressure_ani}

As discussed in sections \ref{sec:bi_observables} and \ref{sec:beyond_mv}, a prominent phenomenon in the early stages of heavy-ion collisions
is the pressure anisotropy and a possible subsequent
isotropization of the system. The main observables in this context
are the transverse and longitudinal pressure components $p_{T}=\e_{L}$
and $p_{L}=\e_{T}-\e_{L}$, averaged over the transverse plane. On the lattice, we compute these observables as described in \cref{sec:em_tensor_latt}.

\begin{figure}[t]
	\centering
	\begin{subfigure}[b]{0.325\textwidth}
		\includegraphics[scale=0.95]{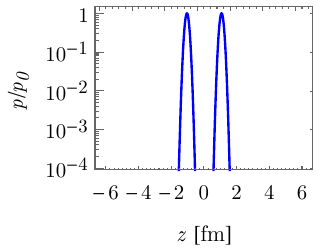}
		\caption{$t=-1\,\fm/c$}
	\end{subfigure}
	\begin{subfigure}[b]{0.325\textwidth}
		\includegraphics[scale=0.95]{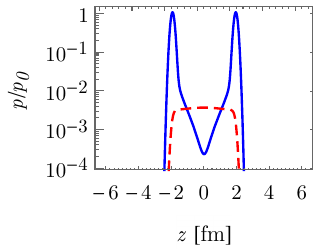}
		\caption{$t=+2\,\fm/c$}
	\end{subfigure}
	\begin{subfigure}[b]{0.325\textwidth}
		\includegraphics[scale=0.95]{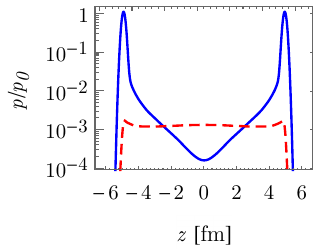}
		\caption{$t=+5\,\fm/c$}
	\end{subfigure}
	
	\caption{Longitudinal and transverse pressure components as functions of the
		longitudinal coordinate $z$ in the laboratory frame at different
		times $t$ before and after the collision. The coordinate origin is
		centered around the collision event at $t=t_c=0$ and $z=0$, where the
		nuclei overlap completely. These plots are taken from \cite{Gelfand:2016yho}.
		Blue solid lines describe $p_L(t,z)$ and red dashed curves are $p_T(t, z)$.
		Both pressures are normalized to the maximum of the initial longitudinal pressure
		$p_L$ before the collision.
		As the Glasma is produced, transverse pressure builds up between the nuclei.
		Longitudinal pressure falls off quickly towards $z = 0$.
		At later times we have $p_T \gg p_L$ in the central region.
		For these plots we use a lattice of $320\times256^{2}$ cells with $a_L=a_T=0.04\,\fm$
		and $\Delta t = a_L / 2$. The thickness parameter
		is set to $\sigma=4a_L$ (which corresponds to a gamma factor of
		$\gamma\approx23$), the IR regulator is set to $m=2\,\gev$
		and the UV cutoff is set to $\Lambda_{\mathrm{UV}}=10\,\gev$. 
		\label{fig:pressures_side}}
\end{figure}

Plots of the (averaged) pressure components in
the laboratory frame at different times are shown in \cref{fig:pressures_side}.
Before the collision, the field strengths of the colliding nuclei are purely transverse and, as such, only contribute to the longitudinal pressure component.
Once the collision starts, longitudinal color-electric and -magnetic fields are produced and transverse pressure builds up. This pressure is much smaller than the longitudinal pressure component, because the large color fields of the nuclei dominate the Glasma fields. After the collision, longitudinal pressure remains largely flat.
On the other hand, $p_L$ falls off exponentially towards $z = 0$. In this central region we observe a much larger $p_T$ compared 
to the $p_L$, which already indicates the characteristic pressure anisotropy of the Glasma.

\begin{figure}
	\centering
	\includegraphics[scale=1.1]{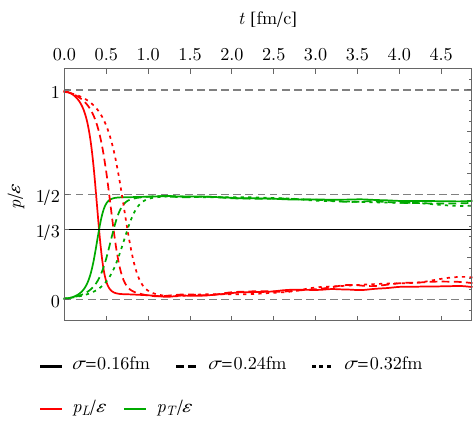}
	\caption{Longitudinal and transverse pressure components in the central region
		$\eta=0$ as a function of time for various nuclear thicknesses $\sigma$ that correspond
		to ``thick" nuclei.
		An IR regulator of $m=2\,\gev$ and a UV cutoff of $\Lambda=10\,\gev$
		has been used. The detailed simulation parameters are explained in
		the main text. This plot is taken from \cite{Gelfand:2016yho}.
		\label{fig:pressure_ratios_1}}
\end{figure}

We now turn towards studying this pressure anisotropy. For the further
analysis it will be sufficient to stay in the central region $\eta=0$ or $z=0$.
Recall from \cref{sec:glasma_initial} that in the boost invariant scenario the field strengths at $\tau=0^{+}$
are purely longitudinal, which leads to maximally
anisotropic initial pressures $\left.p_{T}\right|_{\tau=0^{+}}=\left.\e_{L}\right|_{\tau=0^{+}}$
and $\left.p_{L}\right|_{\tau=0^{+}}=-\left.\e_{L}\right|_{\tau=0^{+}}$.
During the expansion of the Glasma flux tubes, transverse color-electric and -magnetic fields are created until $\e_{L}\simeq\e_{T}$ and $p_{L}\simeq0$. Once the system settles, it will remain in the free streaming limit. As discussed in \cref{sec:glasma_instability}, the free streaming limit is incompatible with the subsequent (isotropic) hydrodynamical evolution of the QGP. Introducing boost invariance breaking fluctuations drive
instabilities in the Glasma, which can move the system towards isotropization
\cite{Fukushima:2011nq,Epelbaum:2013waa,Gelis:2013rba,Berges:2012cj}, albeit more slowly than required by the scenario of fast isotropization.
In the 3+1 dimensional setup
we explicitly violate boost invariance by introducing a finite nucleus
thickness. It is therefore interesting to investigate the effects
of the thickness parameter $\sigma$ on the pressure anisotropy of
the Glasma.

\begin{figure}
	\centering
	\includegraphics[scale=1.1]{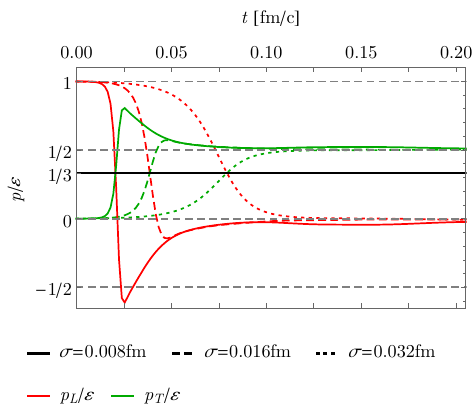}
	\caption{Pressure components in the case of ``thin" nuclei similar to \cref{fig:pressure_ratios_1}. The numerical parameters are explained in the text.
		This plot is taken from \cite{Gelfand:2016yho}.
		\label{fig:pressure_ratios_2}}
\end{figure}

For the numerical study of the anisotropy in the Glasma we use the pressure
to energy density ratios $p_{T} / \e$ and $p_{L} / \e$
with $\e=\e_{L}+\e_{T}$, as introduced in \cref{sec:bi_observables} (see \cref{fig:pressures}). The free streaming limit then corresponds
to $p_{T} / \e \simeq 1/2$ and $p_{L} / \e \simeq0$.
Isotropization would be signaled by $p_{T} / \e \simeq p_{L} / \e \simeq 1/3$.
Both the pressure and energy density components are averaged over
the transverse plane and $32$ events are used for the statistical
sampling. We choose a grid size of $N_L=320$  and $N_T=256$. For
collisions of thick nuclei in \cref{fig:pressure_ratios_1}
we choose a lattice spacing of $a_L = a_T =0.04\fm$, which corresponds to $L_T \approx 10 \fm$.
The transverse area then almost covers the full area $\pi R_{A}^{2} \approx 166 \fm^2$ of
a gold nucleus. 
\interfootnotelinepenalty=10000
For simulations of thin nuclei in \cref{fig:pressure_ratios_2}, we are forced\footnote{These simulations have been carried out in  spring of 2016 with an early build of the Java version. Due to slower performance and high memory requirements, lattice sizes of $320 \times 256^2$ were at the limit of available computational resources (see \cref{sec:implementation}). } to use smaller lattice spacings of $a_L=a_T=0.008\,\fm$
(for $\sigma=0.032\,\fm$), $a_L=a_T=0.004\,\fm$ (for $\sigma=0.016\,\fm$)
and $a_L=a_T=0.002\,\fm$ (for $\sigma=0.008\,\fm$). The transverse area
then only covers $2.5 \%$, $0.64 \%$ and $0.15 \%$ of the full area respectively.
The time step is set to $\Delta t=a_L / 2$.
\interfootnotelinepenalty=100

There are several observations we make regarding the results shown in figs.\ \ref{fig:pressure_ratios_1} and \ref{fig:pressure_ratios_2}:

\begin{itemize}
	\item From \cref{fig:pressure_ratios_1} we see
	that we recover the free streaming limit of the boost invariant case.
	Isotropization is not reached within possible simulation times due
	to limitations from the longitudinal length of the simulation
	box. We observe slight movement of both pressure components
	towards the desired value of $1/3$, but not within any realistic
	time scales.
	\item The initial pressures directly after the collision behave differently
	compared to the boost-invariant case. In our simulations of thick
	nuclei in \cref{fig:pressure_ratios_1} 
	we see that in the beginning $p_{L}$ dominates $p_{T}$ due to the
	presence of the large transverse fields of the colliding nuclei. As the
	nuclei recede from the collision center, the Glasma fields
	have already reached the free streaming limit and therefore no negative
	longitudinal pressures are observed. 
	\item In the results for thin nuclei in \cref{fig:pressure_ratios_2} we recover negative longitudinal pressure. The colliding
	nuclei move away from the collision center fast enough, leaving behind
	longitudinal color flux tubes, which have not decayed yet. The still
	largely longitudinal fields generate negative pressure, which is characteristic
	for the early Glasma phase. We remark that the infrared regulator $m$ only weakly affects the pressure anisotropy.
\end{itemize}

\subsection{Mismatch between electric and magnetic fields} \label{sec:eb_mismatch}

\begin{figure} [t]
	\centering
	\begin{subfigure}[b]{0.45\textwidth}
		\includegraphics[scale=0.9]{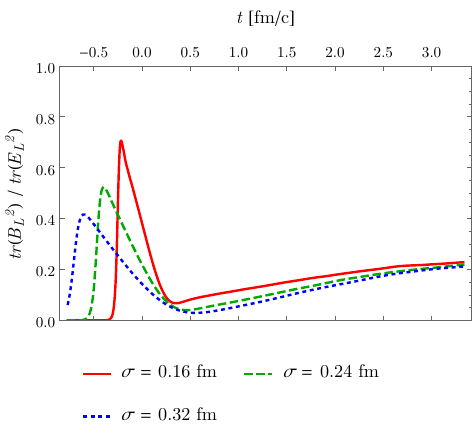}
		\caption{Ratio of longitudinal energy density components
			for thick nuclei.}
	\end{subfigure}
	\qquad
	\quad
	\begin{subfigure}[b]{0.45\textwidth}
		\includegraphics[scale=0.9]{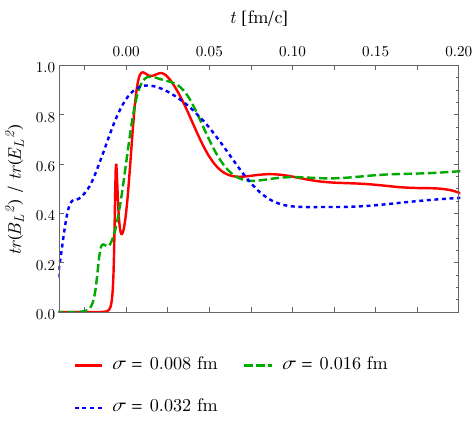}
		\caption{Ratio
			of longitudinal energy density components for thin nuclei.}
	\end{subfigure}
	\caption{Ratio of magnetic to electric longitudinal energy density contributions
		as a function of time for various nuclear thicknesses $\sigma$. The
		ratio increases for thin nuclei, but magnetic flux tubes are still
		heavily suppressed compared to the boost-invariant scenario. The simulation
		parameters are the same as in \cref{sec:scsa_pressure_ani}.
		These plots are taken from \cite{Gelfand:2016yho}.
		\label{fig:BLEL_mismatch_1}}
\end{figure}

\begin{figure} [t]
	\centering
	\includegraphics[scale=1.1]{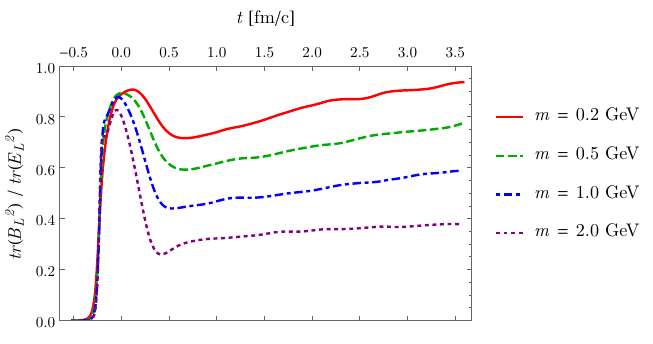}
	\caption{Ratio of magnetic to electric longitudinal energy density contributions
		as a function of time for various values of the IR regulator $m$.
		For this plot we used a grid size of $256^{3}$ cells with a lattice
		spacing $a_L=a_T=0.02\,\fm$ and a statistical average over $32$ events. The
		transverse area covers $25\%$ of the full area of a gold nucleus.
		The thickness parameter is set to $\sigma=0.08\,\fm$, which
		corresponds to $\gamma\approx45$. We find roughly equal contributions for small $m$.
		This plot is taken from \cite{Gelfand:2016yho}.
		\label{fig:BLEL_mismatch_2}}
\end{figure}

In the following, we investigate the production of longitudinal chromo-magnetic
fields $B_{L}$ and -electric fields $E_{L}$ characteristic
for the Glasma at early times. In the boost invariant case the contributions
to the energy density from magnetic and electric color flux tubes
($\tr \left[ B_{L}^{2} \right]$ and $\tr \left[ E_{L}^{2} \right]$ respectively) should
be roughly equal after averaging over initial conditions, see \cref{fig:components_and_density}. In the 3+1D setup
with finite $\sigma$ one observes that this is not the case and there
is a dependency on the thickness parameter $\sigma$ as well as the
IR regulator $m$. The results are presented in figs.\ \ref{fig:BLEL_mismatch_1}
and \ref{fig:BLEL_mismatch_2}.

\Cref{fig:BLEL_mismatch_1} shows the ratio of magnetic and electric
longitudinal fields $\ev{\tr B_{L}^{2}}/\ev{\tr E_{L}^{2}}$
in the central region ($\eta=0$ or $z=0$) for a range of values of the nucleus
thickness $\sigma$ and an IR regulator $m=2\,\gev$. Collisions
of thick nuclei show a very small ratio of about $0.1-0.2$ after
the collision. In the case of thin nuclei in \cref{fig:BLEL_mismatch_1}
(b) the ratio increases to roughly $\sim0.5$, which is
still far away from the ``canonical'' value of $\approx 1$ in the boost invariant
scenario.
One observes that for early times the equal ratio is indeed realized. This is likely due to the fact that the first interactions of color fields (when the nuclei are about to overlap) are very similar to the boost-invariant case. 
Compared to the ultrarelativistic limit, the interaction of nuclei in our simulation takes place over a finite interaction time.
The overlapping of the two colliding nuclei starts with the first few ``layers" of the color fields. 
In this case, the elapsed interaction time is only very short, which could explain why we see similar results to the ultrarelativistic limit. The same effect was observed in \cref{sec:approaching_bi} where the correlation coefficient reached its maximum much earlier than the complete overlap of the two nuclei.
Note that due to the small physical volumes used in the
simulations of thin nuclei it is harder to achieve adequate statistics.
As a result, the curves in \cref{fig:BLEL_mismatch_1} (b) are not as smooth as in  \cref{fig:BLEL_mismatch_1} (a).

The results do not only depend on the thickness $\sigma$. In \cref{fig:BLEL_mismatch_2}
the results are shown for a fixed nuclear thickness $\sigma=0.08\,\fm$
($\gamma\approx45$) and a varying IR regulator. We observe that reducing $m$ to $200\,\mathrm{MeV}$ (which roughly
corresponds to a correlation length of the color fields of the order
of the confinement radius $1\,\fm$) leads to better agreement
with the boost-invariant case with a ratio of $\sim0.8$. This strong dependency of the ratio of magnetic and electric longitudinal
fields on the IR regulator $m$ is not present in the boost invariant
case.

The presented results seemingly suggest a suppression of chromo-magnetic
flux tubes (or an overproduction of chromo-electric flux tubes) in
the Glasma phase when introducing a finite nucleus thickness. However,
the strong dependency of the magnetic to electric longitudinal field
ratio on the IR regulator $m$ leads us to suspect that this discrepancy
between our simulations and the boost invariant case is an artifact
which can be attributed to the single color sheet approximation.
In covariant gauge the three-dimensional color charge density is given by (see \cref{eq:scsa_J})
\begin{equation}
\rho(t,x_T,z) = f(t-z) \rho(x_T).
\end{equation}
The charge density is ``coherent" along $z$ over a length of $\sim \sigma$. Switching to temporal gauge changes the charge density and modifies this longitudinal coherence slightly, but the nucleus largely remains smooth in the direction of $z$. This stands in contrast to the random longitudinal structure of realistic nuclei  and the continuum limit of the generalized MV model as described in \cref{sec:mv_model}. On the other hand, the IR regulator enforces color neutrality on transverse length scales of $m^{-1}$. Consequently, the typical color structures in the initial
conditions have a thickness proportional to $\sigma$ and a transverse
width of the order of $m^{-1}$. To be consistent with the picture
of a highly Lorentz contracted nucleus modeled by classical Yang-Mills
fields, one would demand that $\sigma m\ll1$, such that nucleons within
the nucleus are also contracted to flat ``pancakes''. Therefore,
if we move away from the limit $\sigma m\ll1$, we can expect to see
deviations from the boost invariant case, but these deviations may
very well solely be due to the longitudinal coherence. This reasoning
is consistent with our simulation results: In the case of a very thick nucleus with $\sigma=0.16\,\fm$ and $m=2\,\gev \approx 0.1 \, \fm^{-1}$
the longitudinal magnetic fields are weakened as seen in \cref{fig:BLEL_mismatch_1} (a).
Here we have a value of $\sigma m=1.6$, which corresponds to color
structures which are prolonged in the longitudinal direction. We can
compare this to the case of $\sigma=0.08\,\fm$ and $m=200\,\mathrm{MeV} \approx 1 \, \fm^{-1}$
as presented in \cref{fig:BLEL_mismatch_2}. The ratio
of magnetic to electric fields is closer to $1$ and at the same time
we have $\sigma m=0.08$, which can be considered small. However, this is merely an observation and not a physical explanation for the suppression of the longitudinal chromo-magnetic field.

\subsection{Non-conservation of total energy in the simulation} \label{sec:energy_conservation}

One of the fundamental assumptions made in the CGC framework is the
separation of hard and soft degrees of freedom, which are modeled
as external color charges and classical gauge fields respectively.
As a result of the collision there is an energy exchange between the
charges and the fields. However, since the nuclei are assumed to be
recoilless, the hard sector acts as an inexhaustible energy reservoir
for the gauge fields. The resulting field energy increase can be interpreted
as the work done by the charges against the field. 
In the boost invariant
case this effect is implicitly included in the initial conditions
for the fields at $\tau=0^{+}$. The total increase in energy is less apparent, because the 2+1 dimensional setup is restricted to the future light cone and the charges are not taken into account.

\begin{figure} [t]
	\centering
	\begin{subfigure}[b]{1\textwidth}
		\includegraphics[scale=0.48]{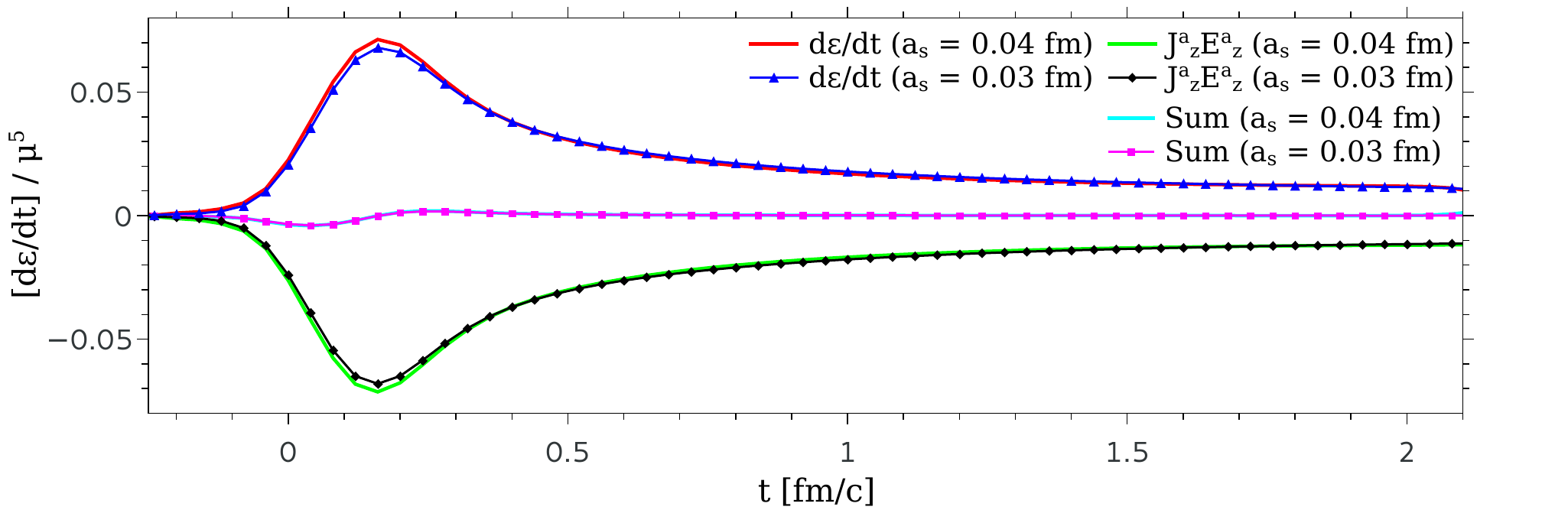}
		\caption{Varying $a_s= a_L=a_T$ while keeping
			$a_t=\Delta t=0.01\:\fm/c$ fixed.}
	\end{subfigure}
	
	\begin{subfigure}[b]{1\textwidth}
		\includegraphics[scale=0.48]{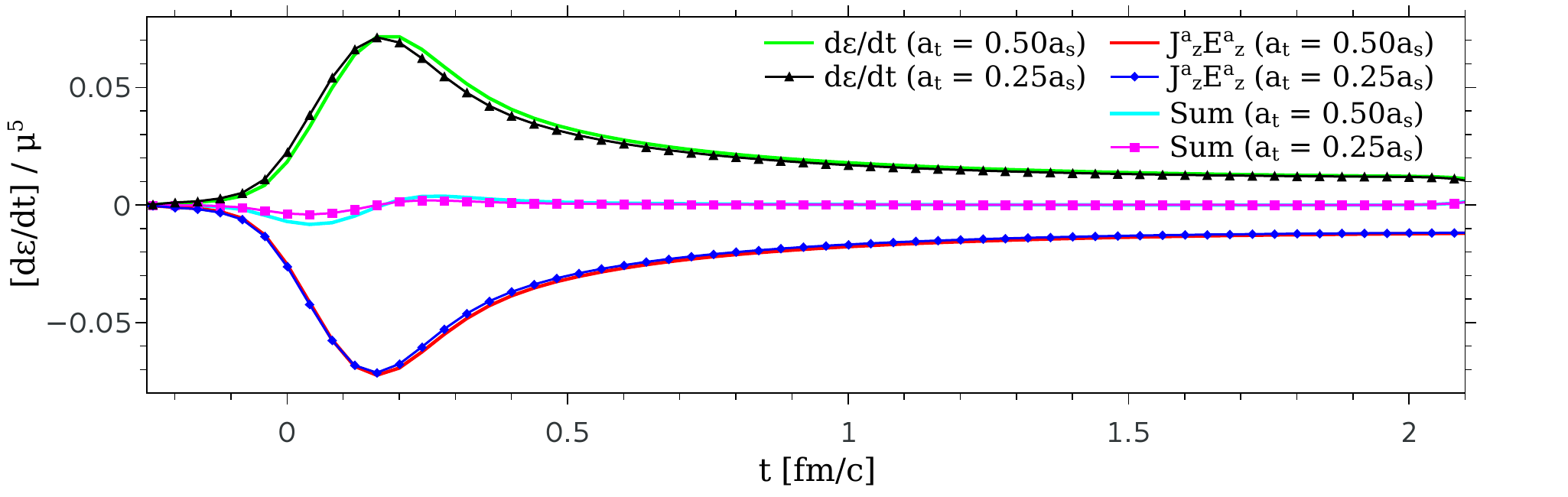}
		\caption{Varying $\Delta t=a_t$ while keeping $a_{s}=a_L=a_T=0.04\:\fm$ fixed.}
	\end{subfigure}
	\caption{Energy production as a function of time with different spatial and
		temporal discretizations from \cite{Gelfand:2016yho}. The ``Sum'' curves correspond to the left-hand
		side of \cref{eq:energy_continuity}. Their deviation from zero is
		a consequence of lattice artifacts and can be reduced by using finer
		time discretizations. The results have been obtained on a cubic lattice
		with a fixed volume of $(5.12\:\fm)^{3}$ with the IR regulator
		set to $m=1\,\gev$ and an UV cutoff $\Lambda=10\,\gev$.
		The nuclear thickness $\sigma$ was set to $0.16\,\fm$. We
		averaged over ten configurations in order to have a sufficient statistical
		sample. 
		\label{fig:poynting_theorem}}
\end{figure}

In our approach we are able to explicitly
compute the energy increase during and after the collision. The change
of the total field energy density $\e$ as a function of time can
be formulated in terms of an energy continuity equation
\begin{equation}
\frac{d\varepsilon}{dt}+\frac{1}{V}\int\partial_{i}S_{i}d^{3}x+\frac{1}{V}\int E_{i}^{a}J_{i}^{a}d^{3}x=0,\label{eq:energy_continuity}
\end{equation}
which is the non-Abelian version of the Poynting theorem. The time
dependence of $\e$ is governed by two terms: the  Poynting vector $S_{i}\equiv2\mathrm{tr}\left(\vec{E}\times\vec{B}\right)_{i}$
and $E_{i}^{a}J_{i}^{a}$. The integral over the divergence
of the Poynting vector can be omitted in the continuum assuming that there is no net energy flux at asymptotic infinity. On the lattice
this term vanishes due to the use of periodic boundary conditions. In the scalar product $E_{i}^{a}J_{i}^{a}$ the only non-vanishing
part of the current $J_{i}^{a}$ is the longitudinal component $J_{z}^{a}$
and therefore the expression reduces to $E_{z}^{a}J_{z}^{a}$. Consequently,
the energy production is caused by longitudinal chromo-electric fields
in the Glasma and must be centered around the collision event and
the boundary of the forward light cone where the color currents are
nonzero.

The energy increase as a function of time is shown in \cref{fig:poynting_theorem}.
We observe that the total energy density is conserved before the onset
of the collision when the external charges and the classical fields
describing both nuclei are propagating through vacuum. There is a strong energy increase during, as well as after the collision.
At later times there is an ongoing, but slowly decreasing energy production,
which finally becomes almost constant. To check the stability of our
results with respect to a change in the spatial and temporal resolution
of the grid, we vary the spatial lattice spacing $a_{s}=a_L=a_T$ and the time step
$\Delta t$. Overall, there is a good agreement between results at different
discretizations. The violation of \cref{eq:energy_continuity} is small
and can be further reduced by using smaller time steps as can be seen in the
lower plot of {\cref{fig:poynting_theorem} (b).}

\section{Rapidity profiles and broken boost invariance} \label{sec:rapidity_profiles}

In \cref{sec:scsa_pressure_energy_density} the focus was on observables evaluated directly in the laboratory frame and particularly at mid-rapidity. In the 3+1 dimensional setup the invariance under boosts is explicitly broken and therefore one expects that observables should depend on rapidity. Transforming quantities, such as the energy momentum tensor, from $\lb t, z\rb$ coordinates to the co-moving frame is conceptually simple, but there are some technical difficulties associated with this procedure that should be addressed properly.

First, in order to define the transformation from $(t, z)$ to $\lb \tau, \eta \rb$, one has to choose a coordinate origin. While this is trivial in the case of infinitely thin nuclei, the introduction of finite thickness makes this choice more ambiguous. The simplest idea is to choose the time and position of complete overlap $(t_c, z_c)$ as we did in the previous sections. However, considering the results of \cref{fig:poynting_theorem}, it is clear that the production of the Glasma fields has not even reached its maximum by that time. One can therefore try a different approach and choose the origin dynamically as the space-time coordinates $\lb t_0, z_0 \rb$ of maximum transverse pressure $p_T$. The reasoning for this is as follows: the transverse pressure is related to the energy density contributions from longitudinal fields $\varepsilon_L$. Therefore, the produced Glasma is the only source of transverse pressure. As the Glasma evolves the initially longitudinal fields decay and $p_T$ decreases. $p_T$ therefore exhibits a maximum at $\lb t_0, z_0 \rb$ close to the overlap event $\lb t_c , z_c \rb$, but $t_0$ is slightly later as the generation of the Glasma is not instantaneous. Furthermore, in symmetric collisions (e.g.\ two gold nuclei in the center-of-mass frame) the spatial coordinate $z_c$ at overlap must be the same as $z_0$ on average. The origin defined by the maximum of $p_T$ is therefore naturally centered, but slightly later than the overlap which accounts for the slowed creation of the Glasma. Having defined the origin, the laboratory frame coordinates can be shifted such that $(t_0, z_0) = (0, 0)$. The future light cone is then parametrized using
\begin{align} 
t = \tau \cosh \eta, \label{eq:cpic_tau_eta_coords_1}\\
z = \tau \sinh \eta. \label{eq:cpic_tau_eta_coords_2}
\end{align}

There is also another problem in the 3+1 dimensional setup: due to the finite thickness of the nuclei, there is no clear distinction between the fields of the Glasma and nuclei. This is particularly problematic when evaluating observables at high values of rapidity at early times $\tau$. Given some $\tau$ directly after the collision, high absolute values of $\eta$ lead to coordinates that can end up within the receding nuclei. Since the color fields of the nuclei are generally much larger than the Glasma fields, getting too close to the nuclei strongly affects the observable one intends to compute. In practice, one is therefore restricted to much smaller, ``safe" rapidity intervals, which depend on $\tau$. We experimented with subtracting the fields of the nuclei in order to circumvent this problem and extend available rapidity intervals. Due to the large difference in size of the nuclear and Glasma fields, we found that this is a highly delicate matter and generally, subtracting did not lead to an overall improvement of results.

A further problem is related to lattice resolution in the laboratory frame: in the 3+1D setup we use a regular hypercuboid grid with lattice spacings $a^\mu$. Observables, such as the energy momentum tensor, are evaluated at discrete positions and times $z_i = i a^3$ and $t_j = j a^0$ (the transverse coordinate $x_T$ is ignored for the moment). Transforming to the co-moving frame involves the smooth mapping of eqs.\ \eqref{eq:cpic_tau_eta_coords_1} and \eqref{eq:cpic_tau_eta_coords_2}, which distorts the regular lattice. Therefore, before performing the coordinate transformation, we compute interpolation functions that can be readily evaluated at arbitrary $z$ and $t$. However, one should be careful when evaluating observables at larger $\abs \eta$ using this method. Finite resolution in the laboratory frame obviously translates to finite resolution in $\tau$ and $\eta$. This implies that for some $\tau > 0$ resolution is highest at mid-rapidity $\eta = 0$ and decreases for higher values of $\abs \eta$. In order to still obtain reasonable numerical results at large $\abs \eta$, the longitudinal coordinate $z$ must be resolved by a very small lattice spacing $a_L$ with $a_L \ll a_T$ and $N_L \gg N_T$.

\subsection{An improved $\sqrt{s_\mathrm{NN}}$ dependent MV model}

Here, we revise the MV model used in sections \ref{sec:approaching_bi} and \ref{sec:scsa_pressure_energy_density}. Until now the only parameter influenced by the collision energy $\sqrt{s_\mathrm{NN}}$ is the longitudinal thickness $\sigma$ via
\begin{equation}
\sigma = \frac{m_N R_A}{\sqrt{s_\mathrm{NN}}},
\end{equation}
where $m_N \approx 1\, \gev$ is the nucleon mass and $R_A = 1.25 A^{1/3} \, \fm$ is the nuclear radius with nucleon number $A$. In particular, we will be interested in Au-Au collisions as performed at RHIC with $\sqrt{s_\mathrm{NN}} = 200 \, \gev$ and $\sqrt{s_\mathrm{NN}} = 130 \, \gev$.
The radius of a gold nucleus is given by $R_A \approx 7.27 \, \fm$. The thickness at the relevant collision energies is $\sigma \approx 0.0364 \, \fm$ for $\sqrt{s_\mathrm{NN}} = 200 \, \gev$ and $\sigma \approx 0.0560 \, \fm$ for $\sqrt{s_\mathrm{NN}} = 130 \, \gev$.

\begin{table} [t]
	\begin{center}
		\begin{tabular}{| c | c | c | c | c |}
			\hline
			$\sqrt{s_\mathrm{NN}}$ [GeV] & $\sigma$ [fm] & $Q_s$ [GeV] & $g$ & $\mu$ [GeV] \\ \hline
			200 & 0.0364 & 1.94 & 2.08 & 0.595 \\ \hline
			130 & 0.0560 & 1.84 & 2.11 & 0.551 \\
			\hline
		\end{tabular}
	\end{center}
	\caption{Parameters for MV model initial conditions for gold nuclei as used in \cite{Ipp:2017lho} and \cite{Ipp:2017uxo}. \label{tab:ic_parameters}}
\end{table}

The saturation momentum $Q_s$ should depend on $\sqrt{s_\mathrm{NN}}$ as well. In order to quickly derive the relation between $Q_s$ and $\sqrt{s_\mathrm{NN}}$ we follow the reasoning presented in \cite{Lappi:2007ku,Lappi:2006hq,Kharzeev:2001gp}: deep inelastic scattering (DIS) experiments at HERA (i.e.\ probing hadrons with leptons, for example using proton-electron collisions) have shown the phenomenon of ``geometric scaling" \cite{Stasto:2000er}. Without going into too much detail, geometric scaling establishes a relationship between the saturation momentum of the proton $Q_{s,p}$ and the longitudinal momentum fraction $x$ of a parton of the proton scattering off an electron:
\begin{equation}
Q^2_{s,p} = Q^2_0 \left( \frac{x_0}{x} \right)^\lambda.
\end{equation}
The values of the parameters $x_0$, $\lambda$ and $Q_0$ can be obtained from fits to experimental data. In \cite{GolecBiernat:1998js} it was found that $x_0 = 3.04 \cdot 10^{-4}$, $\lambda = 0.288$ and $Q_0 = 1\,\gev$. Assuming an average effective longitudinal momentum fraction $x = x_\mathrm{eff}$ given by (see e.g.\ \cite{Lappi:2007ku})
\begin{equation}
x_\mathrm{eff} = \frac{Q_{s,p}}{\sqrt{s_\mathrm{NN}}},
\end{equation}
one can solve for $Q^2_{s,p}$ to obtain
\begin{equation}
Q^2_{s,p}(\sqrt{s_\mathrm{NN}}) \approx 0.13 \cdot \sqrt{s_\mathrm{NN}}^{\overline{\lambda}} \, \gev^2,
\end{equation}
where $\overline{\lambda} = 2 \lambda / \left( 2 + \lambda \right) \approx 0.25$ and $\sqrt{s_\mathrm{NN}}$ is given in GeV.
The saturation momentum $Q_s$ in large, heavy nuclei can be related to the saturation momentum of the proton via
\begin{equation}
Q_s^2 = g(A) Q^2_{s,p},
\end{equation}
where $A$ is the nucleon number and $g(A)$ is a nuclear modification factor for which there exist a number of various models and fits (see the discussion section of \cite{Lappi:2007ku} for a short summary and references). Assuming $g(A) = A^{1/3}$ and $A=197$ for gold nuclei, one finds
\begin{equation} \label{eq:sat_mom_fit}
Q_s^2 \approx 0.76 \cdot \sqrt{s_\mathrm{NN}}^{\overline{\lambda}}  \, \gev^2.
\end{equation}
Depending on which fits are used and what one chooses for $g(A)$, the numerical factors can change slightly.
Similar scaling behavior of $Q_s$ can also be shown theoretically: within the CGC framework both the JIMWLK and the BFKL evolution equations (which are the weak field limit of the JIMWLK equations \cite{JalilianMarian:1997jx}) exhibit geometric scaling and predict energy dependence of $Q_s$ \cite{Weigert:2005us, Iancu:2002tr}.
For simplicity, we use the fitted estimate of \cref{eq:sat_mom_fit} and apply it to the MV model.
 In our publication \cite{Ipp:2017lho} a slightly different estimate was used:
\begin{equation}
Q^2_s \approx \lb \sqrt{s_\mathrm{NN}} \rb^{0.25} \, \gev^2,
\end{equation}
where $\sqrt{s_\mathrm{NN}}$ is given in GeV and the numerical pre-factor was dropped. At RHIC we would therefore have $Q_s \lb 200 \, \gev \rb \approx 1.94 \, \gev$ and $Q_s \lb 130 \, \gev \rb \approx 1.84 \, \gev$. Since the saturation momentum $Q_s$ defines the relevant energy scale for the collision, it also determines the appropriate Yang-Mills coupling constant $g$. Using an approximation of the one-loop running coupling $\alpha_s(Q^2) = g^2 / \lb 4 \pi \rb$ (see \cite{pdg2018} p.~141)
\begin{equation}
\alpha_s(Q^2) = \lb b_0 \ln \lb \frac{Q^2}{\Lambda_\mathrm{QCD}^2} \rb \rb ^{-1},
\end{equation}
where $b_0 = (33 - 3 n_f ) / \lb 12 \pi \rb$ and $n_f = 3$ active fermion flavors, we can estimate the coupling constant $g$ at $Q_s\lb \sqrt{s_\mathrm{NN}}\rb$. With $\Lambda_\mathrm{QCD} \approx 0.2 \, \gev$ this yields $g \approx 2.08$ for $\sqrt{s_\mathrm{NN}} = 200 \, \gev$ and $g \approx 2.11$ for $\sqrt{s_\mathrm{NN}} = 130 \, \gev$. Given $Q_s$ and $g$ we can determine the MV model parameter $\mu$. To do this, we adopt the relationship $Q_s = 0.75 g^2 \mu$ from \cite{Schenke:2012fw}. This yields $\mu \approx 0.595 \, \gev$  for $\sqrt{s_\mathrm{NN}} = 200 \, \gev$ and $\mu = 0.551 \, \gev$ for $\sqrt{s_\mathrm{NN}} = 130 \, \gev$. All parameters are summarized in \cref{tab:ic_parameters}. Regarding infrared and ultraviolet regulation we use $\Lambda_{\mathrm{UV}} = 10 \, \gev$ for all simulations and $m \in \{ 0.2 \, \gev, 0.4 \, \gev, 0.8 \, \gev \}$. For the following simulations we use a numerical grid of size $N_L \times N_T^2 = 2048 \times 192^2$ and a simulation volume $\lb 6\,\fm\rb^3$, which implies $a_L < a_T$. The time step is set to $\Delta t = a_L / 4$.

\begin{figure} [t]
	\centering
	\begin{subfigure}[b]{0.45\textwidth}
		\centering
		\includegraphics[scale=1.0]{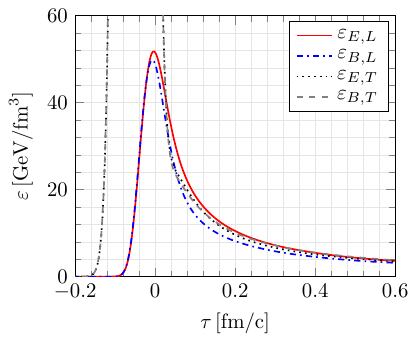}
		\caption{$\sqrt{s_\mathrm{NN}}  = 200 \, \gev$ and $m = 0.2 \, \gev$}
	\end{subfigure}
	\begin{subfigure}[b]{0.45\textwidth}
		\centering
		\includegraphics[scale=1.0]{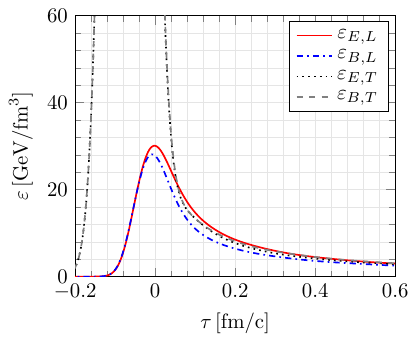}
		\caption{$\sqrt{s_\mathrm{NN}}  = 130 \, \gev$ and $m = 0.2 \, \gev$}
	\end{subfigure}
	\caption{The four contributions to the energy density from longitudinal and transverse chromo-electric and -magnetic fields in a collision of two gold nuclei. For both simulations a statistical average over 15 events was performed. The transverse fields of the nuclei dominate the Glasma fields.
		\label{fig:components_rhic}}
\end{figure}

\begin{figure} [t]
	\centering
	\includegraphics[scale=1.3]{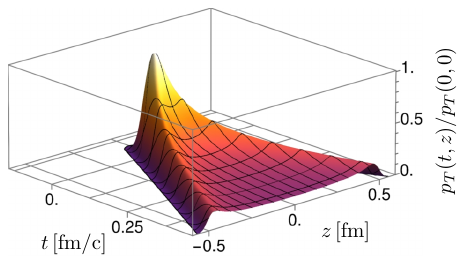}
	\caption{Space-time distribution of the transverse pressure $p_{T}(t,z)$
		normalized to the maximum at the coordinate origin for $\sqrt{s_\mathrm{NN}} = 200 \, \gev$  from \cite{Ipp:2017lho}.
		In the Glasma the transverse pressure is generated by longitudinal magnetic and electric fields
		and it is equivalent to the longitudinal component of the energy density
		$\e_{L}(t,z)$. The drop of $\protect\protect\e_{L}(t,z)$
		along the boundary of the light cone is quite steep and becomes even
		steeper when decreasing the collision energy.
		\label{fig:transverse_pressure}}
\end{figure}

As before, we record the energy density contributions from chromo-electric and -magnetic fields as defined in \cref{sec:em_tensor_latt}. In \cref{fig:components_rhic} we show these quantities evaluated at mid-rapidity or $z = 0$ as a function of time $t$. The coordinate origin is determined by the maximum of transverse pressure $p_T$. These results are for SU(2) and have not been rescaled using $N_c$-dependent factors. In addition, we show the transverse pressure $p_T = \varepsilon_L = \varepsilon_{E,L} + \varepsilon_{B,L}$ in \cref{fig:transverse_pressure}, which shows a pronounced maximum close to the collision event.

\subsection{Energy density rapidity profiles}

During the simulation we record the components of the energy-momentum
tensor $T^{\mu\nu}$ in the laboratory frame and average over a number
of collision events to obtain the expectation value $\ev{T^{\mu\nu}}$.
In the MV model most of components of $T^{\mu\nu}$ vanish after
averaging over all initial conditions due to homogeneity and isotropy
in the transverse plane. Therefore, the energy-momentum tensor reduces
to (see \cref{sec:em_tensor_latt})
\begin{equation}
\ev{T^{\mu\nu}}=\begin{pmatrix}
\ev{\e} & 0 & 0 & \ev{S_{L}}\\
0 & \ev{p_{T}} & 0 & 0\\
0 & 0 & \ev{p_{T}} & 0\\
\ev{S_{L}} & 0 & 0 & \ev{p_{L}}
\end{pmatrix},
\end{equation}
where $\ev{\varepsilon}$ is the energy density, $\ev{p_{L}}$ and
$\ev{p_{T}}$ are the longitudinal and transverse pressure components
and $\ev{S_{L}}$ is the longitudinal component of the Poynting vector.
If we are interested in the energy density in the co-moving frame we perform the projection
\begin{equation}
\ev{T^{\tau\tau}} = u_\mu u_\nu \ev{T^{\mu\nu}},
\end{equation}
where $u^\mu = \lb \cosh \eta, 0, 0, \sinh \eta \rb^\mu$ is the timelike four-velocity of the co-moving observer. Plugging in everything, we find
\begin{equation}
\ev{T^{\tau\tau}} = \cosh^2 \eta \ev{\e} + \sinh^2 \eta \ev{p_L} - 2 \cosh \eta \sinh \eta \ev{S_L}.
\end{equation}
However, this is not the quantity that we will study for two reasons: first, this energy density is sensitive to the exact placement of the coordinate origin. As mentioned in the introduction of this section this choice is slightly ambiguous. Secondly, it is a priori not obvious that the $\lb \tau, \eta \rb$ frame is appropriate for the Glasma with broken boost invariance. A simpler and coordinate-free definition of the Glasma energy density is the local rest frame (LRF) energy density. The LRF corresponds to the frame in which the energy flux (or Poynting vector) vanishes. We can write a general Lorentz transformation of $\ev{T^{\mu\nu}}$ as a similarity transformation via
\begin{equation}
\ev{(T')^\mu_{\,\,\nu}} = \Lambda^\mu_{\,\,\rho} \Lambda_\nu^{\,\,\sigma} \ev{T^\rho_{\,\,\sigma}} = \Lambda^\mu_{\,\,\rho} \ev{T^\rho_{\,\,\sigma}} (\Lambda^{-1})_{\,\,\nu}^{\sigma},
\end{equation}
or using matrix notation
\begin{equation}
\ev{T'} = \Lambda \ev{T} \Lambda^{-1}.
\end{equation}
The matrix $\Lambda$ associated with an arbitrary boost in the $z$ direction is given by
\begin{equation}
\Lambda = \begin{pmatrix}
\gamma & 0 & 0 & -\gamma \beta \\
0 & 1 & 0 & 0 \\
0 & 0 & 1 & 0 \\
- \gamma \beta & 0 & 0 & \gamma \\
\end{pmatrix},
\end{equation}
with $\gamma = 1 / \sqrt{1- \beta^2}$ and $\beta = v_z / c$.
Applying the transformation and requiring that the energy flux in $z$ vanishes in the new frame, we find the velocity of the LRF
\begin{equation} \label{eq:long_velocity}
\beta = v_z / c = \frac{1}{2 \ev{S_L}} \lb \ev{p_L} + \ev{\e} - \sqrt{\lb \ev{p_L} + \ev{\e} \rb^2 - 4 \ev{S_L}^2} \rb.
\end{equation}
The energy density in the LRF is given by
\begin{equation}
\e_{\mathrm{loc}} \equiv \ev{(T')^{00}}  =  \frac{1}{2}\bigg(\ev{\e}-\ev{p_{L}}+\sqrt{\left(\ev{\e}+\ev{p_{L}}\right)^{2}-4\ev{S_{L}}^{2}}\bigg).\label{eq:loc_rest_frame_eng}
\end{equation}
The above result can also be obtained by directly diagonalizing the energy-momentum tensor $\ev T^\mu_{\,\,\nu}$ as $\ev{T'}$ and $\ev{T}$ are related via a similarity transformation.
The LRF energy density $\ev{\e_\mathrm{loc}}$ can be computed at any point $(t, z)$ without explicitly referring to the coordinates origin of the $(\tau, \eta)$ frame. However, since we are specifically interested in the rapidity dependence of observables, we compute $\e_\mathrm{loc}(t, z)$ everywhere in the future lightcone and then, using proper time $\tau=\sqrt{t^{2}-z^{2}}$ and space-time rapidity $\eta_{s}=\frac{1}{2}\ln\left[\left(t-z\right)/\left(t+z\right)\right]$, plot the profile of the LRF energy density as a function of $\eta_{s}$ for various values of $\tau$.

\begin{figure}
	\centering
	\includegraphics[scale=1.1]{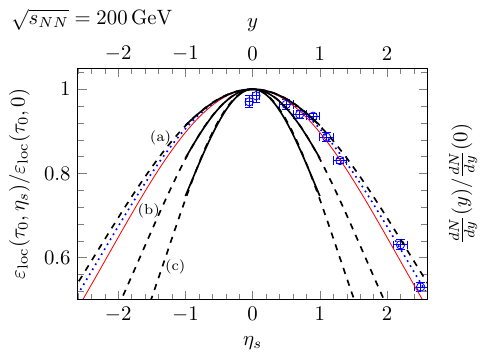}
	\caption{Comparison of the space-time rapidity profile of the local rest frame
		energy density $\protect\e_{\text{loc}}(\tau_{0},\eta_{s})$ for a
		RHIC-like collision (thick solid lines) at $\tau_{0}=1\,\mbox{fm}$/c,
		a measured profile of $\pi^{+}$ multiplicity $dN/dy$ at RHIC (data
		points) and Gaussian fits (dashed and dotted lines) for our simulation
		and experimental data ($\sigma_{\text{exp}}=2.25$). The value of
		the infrared regulator $m$ modifies the width of the profiles: (a)
		$m=0.2\,\mbox{GeV}$ with $\sigma_{\eta}=2.34$, (b) $m=0.4\,\mbox{GeV}$
		with $\sigma_{\eta}=1.66$ and (c) $m=0.8\,\mbox{GeV}$ with $\sigma_\eta=1.28$.
		Data is taken from \cite{Bearden:2004yx}. The thin red line
		corresponds to the profile predicted by the Landau model with $\sigma_{\text{Landau}}=\sqrt{\ln\gamma}\approx2.15$.
		This plot is taken from  \cite{Ipp:2017lho}.
		\label{fig:rhic200}}
\end{figure}

In \cref{fig:rhic200} we see a calculation of the rapidity
profile of the local rest frame energy density for a RHIC-like collision:
choosing parameters $\sqrt{s_{NN}}=200\,\mbox{GeV}$ and $m=0.2\,\text{GeV}$,
we obtain an approximate Gaussian profile of the local rest frame
energy density, i.e.\
\begin{equation}
\e_{\text{loc}}(\tau_{0},\eta_{s})\approx\e_{\text{loc}}(\tau_{0},0)\exp\left(-\frac{\eta_{s}^{2}}{2\sigma_{\eta}^{2}}\right),
\end{equation}
in the range $\eta_{s}\in(-1,1)$. Using a Gaussian fit we compute
$\sigma_{\eta}$ and extrapolate to higher $\eta_{s}$. At higher
values of the IR regulator $m$ the rapidity profiles become more
narrow, which seems to suggest that the transverse size of the color-neutral regions $\sim m^{-1}$ has an effect on broken boost invariance.
In the CGC literature it is well-known that quantities such as the
initial energy density or the gluon multiplicity
can be sensitive to the choice of the infrared regulator \cite{Fukushima:2011nq,Fukushima:2007ki,Lappi:2007ku,Fujii:2008km}.
Here we add another example of a strong dependency on the infrared
regulation.

\begin{figure} [t]
	\centering
	\includegraphics[scale=1.1]{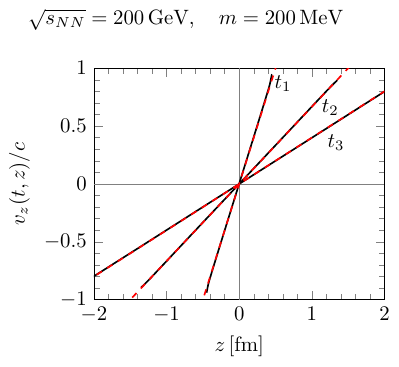}
	\caption{Longitudinal velocity $v_{z}(t,z)$ as a function of the longitudinal coordinate
		$z$ evaluated at different laboratory frame times $t$ from out simulation 
		(black, solid lines) compared to the free streaming case $v_{z}(t,z) = z/t$
		(red, dashed lines). We show three different times:
		$t_1 = 0.5\,\mbox{fm}/c$,
		$t_2 = 1.5\,\mbox{fm}/c$ and
		$t_3 = 2.5\,\mbox{fm}/c$.
		This plot is taken from \cite{Ipp:2017uxo}.
		\label{fig:long_velocity}}
\end{figure}
Even though the results of our model should be regarded as more qualitative
than quantitative, we compare our findings to experimental results:
it is an interesting observation that the rapidity profile of the
energy density for $m=0.2\,\text{GeV}$ agrees with the measured rapidity
profile of pion multiplicities for the most central collisions at
RHIC. Of course, a direct comparison of $\e_{\text{loc}}$ and $dN/dy$
profiles is not strictly valid: the gluon number distribution can
be somewhat broader than the energy density \cite{Hirano:2004en}.
The correct approach would be to use our results as initial conditions
for hydrodynamic simulations in order to make a more direct connection
with measured observables. The subsequent hydrodynamic expansion of
the system likely increases the width of the profiles further as mentioned
in \cite{Schenke:2016ksl},
which would favor the more narrow curves (b) and (c) in \cref{fig:rhic200}.
Under these assumptions, the width of the rapidity profile of measured
charged particle multiplicities can be seen as an upper limit for
realistic rapidity profiles computed from our simulation. We also
compare to the Gaussian rapidity profile of the hydrodynamic Landau
model \cite{Landau:1953gs} with $\sigma_{\text{Landau}}=\sqrt{\ln\gamma}$.

\begin{figure} [t]
	\centering
	\includegraphics[scale=1.1]{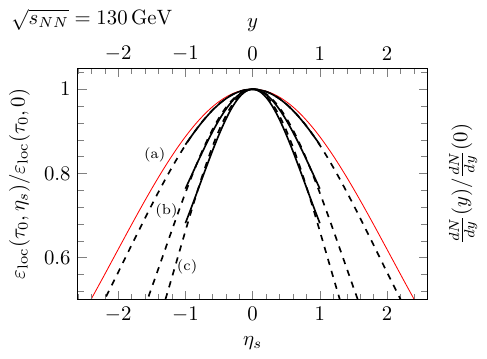}
	\caption{Simulation results of collisions at $\sqrt{s_{NN}}=130\,\text{GeV}$
		at $\tau_{0}=1\,\mbox{fm}$/c from \cite{Ipp:2017lho}. The black solid lines are the computed
		profiles within $\eta_{s}\in(-1,1)$. The dashed lines are fits to
		Gaussian profiles: (a) $m=0.2\,\text{GeV}$ with $\sigma_{\eta}=1.87$,
		(b) $m=0.4\,\text{GeV}$ with $\sigma_{\eta}=1.33$ and (c) $m=0.8\,\text{GeV}$
		with $\sigma_{\eta}=1.10$. The thin red line is the result predicted
		by the Landau model with $\sigma_{\text{Landau}}\approx2.04$. Compared
		to \cref{fig:rhic200}, the profiles are narrower than the
		Landau result. 
		\label{fig:rhic130}}
\end{figure}

The profiles have been computed at $\tau_{0}=1\,\mbox{fm/c}$, which
roughly corresponds to the transition from the Glasma to the QGP.
In general, we observe that the rapidity profiles quickly converge
for $\tau\gtrsim0.3\,\mbox{fm/c}$ and afterwards become independent
of $\tau$. We also observe free streaming behavior signaled by $\e_{\text{loc}}\propto1/\tau$
for a wide range of $\eta_{s}$ and longitudinal velocities $v_{z}\sim z/t$.
The longitudinal velocity can be computed directly from \cref{eq:long_velocity} and
is shown in \cref{fig:long_velocity}, where we plot $v_z(t,z) / c$ in the case of
$\sqrt{s_\mathrm{NN}} = 200 \, \gev $ and $ m=0.2 \, \gev$. We see that the
velocity of the Glasma $v_z/c$ agrees with the longitudinal velocity of
the boost-invariant Glasma $v_z/c = t / z = \tanh \eta_s$. This implies that
there is only negligible flow of energy between different rapidity directions
and that the LRF of the 3+1 dimensional Glasma essentially corresponds to the co-moving frame.
Our results suggest that the rapidity dependence of the Glasma
is fixed early on in the collision and remains
unchanged thereafter. To see the effects of increased longitudinal
thickness, we repeat the calculation of the LRF energy density for $\sqrt{s_{NN}}=130\,\mbox{GeV}$
in \cref{fig:rhic130}. We fit to a Gaussian shape and find
that, as expected, the profiles become narrower as compared to
the $\sqrt{s_\mathrm{NN}} = 200 \, \gev$ case. This time, our results are also narrower than
Landau's prediction.

Investigating the reason behind the non-boost-invariant creation of
the Glasma, we look at the space-time distribution of the transverse
pressure $\ev{p_{T}(z,t)}$ in and along the forward light cone in
\cref{fig:transverse_pressure}. The transverse pressure is solely due to the presence of longitudinal
fields, i.e.\ $\ev{p_{T}(z,t)}$ is equivalent to the energy density
generated by longitudinal fields $\ev{\e_{L}(z,t)}$. In the boost invariant
case the longitudinal fields would be constant along the boundary
of the light cone as determined by the boost invariant Glasma initial
conditions at $\tau=0$. In our simulations we see very different
behavior: the longitudinal fields are mostly centered around the maximum
in the extended collision region $t\sim z\sim0$ and decrease rather
quickly along the $t=\pm z$ boundaries. Since the initially longitudinal
fields are the starting point of the evolution of the Glasma, this
sharp decrease means that for some fixed proper time $\tau$ there
is less Glasma at higher values of rapidity $\eta_{s}$, leading to
the observed Gaussian profiles of the energy density. Furthermore,
we see a mostly constant spatial distribution of $\ev{p_{T}(z,t)}$
in \cref{fig:transverse_pressure} for later times $t$. A similar
distribution has been found in holographic models of heavy-ion collisions
\cite{Casalderrey-Solana:2013aba,vanderSchee:2015rta, Attems:2017zam}.

\subsection{Extension to larger rapidity ranges} \label{sec:extending_rapidity}

In the previous subsections we computed the LRF energy density only within rather restricted
rapidity intervals $\eta_s \in \{ -1, 1\}$. The multiple reasons for this restriction were
already outlined in the beginning of \cref{sec:rapidity_profiles}, but here we present a method that
at least circumvents the problem of reduced lattice resolution at higher values of rapidity.

This method is based on the fact that, at least in principle, the dynamics of Yang-Mills 
fields are invariant under the group of Lorentz transformations, since the underlying action is
written in a Lorentz covariant manner. On the lattice Lorentz symmetry is broken,
 but is restored when taking the continuum limit. Therefore, if the
parameters of the simulation are chosen well such that the numerical results are good
approximations to the continuum system, the following is possible:
first, we perform simulations of a symmetric collision as previously and record observables 
(such as the energy-momentum tensor). Next, we take the same initial conditions as before,
but apply a longitudinal boost of some rapidity $\eta'$, effectively changing the widths of the nuclei
and resulting in an asymmetric scenario.
Again, we simulate the collision and record observables. Finally, due to Lorentz symmetry, it is possible
to undo the Lorentz transformation of the boosted collision and find that the observables
agree in both cases (symmetric and asymmetric collisions) up to small violations due to 
the lattice discretization.   

We can exploit this property to extend the available ``safe" rapidity intervals in the 3+1 dimensional
setup. Consider a simulation of a symmetric collision, where the energy-momentum tensor can be 
computed within $\eta_s \in ( -1, 1 )$ up to sufficient accuracy. Performing a boost by e.g.\ $\eta' = -0.5$,
this interval is shifted to $\eta_s \in ( -0.5, 1.5 )$ (in terms of $\eta_s$ of the original frame). Observables can now be computed in this new interval
with the same resolution as in the un-boosted case. Due to the overlap of intervals in $\eta_s \in ( -0.5, 1 )$, it is possible to
check whether boosting leads to any violations of Lorentz symmetry of the discretized Yang-Mills system.
In the extended interval $\eta_s \in ( 1, 1.5 )$ observables at higher rapidity can be extracted. This process 
can be continued with further boosts $\eta'=-1$, $\eta' =  -1.5$ and so on.

We can perform longitudinal boosts on the initial conditions in a straightforward way using light cone coordinates.
Recall from \cref{sec:mv_model} that the color field of a single nucleus in covariant gauge is given by the solution to the Poisson equation, see \cref{eq:mv_poisson}
\begin{equation} \label{eq:poisson_again}
- \Delta_T A^\pm(x^\mp, x_T) = J^\pm(x^\mp, x_T).
\end{equation}
In light cone coordinates a longitudinal boost along $z$ of rapidity $\eta'$ is given by
\begin{align}
x_T' &= x_T, \\
x'^\pm &= e^{\pm \eta'} x^\pm, \\
J'^\pm(x'^\mp, x_T') &= e^{\pm \eta'} J^\pm(e^{\pm \eta'} x'^\mp, x_T').
\end{align}
In the boost invariant scenario, the current remains unchanged under these boosts (see eqs.\ \eqref{eq:boosted_current_A} and \eqref{eq:boosted_current_B}).
However, this is not the case anymore in the 3+1D setup. For instance, in case of the single color sheet approximation the color current for a right-moving nucleus is given by (see \cref{eq:scsa_J})
\begin{equation}
J^+(x^-, x_T) = \sqrt{2} f(t - z) \rho(x_T), 
\end{equation}
where $f(t - z)$ is the Gaussian profile function of width $\sigma$. Applying the boost, the new current reads
\begin{align}
J'^+(x'^-, x'_T) &= \sqrt{2} e^{\eta'} f\lb e^{\eta'} \lb t' - z' \rb \rb) \rho(x'_T) \nn
&= \sqrt{2} \frac{e^{\eta'}}{\sqrt{2\pi} \sigma} \exp \lb - \frac{\lb t'-z' \rb^2}{2 \sigma^2} e^{2 \eta'} \rb \rho(x'_T) \nn
&= \sqrt{2} \frac{1}{\sqrt{2\pi} \lb e^{-\eta'} \sigma \rb} \exp \lb - \frac{\lb t'-z' \rb^2}{2 \lb e^{-\eta'} \sigma \rb^2} \rb \rho(x'_T) \nn
&= \sqrt{2} f'(t'-z') \rho(x'_T).
\end{align}
In the last two lines we have absorbed the factor $e^{-\eta'}$ into a redefined thickness parameter $\sigma' = e^{-\eta'} \sigma$ and defined a new (boosted) longitudinal profile function $f'(t'-z')$ with thickness $\sigma'$. Boosting in the longitudinal direction is therefore equivalent to simply rescaling the thickness parameter of the nuclei. Analogously, we find $\sigma' = e^{\eta'} \sigma$ for the thickness of a left-moving, boosted nucleus. The associated color field is obtained by solving the Poisson equation \eqref{eq:poisson_again} for the boosted color current.

\begin{figure}[t]
	\centering
	\includegraphics[scale=1]{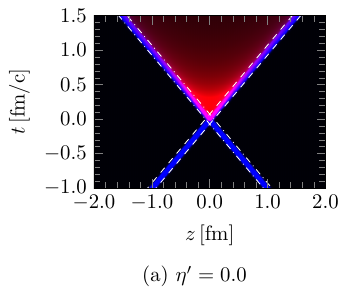}
	\includegraphics[scale=1]{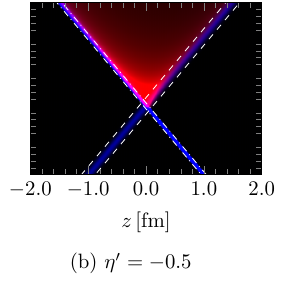}
	\includegraphics[scale=1]{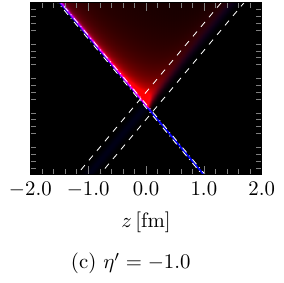}
	
	\caption{Schematic density plots of transverse pressure $p_T$ (red) and longitudinal pressure $p_L$ (blue) in asymmetric collisions of boosted nuclei.
		Different lattices have been used to perform these simulations: (a) and (b) use a simulation volume of $(4\, \fm)^3$ with a lattice of $N_L \times N_T^2 = 1024 \times 128^2$.
		In (c) the size of the lattice was increased to $2048 \times 192^2$ due to the very thin thickness of the right-moving nucleus.
		As $\eta'$ increases, the left-moving nucleus becomes more energetic and starts to ``drag" the Glasma with it. For these simulations a statistical ensemble of 10 collision events was used. The dashed lines indicate the longitudinal diameter $4 \sigma$ (see \cref{eq:scsa_four_sigma}) of the (boosted) nuclei.
		\label{fig:larger_rapidity_pressures}}
\end{figure}

\begin{figure}[t]
	\centering
	\includegraphics[scale=1]{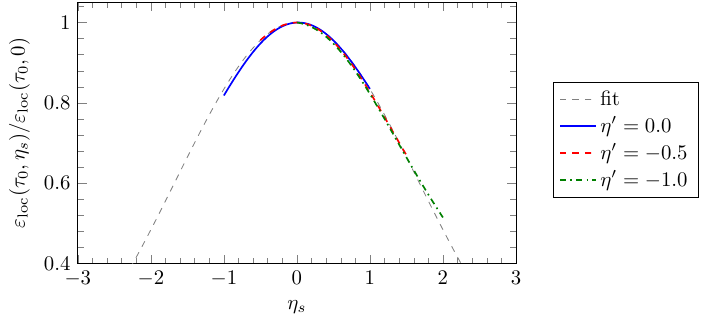}
	\caption{Rapidity profiles corresponding to the simulations shown in \cref{fig:larger_rapidity_pressures} at $\tau = 0.5 \, \fm / c$. The gray dashed line is the Gaussian fit to the numerical results of \cref{fig:rhic200} (b). The blue solid line corresponds to the symmetric collision with $\eta'=0$ restricted to the interval $\eta_s \in (-1, 1)$, the red dashed line corresponds to $\eta'=-0.5$ and the green dot-dashed line corresponds to $\eta'=-1$. We see that the results of the various boosted scenarios agree in the overlapping regions. An extended rapidity profile with range $\eta_s \in (-2, 2)$ can be obtained by combining the results of all simulations. 
		\label{fig:extended_profiles}}
\end{figure}

As a proof-of-concept example we repeat the simulations of the previous subsections for $\sqrt{s_\mathrm{NN}} = 200 \, \gev$ and $m = 0.4 \, \gev$. The results for the two pressure components are shown in \cref{fig:larger_rapidity_pressures}. Here, the effects of the longitudinal boost on the single nucleus fields are clearly visible. For decreasing values of $\eta' $, the width of the left-moving nucleus decreases, while the profile of the right-moving nucleus becomes broader. 
The transverse pressure $p_T$, which is generated by the Glasma fields, is ``dragged along" by the left-moving nucleus towards negative $z$. Finally, the rapidity profiles of the LRF energy density $\e_\mathrm{loc}$ at $\tau = 0.5 \, \fm / c$ are shown in \cref{fig:extended_profiles}. In this plot it is evident that the boosted simulations agree in the overlapping regions, which demonstrates the (approximate) Lorentz symmetry of the numerical method. By combining the results of different boosted collisions, it is possible to extend the available range of observables by one full unit of rapidity. This method is however still limited: by boosting to higher (absolute) rapidities, the thickness of one of the nuclei can become rather small compared to the longitudinal lattice spacing $a_L$. For this reason, the lattice resolution was increased for the highest boost in \cref{fig:larger_rapidity_pressures}.

\chapter{A semi-implicit variational integrator} \label{cha:semi_implicit}

In the previous chapters we presented two slightly different numerical methods to solve the classical Yang-Mills equations in the real-time lattice formalism.
Both the Hamiltonian leapfrog and Lagrangian leapfrog method of \cref{sec:latt_field__eqs} are explicit schemes and accurate up to quadratic error terms in the time step $\Delta t$ and spatial lattice spacings $a^i$.
In terms of accuracy, the two schemes can be considered equivalent.
As already mentioned in \cref{sec:approaching_bi}, we find that using these schemes, our simulations exhibit a numerical instability that results in the production of unphysical spurious fields and an artificial increase of field energy over time.
Generally, increasing lattice resolution dampens this instability up to a point where we can trust our numerical results to be accurate enough at the cost of increased computational resources.
Unfortunately, using the standard schemes, no choice of lattice parameters is able to completely eliminate the instability. 
In this chapter, we identify it as a non-Abelian version of the numerical Cherenkov instability (NCI) \cite{GODFREY1974504} known from Abelian particle-in-cell simulations. Having identified the source of the instability, we are able to develop a new semi-implicit scheme for real-time lattice gauge theory, which almost completely avoids the problems associated with the standard schemes while keeping the same order of accuracy and lattice gauge invariance. This chapter is heavily based on the most recent publication \cite{Ipp:2018hai} with some additional comments and plots added.

\section{Numerical Cherenkov instability and numerical dispersion} \label{sec:nci_dispersion}

We start our discussion by mentioning that the numerical instability already occurs when studying the propagation of single nuclei on the lattice. The problem is therefore not primarily related to the actual collision of nuclei. In principle, the propagation of a single nucleus along the beam axis should be static and stable. In the continuum, both the current and the color fields move at the speed of light while maintaining their shape along the longitudinal coordinate. For observers in the laboratory frame, the transverse color structure of both fields and charges appears static due to time dilation. One encounters  this property when solving the Yang-Mills equations for single nuclei (see \cref{sec:mv_model}): the propagating solutions to the field equations must be independent of one of the light cone coordinates $x^+$ or $x^-$ (at least in covariant and LC gauge).

Ideally, we would like to observe the same properties when solving the Yang-Mills equations numerically on the lattice: both fields and the ensemble of colored particles should move at the speed of light along the beam axis in a stable manner. Let us therefore re-examine the numerical methods of \cref{chap:glasma3d} more closely with regards to propagation at the speed of light. The colored point-like particles which make up the color charge of the nucleus, always move at the speed of light: since we neglect any forces acting on particles, their continuous positions are updated at each time step such that they follow lightlike trajectories. Therefore, in terms of movement, the particle degrees of freedom behave ideally by construction. However, the color charges of particles can be affected by parallel transport. As colored particles move along the beam axis, their color charges are rotated via Wilson lines in the lightlike direction of their trajectory. In temporal gauge, the gauge fields of nuclei are purely transverse and therefore the color charges should ideally remain static. Color charges only rotate if they encounter non-zero longitudinal gauge fields. 

In contrast, the propagation of color fields is governed by the discretized field equations of motion on the lattice. The propagating color fields of nuclei can be decomposed into a superposition of plane wave modes of varying momenta $k$ and associated frequencies $\omega(k)$. In the continuum, plane wave solutions of the Yang-Mills equations have a linear dispersion relation $\omega(k) = \abs k$ and as such they all exhibit the same phase velocity $v_\mathrm{ph} = \omega(k) / \abs k = 1$ equal to the speed of light. This is generally not the case on the lattice. Typically, finite difference schemes such as the Hamiltonian or Lagrangian leapfrog method (see \cref{sec:latt_field__eqs}) show modified dispersion behavior for high momenta $k$. This phenomenon is known as numerical dispersion.

\begin{figure}
	\centering
	\begin{subfigure}[b]{0.48\textwidth}
		\centering
		\includegraphics[scale=1.0]{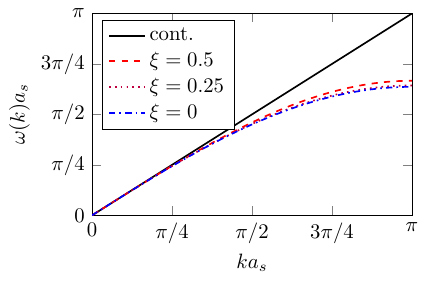}
		\caption{Dispersion relation $\omega(k)$}
	\end{subfigure}
	\begin{subfigure}[b]{0.48\textwidth}
		\centering
		\includegraphics[scale=1.0]{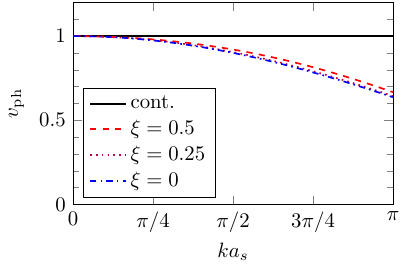}
		\caption{Phase velocity $v_\mathrm{ph} = \omega(k) / k$}
	\end{subfigure}
	\caption{Dispersion relation and phase velocity according to \cref{eq:leapfrog_disp_solved} as a function of momentum $k$ for different values of $\xi = a^0 / a_s$ at fixed lattice spacing $a_s$. At small momenta $k$, the dispersion relation agrees with the linear dispersion of the continuum system and the phase velocity is close to the speed of light. At higher momenta, deviations become more apparent.  
		\label{fig:leapfrog_dispersion}}
\end{figure}

For concreteness, we now derive the wave dispersion relation of the Lagrangian leapfrog method presented in \cref{sec:latt_field__eqs}. Starting from \cref{eq:leapfrog_eom_cpic} without external sources
\begin{equation} \label{eq:semi_eom_intro}
\frac{1}{\lb a^0 a^i \rb^2} \ah{U_{x,i0} + U_{x,i-0}} =
\sum_j \frac{1}{\lb a^i a^j \rb^2} \ah{U_{x,ij} + U_{x,i-j}},
\end{equation}
we first restrict the color field to an Abelian subspace, because plane wave solutions to the Yang-Mills equations are effectively Abelian. As a consequence, all commutator terms can be dropped. Consequently, the plaquette (see \cref{eq:plaquette_definition}) can be written as
\begin{equation} \label{eq:abelian_plaquette}
U_{x,\mu\nu} = \exp \lb i g a^\mu a^\nu F_{x,\mu\nu} \rb,
\end{equation}
where the lattice field strength tensor is given by
\begin{equation}
F_{x,\mu\nu} = \p^F_\mu A_{x,\nu} - \p^F_\nu A_{x,\mu}.
\end{equation}
Here, $\p^F_\mu$ is simply the first order forward finite difference.
In this notation, the lattice gauge field $A_{x,\mu}$ is considered to be defined at the mid-point between $x$ and $x+\hat{\mu}$. The EOM \eqref{eq:semi_eom_intro} are still non-linear due to the use of the matrix exponential in \cref{eq:abelian_plaquette}. Therefore, we go to the limit of small amplitudes and find from \cref{eq:semi_eom_intro} to lowest order in the field strength tensor
\begin{equation}
\p^B_0 F_{x,i0} = \sum_j \p^B_j F_{x,ij},
\end{equation}
which is simply a finite difference approximation of Maxwell's equations. Similarly, we find the discrete Abelian Gauss constraint
\begin{equation}
\sum_i \p^B_i F_{x,i0} = 0.
\end{equation}
The wave dispersion relation can now be derived from these EOM by plugging in a plane wave ansatz
\begin{equation}
A_{x,i} = A_i e^{i \lb \omega x^0 - \sum_i k^i x^i \rb},
\end{equation}
which yields an equation that connects the frequency $\omega$ to the momentum $k$. At this point we simply state the result and postpone the detailed derivation to \cref{sec:abelian_leapfrog}:
\begin{equation} \label{eq:leapfrog_disp_eq}
\sin^2 \lb \frac{\omega a^0}{2} \rb = \sum_j \lb \frac{a^0}{a^j} \rb^2 \sin^2 \lb \frac{k^j a^j}{2} \rb.
\end{equation}
Restricting the momentum to one particular direction $k^i = \delta^{i}_1 k$ and solving for $\omega$ yields the non-linear dispersion relation
\begin{equation} \label{eq:leapfrog_disp_solved}
\omega a^0 = 2 \arcsin \lb \xi \sin \lb \frac{k a_s}{2} \rb \rb,
\end{equation}
where $a_s$ is the spatial lattice spacing and $\xi = a^0 / a_s$. For small momenta $k a_s \ll 1$ we recover the linear result $\omega \simeq \abs k$, but in general, $\omega(k)$ is a non-linear function of $k$. In \cref{fig:leapfrog_dispersion} we plot the dispersion relation and its associated phase velocity for a few parameter choices. In particular, we see that the phase velocity $v_\mathrm{ph}(k)$, i.e.\ the speed at which a plane wave mode with momentum $k$ propagates, is less than the speed of light. This effect is more pronounced at high momenta. Although choosing $\xi = 1$ (or $a^0 = a_s$) would render the dispersion relation linear, this particular parameter choice causes unstable wave propagation: requiring that arbitrary momenta $k$ are associated with real-valued frequencies $\omega(k)$ when solving \cref{eq:leapfrog_disp_eq}, yields the stability criterion (see sections \ref{sec:wave_leapfrog} and \ref{sec:abelian_leapfrog} for a derivation)
\begin{equation}
\sum_i \lb \frac{a^0}{a^i} \rb^2 \leq 1,
\end{equation}
which forbids the time step $a^0$ being equal to any of the lattice spacings for more than one spatial dimension. Violating the above stability constraint leads to modes with imaginary frequencies that can grow exponentially.

Deviations of phase velocity from the speed of light imply that in principle it is impossible to have stable propagation of color fields using the schemes derived in \cref{sec:latt_field__eqs}. Over time, high momentum modes ``fall behind" low momentum modes. This alone would not cause an instability, but merely change the longitudinal shape of the nuclear field as the simulation progresses. A problem arises when we take into account the coupling to the external color currents, which are not affected by numerical dispersion. Consider the Abelian scenario of a highly contracted charge density moving at the speed of light: the longitudinal part of Maxwell's equations then reads
\begin{equation}
\frac{\p E_L}{\p t} = \lb \nabla \times \vec{B} \rb_L - j_L,
\end{equation}
where $E_L$ and $j_L$ are the electric field and current in the longitudinal direction. As in the non-Abelian case, the electric and magnetic field are purely transverse. This implies that there is a precise cancellation of $j_L$ and the curl of the magnetic field. In the continuum we therefore have $\p  E_L / \p t = 0$. Any mismatch between the current and fields leads to the production of longitudinal electric fields. In the discretized scenario, the non-linear dispersion relation of the electromagnetic fields causes this mismatch and leads to a production of spurious field modes. Moving charges seem to be radiating and over time the total field energy of the system increases. In this case electric charges are traveling ``faster than the speed of light", in the sense that the speed of light is defined by the speed of propagation of the electromagnetic field. This is reminiscent of the physical phenomenon of Cherenkov radiation, where relativistic charges move through a medium in which electromagnetic radiation propagates at slower speeds. However, in the case of discretized systems subluminal phase velocities $v_\mathrm{ph}$ are numerical artifacts. This unphysical effect is therefore known as the numerical Cherenkov radiation or instability (NCI) \cite{GODFREY1974504}.
In simulations of Abelian plasmas with dynamical particles, the instability arises because the artificial Cherenkov radiation further affects charged particles through the Lorentz force, which leads to unphysical acceleration of particles.

In contrast, the colored particles of our simulations are not affected by forces, but they interact with the color field via parallel transport. In temporal gauge a non-zero longitudinal color-electric field generally implies that the longitudinal gauge component becomes non-zero as well. The color charges of the nucleus start rotating and the initial mismatch is amplified further. The instability we are facing is therefore due to non-Abelian interactions.

Now that the main culprit is identified, one can think about possible solutions. In the last chapter the instability could be avoided (or rather postponed) by using very large grid sizes and finer lattice spacing. This helped because smaller $a_s$ increases the largest available momentum via $k_\mathrm{max} = \pi / a_s$. In \cref{fig:leapfrog_dispersion} the occupied field modes move towards smaller momenta (compared to $k_\mathrm{\max}$), where the dispersion relation is more linear. As a result, it takes longer for the color fields of the nucleus to disperse and start becoming unstable. However, this does not actually eliminate the instability and the computational cost of our simulations increases quickly. The only way to avoid the instability altogether is to eliminate numerical dispersion in the discretized system. 
A typical example of Abelian PIC simulations being heavily affected by the NCI is the simulation of laser wakefield acceleration \cite{PhysRevSTAB.16.021301}. A number of possible solutions has been developed: for example, instead of using a finite difference scheme in both spatial and temporal directions (finite difference time domain or FDTD), one can solve the equations of motion in Fourier space using so-called ``spectral solvers". In this method the dispersion relation can be modified more directly compared to finite difference schemes. In combination with filtering high momentum modes, the NCI is suppressed successfully \cite{Godfrey:2015sxa}. Another solution is to carefully modify interpolation schemes and finite difference stencils in FDTD approaches, see \cite{GODFREY201333, Godfrey:2014ava} or \cite{GREENWOOD2004665}. In the case of laser wakefield simulations it is in principle not necessary to eliminate the NCI for all propagation directions, but only along a particular beam axis. A numerical scheme based on this idea is developed in \cite{PhysRevSTAB.8.042001}. However, a particularly simple and elegant solution to the problem is the use of semi-implicit schemes to repair the dispersion relation (i.e.\ making it linear) for one direction of propagation \cite{Novokhatski:2012yd}, such that no mismatch between charges and fields can arise. For our 3+1 dimensional setup a corrected dispersion relation along the beam axis would be sufficient to mitigate the NCI. The development of a semi-implicit scheme is therefore the main goal of this chapter.

\section{Lattice gauge invariance} \label{sec:latt_gauge_inv}

One particularly important property of both the Hamiltonian and the Lagrangian leapfrog method for numerically solving the Yang-Mills equations is that there exists an exactly conserved Gauss constraint (GC). Due to this constraint being preserved throughout the time evolution, we are able to derive a covariant conservation equation for the color current and the color charge density on the lattice. Solving the continuity equation for single colored particles as in \cref{sec:latt_particle__eqs} allows us to determine the correct color current necessary for solving the field equations of motion coupled to external sources. Without a preserved GC we have no clear notion of charge conservation on the lattice. It is therefore important that our sought-after semi-implicit scheme also possesses an associated, exactly conserved GC.

Keeping this in mind, it is probably not a good idea to develop new schemes by making arbitrary changes to the EOM, as it is highly unlikely that one ends up with a scheme that preserves the original GC. One then would have to ``guess" a new GC without even knowing if such an equation exists. It is clear that a more systematic approach is necessary. Recall that the Lagrangian leapfrog method of \cref{sec:latt_field__eqs} was derived by first discretizing the Yang-Mills action (yielding the Wilson gauge action) and then varying this action. Analogously to the continuum system, variation with respect to spatial gauge links led to the discrete EOM and through variation with respect to temporal gauge links we found the associated GC. One can then explicitly prove that the GC is identically satisfied, see \cref{app_gauss}. It is no coincidence that the constraint holds exactly, because the action already implies charge conservation through lattice gauge invariance. In the continuum there is a direct connection between gauge invariance and charge conservation through the Noether theorem. Remarkably, this also holds for the discretized case. Consider a general action $S[U]$ formulated in terms of gauge links $U_{x,\mu}$ such that $S[U]$ approximates the Yang-Mills in the continuum limit $a^\mu \rightarrow 0$. Further assume that $S[U]$ is invariant under arbitrary lattice gauge transformations, i.e.\ $S[U'] = S[U]$ with
\begin{equation}
U'_{x,\mu} = \Omega_{x} U_{x,\mu} \Omega^\dg_{x+\mu}.
\end{equation}
Now consider ``small" gauge transformations $\Omega_{x} = \exp \lb i g \alpha_{x} \rb$ with $\alpha_x$ assumed to be infinitesimal. A single gauge link transforms according to
\begin{align}
U'_{x,\mu} &= \Omega_{x} U_{x,\mu} \Omega^\dg_{x+\mu} \nn
&\simeq \lb 1+ i g \alpha_x \rb U_{x,\mu} \lb 1 - i g \alpha_{x+\mu} \rb + \mathcal{O}\lb \alpha^2 \rb \nn
&= \lb 1 - i g a^\mu \lb D^F_\mu \alpha_x \rb \rb U_{x,\mu} + \mathcal{O}\lb \alpha^2 \rb,
\end{align}
where the gauge-covariant forward finite difference is given by (see \cref{eq:gc_finite_forward})
\begin{equation}
D^F_\mu \alpha_x = \frac{1}{a^\mu} \lb U_{x,\mu} \alpha_{x+\mu} U^\dg_{x,\mu} - \alpha_x \rb.
\end{equation}
The above relation can be interpreted as an infinitesimal perturbation of the gauge links, i.e.
\begin{equation}
U'_{x,\mu} = U_{x,\mu} + \dd U_{x,\mu},
\end{equation}
with $\delta U_{x,\mu} = - i g a^\mu D^F_\mu \alpha_x U_{x,\mu}$. This corresponds to a perturbation of gauge fields $A_{x,i} \rightarrow A_{x,i} + \dd A_{x,i}$ by $\delta A_{x,i} = - D^F_\mu \alpha_x$.
Applying this transformation to the action $S[U]$ and expanding in $\alpha$ therefore yields
\begin{align}
S[U'] \simeq S[U] - \sum_{x,\mu, a} \lb D^F_\mu \alpha_x \rb^a \frac{\p S[U]}{\p A^a_{x,\mu}} + \mathcal{O}\lb \alpha^2 \rb,
\end{align} 
where the derivative $\p / \p A^a_{x,\mu}$ acting on gauge links is defined via (see \cref{eq:gauge_link_derivative})
\begin{equation}
\frac{\p U_{x,\mu}}{\p A^a_{x,\mu}} = i g a^\mu t^a U_{x,\mu}.
\end{equation}
The above derivative follows from the variation of gauge links, see \cref{app_var}. Since we assume $S[U]$ to be lattice gauge invariant, the term linear in $\alpha$ must be identically zero (as well as terms of higher order in $\alpha$). This yields
\begin{equation}
\sum_{x,\mu, a} \lb D^F_\mu \alpha_x \rb^a \frac{\p S[U]}{\p A^a_{x,\mu}} = 0.
\end{equation}
Using the adjoint representation of the gauge-covariant finite difference
\begin{equation}
\lb D^F_\mu \alpha_x \rb^a = \frac{1}{a^\mu} \lb U^{ab}_{x,\mu} \alpha^b_{x+\mu} - \alpha^a_x \rb,
\end{equation}
where the gauge link in the adjoint representation is given by $U^{ab}_{x,\mu} = 2 \tr \lb t^a U_{x,\mu} t^b U^\dg_{x,\mu}\rb$, we can use summation by parts to find
\begin{align}
\sum_{x,\mu, a} \lb D^F_\mu \alpha_x \rb^a \frac{\p S[U]}{\p A^a_{x,\mu}} 
&= \sum_{x,\mu, a} \frac{1}{a^\mu} \lb U^{ab}_{x,\mu} \alpha^b_{x+\mu} - \alpha^a_x \rb \frac{\p S[U]}{\p A^a_{x,\mu}} \nn
&= \sum_{x,\mu, a} \frac{1}{a^\mu} \lb U^{\dg ab}_{x-\mu,\mu} \frac{\p S[U]}{\p A^b_{x-\mu,\mu}} - \frac{\p S[U]}{\p A^a_{x,\mu}} \rb \alpha^a_{x} \nn
&= - \sum_{x,\mu, a} \lb \lb D^B_\mu \rb^{ab} \frac{\p S[U]}{\p A^b_{x,\mu}}   \rb \alpha^a_{x}.
\end{align}
This relation must hold for arbitrary $\alpha$ and therefore we have
\begin{equation}
\sum_{\mu, b}\lb D^B_\mu \rb^{ab} \frac{\p S[U]}{\p A^b_{x,\mu}} = 0,
\end{equation}
which is similar to a gauge covariant continuity equation for the ``current" $\mathcal{J}^{a,\mu} = \p S[U]  / \p A^b_{x,\mu}$.
In temporal gauge we have $D^B_0 = \p^B_0$ and can identically rewrite the above equation as
\begin{equation}
\p^B_0 \frac{\p S[U]}{\p A^a_{x,0}} = - \sum_{b, i} \lb D^B_i \rb^{ab} \frac{\p S[U]}{\p A^b_{x,i}}.
\end{equation}
Recalling from \cref{sec:latt_field__eqs} (see also \cref{app_leapfrog}) that ${\p S[U]}/{\p A^a_{x,0}}$ and ${\p S[U]}/{\p A^b_{x,i}}$ yield the GC and the discrete EOM respectively, we see that if the EOM are satisfied, i.e.
\begin{equation}
\frac{\p S[U]}{\p A^b_{x,i}} = 0,
\end{equation}
then the GC must be conserved
\begin{equation}
\p^B_0 \frac{\p S[U]}{\p A^a_{x,0}} = \frac{1}{a^0} \lb \frac{\p S[U]}{\p A^a_{x,0}} - \frac{\p S[U]}{\p A^a_{x-0,0}} \rb = 0,
\end{equation}
or simply
\begin{equation}
\frac{\p S[U]}{\p A^a_{x,0}} = \frac{\p S[U]}{\p A^a_{x-0,0}}.
\end{equation}
This shows that independent of the exact form of $S[U]$, the constraint will be conserved under discrete time evolution if the action is invariant under lattice gauge transformations. In particular, the Lagrangian leapfrog EOM, which are derived from the invariant Wilson gauge action, have this property. In the case of the Hamiltonian leapfrog EOM, which are derived by first taking the partial continuum limit $a^0 \rightarrow 0$ and then re-discretizing the time coordinate, the conservation of the GC is less obvious but still holds. Note however, that the above proof does not apply to the case of the Hamiltonian leapfrog EOM.

A reasonable new numerical scheme for solving the Yang-Mills equations should therefore be derived from a gauge invariant action $S[U]$. Numerical schemes derived from a variational principle are known as variational integrators \cite{marsden_west_2001}. The main advantage of this type of numerical scheme is that any symmetries of the discrete action are inherited by the numerical scheme and any constraints associated with particular symmetries have discrete analogues. In the case of lattice gauge invariance, this leads to gauge covariant discrete EOM and exact constraint conservation. Moreover, variational integrators also tend to have improved energy conservation properties and it is possible to formulate a conserved symplectic structure, i.e.\ there is a notion of conserved phase space volume. For a pedagogical introduction see chapter 5 of \cite{Lew2016}. 

The goal of the rest of this chapter is to formulate a new discrete action $S[U]$ in terms of gauge links, which is invariant under lattice gauge transformations and approximates the Yang-Mills action with (at least) the same accuracy as the Wilson gauge action. Upon variation, this new action should yield a semi-implicit numerical scheme, which allows for stable, dispersion-free propagation along a particular axis on the lattice. In order to gain an understanding of how to formulate such an action, we begin with a simpler model, namely a free scalar field on the lattice. Then we continue with a semi-implicit scheme for Abelian gauge fields on the lattice, where we have a notion of lattice gauge invariance and charge conservation. Finally, we will be able to formulate the non-Abelian generalization of the Abelian semi-implicit scheme. Slightly breaking the conventions of earlier chapters we will choose $x^1$ as the longitudinal (or beam axis) direction.

\section{Scalar field on the lattice}

The basic ideas behind the numerical scheme we are after can be most easily explained using a simple toy model, namely the two-dimensional wave equation. We start by giving a few definitions and then derive three different numerical schemes by discretizing the action of the system in different ways and using a discrete variational principle. We will see how the exact discretization of the action affects the properties of the numerical scheme and in particular how numerical dispersion can be eliminated.

We consider a real-valued scalar field $\phi(x)$ in 2+1D with mostly minuses metric signature $(+1,-1,-1)$ and set the speed of light to $c=1$. The action is given by
\begin{equation} \label{eq:wave_action}
S[\phi] = \intop_x \frac{1}{2} \sum_\mu \p_{\mu}\phi\p^{\mu}\phi,
\end{equation}
which upon demanding that the variation of the action vanishes
\begin{equation}
\dd S = \intop_x \frac{\dd S[\phi]}{\dd \phi(x)} \dd \phi(x) = 0,
\end{equation}
yields the equations of motion (EOM)
\begin{equation} \label{eq:wave_eom}
\p_\mu \p^\mu \phi(x) = \p_0^2 \phi(x) - \sum_i \p_i^2 \phi(x) = 0.
\end{equation}
We use Latin indices $i,j,k,...$ to denote the spatial components, $\p_i^2$ is a shorthand for $\p_i \p_i$ (no sum implied) and $\intop_x = \int dx^0 dx^1 dx^2$. Inserting a plane-wave ansatz 
\begin{equation} \label{eq:plane_wave_ansatz}
\phi(x) = \phi_0 \exp(i \sum_\mu k_\mu x^\mu),
\end{equation}
with $\phi_0 \in \mathbb{R}$ and $k^\mu = (\omega, k^1, k^2)^\mu$ into the EOM gives the dispersion relation
\begin{equation}
\omega = |k| = \sqrt{(k^1)^2+(k^2)^2}.
\end{equation}
Obviously, the phase velocity $v = \omega / |k| = 1$ is constant, i.e.\ there is no dispersion.

\subsection{Leapfrog scheme} \label{sec:wave_leapfrog}
One possible way of discretizing the action is 
\begin{equation} \label{eq:wave_action_d1}
S[\phi] = \frac{1}{2} V \sum_x \lb \lb \p^F_0 \phi_x \rb ^2  - \sum_i (\p^F_i \phi_x)^2\right),
\end{equation}
where $\sum_x$ is the sum over all lattice sites and $V = a^0 a^1 a^2$ is the space-time volume of a unit cell. Introducing small variations $\dd \phi_x$ of the discrete field at each point, the discrete variation of this action reads
\begin{align}
\dd S &= V \sum_x \lb \p^F_0 \phi_x \p^F_0 \dd \phi_x - \sum_i \p^F_i \phi_x \p^F_i \dd \phi_x \rb \nonumber \\
&= - V \sum_x \lb  \p^2_0 \phi_x - \sum_i\p^2_i \phi_x \rb \dd \phi_x
\end{align}
which upon setting it to zero yields the discretized EOM
\begin{equation} \label{eq:wave_eom_d1}
\p^2_0 \phi_x - \sum_i\p^2_i \phi_x = 0,
\end{equation}
where $\p^2_0$ and $\p^2_i$ are second order finite differences
\begin{equation}
\p^2_\mu \phi_x \equiv \p^F_\mu \p^B_\mu \phi_x = \frac{1}{\lb a^\mu \rb^2} \lb \phi_{x+\mu} + \phi_{x-\mu} - 2 \phi_x \rb.
\end{equation}
In the above derivation, we made use of summation by parts, i.e.
\begin{equation}
\sum_x \p^F_0 \phi_x \p^F_0 \dd \phi_x = - \sum_x \p^B_0 \p^F_0 \phi_x \dd \phi_x,
\end{equation}
which is the discrete analogue of integration by parts. If the field is known in two consecutive time slices we can explicitly solve the EOM \eqref{eq:wave_eom_d1} for the field values in the next time slice:
\begin{equation}\label{eq:wave_leapfrog_explicit}
\phi_{x+0} 	= \sum_i \lb\frac{a^0}{a^i}\rb^2 \lb \phi_{x+i} + \phi_{x-i} - 2\phi_x \rb - \phi_{x-0} + 2\phi_x. 
\end{equation}
In fact, this scheme is identical to the explicit leapfrog scheme\footnote{
	The connection to the leapfrog scheme becomes more apparent if we introduce an approximation of the conjugate momentum
	\begin{equation}
	\pi_{x+\frac{0}{2}} \equiv \p^F_0 \phi_x,
	\end{equation}
	which is defined naturally between time slices $x^0$ and $x^0+a^0$ (hence the index ``$x+\frac{0}{2}$" in our notation). The EOM can then be written as
	\begin{align}
	\pi_{x+\frac{0}{2}} &= \sum_i \frac{a^0}{\lb a^i \rb^2} \lb \phi_{x+i} + \phi_{x-i} - 2\phi_x \rb + \pi_{x-\frac{0}{2}}, \\
	\phi_{x+0} &= \phi_x + a^0 \pi_{x+\frac{0}{2}}.
	\end{align}
}, which is accurate up to second order in the time step $a^0$ and spatial lattice spacings $a^i$.
Using the plane-wave ansatz \eqref{eq:plane_wave_ansatz} we find the dispersion relation
\begin{equation} \label{eq:wave_disp_d1}
\sin^2 \lb \frac{\omega a^0}{2} \rb = \sum_i \lb \frac{a^0}{a^i} \rb^2 \sin^2 \lb \frac{k^i a^i}{2} \rb,
\end{equation}
which is in general non-linear and only yields real-valued (stable) frequencies $\omega$ for all wave vectors $k$ if the Courant-Friedrichs-Lewy (CFL) condition holds
\begin{equation} \label{eq:wave_cfl_d1}
\sum_i \lb \frac{a^0}{a^i} \rb^2 \leq 1.
\end{equation}
The discretization errors of this finite difference scheme result in a non-linear dispersion relation, which is usually referred to as numerical dispersion, since this kind of artificial dispersive behavior of plane waves does not show up in the continuum. If it were possible to set $a^0=a^1=a^2$ (the so-called ``magic time-step") the leapfrog scheme would actually be non-dispersive along the lattice axes, but this choice of the parameters is forbidden by the CFL condition in higher dimensions than $1+1$ and would lead to unstable modes as discussed in \cref{sec:nci_dispersion}.

\subsection{Implicit scheme}
Let us consider a different discretization: we define a new action
\begin{equation} \label{eq:wave_action_d2}
S[\phi] = \frac{1}{2} V \sum_x \lb \lb \p^F_0 \phi_x \rb ^2  - \sum_i \p^F_i \phi_x \p^F_i \overline{\phi}_x \rb,
\end{equation}
where $\overline{\phi}_x$ is the temporally averaged field
\begin{equation} 
\overline{\phi}_x \equiv \frac{\phi_{x+0} + \phi_{x-0}}{2} \approx \phi_x +\mathcal{O} \lb (a^0)^2 \rb.
\end{equation}
Note that only one of the spatial finite differences in the squared term is temporally averaged.
Since this action differs from the leapfrog action \eqref{eq:wave_action_d1} only up to an error term quadratic in $a^0$, the numerical scheme derived from this action will have the same accuracy as the leapfrog scheme.

Repeating the steps as before we obtain the discretized EOM
\begin{equation} \label{eq:wave_eom_d2}
\p^2_0 \phi_x - \sum_i\p^2_i \overline{\phi}_x = 0.
\end{equation}
This is an implicit scheme, which is more complicated to solve compared to the explicit leapfrog scheme given by  \cref{eq:wave_leapfrog_explicit}. Here we have to find the solution to a system of linear equations, which can be accomplished using (for instance) iterative methods.

The dispersion relation for this scheme reads
\begin{equation} \label{eq:wave_disp_d2}
\sin^2 \lb \frac{\omega a^0}{2} \rb 	=  \frac{\sum_i \lb \frac{a^0}{a^i} \rb^2 \sin^2 \lb \frac{k^i a^i}{2} \rb}{1 + 2 \sum_i \lb \frac{a^0}{a^i} \rb^2 \sin^2 \lb \frac{k^i a^i}{2} \rb }.
\end{equation}
This relation can always be solved for real-valued frequencies $\omega$ and therefore the implicit scheme is unconditionally stable. Unfortunately, this does not solve the problem of numerical dispersion either, because there is no choice of lattice parameters that results in a linear dispersion relation. 

We quickly summarize: the first action we considered given by \cref{eq:wave_action_d1} gave us the explicit leapfrog scheme, which is rendered non-dispersive but unstable using the ``magic time-step". The second action, \cref{eq:wave_action_d2}, which we obtained by replacing one of the spatial finite differences with a temporally averaged expression, yields an implicit scheme. This scheme is unconditionally stable, but always dispersive. This suggests that a mixture of both discretizations might solve our problem.

\subsection{Semi-implicit scheme} \label{wave_semi_implicit_scheme}
Finally, we consider the action
\begin{equation} \label{eq:wave_action_d3}
S[\phi] = \frac{1}{2} V \sum_x \lb \lb \p^F_0 \phi_x \rb ^2  - \lb \p^F_1 \phi_x \rb^2 -  \p^F_2 \phi_x \p^F_2 \overline{\phi}_x \rb,
\end{equation}
where the derivatives with respect to (w.r.t.)  $x^1$  are treated like in the leapfrog scheme and the derivatives w.r.t.\ $x^2$ involve a temporally averaged expression as in the implicit scheme.
The EOM now read
\begin{equation} \label{eq:wave_eom_d3}
\p^2_0 \phi_x - \p^2_1 \phi_x - \p^2_2 \overline{\phi}_x = 0.
\end{equation}
We call this numerical scheme semi-implicit, because the finite difference equation contains both explicitly and implicitly treated spatial derivatives.
The dispersion relation associated with  \cref{eq:wave_eom_d3} is given by
\begin{equation} \label{eq:wave_disp_d3}
\sin^2 \lb \frac{\omega a^0}{2} \rb 	=  \cfrac{\sum_i \lb \frac{a^0}{a^i} \rb^2 \sin^2 \lb \frac{k^i a^i}{2} \rb}{ 1 + 2\lb \frac{a^0}{a^2} \rb^2 \sin^2 \lb \frac{k^2 a^2}{2} \rb},
\end{equation}
which is stable if
\begin{equation} \label{eq:wave_cfl_d3}
\lb \frac{a^0}{a^1} \rb^2 \leq 1.
\end{equation}
The CFL condition \eqref{eq:wave_cfl_d3} now allows us to set $a^0=a^1$. Looking at the dispersion relation \eqref{eq:wave_disp_d3} we notice that for $k^1 \neq 0$, but $k^2 = 0$ the propagation becomes non-dispersive, i.e.\ $\omega = k^1$. For $k^2 \neq 0$ and $k^1 = 0$ the propagation still exhibits numerical dispersion. The scheme defined by the action \eqref{eq:wave_action_d3} therefore allows for non-dispersive, stable wave propagation along one particular direction on the lattice. This principle also extends to systems with more spatial dimensions, where one treats a preferred direction explicitly and all other spatial directions implicitly.

\begin{figure}
	\centering
	\begin{subfigure}[b]{0.48\textwidth}
		\centering
		\includegraphics[scale=1.0]{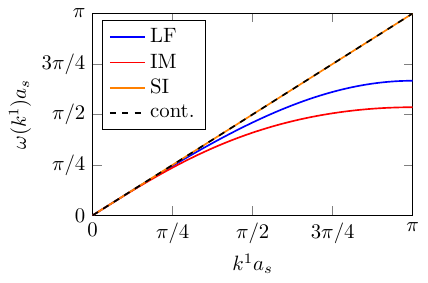}
		\caption{Dispersion relation $\omega(k^1)$ along $x^1$}
	\end{subfigure}
	\begin{subfigure}[b]{0.48\textwidth}
		\centering
		\includegraphics[scale=1.0]{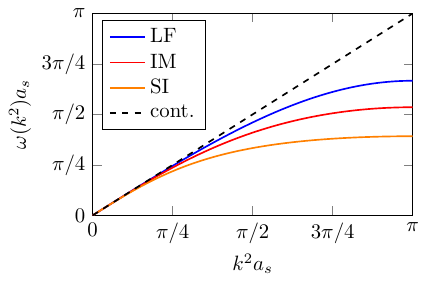}
		\caption{Dispersion relation $\omega(k^2)$ along $x^2$}
	\end{subfigure}
	\caption{Comparison of dispersion relations for the three numerical schemes to the continuum case (cont.): leapfrog (LF), implicit (IM) and semi-implicit (SI). For the leapfrog and the implicit scheme we choose $a^0 = a^1 /2$ and $a^1 = a^2 = a_s$. The dispersion of the semi-implicit scheme is shown for the magic time step $a^0 = a^1$ and $a^1 = a^2 = a_s$. We see that the implicit scheme is generally more dispersive in both directions compared to the leapfrog scheme for the same choice of lattice parameters. The semi-implicit scheme exhibits an ideal dispersion along the $x^1$ direction without numerical dispersion. However, along the orthogonal direction $x^2$ the semi-implicit scheme is less linear than both the leapfrog and the implicit scheme. 
		\label{fig:wave_comparison}}
\end{figure}

A comparison of dispersion relations for all three numerical schemes is shown in \cref{fig:wave_comparison}. Although the semi-implicit scheme is ideal for wave propagation along the $x^1$ direction, it performs worse along the orthogonal direction $x^2$. The semi-implicit scheme therefore offers a trade-off if modified dispersion along other directions is acceptable for the physical situation one is modeling numerically.

\subsection{Solution method and numerical tests}

\begin{figure} [t]
	\centering
	\includegraphics{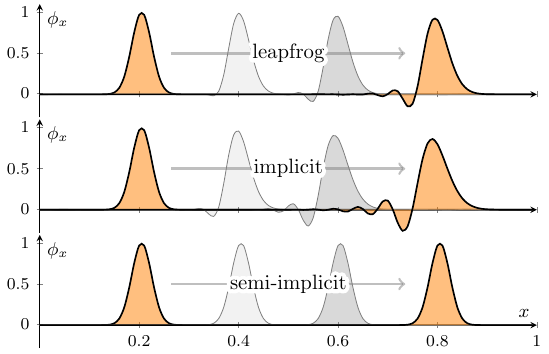}
	\caption{This plot shows a comparison of three different numerical schemes for solving the wave equation in 2+1D: (top) explicit leapfrog scheme \eqref{eq:wave_eom_d1}, (middle) the implicit scheme \eqref{eq:wave_eom_d2}, (bottom) semi-implicit scheme  \eqref{eq:wave_eom_d3} with the ``magic time-step" $a^0=a^1$. The horizontal axis is the $x^1$ coordinate; the vertical axis is the field amplitude $\phi_x$ in arbitrary units. The $x^2$ coordinate is suppressed. The initial condition (seen on the left) is a Gaussian pulse which propagates to the right under time evolution. Due to numerical dispersion of the leapfrog and implicit scheme the original shape of the Gaussian is lost over time. On the other hand, the dispersion-free semi-implicit solver conserves the original pulse shape. This plot is taken from \cite{Ipp:2018hai}. \label{fig:pulse_propagation}}
\end{figure}

To solve the EOM of the implicit or the semi-implicit scheme, one has to solve a linear system of equations. This can be accomplished for instance by inverting a band matrix. Alternatively the equations can also be solved in an iterative manner. Taking the latter approach will be readily applicable to lattice gauge theory.
One example for an iterative method is damped (or relaxed) fixed point iteration: the idea is to first rewrite the EOM \eqref{eq:wave_eom_d3} as a fixed point equation
\begin{equation} \label{eq:wave_fixed_point}
\phi_{x+0} = F \left[ \phi \right] = 2 \phi_x - \phi_{x-0} + \lb a^0 \rb^2 \lb  \p^2_1 \phi_x + \p^2_2 \overline{\phi}_x \rb,
\end{equation}
and then, starting with an initial guess $\phi^{(0)}_{x+0}$ from e.g.\ the explicit leapfrog evolution, use the iteration
\begin{equation}
\phi^{(n+1)}_{x+0} = \alpha \phi^{(n)}_{x+0} + \lb 1 - \alpha \rb F \left[ \phi^{(n)} \right].
\end{equation}
to obtain a new guess $\phi^{(n+1)}_{x+0}$. Here the real-valued parameter $\alpha$ acts as a damping coefficient. Using fixed point iteration might induce other numerical instabilities not covered by the CFL condition \eqref{eq:wave_cfl_d3}. To analyze this we make the ansatz
\begin{equation}
\phi^{(n)}_{x+0} = \lambda^n \varphi_x + \phi^{(\infty)}_{x+0},
\end{equation}
where $\phi^{(\infty)}_{x+0}$ is the true solution to  \cref{eq:wave_fixed_point} and $\varphi_x$ represents a time-independent error term. The growth of the error is determined by the modulus of $\lambda$. Employing a Fourier ansatz $\varphi_x = \exp \lb i \sum_i k^i x^i \rb$ yields
\begin{equation}
\lambda = -2(1-\alpha) \lb \frac{a^0}{a^2} \rb^{2} \sin^{2} \lb \frac{k^2 a^2}{2} \rb +\alpha,
\end{equation}
which is independent of the $k^1$ component and the corresponding lattice spacing $a^1$.
Requiring convergence for high-$k$ modes, i.e.\ $\abs{\lambda} < 1$ for $k^2=\pm \pi / a^2$, we find
\begin{equation} \label{eq:fixed_point_stability}
\frac{2 \dd-1}{2 \dd + 1}<\alpha<1,
\end{equation}
where
\begin{equation}
\dd = \lb \frac{a^0}{a^2} \rb^2.
\end{equation}
In $d+1$ dimensions, where we treat the $i=1$ direction explicitly and all others $2 \leq i \leq d$ implicitly, the stability condition is given by  \cref{eq:fixed_point_stability} with
\begin{equation}
\dd = \sum^d_{i=2} \lb \frac{a^0}{a^i} \rb^2.
\end{equation}
Note that for $\dd < 1/2$ damping is not necessary. A similar stability condition can be derived also for the implicit scheme, which by itself (just from the plane wave analysis) is unconditionally stable. It is important to keep in mind that the use of fixed point iteration can introduce new instabilities depending on the lattice spacing and the time step.

Finally, we perform a crucial numerical test: we compare the propagation of a Gaussian pulse using the three different schemes to show the effects of numerical dispersion and in particular that the semi-implicit scheme is dispersion-free. For simulations using the implicit or semi-implicit method, we solve the equations using damped fixed point iteration. The results are shown in  \cref{fig:pulse_propagation}.

The main insight of this section is that the specific discretization of the action completely fixes the numerical scheme of the discrete equations of motion (and the associated stability and dispersion properties) which one obtains from a discrete variational principle.
The use of temporally averaged quantities in the action leads to implicit schemes. If we treat some derivatives explicitly and some implicitly we can end up with a semi-implicit scheme that can be non-dispersive and still stable for propagation along a single direction on the lattice. As it turns out, this is just what we need to suppress the numerical Cherenkov instability we encountered in our 3+1 dimensional setup for heavy-ion collision simulations. In the next section we will see how we can use the same ``trick" for Abelian gauge fields on the lattice.

\section{Abelian gauge fields on the lattice}\label{abelian_fields}
Before tackling the problem of non-Abelian gauge fields on the lattice, it is instructive to see how we can derive a dispersion-free semi-implicit scheme for discretized Abelian gauge fields. We will approach the problem as before: starting with a discretization of the action, we apply a discrete variational principle to derive discrete equations of motion and constraints. Then we will see what modifications to the action are required to obtain implicit and semi-implicit numerical schemes. Since we are dealing with gauge theory we will take care to retain gauge invariance also for the discretized system.

In the continuum the action for Abelian gauge fields reads
\begin{equation}
S[A] = -\frac{1}{4} \intop_x \sum_{\mu, \nu} F_{\mu \nu}(x) F^{\mu \nu}(x),
\end{equation}
with the field strength tensor given by
\begin{equation}
F_{\mu \nu}(x) = \p_\mu A_\nu(x) - \p_\nu A_\mu(x).
\end{equation}
The field strength and the action are invariant under gauge transformations defined by
\begin{equation}
A_\mu(x) \rightarrow A_\mu(x) + \p_\mu \alpha(x),
\end{equation}
where $\alpha(x)$ is an at least twice differentiable function which defines the gauge transformation.

Varying the action with respect to the gauge field leads to
\begin{equation}
\dd S[A] = \intop_x \sum_{\nu} \lb \sum_\mu \p_\mu F^{\mu\nu}(x) \rb \dd A_\nu(x)=0.
\end{equation}
The term proportional to $\dd A_0(x)$ leads to the GC
\begin{equation}
\sum_i \p_i F_{0i}(x)=0,
\end{equation}
while the term proportional to the variation of spatial components gives the EOM
\begin{equation}
\p_0 F_{i0}(x) = \sum_j \p_j F_{ij}(x).
\end{equation}
It is trivial to see that the EOM imply the conservation of the GC. As we will see next, it is also possible to formulate a discretization of the system where this conservation of the constraint holds exactly.

\subsection{Leapfrog scheme} \label{sec:abelian_leapfrog}
We consider a discretized gauge field $A_{x,\mu}$ at the lattice sites $x$. The field strength tensor $F_{x,\mu\nu}$ at $x$ is defined using forward differences
\begin{equation}
F_{x,\mu\nu} = \p^F_\mu A_{x,\nu} - \p^F_\nu A_{x,\mu},
\end{equation}
which is antisymmetric in the Lorentz index pair $\mu, \nu$ like its continuum analogue.
Furthermore, the lattice field strength is invariant under lattice gauge transformations given by
\begin{equation} \label{eq:lattice_gauge_trans}
A_{x,\mu} \rightarrow A_{x,\mu} + \p^F_\mu \alpha_x,
\end{equation}
where $\alpha_x$ defines the local transformation at each lattice site $x$.
A straightforward discretization of the gauge field action is given by
\begin{equation}\label{eq:abelian_action_d1}
S[A] = \frac{1}{2} V \sum_x \lb \sum_i \ F_{x,0i}^2 - \frac{1}{2} \sum_{i,j} F_{x,ij}^2 \rb,
\end{equation}
where $V=\prod_\mu a^\mu$ is the space-time volume of a unit cell. Due to the invariance of $F_{x,\mu\nu}$ under lattice gauge transformations the action is also gauge invariant.

Performing the variation of \cref{eq:abelian_action_d1} with respect to spatial components $A_{x,i}$ yields the EOM. Using the variation of the magnetic part of the action
\begin{align}
\frac{1}{4} \sum_{x,i,j} \dd \lb F_{x,ij}^2 \rb
&= \frac{1}{2} \sum_{x,i,j} F_{x,ij} \dd F_{x,ij} \nonumber \\
&= \frac{1}{2} \sum_{x,i,j} F_{x,ij} \lb \p^F_i \dd A_{x,j} - \p^F_j \dd A_{x,i} \rb \nonumber \\
&= \sum_{x,i,j} \p^B_j F_{x,ij} \dd A_{x,i},
\end{align}
and the variation of the electric part w.r.t.\ spatial components (denoted by $\dd_s$)
\begin{align}
\frac{1}{2} \sum_{x,i} \dd_s \lb F_{x,0i}^2 \rb
&= \sum_{x,i} F_{x,0i} \p^F_0 \dd A_{x,i} \nonumber \\
&= - \sum_{x,i} \p^B_0 F_{x,i0} \dd A_{x,i},
\end{align}
we find the discrete EOM
\begin{equation} \label{eq:abelian_eom_d1}
\p^B_0 F_{x,i0} = \sum_j \p^B_j F_{x,ij},
\end{equation}
which are of the explicit leapfrog type.
We also obtain a discretized version of the GC by considering the variation w.r.t.\ temporal components $A_{x,0}$ (denoted by $\dd_t$). With
\begin{align}
\frac{1}{2} \sum_{x,i} \dd_t \lb F_{x,0i}^2 \rb
&= - \sum_{x,i} F_{x,0i} \p^F_i \dd A_{x,0} \nonumber \\
&= \sum_{x,i} \p^B_i F_{x,0i} \dd A_{x,0},
\end{align}
we get the constraint
\begin{equation} \label{eq:abelian_constraint_d1}
\sum_i \p^B_i F_{x,0i} = 0.
\end{equation}
Since both the discrete EOM and the constraint follow from the same discretized, gauge-invariant action \eqref{eq:abelian_action_d1}, the GC is guaranteed to be automatically conserved under the EOM. We can show this explicitly via
\begin{align}
\p^B_0 \lb \sum_i \p^B_i F_{x,0i} \rb &= -\sum_i \p^B_i \lb \p^B_0 F_{x,i0} \rb \nonumber \\
&= -\sum_{i,j} \p^B_i \p^B_j F_{x,ij}  = 0,
\end{align}
which is equivalent to
\begin{equation}
\sum_i \p^B_i F_{x,0i} = \sum_i \p^B_i F_{x-0,0i}.
\end{equation}
This means that if the GC is satisfied in one time slice then the EOM will ensure that it remains satisfied in the next time slice.
We can also give a more general proof analogous to the one given in \cref{sec:latt_gauge_inv}: consider an infinitesimal gauge transformation
\begin{equation}
A'_{x,\mu} = A_{x,\mu} + \p^F_\mu \alpha_x,
\end{equation}
and expand the action $S[A']$ for small $\alpha$. We then find
\begin{align}
S[A'] &\simeq S[A] + \sum_{x,y,\mu} \lb \frac{\p S[A']}{\p A'_{x,\mu}} \frac{\p A'_{x,\mu}}{\p \alpha_y} \rb \bigg\rvert_{\alpha=0} \alpha_y +\mathcal{O} \lb \alpha^2 \rb \nonumber \\
&= S[A] + \sum_{x,y,\mu} \frac{\p S[A]}{\p A_{x,\mu}} \p^F_\mu \delta_{xy} \alpha_y +\mathcal{O} \lb \alpha^2 \rb \nonumber \\
&= S[A] - \sum_{x,\mu} \p^B_\mu \frac{\p S[A]}{\p A_{x,\mu}} \alpha_x +\mathcal{O} \lb \alpha^2 \rb.
\end{align}
Since $S[A]$ is invariant for any $\alpha$ it must hold that
\begin{equation}
\sum_{\mu} \p^B_\mu \frac{\p S[A]}{\p A_{x,\mu}} = 0,
\end{equation}
or written slightly differently
\begin{equation}
\p^B_0 \frac{\p S[A]}{\p A_{x,0}} = - \sum_i \p^B_i \frac{\p S[A]}{\p A_{x,i}}.
\end{equation}
If the equations of motion are satisfied in every time slice, i.e.\ ${\p S[A]}/{\p A_{x,i}}=0$, then the GC ${\p S[A]}/{\p A_{x,0}}$ must be conserved from one slice to the next:
\begin{equation}
\p^B_0 \frac{\p S[A]}{\p A_{x,0}} = 0.
\end{equation}
This holds regardless of the exact form of the gauge invariant action $S[A]$. Consequently, it does not matter what kind of discretization of the action we use. As long as $S[A]$ retains lattice gauge invariance in the sense of \cref{eq:lattice_gauge_trans}, we are guaranteed to find that the discrete GC is conserved under the discrete equations of motion.

The EOM given by \cref{eq:abelian_eom_d1} alone are not enough to uniquely determine the time evolution of the field $A_{x,\mu}$: we must specify a gauge condition. As before, we use temporal gauge
\begin{equation}
A_{x,0} = 0,
\end{equation}
which we also use in the case of non-Abelian lattice gauge fields. The EOM then read
\begin{equation} \label{eq:abelian_eom2_d1}
- \p^2_0 A_{x,i} = \sum_j \lb \p^B_j \p^F_i A_{x,j} - \p^2_j A_{x,i} \rb.
\end{equation}
Using a plane wave ansatz 
\begin{equation}
A_{x,i} = A_i e^{i \lb \omega x^0 - \sum_i k^i x^i \rb},
\end{equation}
with amplitude $A_i$, we find the same non-trivial dispersion relation and CFL stability condition as in the case of the leapfrog scheme for the wave equation in 2+1D (see eqs.\ \eqref{eq:wave_disp_d1} and \eqref{eq:wave_cfl_d1}). To show this explicitly we first introduce some notation. Taking a forward (backward) finite difference of the plane wave ansatz yields
\begin{align}
\p^{F}_i A_{x,j} = \frac{e^{-i k^i a^i} - 1}{a^i} A_{x,j}, \\
\p^{B}_i A_{x,j} = \frac{1 - e^{+i k^i a^i}}{a^i} A_{x,j},
\end{align}
which suggests the definition of the forward (backward) lattice momentum
\begin{align}
\kappa^{F}_i = \frac{e^{-i k^i a^i} - 1}{i a^i}, \\
\kappa^{B}_i = \frac{1 - e^{+i k^i a^i}}{i a^i}.
\end{align}
It holds that $\lb \kappa^F_i \rb^\dg = \kappa^B_i$. We define the squared lattice momentum as
\begin{equation}
\kappa^2_i \equiv \kappa^{F}_i \kappa^{B}_i = \lb \frac{2}{a^i} \rb^2 \sin^2 \lb \frac{k^i a^i}{2} \rb.
\end{equation}
Furthermore, we can render the lattice momenta dimensionless by multiplying with $a^0$. We define the dimensionless lattice momentum as
\begin{equation}\label{eq:dimless_lattice_momentum}
\chi^{F/B}_i = \frac{a^0}{2} \kappa^{F/B}_i,
\end{equation}
and
\begin{equation}
\chi^2_i = \chi^F_i \chi^B_i = \lb \frac{a^0}{a^i} \rb^2 \sin^2 \lb \frac{k^i a^i}{2} \rb.
\end{equation}
The factor of $1/2$ in \cref{eq:dimless_lattice_momentum} is introduced for convenience.
For differences with respect to the time coordinate we find $\chi^2_0 = \sin^2 \lb \omega a^0 / 2 \rb$. 
Using these definitions the GC \eqref{eq:abelian_constraint_d1} for the plane wave ansatz can be reduced to
\begin{equation}
\sum_i \chi^B_i A_{i} = 0.
\end{equation}
The discrete EOM \eqref{eq:abelian_eom2_d1} can be written as
\begin{equation}
\chi^2_0 A_{i} = -\sum_j \lb \chi^B_j \chi^F_i A_{j} + \chi^2_j A_{i} \rb.
\end{equation}
One can eliminate the mixed terms $\chi^B_j \chi^F_i A_{j}$ using the GC and find the dispersion relation
\begin{equation}
\chi^2_0 = \sum_j \chi^2_j,
\end{equation}
which is equivalent to the dispersion relation of the wave equation  \eqref{eq:wave_disp_d1}, i.e.\
\begin{equation}
\sin^2 \lb \frac{\omega a^0}{2} \rb = \sum_j \lb \frac{a^0}{a^j} \rb^2 \sin^2 \lb \frac{k^j a^j}{2} \rb.
\end{equation}
\subsection{Implicit scheme}
An implicit scheme analogous to the one we derived for the wave equation (see  \cref{eq:wave_action_d2}) can be found using the action
\begin{equation}\label{eq:abelian_action_d2}
S[A] = \frac{1}{2} V \sum_x \lb \sum_i \ F_{x,0i}^2 - \frac{1}{2} \sum_{i,j} F_{x,ij} M_{x,ij} \rb,
\end{equation}
where we introduce the temporally averaged field-strength
\begin{equation}\label{eq:abelian_M_def}
M_{x,ij} = \avg{F}_{x,ij} =  \frac{1}{2} \lb F_{x+0,ij} + F_{x-0,ij} \rb.
\end{equation}
The averaged field strength $M_{x,ij}$ differs from $F_{x,ij}$ only by an error term proportional to $\lb a^0 \rb^2$.
Replacing one of the field strengths $F_{x,ij}$ in the quadratic term in the action with its temporally averaged expression $M_{x,ij}$ is analogous to replacing the wave amplitude $\phi_x$ with $\overline{\phi}_x=\frac{1}{2} \lb \phi_{x+0} + \phi_{x-0} \rb$ in the action \eqref{eq:wave_action_d2}. The averaged field strength $M_{x,ij}$ is also invariant under lattice gauge transformations:
\begin{equation}
M_{x,ij} \rightarrow M_{x,ij} + \p^F_i \p^F_j \avg{\alpha}_x -  \p^F_j \p^F_i \avg{\alpha}_x = M_{x,ij}.
\end{equation}
Varying the action as we did for the leapfrog scheme we find the EOM
\begin{equation} \label{eq:abelian_eom_d2}
\p^B_0 F_{x,i0} = \sum_j \p^B_j M_{x,ij},
\end{equation}
and employing temporal gauge we have
\begin{equation} \label{eq:abelian_eom2_d2}
- \p^2_0 A_{x,i} = \sum_j \lb \p^B_j \p^F_i \overline{A}_{x,j} - \p^2_j \overline{A}_{x,i} \rb,
\end{equation}
where the fields $A_{x,i}$ have been replaced with temporally averaged expressions $\overline{A}_{x,i}=\frac{1}{2} ( A_{x+0,i} + A_{x-0,i})$ on the right-hand side. 
The GC that arises from varying w.r.t.\ temporal components is simply  \eqref{eq:abelian_constraint_d1}, because we did not modify the term involving $F_{x,0i}$ or introduce new dependencies on $A_{x,0}$. The discrete EOM \eqref{eq:abelian_eom_d2} still conserve the GC due to lattice gauge invariance.
Performing the same steps as in the previous section, we can show that this implicit scheme is unconditionally stable and exhibits the same non-trivial dispersion relation as the implicit scheme for the wave equation, see   \cref{eq:wave_disp_d2}.

\subsection{Semi-implicit scheme} \label{sec:abelian_semi_implicit}

In this section we want to develop a semi-implicit scheme for Abelian gauge fields. Specifically we need an action that allows for dispersion-free propagation of waves in the direction of the $x^1$ coordinate (which we refer to as the longitudinal direction). We call $x^2$ and $x^3$ the transverse coordinates. In this section, Latin indices $i, j, k, ...$ refer to transverse indices and the longitudinal index will always be explicit. Our goal is to define the action in such a way that we obtain equations of motion that include explicit differences in the $x^1$ direction, but temporally averaged finite differences in the $x^2$ and $x^3$ direction. This means that we have to modify the $F_{x,i1}^2$ term of the leapfrog action \eqref{eq:abelian_action_d1} such that it results in terms like $\p^F_i A_{x,1} \p^F_1 \avg{A}_{x,1}$. A first guess for a semi-averaged version of field strength $F_{x,i1}$ that could accomplish this is
\begin{equation}
\p^F_i \avg{A}_{x,1} - \p^F_1 A_{x,i},
\end{equation}
with $\overline{A}_{x,1} = \frac{1}{2} \lb A_{x+0,1} + A_{x-0,1} \rb$. However, it turns out that such a term is not invariant under lattice gauge transformations \eqref{eq:lattice_gauge_trans}. The problem is that $\avg{A}_{x,1}$ transforms differently than $A_{x,i}$. We have
\begin{align}
\avg{A}_{x,1} \rightarrow \avg{A}_{x,1} + \p^F_1 \avg{\alpha}_x, \\
A_{x,i} \rightarrow A_{x,i} + \p^F_i \alpha_x,
\end{align}
which yields
\begin{equation}
\p^F_i \avg{A}_{x,1} - \p^F_1 A_{x,i} \rightarrow \p^F_i \avg{A}_{x,1} - \p^F_1 A_{x,i} + \p^F_i \p^F_1 \lb \avg{\alpha}_x - \alpha_x \rb,
\end{equation}
where the last term $\p^F_i \p^F_1 \lb \avg{\alpha}_x - \alpha_x \rb$ breaks gauge invariance.
To fix this we introduce the ``properly" averaged field strength $\tilde{A}_{x,1}$ given by
\begin{equation} \label{eq:abelian_proper_average}
\tilde{A}_{x,1} \equiv \avg{A}_{x,1} - \frac{1}{2} \lb a^0 \rb^2 \p^F_1 \p^B_0 A_{x,0},
\end{equation}
where the last term transforms as
\begin{align}
\frac{1}{2} \lb a^0 \rb^2 \p^F_1 \p^B_0 A_{x,0} &\rightarrow \frac{1}{2} \lb a^0 \rb^2 \p^F_1 \p^B_0 A_{x,0} + \frac{1}{2} \lb a^0 \rb^2 \p^F_1 \p^2_0 \alpha_{x} \nonumber \\
&= \frac{1}{2} \lb a^0 \rb^2 \p^F_1 \p^B_0 A_{x,0} + \p^F_1 \lb \avg{\alpha}_{x} - \alpha_x \rb. 
\end{align}
Here we made use of the exact relation
\begin{equation}
\frac{1}{2} \lb a^0 \rb^2 \p^2_0 \alpha_x = \avg{\alpha}_x - \alpha_x.
\end{equation}
Therefore, the transformation property of the properly averaged gauge field $\tilde{A}_{x,1}$ is the same as $A_{x,1}$:
\begin{equation}
\tilde{A}_{x,1} \rightarrow \tilde{A}_{x,1} + \p^F_1 \alpha_x.
\end{equation}
It still holds that in the continuum limit the properly averaged gauge field $\tilde{A}_{x,1}$ is the same as $A_{x,1}$ up to an error term quadratic in $a^0$. While the definition \eqref{eq:abelian_proper_average} seems a bit arbitrary at first sight, this way of averaging is more natural using the language of lattice gauge theory: in \cref{semi_scheme} we will find an intuitive picture in terms of Wilson lines that reduces to \cref{eq:abelian_proper_average} in the Abelian limit for small lattice spacings.

The properly semi-averaged gauge-invariant field strength is then given by
\begin{equation} \label{eq:abelian_W_def}
W_{x,i1} \equiv \p^F_i \tilde{A}_{x,1} - \p^F_1 A_{x,i}.
\end{equation}
In order to keep the field strength explicitly antisymmetric we define $W_{x,1i} = - W_{x,i1}$.
Using these definitions we can propose the action
\begin{equation} \label{eq:abelian_action_d3}
S[A] = \frac{1}{2} V \sum_x \bigg(  F_{x,01}^2 + \sum_i F_{x,0i}^2 - \frac{1}{2} \sum_{i,j} F_{x,ij} M_{x,ij}  - \sum_i F_{x,1i} W_{x,1i}\bigg),
\end{equation} 
where the indices $i,j$ denote transverse components. Since we have built the new action from gauge invariant expressions it is also invariant under lattice gauge transformations. The use of $\tilde{A}_{x,i}$ in $W_{x,1i}$ introduces new terms in the action \eqref{eq:abelian_action_d3} dependent on the temporal component of the gauge field. Although these terms disappear in temporal gauge (our preferred choice), they still have an effect on the scheme since we have to perform the variation before choosing a gauge. Therefore, we will obtain a modified GC compatible with the equations of motion derived from the action \eqref{eq:abelian_action_d3} even after setting $A_{x,0} = 0$.

At this point the question might arise if the ``proper" averaging procedure has any effect on the implicit scheme of the previous section as well. It turns out that if one defines the averaged field-strength $M_{x,ij}$ of \cref{eq:abelian_M_def} using the properly averaged gauge field $\tilde{A}_{x,i}$, the action of the implicit scheme (and by extension the EOM and the constraint) remains unchanged. This is due to the fact that the terms proportional to $A_{x,0}$ in \cref{eq:abelian_proper_average} cancel:
\begin{equation}
\p^F_i \tilde{A}_{x,j} - \p^F_j \tilde{A}_{x,i} = \p^F_i \avg{A}_{x,j} - \p^F_j \avg{A}_{x,i}.
\end{equation}
Therefore, no such modification is required in the implicit scheme.

Varying this action w.r.t.\ temporal components $A_{x,0}$ yields the modified GC
\begin{equation} \label{eq:abelian_constraint_d3}
\sum_{i=1}^{d} \p_{i}^{B} F_{x,0i} + \lb \frac{a^{0}}{2} \rb^2 \sum_{i}\p_{i}^{B}\p_{1}^{B}\p_{0}^{F}F_{x,1i} = 0.
\end{equation}
Since $W_{x,1i}$ explicitly depends on $A_{x,0}$, we obtain a correction term to the standard leapfrog GC \eqref{eq:abelian_constraint_d1}. The discrete EOM read
\begin{align} \label{eq:abelian_eom_d3_1}
\p^B_0 F_{x,10} &= \frac{1}{2} \sum_i \p^B_i \lb W_{x,1i} + M_{x,1i} \rb,  \\
\p^B_0 F_{x,i0} &= \sum_{j\neq i} \p^B_j M_{x,ij} + \frac{1}{2} \p^B_1 \lb F_{x,i1} + W_{x,i1} \rb. \label{eq:abelian_eom_d3_2}
\end{align}
We have separate EOM for the longitudinal and transverse components of the gauge field. By replacing the averaged expressions $M$ and $W$ with $F$, the EOM reduce to the leapfrog equations as expected.

The propagation of waves in the semi-implicit scheme turns out to be more complicated compared to the leapfrog or implicit scheme: given a wave vector $k$ and a field amplitude $A_{i}$ (such that the GC \eqref{eq:abelian_constraint_d3} is satisfied) the dispersion relation becomes polarization dependent, i.e.\ the scheme exhibits birefringence. The amplitude $A_i$ of an arbitrary plane wave
\begin{equation}
A_{x,i} = A_i e^{i \lb \omega x^0 - \sum_i k^i x^i \rb},
\end{equation}
has to be split into  a longitudinal and two momentum-dependent transverse components
\begin{equation}
\left\{ \vec{A}_{L},\vec{A}_{T,1},\vec{A}_{T,2}\right\} =
\left\{
\left(
\begin{array}{c}
	1\\
	0\\
	0
\end{array}
\right),
\left(
\begin{array}{c}
	0\\
	-\chi_{3}^{B}\\
	\chi_{2}^{B}
\end{array}
\right),
\left(
\begin{array}{c}
	0\\
	\chi_{2}^{F}\\
	\chi_{3}^{F}
\end{array}
\right)\right\}, 
\end{equation}
where $\chi^{F/B}_i$ are dimensionless lattice momenta given by  \cref{eq:dimless_lattice_momentum}. In \cref{app_abelian_semi} we find that the components $\vec{A}_{L}$ and $\vec{A}_{T,2}$ have the dispersion relation
\begin{equation} \label{eq:semi_dispersion1}
\omega_{1} a^{0}=\arccos\left(\frac{1-\chi_{1}^{2}\left(2+\chi_{2}^{2}+\chi_{3}^{2}\right)}{1+\chi_{2}^{2}\left(2-\chi_{1}^{2}\right)+\chi_{3}^{2}\left(2-\chi_{1}^{2}\right)}\right),
\end{equation}
and the component $\vec{A}_{T,1}$ has a second different dispersion relation
\begin{equation} \label{eq:semi_dispersion2}
\omega_{2}a^{0}=\arccos\left(\frac{1-2\chi_{1}^{2}}{1+2\chi_{2}^{2}+2\chi_{3}^{2}}\right).
\end{equation}
Analyzing the stability of the scheme using the two dispersion relations, yields that it is stable if
\begin{equation}
\chi^2_1 \leq 1,
\end{equation}
which, when requiring stability for all modes, reduces to
\begin{equation}
a^0 \leq a^1.
\end{equation}
If we consider the special case of a plane wave with a  purely transverse amplitude and which propagates in the $x^1$ direction (i.e.\ setting the transverse momenta $\chi_2=\chi_3=0$), we find that both dispersion relations agree:
\begin{equation}
\omega_1 a^0 = \omega_2 a^0 = \arccos \lb 1- 2 \chi^2_1 \rb.
\end{equation}
This dispersion becomes linear if we set $a^0=a^1$. This explicitly shows that the semi-implicit scheme for Abelian gauge fields allows for dispersion-free, stable propagation along the longitudinal direction if we use the ``magic time-step".
The dispersion relations also agree if we set the longitudinal momentum $\chi_1$ to zero for arbitrary transverse momenta. In general however, wave propagation in this scheme is birefractive.
It would be interesting to see if there are alternative discretizations of the action which allow for dispersion-free propagation without being birefractive. 

The main result of this section is the action \eqref{eq:abelian_action_d3} which gives rise to the semi-implicit scheme. Here we used a combination of differently averaged field strengths, $M_{x,ij}$ and $W_{x,1i}$, in the action. Our next goal is to generalize these expressions to non-Abelian gauge fields.

\section{Non-Abelian gauge fields on the lattice}\label{nonabelian_fields}
Recall that the continuum action for non-Abelian Yang-Mills fields is given by
\begin{equation}
S[A] = -\frac{1}{2} \intop_x \sum_{\mu,\nu} \tr \lb F_{\mu\nu}(x) F^{\mu\nu}(x) \rb,
\end{equation}
where the field strength is
\begin{equation}
F_{\mu\nu}(x) = \p_\mu A_\nu(x) - \p_\nu A_\mu(x) + i g \left[ A_\mu(x), A_\nu(x) \right].
\end{equation}
Through variation of the action we obtain the GC and the EOM:
\begin{align}
\sum_i D_i F^{i 0}(x) &= 0, \\
D_0 F^{0 i}(x) &= - \sum_j D_j F^{i j}(x),
\end{align}
where the gauge-covariant derivative acting on an algebra element $\chi$ is given by
\begin{equation}
D_\mu \chi(x) \equiv \p_\mu\chi(x) + i g \left[ A_\mu(x),\chi(x)  \right].
\end{equation}
In real-time lattice gauge theory, instead of gauge fields $A_{\mu}(x)$, we use gauge links (or link variables) $U_{x,\mu}$ as already discussed previously. 

The standard discretization using plaquettes leads to the Wilson gauge action (see \cref{sec:real_time_lattice})
\begin{align} \label{eq:wilson_action_2nd}
S[U] = \frac{V}{g^2} \sum_x \bigg( & \sum_i \frac{1}{\lb a^0 a^i \rb^2} \tr \lb 2 - U_{x,0i} - U^\dg_{x,0i} \rb \nonumber \\
- \frac{1}{2} & \sum_{i,j} \frac{1}{\lb a^i a^j \rb^2}  \tr \lb 2 - U_{x,ij} - U^\dg_{x,ij} \rb \bigg),
\end{align}
where $\sum_i$ denotes a sum over all spatial components. Using
\begin{align} \label{eq:plaquette_trace}
\tr \lb 2 - U_{x,\mu\nu} - U^\dg_{x,\mu\nu} \rb &\simeq \tr \lb F_{x,\mu\nu}^2 \rb \nonumber \\
&\simeq \frac{1}{2}\sum_a \lb g a^\mu a^\nu F^a_{\mu\nu}(x) \rb^2,
\end{align}
it is clear that the action \eqref{eq:wilson_action_2nd} is a discretization of the continuum Yang-Mills action
\begin{equation} \label{eq:ym_action}
S[A] = \frac{1}{2} \intop_x \lb \sum_{a,i} F^a_{0i}(x)F^a_{0i}(x)- \frac{1}{2} \sum_{a,i,j} F^a_{ij}(x)F^a_{ij}(x) \rb,
\end{equation}
where we made the split into temporal and spatial components explicit. Since it is built from gauge-invariant expressions, the Wilson gauge action \eqref{eq:wilson_action_2nd} is invariant under lattice gauge transformations.
\begin{figure}[tbp]
	\centering
	\includegraphics{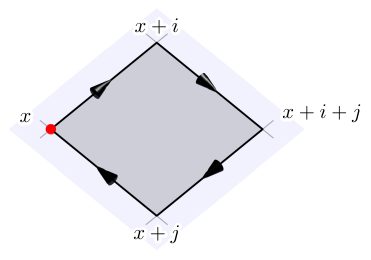} \hspace{25pt}
	\includegraphics{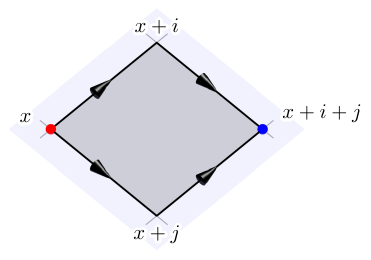}
	
	\caption{ \textit{Left}: the Wilson line associated with the plaquette $U_{x,ij}$. \textit{Right}: the Wilson lines associated with the lattice field strength $C_{x,ij}$.
		Spatial link variables are drawn as solid black arrows.
		While the plaquette starts and ends at the same lattice site $x$ (red dot), the lattice field-strength $C_{x,ij}$ starts at $x$ and ends at $x+i+j$ (blue dot). This figure is taken from \cite{Ipp:2018hai}.
		\label{fig:paths1}}
\end{figure}
At this point we remark that the continuum Yang-Mills action \eqref{eq:ym_action} and its discretization \eqref{eq:wilson_action_2nd} look very different: while the Yang-Mills action is given in terms of squares of the field strength tensor, the Wilson gauge action is linear in plaquette variables. In terms of plaquettes it is not immediately clear how we might generalize our approach from section \ref{abelian_fields}. Fortunately, the action \eqref{eq:wilson_action_2nd} can be written differently so that its functional form is more similar to \cref{eq:ym_action}. We define (see for instance p.\ 94 of \cite{smit_2002})
\begin{equation} \label{eq:lattice_fst}
C_{x,\mu\nu} \equiv U_{x,\mu} U_{x+\mu,\nu} - U_{x,\nu} U_{x+\nu,\mu},
\end{equation}
which transforms non-locally
\begin{equation}
C_{x,\mu\nu} \rightarrow V_{x} C_{x,\mu\nu} V^\dg_{x+\mu+\nu}.
\end{equation}
For comparison to the plaquette $U_{x,ij}$, the path traced by $C_{x,ij}$ is shown in \cref{fig:paths1} on the right. In the continuum limit $C_{x,\mu\nu}$ can be identified (up to constant factors) with the field strength $F_{\mu\nu}(x)$: expanding for small lattice spacing we find
\begin{equation}
C_{x,\mu\nu} \simeq i g a^\mu a^\nu F_{\mu\nu}(x).
\end{equation}
Most noteworthy is the exact relation
\begin{equation}
C_{x,\mu\nu} C^\dg_{x,\mu\nu} = 2 - U_{x,\mu\nu} -U^\dg_{x,\mu\nu},
\end{equation}
with which we can identically rewrite the action as
\begin{equation} \label{eq:wilson_action2}
S[U] = \frac{V}{g^2} \sum_x \bigg( \sum_i \frac{1}{\lb a^0 a^i \rb^2} \tr \lb C_{x,0i} C^\dg_{x,0i} \rb -\frac{1}{2} \sum_{i,j} \frac{1}{\lb a^i a^j \rb^2}  \tr \lb C_{x,ij} C^\dg_{x,ij} \rb \bigg).
\end{equation}
This functional form of the rewritten action \eqref{eq:wilson_action2} is now virtually the same as the continuum case \eqref{eq:ym_action}. We will exploit this when generalizing the implicit \eqref{eq:abelian_action_d2} and semi-implicit schemes \eqref{eq:abelian_action_d3} to non-Abelian gauge fields.

\subsection{Leapfrog scheme} \label{sec:semi_leapfrog_scheme}

For comprehensiveness we summarize the derivation of the Lagrangian leapfrog method. Following the same procedure as in the case of Abelian gauge fields, we vary \cref{eq:wilson_action2} w.r.t.\ spatial components $U_{x,i}$ to obtain the discrete EOM and w.r.t.\ temporal components $U_{x,0}$ to find the GC.

Starting with the constraint we get (see \cref{app_leapfrog_gauss} for details)
\begin{equation} \label{eq:leapfrog_gauss}
\sum_i \frac{1}{\lb a^0 a^i \rb^2} P^a \lb U_{x,0i} + U_{x,0-i}\rb = 0,
\end{equation}
where
\begin{equation}
P^a\lb U \rb = 2 \, \Im \, \tr \lb t^a U \rb.
\end{equation}
Varying the spatial link variables we obtain the EOM (see \cref{app_leapfrog_eom} for more details)

\begin{equation}\label{eq:leapfrog_eom1}
\frac{1}{\lb a^0 a^i \rb^2} P^a \lb U_{x,i0} + U_{x,i-0}\rb =
- \sum_j \frac{1}{\lb a^i a^j \rb^2} P^a \lb U_{x,i} \lb U_{x+i,j} C^\dg_{x,ij} + C^\dg_{x-j,ji} U_{x-j,j}\rb\rb,
\end{equation}
which upon using the definition of $C_{x,ij}$ can be written in the more familiar form
\begin{equation} \label{eq:leapfrog_eom}
\frac{1}{\lb a^0 a^i \rb^2} P^a \lb U_{x,i0} + U_{x,i-0}\rb =
\sum_j \frac{1}{\lb a^i a^j \rb^2} P^a \lb U_{x,ij} + U_{x,i-j}\rb.
\end{equation}
As before, time evolution under the EOM conserves the GC exactly (see \cref{app_gauss}). In order to actually solve the equations we specify the temporal gauge
\begin{equation}
U_{x,0} = \one,
\end{equation}
which enables us to compute the spatial link variables of the next time-slice using past links and the temporal plaquette:
\begin{equation} \label{eq:link_evolve}
U_{x+0,i} = U_{x,0i} U_{x,i}.
\end{equation}
The temporal plaquette $U_{x,0i}$ has to be determined from  \cref{eq:leapfrog_eom}. For SU(2) this can be done explicitly (see \cref{sec:latt_field__eqs}).

\subsection{Implicit scheme} \label{implicit_scheme}

Guided by what we learned from the Abelian case in \cref{abelian_fields}, we would like to replace one of the $C_{x,ij}$ expressions in the Wilson action \eqref{eq:wilson_action2} with a time-averaged equivalent. At the same time we need to retain the gauge invariance of the action. Simply using the temporally averaged expression
\begin{equation} \label{eq:C_simple_avg}
\frac{1}{2} \lb C_{x+0,ij} + C_{x-0,ij} \rb
\end{equation}
is not enough because $C_{x+0,ij}$ and $C_{x-0,ij}$ transform differently. A solution is to include temporal gauge links in order to ``pull back" $C_{x+0,ij}$ and $C_{x-0,ij}$ to the lattice site $x$. This leads us to the definition of the ``properly" averaged field strength
\begin{equation} \label{eq:implicit_M_def}
M_{x,ij} \equiv \frac{1}{2} \lb U_{x,0} C_{x+0,ij} U_{x+i+j,-0} + U_{x,-0} C_{x-0,ij} U_{x+i+j-0,0} \rb,
\end{equation}
which transforms like $C_{x,ij}$, i.e.
\begin{equation}
M_{x,ij} \rightarrow V_{x} M_{x,ij} V^\dg_{x+i+j}.
\end{equation}
This gauge-covariant averaging procedure can be generalized: consider an object $\mathcal{X}_{x,y}$ that transforms like
\begin{equation} \label{eq:X_transf}
\mathcal{X}_{x,y} \rightarrow V_x \mathcal{X}_{x,y} V^\dg_y.
\end{equation}
As an example, $\mathcal{X}_{x,y}$ could be a Wilson line connecting points $x$ and $y$ along some arbitrary path. A time-averaged version of $\mathcal{X}_{x,y}$ is given by
\begin{equation} \label{eq:time_avg}
\avg{\mathcal{X}}_{x,y} = \frac{1}{2} \lb U_{x,0} \mathcal{X}_{x+0,y+0} U_{y+0,-0} 
+ U_{x,-0} \mathcal{X}_{x-0,y-0} U_{y-0,0}\rb,
\end{equation}
where $\mathcal{X}_{x\pm 0,y\pm 0}$ is simply $\mathcal X _{x,y}$ shifted up (or down) by one time step.
It still transforms like  \cref{eq:X_transf}, i.e.
\begin{equation}
\avg{\mathcal{X}}_{x,y} \rightarrow V_x \avg{\mathcal{X}}_{x,y} V^\dg_y.
\end{equation}
Using this we can write
\begin{equation}
M_{x,ij} = \avg{C}_{x,ij}.
\end{equation}
Condensing our notation even further we write
\begin{align}
C^{(+0)}_{x,ij} &= U_{x,0} C_{x+0,ij} U_{x+i+j,-0}, \\
C^{(-0)}_{x,ij} &= U_{x,-0} C_{x-0,ij} U_{x+i+j-0,0},
\end{align}
and
\begin{equation}
M_{x,ij} = \frac{1}{2} \lb C^{(+0)}_{x,ij} + C^{(-0)}_{x,ij} \rb.
\end{equation}
It still holds that
\begin{equation}
M_{x,ij} \simeq C_{x,ij} + \mathcal{O} \lb \lb a^0 \rb^2 \rb,
\end{equation}
so the use of $M_{x,ij}$ (instead of $C_{x,ij}$) does not change the accuracy of the scheme.
Note that temporal gauge renders all temporal link variables trivial and  \cref{eq:time_avg} reduces to the simple time average.

%
\begin{figure}[t]
	\centering
	\includegraphics{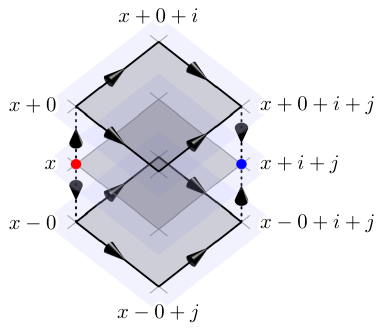} \hspace{25pt}
	\includegraphics{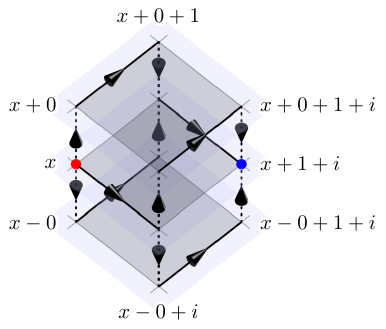}
	\caption{\textit{Left}: the Wilson lines associated with the properly time-averaged field strength $M_{x,ij}$ used in the implicit scheme.
		\textit{Right}: the Wilson lines associated with the partially averaged field-strength $W_{x,1i}$ used in the semi-implicit scheme.
		Spatial link variables are drawn as solid black arrows; temporal links as dashed arrows. The shaded planes represent equal time slices, i.e.\ the spatial lattice (only two dimensions shown) in subsequent time steps.
		The averaged field-strength $M_{x,ij}$ has temporal link connections only at the start and end points. Therefore the diagonal from $x$ to $x+i+j$ is a preferred direction. This asymmetry can be repaired by including $M_{x,i-j}$ terms in the action \eqref{eq:implicit_action}.
		We can also see that in the continuous time limit (the vertically stacked time slices would merge into one) the paths traced by $M_{x,ij}$ and $W_{x,1i}$ would become identical to $C_{x,ij}$. This figure is taken from \cite{Ipp:2018hai}.
		\label{fig:paths2}}
\end{figure}

The inclusion of temporal links in $M_{x,ij}$ breaks a symmetry on the lattice: the diagonal from $x$ to $x+i+j$ is now a preferred direction, which can be seen in \cref{fig:paths2}. The loss of symmetry can be mitigated by also including terms like $M_{x,i-j}$ in the action. Therefore we propose the action

\begin{equation} \label{eq:implicit_action}
S[U] = \frac{V}{g^2} \sum_x \bigg( \sum_i \frac{1}{\lb a^0 a^i \rb^2} \tr \lb C_{x,0i} C^\dg_{x,0i} \rb -\frac{1}{4} \sum_{i, \abs{j}} \frac{1}{\lb a^i a^j \rb^2}  \tr \big( C_{x,ij} M^\dg_{x,ij} \big) \bigg),
\end{equation}
where we explicitly include terms with negative spatial indices using the sum $\sum_{\abs{j}}$ over positive and negative components $j$ to keep the action as symmetric as possible. The action is also invariant under time reversal, real-valued (see \cref{app_implicit} for a proof) and gauge invariant. While we have not made any changes to the terms involving temporal plaquettes, the spatial plaquette terms now include temporal links and therefore we will also obtain a modified GC like in the semi-implicit scheme for Abelian fields.

Performing the variation the same way we did for the leapfrog scheme, we obtain the GC (see  \cref{app_implicit_gauss})
\begin{equation} \label{eq:implicit_gauss}
\sum_{i} \frac{1}{\left(a^{0}a^{i}\right)^{2}} P^a \lb U_{x,0i}+U_{x,0-i} \rb =
- \sum_{\abs i,\abs j} \frac{1}{8} \frac{1}{\lb a^{i}a^{j} \rb^{2}} P^a \lb C_{x,ij}^{(+0)}C_{x,ij}^\dg \rb,
\end{equation}
where $\sum_{\abs i}$ denotes the sum over positive and negative indices $i$. The left hand side (LHS) of \cref{eq:implicit_gauss} is the same as in the leapfrog scheme  \eqref{eq:leapfrog_gauss}, but now there is also a new term on the right hand side (RHS) from varying the spatial part of the action. Performing the continuum limit for the GC (after multiplying both sides with $a^0$), the RHS term vanishes as $\mathcal{O} \lb \lb a^0 \rb^2 \rb$. This shows that the RHS is not a physical contribution, but rather an artifact of the implicit scheme. The correct continuum limit of the constraint (and the EOM) is already guaranteed by the action.

In a similar fashion as before, we perform the variation w.r.t.\ spatial link variables to get the discrete EOM. We find (see \cref{app_implicit_eom})
\begin{equation}\label{eq:implicit_eom}
\frac{1}{\lb a^0 a^i \rb^2} P^a \lb U_{x,i0} + U_{x,i-0}\rb =
- \frac{1}{2} \sum_{\abs j} \frac{1}{\lb a^i a^j \rb^2} P^a \lb U_{x,i} \lb U_{x+i,j} M^\dg_{x,ij} + M^\dg_{x-j,ji} U_{x-j,j}\rb\rb,
\end{equation}
which is formally similar to the leapfrog scheme \eqref{eq:leapfrog_eom1} with $M_{x,ij}$ in place of $C_{x,ij}$, a sum over positive and negative components $j$ (instead of just positive indices) and an additional factor of $1/2$ to avoid overcounting. As a simple check one can replace $M_{x,ij}$ with $C_{x,ij}$ in  \cref{eq:implicit_eom} (only introducing an irrelevant error term quadratic in $a^0$) and recover  \cref{eq:leapfrog_eom1}.

Compared to the leapfrog scheme, solving  \cref{eq:implicit_eom} is more complicated: it is not possible to explicitly solve for the temporal plaquette $U_{x,i0}$ anymore because $M_{x,ij}$ on the RHS involves contributions from both ``past" and ``future" link variables. Completely analogous to the case of Abelian gauge fields on a lattice, we obtain an implicit scheme by introducing time-averaged field strength terms in the action.

Similar to what we did in \cref{wave_semi_implicit_scheme}, we propose to solve  \cref{eq:implicit_eom} iteratively using damped fixed point iteration. Starting from an initial guess for the future link variable $U^{(0)}_{x+0,i}$, for instance by performing a single evolution step using the leapfrog scheme, we iterate
\begin{align}\label{eq:implicit_eom_iteration1}
\frac{1}{\lb a^0 a^i \rb^2} \lb \mathcal{U}^a +  P^a \lb U_{x,i-0}\rb \rb &=
- \frac{1}{2} \sum_{\abs j} \frac{1}{\lb a^i a^j \rb^2} P^a \lb U_{x,i} \lb U_{x+i,j} M^{(n)\dg}_{x,ij} + M^{(n)\dg}_{x-j,ji} U_{x-j,j}\rb\rb, \\ \label{eq:implicit_eom_iteration2}
P^a \lb U^{(n+1)}_{x,i0} \rb &= \alpha P^a \lb  U^{(n)}_{x,i0} \rb + \lb 1 - \alpha \rb \mathcal U^{a},
\end{align}
using $U^{(n)}_{x+0,i}$ in $M^{(n)}_{x,ij}$ from the last iteration step to determine  $U^{(n+1)}_{x,i0}$. In the above equations $P^a \lb U_{x,i0}\rb$ has been replaced with the unknown variable  $\, \mathcal{U}^a$. We first solve  \cref{eq:implicit_eom_iteration1} for  $\,\mathcal U^{a}$ and then update $P^a \lb U^{(n+1)}_{x,i0} \rb$ using  \cref{eq:implicit_eom_iteration2}. The parameter $\alpha$ is used as a damping coefficient to mitigate numerical instabilities induced by fixed point iteration. For SU(2) we construct the temporal plaquette $U^{(n+1)}_{x,i0}$ from $P^a \lb U^{(n+1)}_{x,i0} \rb$ using eqs.\ \eqref{eq:su2_evolve1} and \eqref{eq:su2_evolve2} and using temporal gauge we update the link variables via
\begin{equation}
U^{(n+1)}_{x+0,i} = U^{(n+1)}_{x,0i} U_{x,i}.
\end{equation}
Then we repeat the iteration until convergence.

This iteration scheme can be used to solve the EOM \eqref{eq:implicit_eom} until the GC \eqref{eq:implicit_gauss} is satisfied up to the desired numerical accuracy. Conversely, this means that unlike the leapfrog scheme, where the GC \eqref{eq:leapfrog_gauss} is always satisfied up to machine precision in a single evolution step, the implicit scheme, solved via an iterative scheme, only approximately conserves the GC \cref{eq:implicit_gauss}.
However, in \cref{tests} we will show that using a high number of iterations the constraint can be indeed fulfilled to arbitrary accuracy. In practice however, we find that a lower number of iterations is sufficient for stable and acceptably accurate simulations at the cost of small violations of the constraint. 

It is also immediately obvious that solving the implicit scheme requires higher computational effort compared to the leapfrog scheme. Considering that one has to use the leapfrog scheme for a single evolution step once (as an initial guess) and then use fixed point iteration, where every step is at least as computationally demanding as a single leapfrog step, it becomes clear that the use of an implicit scheme is only viable if increased stability allows the use of coarser lattices while maintaining accurate results.

\subsection{Semi-implicit scheme} \label{semi_scheme}

Using our knowledge from sections \ref{abelian_fields} and \ref{implicit_scheme} we can now generalize the semi-implicit scheme to real-time lattice gauge theory. An appropriate generalization of the semi-averaged field strength \eqref{eq:abelian_W_def} is given by
\begin{equation}
W_{x,1i} = \frac{1}{2} \lb U^{(+0)}_{x,1} +  U^{(-0)}_{x,1}  \rb U_{x+1,i} - \frac{1}{2} U_{x,i} \lb U^{(+0)}_{x+i,1} +  U^{(-0)}_{x+i,1} \rb, 
\end{equation}
where
\begin{align}
U^{(+0)}_{x,\mu} &= U_{x,0} U_{x+0,\mu} U_{x+\mu+0, -0}, \\
U^{(-0)}_{x,\mu} &= U_{x,-0} U_{x-0,\mu} U_{x+\mu-0, 0}.
\end{align}
We also define $W_{x,i1} \equiv - W_{x,1i}$.
Using the time-averaging notation (see  \cref{eq:time_avg}) this can be written more compactly as
\begin{equation}
W_{x,1i} = \avg{U}_{x,1} U_{x+1,i} - U_{x,i} \avg{U}_{x+i,1},
\end{equation}
where
\begin{equation} \label{eq:nonabelian_proper_average}
\avg{U}_{x,1} \equiv \frac{1}{2} \lb U^{(+0)}_{x,1} +  U^{(-0)}_{x,1} \rb.
\end{equation}
Note that
\begin{equation}
\avg{U}_{x,1} \simeq U_{x,1} + \mathcal O \lb \lb a^0 \rb^2 \rb,
\end{equation}
which shows that the semi-averaged field strength $W_{x,1i}$ only differs from $C_{x,1i}$ by an irrelevant error term. Taking the Abelian limit (i.e.\ neglecting commutator terms) of  \cref{eq:nonabelian_proper_average} and expanding for small lattice spacing yields
\begin{equation}
\avg{U}_{x,1} = 1+ia^{1}\left(\avg{A}_{x,1}-\frac{1}{2}\left(a^{0}\right)^{2}\p_{1}^{F}\p_{0}^{B}A_{x,0}\right)+\mathcal{O}\left(\lb a^1 \rb^{2}\right).
\end{equation}
We find that the linear term of the gauge-covariant average \eqref{eq:nonabelian_proper_average} agrees with the expression we constructed in the Abelian semi-implicit scheme \eqref{eq:abelian_proper_average}. While we had to include an ``arbitrary" correction term in the Abelian scheme to fix gauge invariance, the link formalism of lattice gauge theory forces us to only consider closed paths constructed from gauge links in the action and thus naturally leads us to the ``proper" averaging procedure. 

The Wilson line path traced by $W_{x,1i}$, as compared to $M_{x,ij}$, is shown in \cref{fig:paths2} on the right. As before, we keep the action as symmetric as possible by also including terms with negative transverse directions, i.e.\ $W_{x,1-i}$. Inspired by the Abelian semi-implicit case \eqref{eq:abelian_action_d3}, we define the new action as

\begin{align}
\label{eq:semiimplicit_action}
S[U] &= \frac{V}{g^2} \sum_x \bigg( \frac{1}{\lb a^0 a^1 \rb^2} \tr \lb C_{x,01} C^\dg_{x,01} \rb + \sum_i \frac{1}{\lb a^0 a^i \rb^2} \tr \lb C_{x,0i} C^\dg_{x,0i} \rb  \nonumber \\
& \quad -\frac{1}{4} \sum_{i,\abs j} \frac{1}{\lb a^i a^j \rb^2}  \tr \lb C_{x,ij} M^\dg_{x,ij} \rb
-\frac{1}{4} \sum_{\abs j} \frac{1}{\lb a^1 a^j \rb^2}  \tr \lb C_{x,1j} W^\dg_{x,1j} + \hc \rb
\bigg),
\end{align}
where the sum over $i$ and $j$ only run over transverse coordinates and $x^1$ is the longitudinal coordinate. The purely transverse part of the action uses the same terms as the implicit scheme, see  \cref{eq:implicit_action}. The longitudinal-transverse part is now given in terms of $C_{x,1j}$ and $W_{x,1j}$ analogous to  \cref{eq:abelian_action_d2}. We have to explicitly include the hermitian conjugate in order to keep the action real-valued.

Varying with respect to temporal components yields the GC (see \cref{app_semi_gauss})
\begin{align} \label{eq:semi_gauss}
\sum^3_{i=1} \frac{1}{\left(a^{0}a^{i}\right)^{2}} P^a \lb U_{x,0i}+U_{x,0-i} \rb = - \sum_{\abs i,\abs j} \frac{1}{8} \frac{1}{\lb a^{i}a^{j} \rb^{2}} P^a \lb C_{x,ij}^{(+0)}C_{x,ij}^\dg \rb \qquad \qquad \nonumber \\
\qquad \qquad - \frac{1}{8} \frac{1}{\lb a^1 \rb^2} P^a \bigg(U^{(+0)}_{x,1} T^\dg_{x,1} + T^{(+0)}_{x,1} U^\dg_{x,1} + U^{(+0)}_{x-1,1} T^\dg_{x-1,1} + T^{(+0)}_{x-1,1} U^\dg_{x-1,1}  \bigg),
\end{align}
where we use the shorthand
\begin{align}
T_{x,1} &\equiv \sum_{ \abs j} \frac{1}{\lb a^j \rb^2} \lb C_{x,1j} U_{x+1+j,-j} - U_{x,-j} C_{x-j,1j} \rb \nonumber \\
& = \sum_{ \abs j}  \frac{1}{\lb a^j \rb^2} \lb 2 - U_{x,j1} -U_{x,-j1} \rb U_{x,1}.
\end{align}
Varying w.r.t.\ spatial links the discrete semi-implicit EOM read (see \cref{app_semi_eom})
\begin{align}\label{eq:semi_eom_1}
\frac{1}{\lb a^0 a^1 \rb^2} P^a \lb U_{x,10} + U_{x,1-0}\rb =&
- \frac{1}{4} \sum_{ \abs j} \frac{1}{\lb a^1 a^j \rb^2} P^a \bigg( U_{x,1} \bigg( U_{x+1,j} W^\dg_{x,1j} + W^\dg_{x-j,j1} U_{x-j,j} \nonumber \\
& \quad + \avg{ \lb U_{x+1,j} C^\dg_{x,1j} + C^\dg_{x-j,j1} U_{x-j,j}\rb}  \bigg) \bigg),
\end{align}
and for transverse components
\begin{align}\label{eq:semi_eom_i}
\frac{1}{\lb a^0 a^i \rb^2} P^a \lb U_{x,i0} + U_{x,i-0}\rb =&
- \frac{1}{2} \sum_{\abs j} \frac{1}{\lb a^i a^j \rb^2} P^a \lb U_{x,i} \lb U_{x+i,j} M^\dg_{x,ij} + M^\dg_{x-j,ji} U_{x-j,j} \rb \rb \nonumber \\
&- \frac{1}{4} \frac{1}{\lb a^i a^1 \rb^2} \sum_{\abs 1} P^a \bigg( U_{x,i} \bigg( \lb U_{x+i,1} W^\dg_{x,i1} + W^\dg_{x-1,1i} U_{x-1,1} \rb \nonumber \\
& + \lb \avg{U}_{x+i,1} C^\dg_{x,i1} + C^\dg_{x-1,1i} \avg{U}_{x-1,1} \rb \bigg) \bigg),
\end{align}
where $\sum_{\abs 1}$ simply means summing over the terms with positive and negative longitudinal directions.
We now have two sets of equations: one for longitudinal components, and one for the two transverse components.

The equations of motion can be written more compactly by introducing the symbol
\begin{equation}
K_{x,ij}[U, C] = - \frac{1}{2} \frac{1}{\lb a^i a^j \rb^2} \lb U_{x+i,j} C^\dg_{x,ij} - C^\dg_{x-j,ij} U_{x-j,j}\rb,
\end{equation}
where $C$ can be exchanged for corresponding expressions with $M$ or $W$ and $U$ can be exchanged for its temporally averaged version $\avg{U}$. The longitudinal component of the EOM then reads
\begin{align}
\frac{1}{\lb a^{0}a^{1} \rb^{2}} P^{a} \lb U_{x,10}+U_{x,1-0} \rb =  \frac{1}{2} \sum_{\abs i} P^a \lb U_{x,1} \lb K_{x,1i}[U, W] + \avg{K}_{x,1i}[U, C] \rb \rb,
\end{align}
and the transverse components are given by
\begin{align}
\frac{1}{\lb a^{0}a^{i} \rb^{2}} P^{a} \lb U_{x,i0}+U_{x,i-0} \rb &= P^a \bigg( U_{x,i} \bigg( \sum_{\abs j} K_{x,ij}[U, M] \nonumber \\
&\quad + \frac{1}{2} \sum_{\abs 1} \lb K_{x,i1}[U, W] + K_{x,i1}[\avg{U}, C] \rb \bigg) \bigg).
\end{align}
These equations can be solved numerically using damped fixed point iteration completely analogously to \cref{implicit_scheme}. First, one obtains an initial guess $U^{(0)}_{x+0,i}$ for ``future" link variables from a single leapfrog evolution step using eqs.\ \eqref{eq:leapfrog_eom} and \eqref{eq:link_evolve}. Then one iterates from $n=1$ until convergence:
\begin{enumerate} 
	\item Compute the next iteration using damped fixed point iteration: in eqs.\ \eqref{eq:semi_eom_1} and \eqref{eq:semi_eom_i} replace $P^a \lb U_{x,10}\rb \rightarrow \mathcal{U}^a_1$ and $P^a \lb U_{x,i0} \rb \rightarrow \mathcal{U}^a_i$, solve for the unknown $\mathcal U$'s and update the temporal plaquettes using
	\begin{equation}
	P^a \lb U^{(n)}_{x,10} \rb = \alpha P^a \lb  U^{(n-1)}_{x,10} \rb + \lb 1 - \alpha \rb \mathcal U^{a}_1
	\end{equation}
	and analogously for $U^{(n)}_{x,i0}$ and $\mathcal U^{a}_i$, where $\alpha$ is the damping coefficient.
	\item For SU(2) we can reconstruct the full temporal plaquette from its components  $P^a \lb U \rb$ with the identity (see \cref{eq:su2_plaq_from_ah})
	\begin{equation}
	U = \sqrt{1- \frac{1}{4} \sum_a  P^a \lb U \rb ^2} \, \one + \frac{i}{2} \sum_a \sigma^a P^a \lb U \rb,
	\end{equation}
	for $U = U^{(n)}_{x,10}$ and $U = U^{(n)}_{x,i0}$. 
	\item Using $U^{(n)}_{x,10}$ and $U^{(n)}_{x,i0}$, compute the spatial links $U^{(n)}_{x+0,1}$ and $U^{(n)}_{x+0,i}$ via
	\begin{equation}
	U^{(n)}_{x+0,i} = U^{(n)}_{x,0i} U_{x,i}.
	\end{equation}
	\item Repeat with $n\rightarrow n+1$.
\end{enumerate}
As with the implicit scheme, our approach to solving the equations in the semi-implicit scheme is an iterative one. The GC \eqref{eq:semi_gauss} is only approximately satisfied, depending on the degree of convergence\footnote{This is also true for the Abelian semi-implicit scheme. If the equations of motion are solved only approximately using an iterative method, then the conservation of the GC is also only approximate depending on the degree of convergence.}.

\subsection{Coupling to external color currents} \label{external_charges}

Up until now we have only considered pure Yang-Mills fields. In the continuum we can include external color currents by adding a $J \cdot A$ term to the action.
\begin{equation}
S[A] = S_{YM} + S_J = -\frac{1}{2} \intop_x \sum_{\mu, \nu} \tr \lb F_{\mu \nu}(x) F^{\mu \nu}(x) \rb - 2 \intop_x \sum_{\nu} \tr \lb J^\nu(x) A_{\nu}(x) \rb. 
\end{equation}
The EOM then read
\begin{equation}
\sum_\mu D_{\mu} F^{\mu\nu}(x) = J^\nu(x),
\end{equation}
and due to gauge-covariant conservation of charge we have
\begin{equation}
\sum_\mu D_{\mu} J^{\mu}(x)=0,
\end{equation}
which is the non-Abelian continuity equation. On the lattice we can simply add a discrete $J\cdot A$ term to the action as well:
\begin{equation}
S_J = \frac{V}{g^2} \sum_{x,b} \lb - \frac{g}{a^0} \rho^b_x A^b_{x,0} + \sum_{i=1}^3 \frac{g}{a^i} j^b_{x,i} A^b_{x,i} \rb,
\end{equation}
where $A^a_{x,\mu}$ includes a factor of $g a^\mu$ (``lattice units"). We also made the split into 3+1 dimensions explicit using $J^a_0(x) \simeq \rho^a_x$ and $J^a_i(x) \simeq j^a_{x,i}$. The variation of $S_J$ simply reads
\begin{equation}
\dd S_J = \frac{V}{g^2} \sum_{x,b} \lb - \frac{g}{a^0} \rho^b_x \dd A^b_{x,0} + \sum_{i=1}^3 \frac{g}{a^i} j^b_{x,i} \dd A^b_{x,i} \rb,
\end{equation}
which gives the appropriate contributions to the GC and the EOM.
In the leapfrog scheme the GC now reads
\begin{equation} \label{eq:leapfrog_gauss_current}
\sum_i \frac{1}{\lb a^0 a^i \rb^2} P^a \lb U_{x,0i} + U_{x,0-i}\rb = \frac{g}{a^0} \rho^a_x,
\end{equation}
and the EOM read
\begin{equation}\label{eq:leapfrog_eom1_current}
\frac{1}{\lb a^0 a^i \rb^2} P^a \lb U_{x,i0} + U_{x,i-0}\rb =
\sum_j \frac{1}{\lb a^i a^j \rb^2} P^a \lb U_{x,ij} + U_{x,i-j}\rb - \frac{g}{a^i} j^a_{x,i}.
\end{equation}
The constraint taken together with the EOM imply the local conservation of charge (see \cref{app_gauss})
\begin{equation} \label{eq:continuity_discrete}
\frac{\rho_x - \rho_{x-0}}{a^0} = \sum_i \frac{j_{x,i} - U^\dg_{x-i,i} j_{x-i,i} U_{x-i,i}}{a^i},
\end{equation}
which is the discrete version of the continuity equation. For the implicit and semi-implicit schemes the procedure is the same: including the $S_J$ term simply leads to the appearance of $\rho$ on the RHS of the GC (see eqs.\ \eqref{eq:implicit_gauss} and \eqref{eq:semi_gauss}) and $j_{x,i}$ on the RHS of the EOM (see eqs.\ \eqref{eq:implicit_eom} and \eqref{eq:semi_eom_1} -- \eqref{eq:semi_eom_i}). Due to conservation of the GC without external charges, the continuity equation for the implicit and semi-implicit scheme is simply  \cref{eq:continuity_discrete} as well. This implies that our treatment of the external currents in terms of parallel transport does not require any modifications when using the newly derived schemes.

\section{Numerical tests}\label{tests}

In this last section we test the semi-implicit scheme on the propagation of a single nucleus in the CGC framework. For an observer at rest in the laboratory frame, the nucleus moves at the speed of light and consequently exhibits large time dilation. As the nucleus propagates, the interactions inside appear to be frozen and the field configuration is essentially static.
On the lattice we would like to reproduce this behavior as well, but depending on the lattice resolution we run into the numerical Cherenkov instability, which leads to an artificial increase of the total field energy of the system.

\begin{figure}[t]
	\centering
	\includegraphics{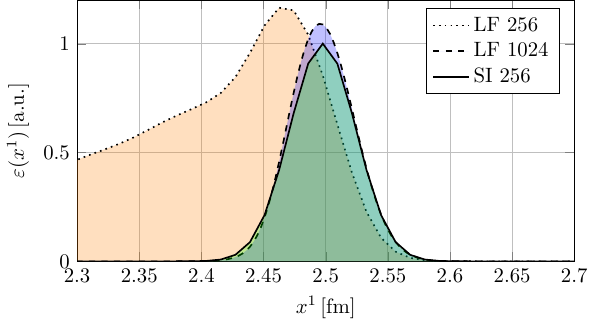}
	\caption{
		The energy density of a right-moving nucleus averaged over the transverse plane as a function of the longitudinal coordinate $x^1$ after $t=2\,\text{fm}/c$. We compare the performance of the leapfrog (LF) scheme (with $N_L=256$ and $N_L=1024$) to the semi-implicit (SI) scheme with $N_L=256$. In the most extreme example (LF 256) the nucleus becomes completely unstable due to the numerical Cherenkov instability. By eliminating numerical dispersion (SI 256) the nucleus retains its original shape almost exactly. This plot is taken from \cite{Ipp:2018hai}.
		\label{fig:profiles}}
\end{figure}

As described in \cref{sec:nci_dispersion}, the root cause of the instability is numerical dispersion: in the CGC framework a nucleus consists of both propagating field modes and a longitudinal current generating the field around it. It is essentially a non-Abelian generalization of the field of a highly relativistic electric charge.
In our simulation the color current is modeled as an ensemble of colored point-like particles moving at the speed of light along the beam axis. The current is unaffected by dispersion, i.e.\ it retains its shape perfectly as it propagates. The field modes suffer from numerical dispersion, which over time leads to a deformation of the original longitudinal profile. The mismatch between the color current and the field leads to the creation of spurious field modes, which interact with the color current non-linearly through parallel transport (color rotation) of the current. This increases the mismatch further and more spurious fields are created. As the simulation progresses, this eventually leads to a large artificial increase of total field energy. The effects of the instability can be quite dramatic as seen in \cref{fig:profiles}.
The main difference to the numerical Cherenkov instability in Abelian PIC simulations is that in electromagnetic simulations the spurious field modes interact with the particles through the Lorentz force \cite{GODFREY1974504}. In our simulations we do not consider any acceleration of the particles (i.e.\ their trajectories are fixed), but interaction is still possible due to non-Abelian charge conservation \eqref{eq:continuity_discrete}, which requires rotating the color charge of the color current. Therefore our type of numerical Cherenkov instability is due to non-Abelian effects.

\begin{figure}[t]
	\centering
	\includegraphics{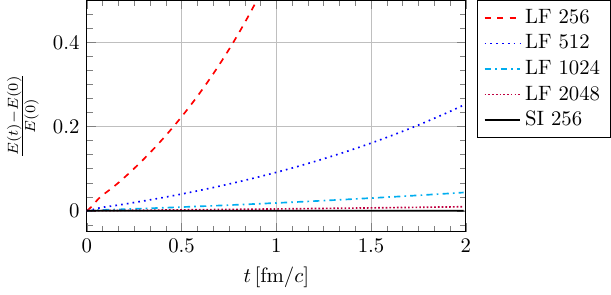}
	\caption{
		The relative increase of the total field energy $E(t)$ as a function of time $t$ for the propagation of a single nucleus in a box of volume $V = \lb 3\,\text{fm}\rb \times \lb 6\,\text{fm}\rb^2$ with a longitudinal length of $\,3\,\text{fm}$ and a transverse area of $\,\lb 6 \, \text{fm}\rb^2$. The lattice size is $N_L \times N_T^2$ with the transverse lattice fixed at $N_T = 256$. Starting with the same initial condition, we evolve forward in time using the leapfrog (LF) and the semi-implicit (SI) method. In the case of the leapfrog scheme we vary the resolution along the beam axis using the number of longitudinal cells $N_L$ of the lattice. For $N_L=256$ the numerical Cherenkov instability leads to catastrophic failure, increasing the energy to many times its original value. The effect is suppressed when increasing the longitudinal resolution, but the instability is still present. In the case of the semi-implicit scheme it is possible to set $N_L=256$ and still obtain (approximate) energy conservation. After $t=2\,\text{fm}/c$ the energy increase for the semi-implicit scheme is roughly $0.02\%$, compared to $1\%$ for the leapfrog with $N_L=2048$. For the simulation using the semi-implicit scheme we used $N_i = 10$ iterations and a damping coefficient of $\,\alpha=0.45$. The time step is set to the longitudinal lattice spacing. In the case of the leapfrog simulation we used $a^0 = a^1 / 4$.  The results were obtained from the same simulations as in \cref{fig:profiles}. This plot is taken from \cite{Ipp:2018hai}.
		\label{fig:increase}}
\end{figure}
We now demonstrate that the instability is cured (or at least highly suppressed for all practical purposes) by using the semi-implicit scheme. We test the scheme's ability to improve energy conservation in the following way: we place a single gold nucleus described by the McLerran-Venugopalan model in a simulation box of volume $V = \lb 3\,\text{fm}\rb \times \lb 6\,\text{fm}\rb^2$ and set the longitudinal extent of the nucleus to roughly correspond to a boosted nucleus with Lorentz factor $\gamma = 100$. The values for $\mu$, $g$ and $\sigma$ are determined from $\gamma$ as detailed in \cref{sec:rapidity_profiles}.
 After setting up the initial condition we let the nucleus freely propagate along the longitudinal axis. As the simulation runs we record the total field energy
\begin{equation}
E(t) = \frac{1}{2} \int_V d^3 x \sum_{i,a} \lb  E^a_i\lb t, \vec{x}\rb^2  +  B^a_i\lb t, \vec{x}\rb^2 \rb,
\end{equation}
where $E^a_i(x)$ and $B^a_i(x)$ are the color-electric and -magnetic fields at each time step. On the lattice the electric and magnetic fields are approximated using plaquettes (see eqs.\ \eqref{sec:em_tensor_latt}, \eqref{eq:electric_plaq} and \eqref{eq:magnetic_plaq}):
\begin{align}
E^a_{i}(x) &\simeq \frac{1}{g a^0 a^i} P^a \lb U_{x,0i} \rb, \\
B^a_{i}(x) &\simeq -\sum_{j, k} \varepsilon_{ijk} \frac{1}{2 g a^j a^k} P^a \lb U_{x,jk} \rb.
\end{align}
We compute the relative change $\lb E(t) - E(0) \rb / E(0)$, which we plot as a function of time $t$. In the continuum we would have $E(t)=E(0)$, but due to numerical artifacts and the Cherenkov instability this is not the case in our simulations.

The numerical results are shown in \cref{fig:increase}. We see that the leapfrog scheme leads to an exponential increase of the total energy over time, which can be suppressed using finer lattices. On the other hand, the semi-implicit scheme leads to better energy conservation even on a rather coarse lattice. Therefore, the resolution that is usually required to obtain accurate, stable results is lowered by using the semi-implicit scheme. However, using the new scheme might not always be economical: finer lattices suppress the instability as well and since the leapfrog scheme is computationally cheaper than the semi-implicit scheme, the leapfrog can be favorable in practice. On our test system (a single $256$ GB node on the VSC 3 cluster) the simulation using the semi-implicit scheme (SI 256) takes $\sim 4$ hours to finish, while the same simulation using the leapfrog with $N_L=1024$ (LF 1024) takes roughly $\sim 2.5$ hours and with $N_L=2048$ (LF 2048) $\sim 10$ hours. Even though energy conservation is not as good as SI 256, the longitudinal resolution is much better in comparison, enabling us to extract observables with higher accuracy. It should be noted however that our implementation of the leapfrog scheme is already highly optimized, while the implementation of the semi-implicit scheme is very basic and should be considered as a proof of concept. Further optimizations and simplifications of the semi-implicit scheme might make it the better choice in many cases. 

\begin{figure}
	\centering
	\includegraphics{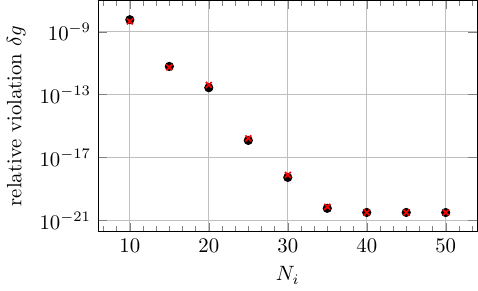}
	\caption{
		The relative GC violation for the semi-implicit scheme as a function of the number of iterations $N_i$ of the damped fixed point iteration. For this plot we use a simulation box of volume $V = \lb 1.5\,\text{fm}\rb \times \lb 6\,\text{fm} \rb^2$ on a lattice with $N_T=256$ points in the transverse directions and $N_L=128$ points in the longitudinal direction. Otherwise, we use the same initial conditions as in figs.\ \ref{fig:profiles} and \ref{fig:increase}. At the beginning of the simulation at $t=0\,\text{fm}/c$ the constraint is conserved up to machine precision by construction. We then let the nucleus propagate until $t=0.5\,\text{fm}/c$ and compute the violation of the GC (black dots). We also compare to the constraint violation after only a single evolution step (red crosses), which does not differ much from the violation after a larger number of time steps. It is evident that the violation systematically converges towards zero (up to machine precision) as we increase the number of iterations $N_i$. This plot is taken from \cite{Ipp:2018hai}.
		\label{fig:gauss}}
\end{figure}

As a second test we look at the violation of the GC. The leapfrog scheme \eqref{eq:leapfrog_eom} conserves its associated GC \eqref{eq:leapfrog_gauss} identically, even for finite time-steps $a^0$. In numerical simulations this conservation is not exact due to floating point number errors, but the violation is zero up to machine precision.
On the other hand, the semi-implicit scheme has to be solved iteratively and therefore the results depend on the number of iterations $N_i$ used in the fixed point iteration method. 
We define the relative violation of the GC as the ratio of the absolute (squared) GC violation to the total (squared) charge on the lattice. In the case of the leapfrog scheme this reads
\begin{equation} \label{eq:rel_gauss}
\dd g(t) = \frac{\sum_{x',a} \lb \sum_i \frac{1}{\lb a^0 a^i \rb^2} P^a \lb U_{x',0i} + U_{x',0-i}\rb - \frac{g}{a^0} \rho^a_{x'}\rb^2}{\sum_{x',a} \lb \frac{g}{a^0} \rho^a_{x'} \rb^2},
\end{equation}
where the sum $\sum_{x'}$ runs over the spatial lattice of a single time slice at $t$. The numerator depends on the GC of the scheme and has to be adjusted according to the implicit or semi-implicit method (either  \cref{eq:implicit_gauss} or \eqref{eq:semi_gauss} including the charge density on the RHS as discussed in \cref{external_charges}). 
In \cref{fig:gauss} we show how the GC violation converges systematically towards zero as we increase the number of iterations. Therefore, even though we can not use an arbitrarily high number of iterations due to limited computational resources, the semi-implicit scheme conserves the GC in principle. The same holds for the purely implicit scheme. In practice it is not necessary to satisfy the constraint up to high precision, as observables such as the energy density seem to converge much faster to satisfying accuracy.
\chapter{Closing remarks}

In this last chapter I present a short summery of my thesis and outline some extensions and generalizations that could be attempted in the future.

\section{Summary}

Starting with \cref{chap:bi_glasma}, I recapitulated the theory and numerical methods for the boost invariant case, where the color currents of the colliding nuclei are assumed to be infinitesimally thin, i.e.
\begin{equation} \label{eq:sum_current_delta}
J^\pm(x^\mp, x_T) = \delta(x^\mp) \rho(x_T),
\end{equation} 
see \cref{eq:mv_color_current2}.
The particular structure of the color currents allows one to unambiguously divide space-time along the light cone centered around the collision event $x^+ = x^- = 0$ into four separate regions. Along the boundary of the future light cone ($x^+ > 0$ and $x^- > 0$), a set of matching conditions can be derived, which turn out to be the initial values for the Glasma fields at proper time $\tau = 0^+$ (see eqs.\ \eqref{eq:glasma_initial_1} and \eqref{eq:glasma_initial_2}).
Due to invariance of \cref{eq:sum_current_delta} under longitudinal boosts, the Glasma fields can be chosen to be independent of space-time rapidity $\eta$, reducing the system to effectively 2+1 dimensions.
Moreover, the color current does not contribute for $\tau > 0$, since its support is limited to the boundary of the light cone.
Therefore, in order to study the evolving Glasma, one has to solve the 2+1 dimensional, non-linear, homogeneous Yang-Mills equations (eqs.\ \eqref{eq:bi_glasma_eom1} and \eqref{eq:bi_glasma_eom2}).   
Although the initial values for the fields in the future light cone are known exactly, the full analytical solution for $\tau > 0$ is not known in general.

Fortunately, approximate solutions can be found numerically using the methods of real-time lattice gauge theory.
One of the main features of the lattice formulation of gauge field theories is that the local fields $A_\mu(x)$ are replaced by non-local gauge variables or links $U_{x,\mu}$, which are Wilson lines connecting nearest neighbors on the underlying rectangular lattice.
This reformulation of Yang-Mills theory allows us to derive discretized actions (the most simple one being the Wilson action, see \cref{eq:wilson_action}), which are exactly invariant under a subset of the full gauge symmetry, namely lattice gauge transformations.
The most important symmetry characterizing Yang-Mills theories is therefore also realized when computing numerical approximations of its solutions.

As a model for the Glasma initial conditions we employed the McLerran-Venugopalan (MV) model. The MV model approximates relativistic nuclei as longitudinally infinitesimally thin, transversally infinitely extended walls of random, homogeneously distributed color charge.
Neither a finite transverse radius nor nucleonic structure are described by the original MV model, although in principle such features can be added as well. 

Equipped with these tools, numerical simulations were performed to study the boost invariant Glasma.
In particular we studied the components of the energy-momentum tensor $T_{\mu\nu}$ of the Glasma, where a number of observations can be made (see figs.\ \ref{fig:density2d_plots}, \ref{fig:components_and_density} and \ref{fig:pressures} in \cref{sec:bi_observables}):
\begin{itemize}
\item Initially, the Glasma consists of purely longitudinal color-electric and color-magnetic fields in the form of color flux tubes. These fields are correlated over lengths inversely proportional to the saturation momentum $Q_s$, which also roughly fixes the radius of the flux tubes. This initial state is highly anisotropic. The longitudinal pressure is negative with $p_L = - \varepsilon$ and the transverse pressure positive with $p_T = + \varepsilon$, where $\varepsilon$ is the energy density of the Glasma. The pressure anisotropy is maximal in the sense of $p_L / \varepsilon = - 1$  and $p_T / \varepsilon = + 1$.
\item As the system evolves the flux tubes start to expand in the transverse plane and transverse (color)-electric and -magnetic fields are generated. This leads to an increase of $p_L$ and a decrease of $p_T$. Nevertheless, the system never becomes isotropic ($p_L \approx p_T$) as the change in pressures stops when the longitudinal pressure reaches zero. The system ends up in an anisotropic state where $p_L \approx 0$ and $p_T \approx \varepsilon / 2$.
\item This anisotropic late time behavior is reached after $\tau > Q_s^{-1}$ and is referred to as free streaming. In this case the energy density of the system exhibits characteristic behavior given by $\varepsilon(\tau) \propto 1 / \tau$.
\item Obviously, by construction the system and all of its observables are independent of space-time rapidity $\eta$.
\end{itemize}

Although the 2+1D framework can be extended to include the rapidity coordinate $\eta$ in a straightforward way, it is less clear how to formulate rapidity dependent initial conditions that satisfy the Gauss constraint (GC). Previous 3+1 dimensional approaches therefore focus on studying non-Abelian plasma instabilities due to small fluctuations on top of the boost invariant Glasma background field  \cite{Romatschke:2005ag, Romatschke:2006nk,Fukushima:2011nq,Epelbaum:2013waa,Gelis:2013rba, Fukushima:2006ax, Dusling:2010rm, Berges:2012cj}.
If one is interested in a generalization of \cref{eq:sum_current_delta}, i.e.\ color currents of the form
\begin{equation} \label{eq:sum_current_general}
J^\pm(x^\mp, x_T) = \rho(x^\mp, x_T),
\end{equation}
then the above outlined approach is too constraining. Due to a small, but finite longitudinal support of the color charge density of the nuclei, one cannot rely on a well-separated future light cone anymore. Therefore, in \cref{chap:glasma3d} a numerical method is developed to simulate this generalized case. The most obvious difference to the 2+1 dimensional method is that in order to accommodate finite width currents \eqref{eq:sum_current_general}, the whole Minkowski space-time must be considered and color currents have to be explicitly taken into account when solving the Yang-Mills equations on the lattice. This generalization implies a number of differences:
\begin{itemize}
\item The Yang-Mills equations are written in terms of flat coordinates using the laboratory frame time $t$ and the longitudinal coordinate $z$ instead of formulating them in the curved coordinate patch spanned by $\tau$ and $\eta$. As a result, the field equations simplify due to lack of explicit $\tau$-dependent terms. The time evolution in this frame is an evolution in $t$ instead of $\tau$. Therefore, the natural choice of gauge for this case is the laboratory frame temporal gauge $A^0 = 0$. The standard leapfrog equations of motion in the lattice formalism are derived in \cref{sec:latt_field__eqs} for this purpose.

\item Instead of specifying the Glasma fields at $\tau = 0^+$, one has to set the fields of single nuclei at some initial time $t_0$ before the collision has happened. Analytically one can find the solution to the Yang-Mills equations, if at time $t_0$ both nuclei are spatially separated such that their color fields do not interact and each single nucleus can be treated independently. Furthermore, the initial conditions have to be compatible with the choice of gauge. The temporal gauge initial conditions are derived in \cref{sec:coll_with_finite} for the continuum case where it is shown that the color field of a single nucleus in temporal gauge is identical to its light cone gauge version. The lattice discretization of these initial conditions is derived in \cref{sec:initial_fields_cpic}.

\item The 3+1D method takes finite width color currents explicitly into account. In order to describe these currents numerically, one can make use of the colored particle-in-cell method originating from non-Abelian plasma simulations. The main idea is to replace the smooth color charge density of a nucleus with a large number of densely placed colored point particles. The movement of this ensemble of particles approximates the smooth color current of the nucleus. Although each particle's trajectory is known in advance as all forces acting on them are neglected (the eikonal approximation), their color charge precesses due to being parallel transported in the non-Abelian color field. This color charge precession or rotation must be accounted for in order to satisfy the Gauss constraint. All the necessary equations to perform the numerical time evolution of fields and currents in this manner have been derived in \cref{sec:cpic}.
\end{itemize}

In \cref{chap:single_color_sheet} the previously developed method is applied to study collisions of nuclei with color currents of the form
\begin{equation} \label{eq:sum_current_sca}
J^\pm(x^\mp, x_T) = f(x^\mp) \rho(x_T),
\end{equation} 
where $f(x^\mp)$ is the longitudinal profile function of the nucleus and $\rho(x_T)$ determines the distribution of color charge in the transverse plane.
\Cref{eq:sum_current_sca} can be viewed as a regularization of \cref{eq:sum_current_delta} and a special case of the more general form \cref{eq:sum_current_general}. The color charges are correlated along the longitudinal extent of the nucleus, i.e.\ the nucleus only consists of a single sheet of color charge with some finite width. This specific case was therefore referred to as the single color sheet approximation. The width of the profile function $f(x^\mp)$ is set by the Lorentz-contracted longitudinal extent of the nucleus, i.e.\ $2 R / \gamma$, where $R$ is the nuclear radius and $\gamma$ is the Lorentz factor. 

First, the 3+1D method was put to the test by directly comparing to the boost invariant case in \cref{sec:approaching_bi}. Immediately after the collision, the longitudinal color-electric field in the transverse plane was compared to the Glasma initial conditions. It was demonstrated that when the profile function $f(x)$ approaches the delta distribution $\delta(x)$, the results from the 3+1D method and the boost invariant initial conditions agree, see \cref{fig:correlations_initial}. Next, the pressure components were studied more closely. The 3+1D Glasma behaves very similar to the boost invariant case: the system exhibits free streaming and the pressure anisotropy remains large, see figs.\ \ref{fig:pressure_ratios_1} and \ref{fig:pressure_ratios_2}. However, a subtle difference which strongly depends on the longitudinal width of the colliding nuclei, can be observed: in the boost invariant case the magnetic and electric components roughly contribute equally to the energy density of the Glasma. This does not appear to be the case for three-dimensional collisions. Immediately after the collision event the longitudinal magnetic field appears to be reduced compared to the electric component. This mismatch grows with larger widths and larger values of the infrared regulator and is only slowly cured over time, see figs.\ \ref{fig:BLEL_mismatch_1} and \ref{fig:BLEL_mismatch_2}. As a further test of the validity of the three-dimensional simulations, local energy conservation was explicitly checked using the non-Abelian version of Poynting's theorem, see \cref{fig:poynting_theorem}.

Having established that the 3+1D method is able to produce valid results, initial conditions were chosen that resemble the collisions at RHIC more closely. This was done to demonstrate the main feature of the method, namely the explicit breaking of boost invariance.
In \cref{sec:rapidity_profiles} we extracted the rapidity profile of the energy density of the Glasma in its rest frame.
For certain choices of parameters, the obtained profiles are roughly in agreement with experimental results at RHIC, see figs.\ \ref{fig:rhic200} and \ref{fig:rhic130}. Surprisingly, infrared regulation seems to have a large effect on the widths of these profiles. It can also be shown that the longitudinal velocity of the Glasma is very close to the free streaming result, see \cref{fig:long_velocity}. These findings indicate that the rapidity profile of the energy density develops very shortly after the collision. Afterwards, the Glasma expands in a free streaming manner. Although the range of rapidity that can be probed with sufficient numerical accuracy is rather limited, a workaround method has been developed to extend this interval slightly (see \cref{sec:extending_rapidity}).

Unfortunately, the original numerical approach presented in chapters \ref{chap:glasma3d} and \ref{chap:single_color_sheet}, which is based on the standard Wilson action, suffers from a non-Abelian numerical Cherenkov instability. This instability leads to an artificial increase of field energy over time even when only considering the propagation of a single nucleus. In \cref{cha:semi_implicit} the main culprit was identified as the different numerical dispersion behaviors of the two types of degrees of freedom in the simulation: while the color current, composed of many colored point particles, keeps its shape as it moves, the discretized gauge field suffers from numerical dispersion. As a result, the growing mismatch between the color fields of nuclei and their color currents, drives an instability that is only cured when taking the continuum limit. To resolve this issue, an alternative lattice discretization of the Yang-Mills action was proposed, which eliminates the problematic dispersion behavior along the trajectories of the nuclei. It was shown that this action is invariant under lattice gauge transformations and exhibits the same order of accuracy as the original Wilson action. Varying this action led to semi-implicit equations of motion for the gauge field, which can still be solved numerically, see \cref{semi_scheme}. Compared to the leapfrog method, the semi-implicit method requires higher computational effort, but was shown to successfully suppress the numerical Cherenkov instability (see \cref{tests}). 

\section{Outlook}

A straightforward continuation of the work presented in this thesis
is the study of random fluctuations within the longitudinal extent of
nuclei and their effects on the Glasma.
Most of the focus in the CGC approach to the initial state in heavy-ion collisions
 has been on realistically modeling the transverse structure
of nuclei. For instance, the IP-Glasma model includes impact parameter
dependence for non-central collisions and nucleonic and sub-nucleonic
fluctuations \cite{Schenke:2012wb}. On the other hand, the random longitudinal
structure of nuclei has been largely neglected, except for the approximation
of the path ordered Wilson line as an infinitely thin stack of uncorrelated
sheets of color charge \cite{Fukushima:2007ki} (see also \cref{sec:mv_model}).
The 3+1D setup using the numerical improvements of \cref{cha:semi_implicit}
should make the study of initial conditions including these fluctuations feasible.

In the spirit of \cite{Fukushima:2007ki}, the starting point of this study
could be a generalization of the MV model including random fluctuations in the
longitudinal extent.
Extending \cref{eq:mv2_twop}, the general charge density two-point function of this model (in covariant gauge, for a right-moving nucleus) can be expressed as
\begin{equation} \label{eq:closing_mv1}
\left\langle \rho^{a}(x^{-},x_{T})\rho^{b}(y^{-},y_{T})\right\rangle = g^{2}\mu^{2}( \frac{x^- + y^-}{2}, x^- - y^-) \delta^{ab}
\delta^{(2)}(x_{T}-y_{T}),
\end{equation}
where $x_{T}$ and $x^{-}$ refer to transverse and light cone coordinates,
$g$ is the coupling constant, and $\mu^2$ is a generalization of the phenomenological MV
model parameter.
In this model, $\mu^2$ is considered to be a function of $(x^- + y^-) / 2$ and $x^- - y^-$. The first argument $(x^- + y^-) / 2$ defines the overall longitudinal shape of the nucleus, while the second argument 
$x^- - y^-$ controls the correlation of random color charges along the longitudinal extent. The one-point function is assumed to be identically zero
to implement color neutrality on average.
A specific realization of \cref{eq:closing_mv1} is given by (see \cite{Fukushima:2007ki})
\begin{equation} \label{eq:closing_mv2}
\left\langle \rho^{a}(x^{-},x_{T})\rho^{b}(y^{-},y_{T})\right\rangle =g^{2}\mu^{2} f_{\sigma}(\frac{x^{-}+y^{-}}{2})f_{\epsilon}(x^{-}-y^{-}) \delta^{ab} \delta^{(2)}(x_{T}-y_{T}),
\end{equation}
where the functions $f_{\sigma}(x)$ and $f_{\epsilon}(x)$
represent Gaussian profile functions with widths $\sigma$ and $\epsilon$
respectively and $\mu^2$ is a constant. The parameter $\sigma$ defines the longitudinal
width (as in the single color sheet approximation, see \cref{chap:single_color_sheet}) and $\epsilon$ is a new parameter
to control the amount of random longitudinal fluctuations in the nucleus. The two-point function can be
interpreted as a regularization of \cref{eq:mv2_twop}.

The Wilson line associated with the transformation to LC gauge
is given by (see \cref{eq:lc_gauge_wilson_line})
\begin{equation}
W^\dg(x^{-},x_{T})=\mathcal{P}\exp\left(ig\intop_{-\infty}^{x^{-}}dz^{-}\,\,\frac{\rho^{a}(z^{-},x_{T})}{\nabla_{T}^{2}-m^{2}}t^{a}\right),
\end{equation}
with $m\sim\Lambda_{QCD}$ as the infrared regulator to control the
divergence in the model. For $\epsilon\sim\sigma$ we obtain the MV
model with finite width $\sigma$ in the single color sheet approximation
which has been thoroughly investigated in \cref{chap:single_color_sheet}.
The boost invariant limit is achieved by taking the limit $\sigma\rightarrow0$
and $\epsilon\rightarrow0$ with the order of limits being $\sigma>\epsilon$.
The color field of the nucleus is then treated
as a shock wave and the $x^{-}$ dependence of $W(x^{-},x_{T})$ becomes
trivial as detailed in \cref{sec:mv_model}. Random longitudinal structure
is accounted for using the color sheet regularization, see eqs.\ \eqref{eq:mv2_twop_reg}, 
\eqref{eq:mv2_poisson_reg} and \eqref{eq:asym_Wilson_line_reg}.

For the present purpose it would be interesting to study a different limit: in
the case of $\epsilon\rightarrow0$ while keeping $\sigma > 0$, we have $f_{\epsilon}(x)\rightarrow\delta(x)$
and the generalized model reduces to the three-dimensional MV model with finite longitudinal
width $\sigma$, see \cref{eq:mv2_twop}. In this case the nucleus exhibits random longitudinal
structure and it is this limit which has not been studied previously.

In principle, the model for initial conditions described above is simple to implement and simulate with
the 3+1D method presented in this thesis. The procedure for setting
up initial conditions discussed in \cref{sec:initial_lattice} is general enough
to work with an arbitrary charge density $\rho^a(x^-, x_T)$ regardless of the 
exact dependence on $x^-$. It would then be interesting to study the Glasma
created from collisions of such nuclei, in particular with regards to the breaking of boost invariance
and the problem of pressure isotropization and (physical) instabilities. The effects of random variation
in the longitudinal extent should be compared to the setting that has been
thoroughly investigated in the past decade, namely rapidity-dependent fluctuations
on top of a boost invariant background field \cite{Romatschke:2006nk,Epelbaum:2013waa,Gelis:2013rba,Fukushima:2007ja,Berges:2012cj}.
The main difference to previous works is that the rapidity-dependent fluctuations
which seed the Glasma instability are of different origin than usually assumed.
In the proposed approach, deviations from boost invariance are due to the color structure
of the colliding nuclei and not of quantum origin as e.g.\ in \cite{Berges:2012cj}.
Furthermore, instead of having to model the Glasma by superimposing fluctuations and background fields, the Glasma field in the 3+1D method arises directly from
solving the Yang-Mills equations. By construction, these fields fulfill the Gauss constraint.
The 3+1D setting also differs from previous approaches in more subtle ways: in the ``background plus fluctuations"
setup, which is generally formulated in the future light cone using $(\tau, \eta)$ coordinates,
the axis along the rapidity coordinate $\eta$ is assumed to have periodic boundary conditions. In
the 3+1D setup with $(t, z)$ coordinates, the boundary condition in $\eta$ is more similar to an open boundary. 

One can even conceive a slightly artificial, but more direct way of showing how much
the variation in the longitudinal extent affects the Glasma: starting with nuclei in
the single color sheet approximation, one could make arbitrary changes to the Wilson line
$W(x^-, x_T)$ inside the nucleus without affecting the asymptotic value 
\begin{equation}
W(x_T) = \lim_{x^- \rightarrow \infty} W(x^-, x_T).
\end{equation}
In the boost invariant scenario, these different initial conditions would be considered identical,
as only the asymptotic Wilson line is used as input for the initial Glasma fields.
In contrast, using the 3+1D setup it should be possible to differentiate between different initial conditions which share the same asymptotic Wilson line, if random variations within the longitudinal extent strongly affect the produced Glasma.
Therefore, a more thorough study of these effects and more direct comparisons to the boost invariant case
could prove interesting and provide new insights into the creation and the dynamics of the Glasma with broken boost invariance.

Another phenomenon that calls for further investigation using 3+1D
simulations is the production of axial charge in the earliest
stages of the collision, which is related to the chiral magnetic effect
\cite{Fukushima:2008xe,Skokov:2016yrj}.
The chiral magnetic effect predicts an anomalous separation of charges
in non-central collisions where large magnetic fields are
present. This anomalous transport phenomenon is due to a chirality
imbalance or axial charge, i.e.\ a difference in the numbers of
left-handed and right-handed fermions in the system.
Both the theoretical
aspects and the experimental signatures of the chiral magnetic effect
are so far inconclusive and active topics of current research. The
importance of the chiral magnetic effect can be seen by the fact that
in 2018 a new kind of experiment within Run-18 at RHIC has
been devoted exclusively to the exploration of this effect: by performing
collisions of $^{96}{\rm \mathbf{Zr}}^{40+}$ nuclei pairs and comparing
to $^{96}{\rm \mathbf{Ru}}^{44+}$ isobars, which are nuclei with
the same mass but different number of protons, one can perform controlled
experiments where the only changing parameter is the electric charge
and thus a 10\% change in the magnetic field induced by its current
\cite{Skokov:2016yrj}.

A theoretical
uncertainty in the chiral magnetic effect is how large the produced axial charge is and
how it is created. Locally the generation of axial charge is due to
the non-zero covariant divergence of the axial charge current $j_{(5)}^{\mu}$
which exhibits the axial anomaly
\begin{equation}
\partial_{\mu}j_{(5)}^{\mu}=-\frac{g^{2}N_{f}}{8\pi^{2}}\mbox{tr}\left(F_{\mu\nu}\tilde{F}^{\mu\nu}\right)\propto\sum_{a}E_{i}^{a}B_{i}^{a}.
\end{equation}
Here $N_{f}$ refers to the number of fermion flavors and $\tilde{F}^{\mu\nu}$
is the dual to the non-Abelian field strength tensor $F_{\mu\nu}$.
Because of this anomaly the axial charge density
$j_{5}^{0}$ is not conserved, but arises due to parallel (color-)electric
and magnetic fields. The right hand side $\sum_{a}E_{i}^{a}B_{i}^{a}$
is readily computable in Yang-Mills simulations and has been investigated
in boost invariant, Glasma-like scenarios 
using numerical and analytical methods \cite{Kharzeev:2001ev,Mace:2016svc,Lappi:2017skr}.
While the boost invariant scenario allows for strong local axial charge
production due to the nature of longitudinal (and therefore parallel)
color-electric and magnetic flux tubes early in the evolution of the
Glasma, there are topological arguments that disallow a net change
in the total axial charge \cite{Lappi:2006fp,Kharzeev:2001ev}. However,
this no-go theorem only strictly applies to the boost invariant case
where the system is effectively 2+1 dimensional, so it might be 
possible to find different behavior in our 3+1 dimensional setup.
Even if no net axial charge will be found, one could still study local variations 
of axial charge density, make direct comparisons
to boost invariant simulations and investigate how the mismatch of
color-electric and color-magnetic fields (see \cref{sec:eb_mismatch}) affects axial charge production.

An important quantity that has not been computed yet in the 3+1D setup
is the gluon occupation number. In the boost invariant case this quantity
can be readily computed as mentioned in \cref{sec:bi_observables}.
It has been used to study the gluon spectrum in the Glasma \cite{Lappi:2009xa},
to fix phenomenological parameters of models (such as $\mu$ in the
MV model) and to determine the average transverse momentum of gluons
$\left\langle p_{T}^{2}\right\rangle $ \cite{Lappi:2004sf,Lappi:2003bi}.
Furthermore it can be used to directly compute flow coefficients from
the initial state from a Fourier decomposition with respect to the
azimuthal angle \cite{Schenke:2015aqa,Schlichting:2016sqo}.

There are two technical problems that complicate the extraction of
gluon numbers in 3+1 dimensions. The first problem is related to gauge fixing:
in the boost invariant case the common procedure is to use the transverse
Coulomb gauge condition
\begin{equation} \label{eq:coulomb_2d}
\sum_{i=1,2} \p_i A^{i,a}_c(\tau_0, x_T) = 0
\end{equation}
at some proper time $\tau_0$.
In the 3+1D method, which is formulated in the laboratory
frame of the collision, the analogue to this gauge condition would be the
three-dimensional Coulomb gauge
\begin{equation} \label{eq:coulomb_3d}
\sum_{i=1,2,3} \p_i A^{i,a}_c(t_0, x_T, z) = 0,
\end{equation}
fixed at some time $t_0$. \Cref{eq:coulomb_3d} does not necessarily
imply \cref{eq:coulomb_2d} and therefore it is possible that the gluon numbers
extracted in 3+1 dimensions differ from the definition used in the boost invariant
scenario. It might be necessary to choose a different gauge condition than the
obvious choice of \cref{eq:coulomb_3d}. 

A second problem is related to the choice of frame: as the $(\tau, \eta)$
frame only covers the future light cone, one only obtains the gluon numbers of
the Glasma. In the 3+1D method, both the Glasma and the color fields of the nuclei
are part of the simulation. Therefore, unwanted contributions from the nuclei
would be present in the gluon occupation numbers. Direct comparisons between
the boost invariant and the 3+1D method are only possible if these contributions
can be somehow canceled or filtered out.

Once these issues are solved, it would be interesting to revisit the study of 
rapidity profiles from \cref{sec:rapidity_profiles}. Instead of using the
rest-frame energy density, one could use the gluon occupation number as
a function of momentum rapidity, which would allow for better comparison
to experimental data.

Finally, a much harder problem to tackle would be to study the production
of quarks in the background of fully non-perturbative Glasma fields in 3+1D.
The main idea
is to solve the Dirac equation coupled to an external non-Abelian
gauge field, which is treated as a background field (i.e.$\ $no backreaction
from the fermion fields to the gauge fields) and then compute the
number of quark-antiquark pairs that arises from the interaction using
the solution of the Dirac equation. This problem can be solved analytically
for Abelian fields \cite{Baltz:2001dp} and in the special case of
proton-nucleus collisions \cite{Blaizot:2004wv,Fujii:2006ab}, where
one can consider the color field of the proton to be weak compared
to the color field of the nucleus.

The fully non-perturbative problem for nucleus-nucleus collisions
can only be solved numerically and turns out to be quite complicated:
early numerical investigations have been performed first in a 1+1
dimensional toy model \cite{Gelis:2004jp} and later in boost invariant
Glasma background fields \cite{Gelis:2005pb}. However, there are
multiple numerical difficulties arising in these earlier approaches:
it turns out that even though the color fields are boost invariant
(effectively 2+1 dimensional) one has to solve the Dirac equation
in 3+1 dimensions. A further complication is the choice of coordinate
systems. The boost invariant Glasma fields are naturally solved in
co-moving coordinates in the future light cone, but the solution
of the Dirac equation also depends on information about the field
in other parts of the light cone. Consequently due to mixing of different
reference frames, one has to transform the Glasma fields into a more
suitable coordinate system and special care has to be taken with coordinate
singularities associated with the co-moving frame. Although there
has been progress in improving the mathematical formalism \cite{Gelis:2015eua},
the fully numerical calculation with Glasma fields
in 3+1D could still provide new results.

In the light of these technical issues, the 3+1D approach
using laboratory frame (Cartesian) coordinates seems exceptionally well-suited
to revisit the problem of quark pair production in heavy-ion collisions:
the simulations presented in this thesis already provide information about the color fields
in all parts of the light cone and by construction there are no
ambiguities with coordinate singularities. Thus, even though it would be
a difficult task to undertake due to the large computational effort,
 the full numerical solution could be attainable.

\section{Conclusions}

In conclusion, the 3+1D CPIC method for simulating collisions within the CGC framework developed in this thesis was shown to be viable and in agreement with the boost invariant scenario as a limiting case. It genuinely extends the boost invariant method by including the longitudinal coordinate and is able to describe the fully rapidity-dependent dynamics of the Glasma. In particular, it allows for studying the effects of finite longitudinal width of the colliding nuclei and their longitudinal structure. Other extensions, such as axial charge production, the rapidity dependence of gluon occupation numbers, and quark pair production in the Glasma background field, could provide interesting new insights into the pre-equilibrium stage of heavy-ion collisions in future works.

\appendix

\chapter{Notation and conventions} \label{cha:notation}

In this chapter of the appendix we summarize special notation and conventions that are used throughout the thesis.

\section{Natural units} \label{sec:unit_conventions}

In this thesis we use natural units as defined by $\hbar = 1$ and $c = 1$. This allows us to express particle masses and momenta in units of energy. In particle physics the preferred unit of energy is the electron volt denoted by eV.
 For example, the mass of the proton is ${m_p \approx 0.938 \, \gev}$. Length and time scales are given in inverse units of energy, i.e.\ $[l] = [t] = \gev^{-1}$. It is convenient to also use femtometers (or Fermi) for lengths, e.g.\ the radius $R_A$ of a nucleus with mass number $A$ is given by $R_A \approx 1.25 A^\frac{1}{3} \, \fm$.
A gold nucleus with mass number $A=197$ has a radius of $R_A \approx 7.27 \, \fm$.

Using the relation
\begin{equation}
\hbar c \approx 0.197327 \, \gev \, \fm,
\end{equation}
one can translate between inverse energy units and Fermi. For a gold nucleus one finds ${R_A \approx 36.9 \, \gev^{-1}}$. It is common to express time scales in Fermi divided by the speed of light $c$, e.g.\ $t = 1 \, \fm / c \approx  5.07 \, \gev^{-1}$. In some cases, such as energy densities $\varepsilon$, we use a mix of Fermi and electron volts:
\begin{equation}
\varepsilon = 100 \, \frac{\gev}{\fm^3} \approx 0.768 \, \gev^4.
\end{equation}

\section{Yang-Mills theory} \label{sec:ym_conventions}

Throughout the thesis we use the mostly minus metric convention
\begin{equation}
g^{\mu\nu} = \mathrm{diag} \lb 1, -1, -1 -1\rb^{\mu\nu},
\end{equation}
where Greek indices run from $0$ to $3$. The symbol $\p_\mu$ is used as a shorthand for the partial derivative $\p / \p x^\mu$ with respect to the coordinate $x^\mu$.

The Lagrangian density of $\SUN$ Yang-Mills theory with $N_c$ colors is given by
\begin{equation}
\mathcal{L} = - \frac{1}{4} F^a_{\mu\nu} F^{a,\mu\nu}.
\end{equation}
$F^a_{\mu\nu}$ are the components of the non-Abelian field strength tensor given by
\begin{equation}
F^a_{\mu\nu} = \p_\mu A^a_\nu - \p_\nu A^a_\mu - g f^{abc} A^b_\mu A^c_\nu,
\end{equation}
where $A^a_\mu$ is a non-Abelian gauge field and $g$ is the Yang-Mills coupling constant. $f^{abc}$ are the real-valued, totally antisymmetric structure constants of the Lie algebra $\sun$ of the Lie group $\SUN$. The color component index $a$ runs from $1$ to $N_c^2 - 1$.

Often it is convenient to work with algebra elements instead of color components $A^a_\mu$. The gauge field can be written as
\begin{equation}
A_\mu = A^a_\mu t^a,
\end{equation}
where $t^a$ are the traceless, hermitian generators of the gauge group $\SUN$. In the fundamental representation of $\sun$, we use the normalization
\begin{equation} \label{eq:generator_normalization}
\tr [t^a t^b ] = \frac{1}{2} \delta^{ab}.
\end{equation}
The commutator of two generators is related to the antisymmetric structure constants $f^{abc}$ via
\begin{equation}
\cm{t^a}{t^b} = i f^{abc} t^c.
\end{equation}
The anti-commutator defines the real-valued, symmetric structure constants $d^{abc}$
\begin{equation}
\{ t^a, t^b \} = \frac{1}{N_c} \delta^{ab} + d^{abc} t^c.
\end{equation}
Using generators we can write the field strength tensor as an algebra element
\begin{equation}
F_{\mu\nu} = F^a_{\mu\nu} t^a = \p_\mu A_\nu - \p_\nu A_\mu + i g \cm{A_\mu}{A_\nu}.
\end{equation}
The Lagrangian density can be expressed as
\begin{equation}
\mathcal{L} = - \frac{1}{2} \tr [ F_{\mu\nu} F^{\mu\nu} ].
\end{equation}

The Lagrangian density of Yang-Mills theory is invariant under local $\SUN$ gauge transformations. The gauge field $A_\mu$ transforms according to
\begin{equation} \label{eq:gf_gauge_transformation}
A'_\mu(x) = V(x) \lb A_\mu(x) + \frac{1}{ig} \p_\mu \rb V^\dg(x), 
\end{equation}
where $V(x) \in \SUN$ is a space-time dependent gauge transformation. It holds that
\begin{align}
V V^\dg &= V^\dg V = \one, \\
\det V &= 1.
\end{align}
The gauge transformation of the field strength tensor is given by
\begin{equation}
F'_{\mu\nu} = V F_{\mu\nu} V^\dg.
\end{equation}
Compared to \cref{eq:gf_gauge_transformation} the transformation of $F_{\mu\nu}$ does not involve any partial derivatives acting on $V(x)$.
The invariance of $\mathcal{L}$ is easy to show using the above transformation law and the cyclicity of the trace
\begin{align}
\tr [ F'_{\mu\nu} F'^{\mu\nu} ] &= \tr [V F_{\mu\nu} V^\dg V F^{\mu\nu} V^\dg] \nn
&= \tr [ F_{\mu\nu} F^{\mu\nu} ].
\end{align}

The gauge-covariant derivative $D_\mu$, acting on an algebra element $\chi$, is given by
\begin{equation}
D_\mu (\chi) = \p_\mu \chi + i g \cm{A_\mu}{\chi}.
\end{equation}
If $\chi$ transforms like
\begin{equation}
\chi' = V \chi V^\dg,
\end{equation}
then $D_\mu \chi$ transforms in a similar fashion
\begin{align}
D'_\mu \chi' &=\p_\mu \chi' + i g \cm{A'_\mu}{\chi'} \nn
&= V \p_\mu \chi V^\dg - \cm{V \p_\mu V^\dg}{\chi'} + i g V \cm{A_\mu}{\chi} V^\dg + \cm{V \p_\mu V^\dg}{\chi'}\nn
&= V \lb \p_\mu \chi + i g \cm{A_\mu}{\chi} \rb V^\dg \nn
&= V D_\mu \chi V^\dg,
\end{align}
where we have used $\p_\mu (V V^\dg) = \p_\mu V V^\dg + V \p_\mu V^\dg = 0$.

The anti-hermitian, traceless part $\ah{V}$ of a  matrix $V$ in the fundamental representation is given by
\begin{equation}  \label{eq:ah_definition_2}
\ah{V} \equiv \frac{1}{2i} \lb V -V^\dg\rb - \frac{1}{N_c} \tr \left[ \frac{1}{2i} \lb V -V^\dg\rb \right] \, \one.
\end{equation}
It holds that
\begin{equation}
\ah{V^\dg} = - \ah{V},
\end{equation}
and
\begin{equation}
\tr \left( \ah{V} \right) = 0.
\end{equation}
If $V$ is hermitian, i.e.\ $V^\dg = V$, then $\ah{V} = 0$.
The anti-hermitian traceless part $\ah{V}$ is related to the components $P^a(V)$. They are defined by
\begin{equation}
P^a(V) \equiv 2 \, \Im \, \tr \left[ t^a V \right],
\end{equation}
where ``$\Im$" denotes the imaginary part.
For the fundamental representation of $\SUN$, it holds that
\begin{align} 
\sum_a t^{a}P^{a}\left(V\right) &= \frac{1}{2i}\left(V-V^{\dg}\right)-\frac{1}{N_c}\tr\left(\frac{1}{2i}\left(V-V^{\dg}\right)\right) \one, \nn
&= \ah{V} \label{eq:Pa_and_ah}
\end{align}
which can be shown using the Fierz identity for the generators of $\sun$
\begin{equation}
\sum_a t_{ij}^{a}t_{kl}^{a}=\frac{1}{2}\left(\delta_{il}\delta_{jk}-\frac{1}{N_c}\delta_{ij}\delta_{kl}\right),
\end{equation}
where $i,j,k,l$ are fundamental representation matrix indices.

\section{Lattice notation} \label{sec:lattice_notation}

In order to make calculations in discretized systems less cumbersome, we use a compact notation for discretized fields on lattices. We define the four-dimensional, regular hypercubic lattice $\Lambda^4$ as the set of lattice sites:
\begin{equation} \label{eq:lambda4_def}
\Lambda^4 = \left\{ x \, | \, x = \sum_{\mu=0}^3 n_\mu \hat{a}^\mu, \quad n_\mu \in \mathbb{Z}  \right\},
\end{equation}
where $\hat{a}^\mu = a^\mu \hat{e}_\mu$ (no sum implied) with unit vectors $\hat{e}_\mu$. The time step is $a^0$ (sometimes $\Delta t$) and the spatial lattice spacings are $a^i$. We use a shorthand notation to denote the neighbors of a lattice site $x$. For example, $x+\mu$ corresponds to the lattice site $x$ shifted by $\hat{a}^\mu$. A set of points $x \in \Lambda^4$ which share the same time coordinate $x^0$ are referred to as a time slice.

A discretized field $\{ \phi_x \}$ is the set of field values defined at every lattice site $x \in \Lambda^4$. The neighbors of $\phi_x$ are denoted as $\phi_{x+\mu}$. For example, $\phi_{x+0}$ is the field $\phi_x$ shifted by one time step. $\phi_{x+1}$ is $\phi_x$ shifted by one lattice spacing $a^1$ along the $x^1$ axis.

Forward and backward finite differences are defined as
\begin{align}
\p^F_\mu \phi_x &\equiv \frac{\phi_{x+\mu} - \phi_x}{a^\mu}, \\
\p^B_\mu \phi_x &\equiv \frac{\phi_x - \phi_{x-\mu}}{a^\mu}.
\end{align}

We use the ``$\sim$" symbol to denote equivalence under the sum over all lattice sites. For example, using explicit sums it holds that
\begin{equation}
\sum_x \phi_x \phi_{x+\mu} = \sum_x \phi_{x-\mu} \phi_{x},
\end{equation}
where we have simply used the shift $x \rightarrow x - \mu$. We write ``$\sim$" when we omit the explicit sum, i.e.\
\begin{equation}
\phi_x \phi_{x+\mu} \sim \phi_{x-\mu} \phi_{x}.
\end{equation} 
If $\phi_x$ is matrix valued (e.g.\ when working with real-time lattice gauge theory) and an expression involves a trace, ``$\sim$" also implies equivalence under the trace. For example the expression
\begin{equation}
\phi_x \phi_{x+\mu} \sim \phi_{x} \phi_{x-\mu}
\end{equation}
is shorthand for
\begin{equation}
\sum_x \tr \left[ \phi_x \phi_{x+\mu} \right] = \sum_x \tr \left[ \phi_{x} \phi_{x-\mu} \right],
\end{equation}
where we have used the cyclic property of the trace.

As an example of a derivation using this shorthand notation, we prove a formula for summation by parts, which is the discrete analogue of integration by parts. It holds that
\begin{equation}
\sum_x \phi_x \p^F_\mu \phi_x = - \sum_x \lb \p^B_\mu \phi_x \rb \phi_x.
\end{equation}
To see this, we write first write the LHS of the above equation as
\begin{equation}
\phi_x \p^F_\mu \phi_x,
\end{equation}
and then perform simple manipulations using the ``$\sim$" symbol:
\begin{align}
\phi_x \p^F_\mu \phi_x &= \frac{1}{a^\mu} \lb \phi_x   \phi_{x+\mu} - \phi_x^2 \rb \nn
&\sim \frac{1}{a^\mu} \lb  \phi_{x-\mu} \phi_x - \phi_x^2 \rb \nn
&= - \lb \p^B_\mu \phi_x\rb \phi_x,
\end{align}
or simply
\begin{equation}
\phi_x \p^F_\mu \phi_x \sim - \lb \p^B_\mu \phi_x \rb \phi_x.
\end{equation}

\chapter{Boost invariant Yang-Mills theory}

\section{Yang-Mills theory in $\tau$, $\eta$ coordinates} \label{sec:ym_tau_eta}
The Yang-Mills action in flat coordinates (e.g.\ laboratory frame coordinates $x^\mu = (t,x,y,z)^\mu$) is given by
\begin{equation}
S = \intop d^4x \lb - \frac{1}{2} \tr \left[ F_{\mu\nu} F^{\mu\nu} \right] \rb .
\end{equation}
The goal is to rewrite this action in terms of the $(\tau, x, y, \eta)$ coordinate set, which is defined via
\begin{align}
\tau &= \sqrt{t^2 - z^2}, \\
\eta &= \frac{1}{2} \ln \left( \frac{t+z}{t-z} \right).
\end{align}
In terms of light cone coordinates
\begin{equation}
x^\pm = \frac{t \pm z}{\sqrt{2}},
\end{equation}
one finds
\begin{align}
\tau &= \sqrt{2 x^+ x^-}, \\
\eta &= \frac{1}{2} \ln \left( \frac{x^+}{x^-} \right).
\end{align}
The gauge field $A^\mu$ transforms as a four-vector:
\begin{equation}
A'^\mu(x') = \frac{\p x'^\mu}{\p x^\nu} A^\nu(x).
\end{equation}
The temporal component $A^\tau$ expressed in terms of $A^+$ and $A^-$ is therefore
\begin{align}
A^\tau &= \frac{\p \tau}{\p x^+} A^+ + \frac{\p \tau}{\p x^-} A^- \nn
&= \frac{1}{\tau} \left( x^- A^+ + x^+ A^-\right).
\end{align}
Similarly, one finds
\begin{equation}
A^\eta = \frac{1}{\tau^2}  \left( x^- A^+ - x^+ A^-\right).
\end{equation}
In $\tau$, $\eta$ coordinates we have the infinitesimal line element
\begin{equation}
ds^2 = d\tau^2 - dx_T^2 - \tau^2 d\eta^2,
\end{equation}
which implies the metric
\begin{equation}
g_{\mu \nu} = \mathrm{diag} \lb 1, -1, -1, -\tau^2 \rb.
\end{equation}
The infinitesimal space-time volume is given by $d^4x = \tau d\tau d^2 x_T d\eta$. Therefore we can write
\begin{equation}
S = - \intop d\tau d^2x_T d\eta \, \tr \left[ \tau \lb F_{\tau i} F^{\tau i} + F_{\tau \eta} F^{\tau \eta} + \frac{1}{2} F_{ij} F^{ij} + F_{\eta i} F^{\eta i} \rb \right].
\end{equation}
Lowering all indices we find
\begin{equation} \label{eq:taueta_action}
S = + \intop d\tau d^2x_T d\eta \, \tr \left[  \tau F_{\tau i} F_{\tau i} + \frac{1}{\tau} F_{\tau \eta}^2 - \frac{\tau}{2} F_{ij} F_{ij} - \frac{1}{\tau} F_{\eta i} F_{\eta i} \right].
\end{equation}
One could define the electric fields in the $\tau, \eta$ frame as
\begin{align}
E_i &= F_{\tau i}, \\
E_\eta &= F_{\tau \eta},
\end{align}
and thus we can rewrite the action as
\begin{equation}
S = + \intop d\tau d^2x_T d\eta \, \tr \left[  \tau E_i E_i + \frac{1}{\tau} E_\eta^2 - \frac{\tau}{2} F_{ij} F_{ij} - \frac{1}{\tau} F_{\eta i} F_{\eta i} \right].
\end{equation}
However, interpreting $E_\eta$ as the actual longitudinal electric field is problematic, since it only has units of energy instead of units of energy squared. Relating the field strength tensor in $\tau, \eta$ coordinates to the laboratory frame field strengths one finds \cite{Fujii:2008dd}
\begin{align}
E_i &= F_{\tau i} \cosh(\eta) - \frac{1}{\tau} F_{\eta i} \sinh(\eta), \label{eq:lab_to_taueta_1} \\
E_3 &= \frac{1}{\tau} F_{\tau \eta}, \label{eq:lab_to_taueta_2} \\
B_i &= \epsilon_{ij} \lb F_{j\tau} \sinh(\eta) + \frac{1}{\tau} F_{\eta j} \cosh (\eta) \rb, \label{eq:lab_to_taueta_3} \\
B_3 &= - F_{12}. \label{eq:lab_to_taueta_4}
\end{align}
The additional factor $\tau^{-1}$ in the relation for $E_3$ restores the correct units. We also define the canonical momenta using functional derivatives
\begin{align}
P^{a,\eta} &= \frac{\dd S}{\dd \lb \p_\tau A^a_\eta \rb} = \frac{1}{\tau} F^a_{\tau \eta}, \\
P^{a,i} &=  \frac{\dd S}{\dd \lb \p_\tau A^a_i \rb} = \tau F^a_{\tau i}.
\end{align}

We start varying the action \eqref{eq:taueta_action} with respect to $A_\tau$ without specifying any gauge condition. This yields
\begin{align}
\dd_\tau S &= 2 \int d\tau d^2x_T d\eta \, \tr \bigg[ \tau F_{\tau i} \lb - \p_i \dd A_\tau + i g \cm{\dd A_\tau}{A_i} \rb \nn
& \quad + \frac{1}{\tau} F_{\tau \eta} \lb - \p_\eta \dd A_\tau + i g \cm{\dd A_\tau}{A_\eta} \rb\bigg].
\end{align}
Integrating by parts and making use of $\tr \lb A \cm{B}{C} \rb = \tr \lb \cm{A}{B} C \rb$ gives
\begin{equation}
\dd_\tau S = 2 \int d\tau d^2x_T d\eta \, \tr \bigg[ \lb \tau D_ i F_{\tau i} + \frac{1}{\tau} D_\eta F_{\tau \eta} \rb \delta A_\tau \bigg],
\end{equation}
which by requiring $\dd S = 0$ yields the Gauss constraint
\begin{equation}
\tau D_ i F_{\tau i} + \frac{1}{\tau} D_\eta F_{\tau \eta} = 0,
\end{equation}
or in terms of canonical momenta
\begin{equation} \label{eq:taueta_gauss}
D_i P^i + D_\eta P^\eta = 0.
\end{equation}
The only remaining variations to be done are with respect to $A_i$ and $A_\eta$. We perform both already employing Fock-Schwinger gauge
\begin{equation}
x^+ A^- + x^- A^+ = \tau A^\tau = 0,
\end{equation}
which is equivalent to temporal gauge $A^\tau = 0$. Furthermore, we require that the gauge fields are boost invariant, i.e.\ they do not depend on $\eta$. The action then simply reads
\begin{equation} \label{eq:taueta_action2}
S = \intop d\tau d^2x_T d\eta \, \tr \left[  \tau  \p_\tau A_i \p_\tau A_i  + \frac{1}{\tau} \lb \p_\tau A_\eta \rb^2 - \frac{\tau}{2} F_{ij} F_{ij}  - \frac{1}{\tau} D_i A_\eta D_i A_\eta  \right].
\end{equation}
In order to always guarantee that the fields are independent of $\eta$, gauge transformations must be independent of $\eta$ as well. This restriction affects the transformation behavior of $A_\eta$, which now transforms as a scalar
\begin{equation}
A_\eta(\tau,x_T) \rightarrow \Omega(x_T) A_\eta(\tau,x_T) \Omega^\dg(x_T).
\end{equation}
In fact, the action \eqref{eq:taueta_action2} is equivalent to the action of a boost invariant, color charged scalar:
\begin{equation}
S = \intop d\tau d^2x_T d\eta \, \tau \, \tr \left[ \p_\tau A_i  \p_\tau A_i - \frac{1}{2} F_{ij} F_{ij} + \frac{1}{\tau^2} \lb \p_\tau \phi \rb^2- \frac{1}{\tau^2} D_i \phi D_i \phi \right],
\end{equation}
with $A_\eta = \phi$.

Varying the action \eqref{eq:taueta_action2} with respect to $A_i$ yields
\begin{equation}
\dd_i S = - 2 \intop d\tau d^2x_T d\eta \, \tr \left[ \lb \p_\tau \lb \tau \p_\tau A_i \rb  - \tau D_j F_{ji} + \frac{ig}{\tau}  \cm{A_\eta}{D_i A_\eta} \rb \dd A_i \right],
\end{equation}
while varying with respect to $A_\eta$ gives
\begin{equation}
\dd_\eta S = - 2 \intop d\tau d^2x_T d\eta \, \tr \left[ \lb \p_\tau \lb \frac{1}{\tau} \p_\tau A_\eta \rb  - \frac{1}{\tau} D_i \lb D_i A_\eta \rb \rb \dd A_\eta \right].
\end{equation}
The equations of motion therefore read
\begin{align}
\p_\tau \lb \tau \p_\tau A_i \rb &=  \tau D_j F_{ji} - \frac{ig}{\tau}  \cm{A_\eta}{D_i A_\eta}, \\
\p_\tau \lb \frac{1}{\tau} \p_\tau A_\eta \rb &= \frac{1}{\tau} D_i \lb D_i A_\eta \rb,
\end{align}
or in terms of canonical momenta
\begin{align}
\p_\tau P^i &=  \tau D_j F_{ji} - \frac{ig}{\tau}  \cm{A_\eta}{D_i A_\eta}, \label{eq:bi_eq_i} \\
\p_\tau P^\eta &= \frac{1}{\tau} D_i \lb D_i A_\eta \rb, \label{eq:bi_eq_eta}
\end{align}
with
\begin{align}
\p_\tau A_\eta &= \tau P^\eta, \\
\p_\tau A_i &= \frac{1}{\tau} P^i. \label{eq:canon_momentum_i}
\end{align}
Under the assumption of boost invariance the Gauss constraint \eqref{eq:taueta_gauss} simplifies to
\begin{equation} \label{eq:bi_gauss}
D_i P^i + i g \cm{A_\eta}{P^\eta} = 0.
\end{equation} 

\section{Boost invariant real-time lattice gauge theory} \label{app_bi_latt}

In order to solve the boost invariant Yang-Mills equations numerically, we employ real-time lattice gauge theory techniques. In this section we present a non-rigorous, but quick derivation of the discretized equations. A more thorough discussion of real-time lattice gauge theory techniques follows in \cref{cha:rtlgt}. 

Starting from the action \eqref{eq:taueta_action2}, we replace the continuous transverse plane with a finite, rectangular grid while keeping proper time $\tau$ continuous. Instead of the transverse gauge field $A_i(\tau, x_T)$, we use the transverse gauge links
\begin{equation}
U_{x,i}(\tau) \simeq \exp \lb i g a^i A^a_i(\tau, x_T) t^a\rb,
\end{equation}  
where the index $x$ refers to the coordinate in the transverse plane $x_T$ and $a^i$ is the lattice spacing in the $x^i$ direction. The gauge links are elements of the gauge group in the fundamental representation, i.e.\ unitary $N_c \times N_c$ matrices.
As a conjugated momentum we use $P^i_x(\tau)$ (instead of the continuous canonical momentum $P^i(\tau, x_T)$) which we define to be acting on the left via
\begin{equation} \label{eq:Ui_update_cont}
\p_\tau U_{x,i}(\tau) = \frac{i g a^i}{\tau} P^i_x(\tau) U_{x,i}(\tau),
\end{equation}
which reduces to \cref{eq:canon_momentum_i} in the continuum limit $a^i \rightarrow 0$. Note that $P^i_x(\tau)$ is an algebra element of the gauge group in the fundamental representation, i.e.\ a hermitian, traceless $N_c \times N_c$ matrix. Consequently, it is guaranteed that the matrices $U_{x,i}(\tau)$ remain unitary under time evolution, since
\begin{equation}
\p_\tau \lb U_{x,i}(\tau) U^\dg_{x,i}(\tau) \rb = \p_\tau U_{x,i}(\tau) U^\dg_{x,i}(\tau) + U_{x,i}(\tau) \p_\tau U^\dg_{x,i}(\tau) \equiv 0,
\end{equation}
and similarly $\p_\tau \det U_{x,i}(\tau) \equiv 0$.

The gauge links are to be interpreted as Wilson lines connecting nearest neighbors on the rectangular lattice. Under discrete gauge transformations (i.e.\ gauge transformations defined at the discrete lattice sites $x$), the gauge links transform according to
\begin{equation}
U_{x,i}(\tau) \rightarrow \Omega_x U_{x,i}(\tau) \Omega^\dg_{x+i},
\end{equation}
where $x+i$ refers to the lattice site $x$ shifted by one cell in the direction of $x^i$. Since we are employing temporal gauge $A_\tau = 0$, the discrete gauge transformations $\Omega_x$ must be independent of $\tau$.
Furthermore, since we implement boost invariance at the level of gauge fields, gauge transformations are also independent of $\eta$.

The smallest possible Wilson loops that can be formed on the lattice are the transverse plaquettes
\begin{equation}
U_{x,ij}(\tau) \equiv U_{x,i}(\tau) U_{x+i,j}(\tau) U_{x+i+j,-i}(\tau) U_{x+j,-j}(\tau),
\end{equation}
where we define $U_{x,-i}(\tau) \equiv U^\dg_{x-i,i}(\tau)$. The plaquette at $x$ transforms locally, i.e.\
\begin{equation}
U_{x,ij}(\tau) \rightarrow \Omega_x U_{x,ij}(\tau) \Omega^\dg_x.
\end{equation}
In the continuum limit the plaquette $U_{x,ij}(\tau)$ can be related to the field strength tensor (see \cref{eq:plaquette_exponential})
\begin{equation}
U_{x,ij}(\tau) \simeq \exp \lb i \lb g a^i a^j F_{ij}(\tau, x_T) +  \mathcal{O}(a^3) \rb \rb,
\end{equation}
which yields the approximate, gauge invariant relation (see \cref{eq:tr_umunu})
\begin{equation}
\tr \lb 2 - U_{x,ij}(\tau) - U^\dg_{x,ij} (\tau) \rb \simeq \lb g a^i a^j\rb^2 \tr \lb F^2_{ij}(\tau, x_T) \rb + \mathcal{O}(a^6).
\end{equation}

On the other hand, the gauge field component $A_\eta(\tau, x_T)$ does not need to be replaced by a link, since the $\eta$ direction is kept continuous. It is simply replaced by $A_{x,\eta} (\tau)$ defined at the transverse lattice sites $x$. Under gauge transformations, this field transforms locally
\begin{equation}
A_{x,\eta}(\tau) \rightarrow \Omega_x A_{x,\eta}(\tau) \Omega^\dg_x,
\end{equation} 
since the transformations are by definition independent of $\eta$.
The discretized canonical momentum of $A_{x,\eta} (\tau)$ is simply defined by
\begin{equation} \label{eq:Aeta_update_cont}
P^\eta_x(\tau) \equiv \frac{1}{\tau} \p_\tau A_{x,\eta}(\tau).
\end{equation}
Gauge-covariant derivatives acting on $A_{x,\eta}(\tau)$ in the transverse plane have to be replaced by gauge-covariant finite difference expressions. We define the first-order forward and backward derivatives
\begin{align}
D^F_i A_{x,\eta}(\tau) & \equiv \frac{1}{a^i} \lb {U_{x,i}(\tau) A_{x+i,\eta}(\tau) U_{x+i,-i}(\tau) - A_{x,\eta}} \rb, \\
D^B_i A_{x,\eta}(\tau) & \equiv \frac{1}{a^i} \lb {A_{x,\eta}(\tau) - U_{x,-i}(\tau) A_{x-i,\eta} U_{x-i,i}(\tau)} \rb.
\end{align}

The (semi-)discretized, boost invariant action in temporal gauge can now be defined as
\begin{align} \label{eq:taueta_action_disc}
S & = \intop d\tau d\eta \sum_x \lb \prod_k a^k \rb \, \tr \bigg[ \sum_i \frac{1}{\tau} \lb P^i_x(\tau) \rb^2 + \tau \lb P^\eta_x(\tau) \rb^2 \nn
& \quad - \sum_{i,j} \frac{\tau}{2 \lb g a^i a^j \rb^2} \lb 2 - U_{x,ij}(\tau) - U^\dg_{x,ij} (\tau) \rb - \sum_i \frac{1}{\tau} \lb D^F_i A_{x,\eta}(\tau) \rb^2  \bigg],
\end{align}
and the discretized Gauss constraint reads
\begin{equation} \label{eq:bi_latt_gauss}
\sum_i D^B_i P^i_x(\tau) + i g \cm{A_{x,\eta}(\tau)}{P^\eta_x(\tau)} = 0.
\end{equation}
In order to derive the discretized equations of motion we have to vary the above action with respect to $A_{x,\eta}(\tau)$ and $U_{x,i}(\tau)$. The variation with respect to the scalar component $A_{x,\eta}$ is performed in a straightforward way with $\dd P^\eta_{x}(\tau) = \frac{1}{\tau} \p_\tau \dd A_{x,\eta}(\tau)$. The variation then reads
\begin{equation}
\dd_\eta S = 2 \intop d\tau d\eta \sum_x \lb \prod_k a^k \rb \, \tr \bigg[ P^\eta_x \p_\tau \dd A_{x,\eta} - \frac{1}{\tau} D^F_i A_{x,\eta} D^F_i \dd A_{x,\eta} \bigg].
\end{equation}
The terms involving the $\tau$ derivative can be integrated by parts in the usual way. For the discrete gauge-covariant differences we can perform a summation by parts, i.e.
\begin{equation}
\sum_x D^F_i A_{x,\eta} D^F_i \dd A_{x,\eta} = - \sum_x D^B_i D^F_i A_{x,\eta} \dd A_{x,\eta},
\end{equation}
where the forward difference turns into a backward difference. Defining $D^2_i = D^B_i D^F_i$ (no sum implied), the variation then reads
\begin{equation}
\dd_\eta S = - 2 \intop d\tau d\eta \sum_x \lb \prod_k a^k \rb \, \tr \left[ \lb \p_\tau P^\eta_x  - \sum_i \frac{1}{\tau} D^2_i A_{x,\eta} \rb \dd A_{x,\eta} \right],
\end{equation}
and with $\dd_\eta S = 0$ we can read off the discrete equation of motion
\begin{equation} \label{eq:bi_latt_eta}
\p_\tau P^\eta_x  = \sum_i \frac{1}{\tau} D^2_i A_{x,\eta},
\end{equation}
which is completely analogous to the continuous equation \eqref{eq:bi_eq_eta}.

The variation of \cref{eq:taueta_action_disc} with respect to the transverse gauge links $U_{x,i}(\tau)$ is slightly more involved. It is insufficient to simply vary the matrix elements of $U_{x,i}(\tau)$ independently, because gauge links are elements of $\textrm{SU}(N_c)$, i.e.\ they satisfy the constraints $U_{x,i}(\tau) U^\dg_{x,i}(\tau) = \one$ and $\det U_{x,i}(\tau) = 1$. However, we can perform a variation that leaves these constraints fulfilled (see \cref{app_var}):
\begin{equation}
\dd U_{x,i}(\tau) = i g a^i \dd A_{x,i}(\tau) U_{x,i}(\tau).
\end{equation}
Starting with the kinetic term  of \cref{eq:taueta_action_disc}, we find after some algebra and integration by parts
\begin{equation}
\dd_i \frac{1}{\tau} \tr \bigg[ \lb P^i_{x}(\tau) \rb^2 \bigg] = - 2 \, \tr \bigg[ \p_\tau P^i_x(\tau) \dd A_{x,i}(\tau) \bigg] = - \p_\tau P^{a,i}_x \dd A^a_{x,i}.
\end{equation}
Varying the plaquette term  of \cref{eq:taueta_action_disc} we find
\begin{align}
\sum_{x,i,j} \tr \bigg[ 2 - U_{x,ij} - U^\dg_{x,ij} \bigg] & = - 2 \sum_{x,i,j} g a^i \tr \bigg[ i \dd A_{x,i} \lb U_{x,ij} + U_{x,i-j} \rb + \hc \bigg] \nn
& = - 2 i \sum_{x,i,j} g a^i \tr \bigg[ \lb U_{x,ij} + U_{x,i-j} - \hc \rb t^a \bigg] \dd A^a_{x,i}. 
\end{align}
With the shorthand
\begin{equation}
P^a \lb U \rb \equiv 2 \, \Im \, \tr \lb t^a U \rb = - i \,  \tr \lb t^a \lb U- U^\dg \rb \rb,
\end{equation}
we can write this as
\begin{equation}
\sum_{x,i,j} \tr \bigg[ 2 - U_{x,ij} - U^\dg_{x,ij} \bigg] = + 2 \sum_{x,i,j} g a^i P^a \lb U_{x,ij} + U_{x,i-j}  \rb \dd A^a_{x,i}.
\end{equation}
Varying the term involving $D^F_i A_{x,\eta}$  of \cref{eq:taueta_action_disc} with respect to $U_{x,i}$ we find
\begin{equation}
\dd_i \sum_{x} \tr \bigg[ \lb D^F_i A_{x,\eta} \rb^2 \bigg] = - g P^a \lb \cm{A^{(+i)}_{x,\eta}}{D^F_i A_{x,\eta}} \rb \dd A^a_{x,i},
\end{equation}
where we used the shorthand notation $A^{(+i)}_{x,\eta} = U_{x,i} A_{x+i,\eta} U_{x+i,-i}$.

Combining all terms, the variation of the action reads
\begin{align}
\dd_i S & = \intop d\tau d\eta \sum_x \lb \prod_k a^k \rb \, \bigg[ - \p_\tau P^{a,i}_x
 - \sum_j \frac{\tau}{g a^i \lb a^j \rb^2} P^a \lb U_{x,ij} + U_{x,i-j} \rb \nn
& \quad  + \frac{g}{\tau} P^a \lb \cm{A^{(+i)}_{x,\eta}}{D^F_i A_{x,\eta}} \rb \bigg] \dd A^a_{x,i}.
\end{align}
With $\dd_i S = 0$ the equations of motion for $U_{x,i}$ read
\begin{equation} 
\p_\tau P^{a,i}_x = - \sum_j \frac{\tau}{ g a^i \lb a^j \rb^2} P^a \lb U_{x,ij} + U_{x,i-j} \rb - \frac{g}{\tau} P^a \lb \cm{A^{(+i)}_{x,\eta}}{D^F_i A_{x,\eta}} \rb.
\end{equation}
Multiplying everything with $t^a$ and using (see \cref{eq:Pa_and_ah})
\begin{align}
\sum_a t^a P^a \lb U \rb &= \ah{U} = \frac{1}{2 i} \lb \lb U- U^\dg \rb - \frac{1}{N_c }  \tr  \lb U- U^\dg \rb \one \rb,
\end{align}
and
\begin{align}
\sum_a t^a P^a \lb \cm{A}{B} \rb
&= 2 \sum_{a, b, c, d} t^a f^{bcd} A^b B^c \tr \lb t^a t^d \rb  \nn 
&= \sum_{b,c,d} f^{bcd} A^b B^c t^d \nn
&= \frac{1}{i} \cm{A}{B},
\end{align}
we can write this equation directly in terms of algebra and group elements
\begin{equation} \label{eq:bi_latt_i}
\p_\tau P^i_x = - \sum_j \frac{\tau}{g a^i \lb a^j \rb^2} \ah{ U_{x,ij} + U_{x,i-j}}
- \frac{i g}{\tau} \cm{A^{(+i)}_{x,\eta}}{D^F_i A_{x,\eta}}.
\end{equation}
While the continuum limit of the commutator term in the above equation on the right is obvious, the term involving plaquettes is more subtle. Using $U_{x,ij} \simeq \exp \lb i g a^i a^j F_{ij}(x_T) \rb$ and $U_{x,i-j} = U_{x,-j} U_{x-j,ji} U_{x-j,j}$, we can expand for small lattice spacing to arrive at
\begin{align}
\sum_j \frac{\tau}{g a^i \lb a^j \rb^2} \ah{ U_{x,ij} + U_{x,i-j}} & \simeq \sum_j \frac{\tau}{a^j} \lb F_{ij}(x_T) -  U_{x,-j} F_{ij}(x_T-a^j \hat{e}_j) U_{x-j,j}\rb \nn
& = - \sum_j \tau D^B_j F_{x,ji}(x_T).
\end{align}
Therefore the equations of motion for $U_{x,i}$ reduce to \cref{eq:bi_eq_i} in the continuum limit $a^i \rightarrow 0$.

Finally, in order to solve the time evolution numerically, the time coordinate $\tau$ also has to be discretized in eqs.\ \eqref{eq:Ui_update_cont}, \eqref{eq:Aeta_update_cont}, \eqref{eq:bi_latt_gauss}, \eqref{eq:bi_latt_eta} and \eqref{eq:bi_latt_i}. A simple approach is to use a leapfrog time integrator. We introduce the finite time step $\Delta \tau$ and discrete times $\tau_n = n \Delta \tau$. We then define that the scalar component $A_{x,\eta}$ and the transverse gauge links $U_{x,i}$ are evaluated at whole-numbered time steps $\tau_n$, while the canonical momenta $P^\eta_x$ and $P^i_x$ are evaluated in-between these steps, i.e.\ at $\tau_{n+1/2}$. The discretized equations (i.e.\ update equations) then read
\begin{align} \label{eq:bi_leapfrog_1}
P^\eta_x(\tau_{n+\frac{1}{2}}) &= P^\eta_x(\tau_{n-\frac{1}{2}}) + \frac{\Delta \tau}{\tau_n} \sum_i D^2_i A_{x,\eta}(\tau_n), \\
\label{eq:bi_leapfrog_2}
P^i_x(\tau_{n+\frac{1}{2}}) &= P^i_x(\tau_{n-\frac{1}{2}}) - \sum_j \frac{\Delta \tau \tau_n}{g a^i \lb a^j \rb^2} \ah{ U_{x,ij}(\tau_n) + U_{x,i-j}(\tau_n)} \nn
& \quad - \frac{i g \Delta \tau}{\tau_n} \cm{A^{(+i)}_{x,\eta}(\tau_n)}{D^F_i A_{x,\eta}(\tau_n)}, \\
\label{eq:bi_leapfrog_3}
A_{x,\eta}(\tau_{n+1}) &= A_{x,\eta}(\tau_{n}) + \Delta \tau \tau_{n+\frac{1}{2}} P^\eta_x(\tau_{n+\frac{1}{2}}), \\
\label{eq:bi_leapfrog_4}
U_{x,i}(\tau_{n+1}) &= \exp \lb \frac{i g a^i \Delta \tau}{\tau_{n+\frac{1}{2}}} P^i_x(\tau_{n+\frac{1}{2}}) \rb U_{x,i}(\tau_n),
\end{align}
and the discrete Gauss constraint consistent with these update equations is given by
\begin{equation}
\sum_i D^B_i P^i_x(\tau_{n+\frac{1}{2}}) + i g \cm{A_{x,\eta}(\tau_n)}{P^\eta_x(\tau_{n+\frac{1}{2}})} = 0.
\end{equation}
The Gauss constraint contains momenta evaluated at $\tau_{n+1/2}$, while gauge fields and links are evaluated at $\tau_n$. It is straightforward to prove that the Gauss constraint is conserved under the update equations.

\chapter{Real-time lattice gauge theory} \label{cha:rtlgt}

\section{Parametrization for $\mathrm{SU(2)}$} \label{app_su2}

In this section of the appendix we list a few formulae and identities for the parametrization of the SU(2) group that is being used in the simulation. This parametrization can also be used for elements of $U(2)$ and anti-hermitian, traceless matrices in $\mathbb{C}^2$. 


We can represent any SU(2) group element using the exponential
map applied to an algebra element. The dimension of the Lie algebra $\mathfrak{su}$(2) of SU(2) is $3$. Given the three-component vector $A_{a}$
and the generators $t_{a}={\sigma_{a}} / {2}$ with the Pauli matrices
$\sigma_{a}$, we can write any element $U\in\mathrm{SU(2)}$ in the fundamental representation as

\begin{equation}
	U=\exp(i \sum_a A_{a}t_{a}).
\end{equation}
The matrix $U$ is unitary and has determinant one. Exploiting the
property of the Pauli matrices (no sum over $a$ implied)
\begin{align}
	\sigma_{a}^{2}&=\one, \\
	t_{a}^{2}&=\frac{1}{4} \, \one,
	\end{align}
we write 
\begin{align}
	U  = & \exp (i \sum_a A_{a} t_{a} )\nonumber \\
	 = & \sum_{n=0}^{\infty}\frac{(i \sum_aA_{a}t_{a})^{n}}{n!}\nonumber \\
	 = & \sum_{n=0}^{\infty}\frac{(-1)^{n}(\sum_aA_{a}t_{a})^{2n}}{2n!}+i\sum_{n=0}^{\infty}\frac{(-1)^{n}}{(2n+1)!}(\sum_a A_{a}t_{a})^{2n} \sum_aA_{a}t_{a}\nonumber \\
	 = & \sum_{n=0}^{\infty}\frac{(-1)^{n}\left(\frac{A}{2}\right)^{2n}}{2n!} \, \one+i\sum_{n=0}^{\infty}\frac{(-1)^{n}\left(\frac{A}{2}\right)^{2n+1}}{(2n+1)!}\sum_a e_{a}\sigma_{a}\nonumber \\
	 = & \cos(\frac{A}{2}) \, \one+i\sin(\frac{A}{2})\sum_a e_{a}\sigma_{a},
\end{align}
where we used $A=\sqrt{\sum_a A_{a}A_{a}}$ and $e_{a}={A_{a}}/ {A}$.
The above exact relation provides a simple way of computing the exponential
map in SU(2).
Moreover, it allows us to represent any group element in SU(2) with only 
three real-valued parameters $A_a$.
Since computing trigonometric functions can be numerically costly, one can switch to a
slightly different representation. By introducing the four real-valued parameters
\begin{eqnarray}
	u_{0} & = & \cos(\frac{A}{2}), \label{eq:su2_param_cos}\\
	u_{a} & = & \sin(\frac{A}{2})e_{a}, \label{eq:su2_param_sin}
\end{eqnarray}
the group element can be written as
\begin{equation}
	U=u_{0}\, \one+i \sum_a u_{a}\sigma_{a}.
\end{equation}
The above is only a group element if the constraint
\begin{equation} \label{eq:app_su2_constraint}
u_{0}^{2}+ \sum_a \lb u_a \rb^2 = 1,
\end{equation}
holds. This parametrization is particularly useful, since it also allows us to represent other complex matrices as well, such as sums of unitary matrices. In the following sections we use more compact notation and drop the sum symbol.

\subsection{Multiplication}

Scalar multiplication is very simple in terms of parameters $u_0$ and $u_a$. Let $\lambda\in\mathbb{R}$
be a real number. We then have
\begin{eqnarray}
	U' & = & \lambda U\nonumber \\
	& = & \left(\lambda u_{0}\right)\one+i\left(\lambda u_{i}\right)\sigma_{i}\nonumber \\
	& = & u_{0}' \, \one+iu_{i}'\sigma_{i}.
\end{eqnarray}
We just have to multiply the parameter vector $u$ by the scalar $\lambda$.
Note that the norm of the parameter vector changes to
\begin{equation}
	(u')^{2}=\lambda^{2}u^{2}.
\end{equation}

Matrix multiplication in terms of parameter vectors is more complicated.
Let $U_{1}$ and $U_{2}$ be two group elements with parameter vectors
$a$ and $b$ respectively. Using the property
\[
\sigma_{i}\sigma_{j}=\delta_{ij}\one+i\e_{ijk}\sigma_{k}
\]
we find
\begin{align}
	U_{3}=U_{1}U_{2} & = \left(a_{0}b_{0}-a_{j}b_{j}\right)\one\nonumber \\
	  & \quad +i\sigma_{k}\left(a_{0}b_{k}+a_{k}b_{0}-\e_{ijk}a_{i}b_{j}\right).
\end{align}
Let $c$ be the parameter vector for $U_{3}$. We then find
\begin{align}
	c_{0} & = a_{0}b_{0}-a_{j}b_{j},\\
	c_{1} & = a_{0}b_{1}+a_{1}b_{0}-a_{2}b_{3}+a_{3}b_{2},\\
	c_{2} & = a_{0}b_{2}+a_{2}b_{0}-a_{3}b_{1}+a_{1}b_{3},\\
	c_{3} & = a_{0}b_{3}+a_{3}b_{0}-a_{1}b_{2}+a_{2}b_{1}.
\end{align}
The matrix $U_{3}$ is again a $SU(2)$ matrix due to
\begin{align}
	c_{0}^{2}+c_{k}^{2} & =  \left(a_{0}b_{0}-a_{j}b_{j}\right)^{2}+\left(a_{0}b_{k}+a_{k}b_{0}-\e_{ijk}a_{i}b_{j}\right)^{2}, \\
	\left(a_{0}b_{0}-a_{j}b_{j}\right)^{2} & =  a_{0}^{2}b_{0}^{2}-2a_{0}b_{0}\vec{a}\cdot\vec{b}+\left(\vec{a}\cdot\vec{b}\right)^{2},\\
	\left(a_{0}b_{k}+a_{k}b_{0}-\e_{ijk}a_{i}b_{j}\right)^{2} & =  a_{0}^{2}\vec{b}^{2}+b_{0}^{2}\vec{a}^{2}+2a_{0}b_{0}\vec{a}\cdot\vec{b}\nonumber \\
	  & \quad +\vec{a}^{2}\vec{b}^{2}-\left(\vec{a}\cdot\vec{b}\right)^{2},
\end{align}
and consequently

\begin{align}
	c_{0}^{2}+c_{k}^{2} & = a_{0}^{2}b_{0}^{2}+a_{0}^{2}\vec{b}^{2}+b_{0}^{2}\vec{a}^{2}+\vec{a}^{2}\vec{b}^{2}\nonumber \\
	& = \left(a_{0}^{2}+\vec{a}^{2}\right)\left(b_{0}^{2}+\vec{b}^{2}\right).
\end{align}
If $U_{1}$ and $U_{2}$ are elements of $SU(2)$ (i.e.\ $a_{0}^{2}+\vec{a}^{2}=1$
and $b_{0}^{2}+\vec{b}^{2}=1$), then $U_{3}$ will also be in $SU(2)$
(i.e.\ $c_{0}^{2}+c_{k}^{2}=1$). This multiplication rule also works for general complex matrices using this parametrization.

\subsection{Inverse of the exponential map}

If we want to extract the original algebra vector $A_{a}$ from the
parameters $u$ (see eqs.\ \eqref{eq:su2_param_cos} and \eqref{eq:su2_param_sin}), i.e.\ perform the inverse exponential map, then we have to do the following: First, we find the normalized
vector using
\begin{equation}
	e_{i}=\frac{u_{i}}{\sqrt{u_{i}u_{i}}}=\frac{A_{i}}{A}.
\end{equation}
We can compute the norm $A=\sqrt{A_{i}A_{i}}$ from eq.\ \eqref{eq:su2_param_sin}
\begin{equation}
	A=2\arcsin\left(\sqrt{u_{i}u_{i}}\right),
\end{equation}
as long as $\sqrt{u_{i}u_{i}} < 1$. The original components $A_i$ are then given by
\begin{equation}
	A_{i}=2\arcsin\left(\sqrt{u_{i}u_{i}}\right)\frac{u_{i}}{\sqrt{u_{i}u_{i}}}.
\end{equation}
If $\sqrt{u_i u_i} \ll 1$, one can also linearize:
\begin{equation}
	u_{i}=\sin\frac{A}{2}e_{i}\simeq\left(\frac{A}{2}+\mathcal{O}(A^{3})\right)e_{i}\approx\frac{A}{2}e_{i}.
\end{equation}
We then get
\begin{equation}
	A_{i}=Ae_{i}\approx2u_{i}.
\end{equation}

\subsection{Anti-hermitian traceless part of group elements}

The anti-hermitian traceless part of $U \in \SUN$ is given by (see \cref{eq:ah_definition_2})
\begin{equation}
\ah{U} = \frac{1}{2i} \lb U - U^\dg \rb - \frac{1}{N_c} \tr \lb \frac{1}{2i} \lb U - U^\dg \rb  \rb \one.
\end{equation}
Using $U = u_0 \, \one + i u_a \sigma_a$, we find
\begin{equation}
\tr \lb \frac{1}{2i} \lb U - U^\dg \rb  \rb = 0,
\end{equation}
and therefore
\begin{equation}
\ah{U} = \sigma_a u_a = 2 t_a u_a.
\end{equation}
Similarly, we find
\begin{equation}
P^a \lb U \rb = 2 \, \Im \,  \tr \lb t^a U \rb = 2 u^a.
\end{equation}
It holds that
\begin{equation}
t^a P^a \lb U \rb = \ah{U}.
\end{equation}
This is an exact relation, which generally holds for the fundamental representations of $\SUN$ (see \cref{app_gauss}).

In the case of SU(2), the relation between $P^a(U)$ and $u^a$ also allows us to write $U$ in terms of $P^a(U)$. Using $u^a u^a = \frac{1}{4} \lb P^a(U) \rb^2$ and assuming $u^0 = \sqrt{1- u^a u^a} > 0$ we find from \cref{eq:app_su2_constraint}
\begin{equation} \label{eq:su2_plaq_from_ah}
U = \sqrt{1- \frac{1}{4} \lb P^a (U) \rb^2} \, \one + \frac{i}{2} \sigma_a P^a (U).
\end{equation}
\subsection{Inverse of matrices} \label{sec:su2_inverse}

If $U \in$ SU(2), then $U$ is unitary and therefore $U^{-1} = U^\dg$. In terms of our parametrization we then have
\begin{align}
U &= u_0 \, \one + i \sigma_a u_a, \\
U^\dg &= u_0 \, \one - i \sigma_a u_a.
\end{align} 
However, if $U$ is not an element of SU(2) (i.e.\ $u^2_0 + u_a u_a \neq 1$), then we can still use the parametrization to compute the inverse. Writing
\begin{equation}
U^{-1} = v_0 \, \one + i \sigma_b v_b,
\end{equation}
and requiring $U U^{-1} = \one$ we find
\begin{align}
v_0 &= \alpha u_0, \\ 
v_a &= - \alpha u_a, 
\end{align}
with $\alpha = \lb u_0^2 + u_a u_a \rb^{-1}$.

This is mainly used for the Glasma initial conditions on the lattice, see sections \ref{sec:initial_lattice} and \ref{eq:Ui_initial_su2_sol}.

\section{Variation of gauge links}\label{app_var}

When performing the variation of e.g.\ the Wilson action $S[U]$ (see \cref{eq:wilson_action}) with respect to gauge links $U_{x,\mu}$, we have to keep in mind that one considers a variation w.r.t.\ special unitary matrices, i.e.\ it holds that 
\begin{align}
\det U_{x,\mu} &= 1, \label{eq:su_constraint_1}\\
U_{x,\mu} U^\dg_{x,\mu} &= \one. \label{eq:su_constraint_2}
\end{align}
It is therefore incorrect to simply vary $S[U]$ as a function of the matrix elements of $U_{x,\mu}$. Instead, we must make sure that the constraints in eqs.\ \eqref{eq:su_constraint_1} and \eqref{eq:su_constraint_2} remain fulfilled.

For instance, imagine writing the Wilson action $S[U]$ for $N_c = 2$ in terms of the parametrization of SU(2) introduced in \cref{app_su2}, such that every gauge link is given in terms of the four real-valued parameters $u_0$ and $u_a$ ($a \in \{1,2,3 \}$) with $U = u_0 \one + i \sigma_a u_a$. One could then try to find the variation of $S[U]$ by taking the independent derivatives w.r.t.\ the parameters $u_0$ and $u_a$ for each link. This would yield the wrong result, because the parametrization using four parameters is in fact overdetermined. In order for $U$ to be an element of SU(2), the parameters satisfy \cref{eq:app_su2_constraint}
\begin{equation}
u_0^2 + \sum_a \lb u_a \rb^2 = 1,
\end{equation}
which describes a three-dimensional sphere embedded in $\mathbb{R}^4$. Taking independent derivatives w.r.t.\ $u_0$ and $u_a$ generally violates this constraint. Geometrically speaking, one would ``leave" the sphere instead of ``moving along" the sphere while performing the variation.  

In order to solve this problem, we could implement the constraints using the method of Lagrange multipliers. However, this can be quite cumbersome: for each link we would need to introduce independent multipliers for eqs.\ \eqref{eq:su_constraint_1} and \eqref{eq:su_constraint_2} in the Wilson action and then additionally solve for the unknown multipliers. 

In order to avoid the use of Lagrange multipliers, we use the method of constrained variation, which turns out to be much easier to handle.
We introduce the infinitesimal variation of a gauge link variable
\begin{equation}
\dd U_{x,\mu} = i g a^\mu \dd A_{x,\mu} U_{x,\mu},
\end{equation}
or written shorter
\begin{equation}
\dd U_{x,\mu} = i \dd A_{x,\mu} U_{x,\mu},
\end{equation}
where the variation of the gauge field $\dd A_{x,\mu}$ is traceless and hermitian and the factor $g a^\mu$ is absorbed in a re-definition of $\dd A_{x,\mu}$. In the continuum limit $\dd A_{x,\mu}$ becomes the infinitesimal variation of the gauge field $A_{\mu}(x)$.
The infinitesimal variation $\dd U_{x,\mu}$ preserves the unitarity of gauge links. Let $U'_{x,\mu} = U_{x,\mu} + \dd U_{x,\mu}$. We then find
\begin{align}
U'_{x,\mu} U'^\dg_{x,\mu} &= \lb U_{x,\mu} + \dd U_{x,\mu} \rb \lb U^\dg_{x,\mu} + \dd U^\dg_{x,\mu} \rb \nonumber \\
& \simeq \one + \dd U_{x,\mu} U^\dg_{x,\mu} + U_{x,\mu} \dd U^\dg_{x,\mu} +\mathcal O \lb \dd A^2 \rb \nonumber \\
& \simeq \one + i \dd A_{x,\mu} - i \dd A_{x\,\mu } +\mathcal O \lb \dd A^2 \rb \nonumber \\
& \simeq \one + \mathcal O \lb \dd A^2 \rb.
\end{align}
The determinant is also unaffected for infinitesimal variations. Using Taylor expansion of the determinant we find
\begin{align} \label{eq:det_constraint_proof}
\det U'_{x,\mu} &= \det \lb U_{x,\mu} + \dd U_{x,\mu} \rb  \nn
& \simeq 1 + \tr \lb \adj \lb U_{x,\mu} \rb \dd U_{x,\mu} \rb + \mathcal O \lb \dd A^2 \rb \nn
& \simeq 1 + \tr \lb U^\dg_{x,\mu} \dd U_{x,\mu} \rb + \mathcal O \lb \dd A^2 \rb \nn
& \simeq 1 + \mathcal O \lb \dd A^2 \rb,
\end{align}
where ``$\adj$" refers to the adjugate which is given by $\adj \lb U_{x,\mu} \rb = \det U_{x,\mu} U^{-1}_{x,\mu} = U^\dg_{x,\mu}$. In the second line of \cref{eq:det_constraint_proof} we have made use of Jacobi's formula for the derivative of the determinant.
The variation $\dd U_{x,\mu}$ therefore preserves the constraints and allows us to vary the action without the use of Lagrange multipliers, which dramatically simplifies the derivation of equations of motion.

The constrained variation of the gauge link further allows us to define the derivative
\begin{equation} \label{eq:gauge_link_derivative}
\frac{\p U_{x,\mu}}{\p A^a_{x,\mu}} = i g a^\mu t^a U_{x,\mu}.
\end{equation}
This follows from
\begin{align}
\frac{\p U_{x,\mu}}{\p A^a_{x,\mu}} &\equiv \lim_{\delta A^a_{x,\mu} \rightarrow 0} \frac{U_{x,\mu} \left[ A_\mu + \delta A_\mu \right] - U_{x,\mu} [A_\mu]}{\delta A^a_{x,\mu}} \nn
&= \lim_{\delta A^a_{x,\mu} \rightarrow 0} \frac{i g a^\mu \dd A^a_{x,\mu} t^a U_{x,\mu}}{\delta A^a_{x,\mu}} \nn
&= i g a^\mu t^a U_{x,\mu}.
\end{align}
\section{Variation of the leapfrog action} \label{app_leapfrog}
In this section of the appendix we derive the discrete equations of motion (EOM) (as used in sections \ref{sec:latt_field__eqs} and \ref{sec:semi_leapfrog_scheme}) obtained from the standard Wilson action \eqref{eq:wilson_action} (or \cref{eq:wilson_action2}) using the constraint preserving variation of link variables of the previous section. To make the calculation more organized, we first split the action into two parts: a part containing temporal plaquettes $S_E[U]$ (``E" for electric) and a part containing spatial plaquettes $S_B[U]$ (``B" for magnetic). We write
\begin{equation}
S[U] = S_E[U] - S_B[U],
\end{equation}
where
\begin{align}
	S_E[U] &= \frac{V}{g^2} \sum_{x,i} \frac{1}{\lb a^0 a^i \rb^2} \tr \lb 2- U_{x,0i} - U^\dg_{x,0i} \rb,\\
	S_B[U] &= \frac{V}{g^2} \sum_{x,i,j} \frac{1}{2} \frac{1}{\lb a^i a^j \rb^2} \tr \lb 2- U_{x,ij} - U^\dg_{x,ij}  \rb.
\end{align}
Furthermore, we use the identity introduced in \cref{nonabelian_fields}
\begin{equation}
C_{x,\mu\nu} C^\dg_{x,\mu\nu} = 2 - U_{x,\mu\nu} -U^\dg_{x,\mu\nu},
\end{equation}
with
\begin{equation}
C_{x,\mu\nu} \equiv U_{x,\mu} U_{x,\nu} - U_{x,\nu} U_{x,\mu}.
\end{equation}
This allows us to rewrite the electric and magnetic part as
\begin{align}
S_E[U] &= \frac{V}{g^2} \sum_{x,i} \frac{1}{\lb a^0 a^i \rb^2} \tr \lb C_{x,0i} C^\dg_{x,0i} \rb, \label{eq:wilson_action_electric}\\
S_B[U] &= \frac{V}{g^2} \sum_{x,i,j} \frac{1}{2} \frac{1}{\lb a^i a^j \rb^2} \tr \lb C_{x,ij} C^\dg_{x,ij} \rb.
\end{align}
\subsection{Gauss constraint} \label{app_leapfrog_gauss}
For the Gauss constraint (GC) in the leapfrog scheme we only have to consider the variation $\dd_t S_E[U]$, since $S_B[U]$ does not contain any temporal links.
In the following sections we make use of the ``$\sim$" symbol, denoting equality under the sum over lattice sites $x$ and under the trace (see \cref{sec:lattice_notation}).
We then have
\begin{align}
\dd_t \lb C_{x,0i} C^\dg_{x,0i} \rb &= \lb \dd U_{x,0} U_{x+0,i} - U_{x,i} \dd U_{x+i,0} \rb C^\dg_{x,0i} + \hc \nonumber \\
&\sim \dd U_{x,0} \lb U_{x+0,i} C^\dg_{x,0i} - C^\dg_{x-i,0i} U_{x-i,i}\rb + \hc \nonumber \\
&=i \dd A_{x,0} \lb U_{x,0} \lb U_{x+0,i} C^\dg_{x,0i} - C^\dg_{x-i,0i} U_{x-i,i} \rb \rb + \hc
\end{align}
To go from the first to the second line, we applied a shift $x \rightarrow x-i$ in the right term of the first line and made use of the cyclicity of the trace. In the third line we simply used the definition of the variation of gauge links.
The variation of the action therefore reads
\begin{equation}
\dd_t S_E[U] = -\frac{V}{g^{2}}\sum_{x,i,a}\frac{1}{\left(a^{0}a^{i}\right)^{2}}P^{a}\left(U_{x,0}\left(U_{x+0,i}C_{x,0i}^{\dg}-C_{x-i,0i}^{\dg}U_{x-i,i}\right)\right)\delta A_{x,0}^{a},
\end{equation}
where we used
\begin{equation}
P^a\lb U \rb \equiv 2 \Im \, \tr \lb t^a U \rb = -i \tr \lb t^a \lb U - U^\dg \rb \rb.
\end{equation}
Replacing the $C_{x,ij}$ terms with link variables we find
\begin{equation}
U_{x,0}\left(U_{x+0,i}C_{x,0i}^{\dg}-C_{x-i,0i}^{\dg}U_{x-i,i}\right) = 2-U_{x,0i}-U_{x,0-i},
\end{equation}
and subsequently
\begin{equation}
\dd_t S_E[U] = \frac{V}{g^{2}}\sum_{x,i,a} \frac{1}{\left(a^{0}a^{i}\right)^{2}}P^{a}\left(U_{x,0i}+U_{x,0-i}\right)\delta A_{x,0}^{a}.
\end{equation}
We require that the variation vanishes, i.e.\ $\dd S[U] = 0$. Since all gauge links can be varied independently we find
\begin{equation} \label{eq:app_lf_gc}
\sum_{i}\frac{1}{\left(a^{0}a^{i}\right)^{2}}P^{a}\left(U_{x,0i}+U_{x,0-i}\right)=0.
\end{equation}
\subsection{Equations of motion} \label{app_leapfrog_eom}

First, we consider the variation of $S_E[U]$ w.r.t.\ spatial links. We find a result that is similar to the expression for the GC
\begin{equation}
\delta_{s}\left(C_{x,0i}C_{x,0i}^{\dg}\right) \sim i\delta A_{x,i}\left(U_{x,i}\left(C_{x-0,0i}^{\dg} U_{x-0,0}-U_{x+i,0} C_{x,0i}^{\dg}\right)\right)+ \hc
\end{equation}
and
\begin{align}
\dd_s S_E[U] &= -\frac{V}{g^{2}}\sum_{x,i,a}\frac{1}{\left(a^{0}a^{i}\right)^{2}} \dd A_{x,i}^{a}  P^{a}\left(U_{x,i}\left(C_{x-0,0i}^{\dg}U_{x-0,0}-U_{x+i,0}C_{x,0i}^{\dg}\right)\right)\nonumber \\
&= \frac{V}{g^{2}}\sum_{x,i,a}\frac{1}{\left(a^{0}a^{i}\right)^{2}} \dd A_{x,i}^{a}  P^{a}\left(U_{x,i0}+U_{x,i-0}\right).
\end{align}
For the variation of $S_B[U]$ we use
\begin{align}
\sum_{i,j} \dd_s \lb C_{x,ij} C^\dg_{x,ij} \rb & \sim \sum_{i, j} \left(\delta U_{x,i}U_{x+i,j}+U_{x,i}\delta U_{x+i,j} - \dd U_{x,j} U_{x+j,i} - U_{x,j} \dd U_{x+j,i}\right)C_{x,ij}^{\dg}+\hc \nonumber \\
& \sim i \sum_{i, j} 2 \delta A_{x,i}U_{x,i}\left(U_{x+i,j}C_{x,ij}^{\dg}+C_{x-j,ji}^{\dg}U_{x-j,j}\right)+\hc
\end{align}
The variation then reads
\begin{equation}
\dd_s S_B[U] = -\frac{V}{g^{2}}\sum_{x,i,j,a}\frac{1}{\left(a^{i}a^{j}\right)^{2}} \dd A_{x,i}^{a} P^{a}\left(U_{x,i}\left(U_{x+i,j}C_{x,ij}^{\dg}+C_{x-j,ji}^{\dg}U_{x-j,j}\right)\right).
\end{equation}
We set $\dd S = 0$ and, after canceling some constants, we find the discrete EOM
\begin{equation}
\frac{1}{\left(a^{0}a^{i}\right)^{2}}P^{a}\left(U_{x,i0}+U_{x,i-0}\right)=-\sum_{j}\frac{1}{\left(a^{i}a^{j}\right)^{2}}P^{a}\left(U_{x,i}\left(U_{x+i,j}C_{x,ij}^{\dg}+C_{x-j,ji}^{\dg}U_{x-j,j}\right)\right),
\end{equation}
which can also be written as
\begin{equation} \label{eq:app_lf_eom}
\frac{1}{\lb a^0 a^i \rb^2} P^a \lb U_{x,i0} + U_{x,i-0}\rb =
\sum_j \frac{1}{\lb a^i a^j \rb^2} P^a \lb U_{x,ij} + U_{x,i-j}\rb.
\end{equation}
\section{Conservation of the Gauss constraint}\label{app_gauss}
We now explicitly show that the leapfrog EOM \eqref{eq:app_lf_eom} preserve the associated GC \eqref{eq:app_lf_gc}. We use the identity for the fundamental representation of $\SUN$ (see \cref{eq:Pa_and_ah})
\begin{equation}
\sum_a t^{a}P^{a}\left(X\right) = \ah{X},
\end{equation}
where ``ah" denotes the anti-hermitian traceless part of $X$, see \cref{eq:ah_definition_2}. The constraint and the equations of motion (including external charges) then read 
\begin{align}
\sum_{i}\frac{1}{\left(a^{0}a^{i}\right)^{2}}\ah{U_{x,0i}+U_{x,0-i}} &= \frac{g}{a^0} \rho_x,\label{eq:app_gauss_ah} \\
\frac{1}{\lb a^0 a^i \rb^2} \ah{U_{x,i0} + U_{x,i-0}} &=
\sum_j \frac{1}{\lb a^i a^j \rb^2} \ah{U_{x,ij} + U_{x,i-j}} - \frac{g}{a^i} j_{x,i}. \label{eq:app_eom_ah}
\end{align}
We take \cref{eq:app_eom_ah} and sum over $i$. Due to $\ah{U_{x,ij}}$ being antisymmetric in the index pair $i,j$ we find
\begin{equation}
\sum_i \frac{1}{\lb a^0 a^i \rb^2} \ah{U_{x,i0} + U_{x,i-0}} =
\sum_{i,j} \frac{1}{\lb a^i a^j \rb^2} \ah{U_{x,i-j}} - \sum_i \frac{g}{a^i} j_{x,i}.
\end{equation}
Doing the same at $x-i$ and parallel transporting from $x-i$ to $x$ yields
\begin{equation}
\sum_i \frac{1}{\lb a^0 a^i \rb^2} \ah{U_{x,0-i} + U_{x,-0-i}} =
\sum_{i,j} \frac{1}{\lb a^i a^j \rb^2} \ah{U_{x,j-i}} - \sum_i \frac{g}{a^i} U^\dg_{x-i,i} j_{x-i,i} U_{x-i,i}.
\end{equation}
Subtracting the above two equations gives
\begin{equation}
\sum_i \frac{1}{\lb a^0 a^i \rb^2} \ah{U_{x,i0} + U_{x,i-0} - U_{x,0-i} - U_{x,-0-i}} =
- \sum_i \frac{g}{a^i} \lb j_{x,i} - U^\dg_{x-i,i} j_{x-i,i} U_{x-i,i} \rb.
\end{equation}
Using antisymmetry we have 
\begin{equation}
\ah{U_{x,i0}} = - \ah{U_{x,0i}}
\end{equation}
and
\begin{equation}
\ah{U_{x,-0-i}} = - \ah{U_{x,-i-0}}.
\end{equation}
Moreover, in temporal gauge we have $U_{x,i-0} = U_{x-0,0i}$ and $U_{x,-i-0}=U_{x-0,0-i}$, which leads to
\begin{align}
\sum_i \frac{1}{\lb a^0 a^i \rb^2} \ah{U_{x,i0} + U_{x,i-0} - U_{x,0-i} - U_{x,-0-i}} &= \nonumber \\
\sum_i \frac{1}{\lb a^0 a^i \rb^2} \ah{-\lb U_{x,0i} + U_{x,0-i} \rb + \lb U_{x-0,0i} + U_{x-0,0-i} \rb} &= \nonumber \\
-\frac{g}{a^0} \lb \rho_x -\rho_{x-0} \rb, &
\end{align}
where we used the GC in the last line to replace the temporal plaquette terms with charge densities.
This yields the gauge-covariant continuity equation
\begin{equation}
\frac{1}{a^0} \lb \rho_x -\rho_{x-0} \rb = \sum_i \frac{1}{a^i} \lb j_{x,i} - U^\dg_{x-i,i} j_{x-i,i} U_{x-i,i} \rb.
\end{equation}
If there are no external charges, we simply have the conservation of the GC: assuming that the constraint in the previous time slice holds, i.e.
\begin{equation}
\sum_i \frac{1}{\lb a^0 a^i \rb^2} \ah{U_{x-0,0i} + U_{x-0,0-i}} = 0,
\end{equation}
then the EOM guarantee that it will also hold in the next one, i.e.
\begin{equation}
\sum_i \frac{1}{\lb a^0 a^i \rb^2} \ah{U_{x,0i} + U_{x,0-i}} = 0.
\end{equation}

\section{Variation of the implicit action}\label{app_implicit}

We now consider the action introduced in \cref{implicit_scheme}
\begin{equation}
S[U] = S_E[U] - S_B[U],
\end{equation}
where
\begin{align}
S_E[U] &= \frac{V}{g^2} \sum_{x,i} \frac{1}{\lb a^0 a^i \rb^2} \tr \lb C_{x,0i} C^\dg_{x,0i} \rb,\\
S_B[U] &= \frac{V}{g^2} \sum_{x,i,\abs j} \frac{1}{4} \frac{1}{\lb a^i a^j \rb^2} \tr \lb C_{x,ij} M^\dg_{x,ij} \rb.
\end{align}
The electric part is the same as in the leapfrog scheme. However, the magnetic part $S_B[U]$ now contains temporal gauge links and gives a new contribution to the GC. 

Before we vary this action, we must verify that $S_B[U]$ is indeed real-valued. While it is easy to see that the original leapfrog action is real-valued because of the obvious hermicity of $C_{x,ij} C^\dg_{x,ij}$, the term $C_{x,ij} M^\dg_{x,ij}$ is not hermitian in general. Still, we can show that the action is real: we have
\begin{equation}
C_{x,ij} M^\dg_{x,ij} = \frac{1}{2} C_{x,ij} \lb C^{(+0)}_{x,ij} + C^{(-0)}_{x,ij} \rb^\dg,
\end{equation}
where we can rewrite
\begin{align}
C_{x,ij} C^{(+0)\dg}_{x,ij} &= C_{x,ij} \lb U_{x,0} C_{x+0,ij} U^\dg_{x+i+j,0}\rb^\dg \nonumber \\
&= C_{x,ij} U_{x+i+j,0} C^\dg_{x+0,ij} U^\dg_{x,0} \nonumber \\
&\sim U^\dg_{x,0} C_{x,ij} U_{x+i+j,0} C^\dg_{x+0,ij} \nonumber \\
&= C^{(-0)}_{x+0,ij} C^\dg_{x+0,ij}.
\end{align}
Here we used the cyclicity of the trace. Then using a shift $x \rightarrow x-0$ we have 
\begin{equation}
C_{x,ij} C^{(+0) \dg}_{x,ij} \sim C^{(-0)}_{x,ij} C^\dg_{x,ij}.
\end{equation}
Likewise we have 
\begin{equation}
C_{x,ij} C^{(-0) \dg}_{x,ij} \sim C^{(+0)}_{x,ij} C^\dg_{x,ij},
\end{equation}
which leads to
\begin{equation}
C_{x,ij} M^\dg_{x,ij} \sim M_{x,ij} C^\dg_{x,ij}.
\end{equation}
Incidentally, the RHS term is exactly the hermitian conjugate of the LHS term. In other words
\begin{equation}
\lb C_{x,ij} M^\dg_{x,ij} \rb^\dg = M_{x,ij} C^\dg_{x,ij} \sim C_{x,ij} M^\dg_{x,ij}.
\end{equation}
This means that under the sum over $x$ and the trace, the expression $C_{x,ij} M^\dg_{x,ij}$ is indeed real-valued and by extension $S_B[U]$ is real-valued as well. We have also shown that the time-average in $M_{x,ij}$ can be ``shifted" to the other term $C_{x,ij}$ under the sum and trace, i.e.
\begin{equation}
C_{x,ij} \avg{C}^\dg_{x,ij} \sim \avg{C}_{x,ij} C^\dg_{x,ij}.
\end{equation}
This is a useful property that we will need in the following derivation.

\subsection{Gauss constraint} \label{app_implicit_gauss}
We perform the variation of $S_E[U]$ and $S_B[U]$ w.r.t.\ temporal links to derive the GC in the implicit scheme. Since $S_E[U]$ is the same for all schemes, we do not have to repeat it. On the other hand, the variation of the magnetic part involves terms like $\dd_t \lb C_{x,ij} M^\dg_{x,ij} \rb$, which we now discuss explicitly. First, we make use of
\begin{equation}
\dd_t \lb C_{x,ij} M^\dg_{x,ij} \rb \sim \dd_t M_{x,ij} C^\dg_{x,ij} = \frac{1}{2} \lb \dd_t C^{(+0)}_{x,ij} + \dd_t C^{(-0)}_{x,ij} \rb C^\dg_{x,ij}.
\end{equation}
Then, after some algebra we find
\begin{equation}
\dd_t C^{(+0)}_{x,ij} C^\dg_{x,ij} \sim i \dd A_{x,0} \lb C^{(+0)}_{x,ij} C^{\dg}_{x,ij} - C_{x,-i-j} \lb C^{(+0)}_{x,-i-j} \rb^\dg \rb
\end{equation}
and 
\begin{align}
\dd_t C^{(-0)}_{x,ij} C^\dg_{x,ij} &\sim i \dd A_{x,0} \lb - C_{x,ij} \lb C^{(+0)}_{x,ij} \rb^\dg + C^{(+0)}_{x,-i-j} C^\dg_{x,-i-j} \rb \nonumber \\
&= \lb \dd_t C^{(+0)}_{x,ij} C^\dg_{x,ij} \rb^\dg,
\end{align}
which yields
\begin{equation}
\dd_t M_{x,ij} C^\dg_{x,ij} \sim \frac{i}{2} \dd A_{x,0} \lb \left[ C^{(+0)}_{x,ij} C^{\dg}_{x,ij} - \hc \right] + \left[  C^{(+0)}_{x,-i-j} C^\dg_{x,-i-j} -\hc \right] \rb.
\end{equation}
The variation of the magnetic part therefore reads
\begin{align} \label{eq:var_imp_SB_t}
\dd_t S_B[U] =& - \frac{V}{g^2} \sum_{x,a,i,\abs j} \frac{1}{8} \frac{1}{\lb a^i a^j \rb^2} \dd A^a_{x,0} P^a \lb
C^{(+0)}_{x,ij} C^{\dg}_{x,ij} + C^{(+0)}_{x,-i-j} C^\dg_{x,-i-j}
\rb \nonumber \\
&= - \frac{V}{g^2} \sum_{x,a,\abs i,\abs j} \frac{1}{8} \frac{1}{\lb a^i a^j \rb^2} \dd A^a_{x,0} P^a\lb
C^{(+0)}_{x,ij} C^{\dg}_{x,ij} 
\rb.
\end{align}
In the last line we consolidated terms with index $i$ and $-i$ into a single term using the sum $\sum_{\abs i}$.
Taking the result for $\dd_t S_E[U]$ from the leapfrog scheme, we find the GC
\begin{equation}
\sum_{i}\frac{1}{\lb a^{0}a^{i} \rb^{2}}P^{a}\lb U_{x,0i}+U_{x,0-i}\rb = 
- \sum_{\abs i,\abs j} \frac{1}{8} \frac{1}{\lb a^i a^j \rb^2} P^a\lb C^{(+0)}_{x,ij} C^{\dg}_{x,ij} \rb.
\end{equation}
\subsection{Equations of motion} \label{app_implicit_eom}
For the EOM we vary $S[U]$ w.r.t.\ spatial links. Again, we already have the result for $\dd_s S_E[U]$ from the leapfrog scheme and only need to calculate $\dd_s S_B[U]$. In particular we consider the term 
\begin{equation}
\dd_s \lb C_{x,ij} M^\dg_{x,ij} \rb = \dd_s  C_{x,ij} M^\dg_{x,ij} + C_{x,ij} \dd_s M^\dg_{x,ij}.
\end{equation}
Since the variation only acts on spatial links we can shift the time-average of the right term from $M^\dg_{x,ij}$ to $C_{x,ij}$. This gives
\begin{align}
\dd_s \lb C_{x,ij} M^\dg_{x,ij} \rb &\sim \dd_s C_{x,ij} M^\dg_{x,ij} + M_{x,ij} \dd_s C^\dg_{x,ij} \nonumber \\
&= \dd_s C_{x,ij} M^\dg_{x,ij} + \hc
\end{align}
The variation then proceeds analogously to the derivation of the leapfrog scheme in \cref{app_leapfrog_eom}. We find
\begin{equation} \label{eq:var_imp_SB_s}
\dd_s S_B[U] =-\frac{V}{g^2} \sum_{x,i,\abs j} \frac{1}{2} \frac{1}{\lb a^{i}a^{j}\rb^{2}} \dd A_{x,i}^{a} P^{a}\lb U_{x,i}\lb U_{x+i,j} M_{x,ij}^{\dg}+M_{x-j,ji}^{\dg} U_{x-j,j}\rb \rb.
\end{equation}
With the result for $\dd_s S_E[U]$ we obtain the discrete EOM
\begin{equation}
\frac{1}{\lb a^0 a^i \rb^2} P^a \lb U_{x,i0} + U_{x,i-0}\rb =
- \frac{1}{2} \sum_{\abs j} \frac{1}{\lb a^i a^j \rb^2} P^a \lb U_{x,i} \lb U_{x+i,j} M^\dg_{x,ij} + M^\dg_{x-j,ji} U_{x-j,j}\rb\rb.
\end{equation}
Introducing the shorthand
\begin{equation}
K_{x,ij}[U, M] = - \frac{1}{2} \frac{1}{\lb a^i a^j \rb^2} \lb U_{x+i,j} M^\dg_{x,ij} - M^\dg_{x-j,ij} U_{x-j,j}\rb,
\end{equation}
allows us to write the EOM rather compactly as
\begin{equation}
\frac{1}{\lb a^0 a^i \rb^2} P^a \lb U_{x,i0} + U_{x,i-0}\rb =
\sum_{\abs j} P^a \lb U_{x,i} K_{x,ij}[U,M] \rb.
\end{equation}
\section{Variation of the semi-implicit action}\label{app_semi}

In the semi-implicit scheme (see \cref{semi_scheme}) the action reads
\begin{equation}
S[U] = S_E[U] - S_B[U],
\end{equation}
where $S_E[U]$ is given by \cref{eq:wilson_action_electric}. The magnetic part comprises of $S_B[U] = S_{B,M}[U] + S_{B,W}[U]$, where $S_{B,M}[U]$ is the same as $S_B[U]$ from the implicit scheme except that the indices $i$ and $j$ in $\sum_{i, \abs j}$ only run through transverse components:
\begin{equation}
S_{B,M}[U] = \frac{V}{g^2} \sum_{x,i,\abs j} \frac{1}{4} \frac{1}{\lb a^i a^j \rb^2} \tr \lb C_{x,ij} M^\dg_{x,ij} \rb.
\end{equation}
Therefore we can take the results from the previous section for $\dd S_{B,M}[U]$, eqs.\ \eqref{eq:var_imp_SB_t} and \eqref{eq:var_imp_SB_s}. The new part is given by
\begin{equation}
S_{B,W}[U] = \frac{V}{g^2} \sum_{x, \abs j} \frac{1}{4} \frac{1}{\lb a^1 a^j \rb^2}  \tr \lb C_{x,1j} W^\dg_{x,1j} + \hc \rb.
\end{equation}
\subsection{Gauss constraint} \label{app_semi_gauss}
We already know $\dd_t S_E[U]$ and $\dd_t S_{B,M}[U]$ from sections \ref{app_leapfrog_gauss} and \ref{app_implicit_gauss}, so we only have to compute $\dd_t S_{B,W}[U]$. The relevant terms are
\begin{align}
\sum_{\abs j} \frac{1}{\lb a^j \rb^2}\dd_t W_{x,1j} C^\dg_{x,1j} &= \sum_{\abs j} \frac{1}{\lb a^j \rb^2} \lb \dd_t \avg{U}_{x,1} U_{x+1,j} - U_{x,j} \dd_t \avg{U}_{x+j,1} \rb C^\dg_{x,1j} \nonumber \\
&\sim \dd_t \avg{U}_{x,1} \sum_{\abs j} \frac{1}{\lb a^j \rb^2} \lb U_{x+1,j} C^\dg_{x,1j} - C^\dg_{x-j,1j} U_{x-j,j} \rb \nonumber \\
&= \dd_t \avg{U}_{x,1} T^{\dg}_{x,1},
\end{align}
where we defined
\begin{equation}
T^{\dg}_{x,1} = \sum_{\abs j} \frac{1}{\lb a^j \rb^2} \lb U_{x+1,j} C^\dg_{x,1j} - C^\dg_{x-j,1j} U_{x-j,j} \rb.
\end{equation}
Using the same techniques as before, we find 
\begin{align}
\dd_t \avg{U}_{x,1} T^{\dg}_{x,1} \sim \frac{i}{2} \dd A_{x,0} 
\bigg( & U^{(+0)}_{x,1} T^\dg_{x,1} - U_{x,1} T^{(+0) \dg}_{x,1} \nonumber \\
- &  T^\dg_{x-1,1} U^{(+0)}_{x-1,1} + T^{(+0) \dg}_{x-1,1} U_{x-1,1}  \bigg),
\end{align}
and the variation of $S_{B,W}[U]$ reads
\begin{align}
\dd_t S_{B,W}[U] = - \frac{V}{g^2} \sum_{x,a} \frac{1}{8\lb a^1 \rb^2} \dd A^a_{x,0} P^a \bigg(
&U^{(+0)}_{x,1} T^\dg_{x,1} + T^{(+0)}_{x,1} U^\dg_{x,1} \nonumber \\
+ &U^{(+0)\dg}_{x-1,1} T_{x-1,1} +  T^{(+0) \dg}_{x-1,1} U_{x-1,1}
\bigg),
\end{align}
which (with the previous results taken into account) yields the GC in the semi-implicit scheme, see  \cref{eq:semi_gauss}.

\subsection{Equations of motion} \label{app_semi_eom}

For the variation w.r.t.\ spatial links we have to distinguish two cases: the longitudinal and the transverse links. Starting with the variation of longitudinal links we find
\begin{align}
\dd_1 \lb C_{x,1j} W^\dg_{x,1j} \rb +\hc &=
\dd_1 C_{x,1j} W^\dg_{x,1j} + C_{x,1j} \dd_1 W^\dg_{x,1j} + \hc \nonumber \\
&\sim i \dd A_{x,1} U_{x,1} \bigg( \lb U_{x+1,j} W^\dg_{x,1j} + W^\dg_{x-j,j1} U_{x-j,j} \rb \nonumber \\
& + \avg{\lb U_{x+1,j} C^\dg_{x,1j} + C^\dg_{x-j,j1} U_{x-j,j} \rb}\bigg) + \hc
\end{align}
Again we used the fact that the time-average can be shifted to other terms by exploiting the sum over $x$ and the cyclicity of the trace. The variation of $S_{B,W}[U]$ then reads
\begin{align}
\dd_1 S_{B,W}[U] = -\frac{V}{g^2} \sum_{x,a,j} \frac{1}{4} \frac{1}{\lb a^1 a^j \rb^2} \dd A^a_{x,1} P^a \bigg(
U_{x,1} \bigg( &\lb U_{x+1,j} W^\dg_{x,1j} + W^\dg_{x-j,j1} U_{x-j,j} \rb \nonumber \\
+&\avg{\lb U_{x+1,j} C^\dg_{x,1j} + C^\dg_{x-j,j1} U_{x-j,j} \rb}\bigg) \bigg).
\end{align}
Combining the above with $\dd_1 S_E[U]$ yields the longitudinal EOM \eqref{eq:semi_eom_1}. No contributions from $S_{B,M}[U]$ are necessary because it does not include any longitudinal links.

For the transverse components of the EOM we vary w.r.t.\ $U_{x,j}$, where $j$ is a transverse index. The relevant terms for $j>0$ are
\begin{align}
\dd_j W_{x,1j} C^\dg_{x,1j} \sim i \dd A_{x,j} U_{x,j} \lb \avg{U}_{x+j,1} C^\dg_{x,j1} - C^\dg_{x-1,j1} \avg{U}_{x-1,1} \rb,
\end{align}
and
\begin{align}
\dd_j C_{x,1j} W^\dg_{x,1j} \sim i \dd A_{x,j} U_{x,j} \lb U_{x+j,1} W^\dg_{x,j1} - W^\dg_{x-1,j1} U_{x-1,1} \rb.
\end{align}
For terms with negative component indices $j<0$ we can show that they are identical (under the sum over $x$ and the trace) to the last two terms except for the substitution $1 \rightarrow -1$:
\begin{align}
\dd_j \sum_{-j}  \lb W_{x,1j} C^\dg_{x,1j} + \hc \rb &= \dd_j \sum_{j} \lb W_{x,1-j} C^\dg_{x,1-j} + \hc \rb \nonumber \\
&\sim \sum_{j} \lb \dd_j W_{x,-1j} C^\dg_{x,-1j}  + \dd_j C_{x,-1j} W^\dg_{x,-1j} + \hc \rb.
\end{align}
This allows us to write
\begin{align}
\sum_{\abs j} \dd_j W_{x,1j} C^\dg_{x,1j} &\sim \sum_{\abs 1} i \dd A_{x,j} U_{x,j} \lb \avg{U}_{x+j,1} C^\dg_{x,j1} - C^\dg_{x-1,j1} \avg{U}_{x-1,1} \rb, \\
\sum_{\abs j} \dd_j C_{x,1j} W^\dg_{x,1j} &\sim \sum_{\abs 1} i \dd A_{x,j} U_{x,j} \lb U_{x+j,1} W^\dg_{x,j1} - W^\dg_{x-1,j1} U_{x-1,1} \rb,
\end{align}
where $\sum_{\abs 1}$ stands for summing over terms with component indices $1$ and $-1$.
The variation of $S_{B,W}[U]$ then reads
\begin{align}
\dd_j S_{B,W}[U] =& \frac{V}{g^2} \sum_{x, \abs{j}} \frac{1}{4} \frac{1}{\lb a^1 a^j \rb^2}  \dd_j \tr \lb C_{x,1j} W^\dg_{x,1j} + \hc \rb & \nonumber \\
=& - \frac{V}{g^2} \sum_{x,j,a} \frac{1}{4} \frac{1}{\lb a^1 a^j \rb^2} \dd A^a_{x,j} P^a \bigg( U_{x,j} \sum_{\abs 1}\bigg( \lb \avg{U}_{x+j,1} C^\dg_{x,j1} - C^\dg_{x-1,j1} \avg{U}_{x-1,1} \rb \nonumber \\
& \qquad\qquad+ \lb U_{x+j,1} W^\dg_{x,j1} - W^\dg_{x-1,j1} U_{x-1,1} \rb \bigg) \bigg).
\end{align}
With the expressions for $\dd_j S_E[U]$ and $\dd_j S_{W,M}[U]$ we find the transverse components of the EOM \eqref{eq:semi_eom_i}.

\chapter{Stability of the semi-implicit Abelian scheme}\label{app_abelian_semi}

In this section of the appendix we prove stability for the semi-implicit scheme for Abelian fields derived in \cref{sec:abelian_semi_implicit}. Using temporal gauge, $A_{x,0}=0$, the EOM \eqref{eq:abelian_eom_d3_1} and \eqref{eq:abelian_eom_d3_2} read
\begin{align}
-\p_{0}^{2}A_{x,1}	&=	\frac{1}{2}\sum_{i}\p_{i}^{B}\left(W_{x,1i}+M_{x,1i}\right), \\
-\p_{0}^{2}A_{x,i}	&=	\sum_{j\neq i}\p_{j}^{B}M_{x,ij}+\frac{1}{2}\p_{1}^{B}\left(W_{x,i1}+F_{x,i1}\right),
\end{align}
where $W_{x,1i}$ reduces to
\begin{equation}
W_{x,1i}=\p_{1}^{F}A_{x,i}-\p_{i}^{F}\bar{A}_{x,1}.
\end{equation}
The Gauss constraint \eqref{eq:abelian_constraint_d3} reads
\begin{equation}
\sum_{i=1}^{d}\p_{i}^{B}\p_{0}^{F}A_{x,i}+\left(\frac{a^{0}}{2}\right)^{2}\sum_{i}\p_{1}^{B}\p_{i}^{B}\p_{0}^{F}\left(\p_{1}^{F}A_{x,i}-\p_{i}^{F}A_{x,1}\right)=0.
\end{equation}
Splitting the equations of motion into Laplacian terms and mixed derivative terms we find
\begin{align}
-\p_{0}^{2}A_{x,1}	&=	-\sum_{i}\p_{i}^{2}\avg{A}_{x,1}
+\frac{1}{2}\sum_{i}\p_{i}^{B}\p_{1}^{F}\left(A_{x,i}+\avg{A}_{x,i}\right), \\
-\p_{0}^{2}A_{x,i}	&=	-\sum_{j\neq i}\p_{j}^{2}\avg{A}_{x,i}-\p_{1}^{2}A_{x,i}
+\sum_{j\ne i}\p_{j}^{B}\p_{i}^{F}\avg{A}_{x,j}+\frac{1}{2}\p_{1}^{B}\p_{i}^{F}\left(\avg{A}_{x,1}+A_{x,1}\right).
\end{align}
Using a plane wave ansatz
\begin{equation}
A_{x,i} = A_i e^{i \lb \omega x^0 - \sum_i k^i x^i \rb},
\end{equation}
we will use the Gauss constraint to first reduce the number of degrees of freedom and then compute the dispersion relation $\omega(k)$. Inserting the ansatz into the Gauss constraint we find
\begin{equation}
\left(1+\sum_{i}\chi_{i}^{2}\right)\chi_{1}^{B}A_{1}+\left(1-\chi_{1}^{2}\right)\sum_{i}\chi_{i}^{B}A_{i}=0,
\end{equation}
where we use the dimensionless lattice momenta \eqref{eq:dimless_lattice_momentum}. The constraint equation can alternatively be written as
\begin{equation}
\chi_{1}^{B}A_{1}=-\beta\sum_{i}\chi_{i}^{B}A_{i},
\end{equation}
where $\beta$ is a momentum-dependent factor given by
\begin{equation}
\beta = \frac{1-\chi_{1}^{2}}{1+\sum_{i}\chi_{i}^{2}}.
\end{equation}
The temporal average $\avg{A}_{x,i}$ reduces to a multiplication with a frequency dependent factor:
\begin{equation}
\avg{A}_{x,i} = \cos \lb \omega a^0 \rb A_{x,i} = c A_{x,i},
\end{equation}
where we used the shorthand $c = \cos \lb \omega a^0 \rb$. Inserting the plane wave ansatz into the EOM yields
\begin{align}
\chi_{0}^{2}A_{1} &= c \sum_{i} \chi_{i}^{2} A_{1} - \frac{1}{2} \left( 1+c \right) \sum_{i} \chi_{i}^{B} \chi_{1}^{F} A_{i}, \\
\chi_{0}^{2}A_{i} &= c \sum_{j\neq i} \chi_{j}^{2} A_{i} + \chi_{1}^{2} A_{i} - c \sum_{j\ne i} \chi_{j}^{B} \chi_{i}^{F} A_{j} - \frac{1}{2}\left(1+c\right)\chi_{1}^{B}\chi_{i}^{F}A_{1}.
\end{align}
Note that both $\chi_0$ and $c$ depend on $\omega$. After making use of the Gauss constraint the longitudinal EOM reads
\begin{equation}
\chi_{0}^{2}A_{1}=c\left(\chi_{2}^{2}+\chi_{3}^{2}\right)A_{1}+\frac{1}{2}\left(1+c\right)\beta^{-1}\chi_{1}^{2}A_{1},
\end{equation}
and the two transverse equations read
\begin{align}
\chi_{0}^{2}A_{2}	&=	c\chi_{3}^{2}A_{2}+\chi_{1}^{2}A_{2}-c\chi_{3}^{B}\chi_{2}^{F}A_{3}-\frac{1}{2}\left(1+c\right)\chi_{1}^{B}\chi_{2}^{F}A_{1} \nonumber \\
&=	c \chi_{3}^{2}A_{2}+\chi_{1}^{2}A_{2}+\frac{1}{2}\left(1+c\right)\beta\chi_{2}^{2}A_{2}+\left(\frac{1}{2}\left(1+c\right)\beta-c\right)\chi_{2}^{F}\chi_{3}^{B}A_{3}, \\
\chi_{0}^{2}A_{3}	&=	c\chi_{2}^{2}A_{3}+\chi_{1}^{2}A_{3}-c\chi_{2}^{B}\chi_{3}^{F}A_{2}-\frac{1}{2}\left(1+c\right)\chi_{1}^{B}\chi_{3}^{F}A_{1} \nonumber \\
&=	c\chi_{2}^{2}A_{3}+\chi_{1}^{2}A_{3}+\frac{1}{2}\left(1+c\right)\beta\chi_{3}^{2}A_{3}+\left(\frac{1}{2}\left(1+c\right)\beta-c\right)\chi_{2}^{B}\chi_{3}^{F}A_{2}.
\end{align}
This system of equations can be written in matrix notation as an eigenvalue problem
\begin{equation}
M \vec{A} = \chi^2_0 \vec{A},
\end{equation}
where the coefficient matrix $M$ is given by
\begin{equation}
M = \begin{pmatrix}
	\frac{1}{2}\left(1+c\right)\beta^{-1}\chi_{1}^{2}+c\left(\chi_{2}^{2}+\chi_{3}^{2}\right) & 0 & 0\\
	0 & \chi_{1}^{2}+\frac{1}{2}\left(1+c\right)\beta\chi_{2}^{2}+c\chi_{3}^{2} & \left(\frac{1}{2}\left(1+c\right)\beta-c\right)\chi_{2}^{F}\chi_{3}^{B}\\
	0 & \left(\frac{1}{2}\left(1+c\right)\beta-c\right)\chi_{2}^{B}\chi_{3}^{F} & \chi_{1}^{2}+c\chi_{2}^{2}+\frac{1}{2}\left(1+c\right)\beta\chi_{3}^{2}
\end{pmatrix},
\end{equation}
and the vector $\vec{A}$ is simply
\begin{equation}
\vec{A} =  \lb \begin{array}{c}
	A_{1}\\
	A_{2}\\
	A_{3}
\end{array} \rb.
\end{equation}
The eigenvectors of $M$ are
\begin{equation}
\left\{ \vec{A}_{L},\vec{A}_{T,1},\vec{A}_{T,2}\right\} =
\left\{
\left(
\begin{array}{c}
	1\\
	0\\
	0
\end{array}
\right),
\left(
\begin{array}{c}
	0\\
	-\chi_{3}^{B}\\
	\chi_{2}^{B}
\end{array}
\right),
\left(
\begin{array}{c}
	0\\
	\chi_{2}^{F}\\
	\chi_{3}^{F}
\end{array}
\right)\right\}, 
\end{equation}
where we find two transverse, momentum dependent eigenvectors $\vec{A}_{T,1}$, $\vec{A}_{T,2}$ and the longitudinal unit vector $\vec{A}_L$. The two transverse vectors $\vec{A}_{T,1}$ and $\vec{A}_{T,2}$ can be interpreted as transverse polarization modes and are orthogonal in the sense of $\left(\vec{A}_{T,1}\right)\cdot\left(\vec{A}_{T,2}\right)^{\dg}=0$. The three eigenvectors yield three different equations for the eigenvalue problem, namely
\begin{align}
M \vec{A}_L &= \lambda_L(c, \chi_i) \vec{A}_L = \chi^2_0 \vec{A}_L, \\
M \vec{A}_{T,l} &= \lambda_{T,l}(c, \chi_i) \vec{A}_{T,l} = \chi^2_0 \vec{A}_{T,l}, \quad l \in {1, 2},
\end{align}
where $\lambda_L(c, \chi_i)$ and $\lambda_{T,k}(c, \chi_i)$ are expressions which depend on the momenta $\chi_i$ and the frequency $\omega$ via $c=\cos{\lb \omega a^0 \rb}$.
Solving the first equation $\lambda_L(c, \chi_i) = \chi^2_0$ for $\omega$ yields the dispersion relation of the longitudinal component:
\begin{equation} \label{eq_app_semi_freq1}
\omega_{L}a^{0}=\arccos\left(\frac{1-\chi_{1}^{2}\left(2+\chi_{2}^{2}+\chi_{3}^{2}\right)}{1+\chi_{2}^{2}\left(2-\chi_{1}^{2}\right)+\chi_{3}^{2}\left(2-\chi_{1}^{2}\right)}\right).
\end{equation}
Solving the two transverse equations $\lambda_{T,l}(c, \chi_i) = \chi^2_0$ yields
\begin{equation} \label{eq_app_semi_freq2}
\omega_{T,1}a^{0}=\arccos\left(\frac{1-2\chi_{1}^{2}}{1+2\chi_{2}^{2}+2\chi_{3}^{2}}\right),
\end{equation}
and
\begin{equation}
\omega_{T,2}a^{0}=\arccos\left(\frac{1-\chi_{1}^{2}\left(2+\chi_{2}^{2}+\chi_{3}^{2}\right)}{1+\chi_{2}^{2}\left(2-\chi_{1}^{2}\right)+\chi_{3}^{2}\left(2-\chi_{1}^{2}\right)}\right)=\omega_{L}a^{0}.
\end{equation}
It turns out that the expressions for $\omega$ associated with the different eigenvectors are not the same, except that $\vec{A}_L$ and $\vec{A}_{T,2}$ share the same dispersion relation, that is to say  $\omega_{T,1} \neq \omega_L = \omega_{T,2}$ in general.
This implies that for a given momentum $k$, the amplitude of an arbitrary wave has to be split into two components: a part which is projected into the plane spanned by $\vec{A}_L$ and $\vec{A}_{T,2}$ which oscillates with $\omega_L=\omega_{T,2}$ and a part parallel to $\vec{A}_{T,1}$ which oscillates with frequency $\omega_{T,1}$.
We interpret this as numerical (or artificial) birefringence. 

We require the propagation of a wave to be stable, i.e.\ we require the frequencies $\omega$ to be real-valued. This is guaranteed if the arguments of the $\arccos$ expressions in eqs.\ \eqref{eq_app_semi_freq1} and \eqref{eq_app_semi_freq2} are restricted to $[-1,1]$. Both dispersion relations remain stable if the CFL condition
\begin{equation}
\chi^2_1 \leq 1,
\end{equation}
holds. Using 
\begin{equation}
\chi^2_1 = \lb \frac{a^0}{a^1} \rb \sin^2 \lb \frac{k^1 a^1}{2} \rb,
\end{equation}
and requiring stability for all values of $k^1$ yields the constraint
\begin{equation}
a^0 \leq a^1.
\end{equation}
This concludes the proof that the semi-implicit scheme, even though exhibiting peculiar wave propagation phenomena, is stable.

\chapter{Abbreviations}

\begin{description}[labelsep=4em, align=left, labelwidth=1in]
\item [2+1D]		2+1 dimensions / dimensional
\item [3+1D]		3+1 dimensions / dimensional
\item [BFKL]		Balitsky, Fadin, Kuraev, Lipatov 
\item [CFL]			Courant, Friedrichs, Levy
\item [CGC]			Color Glass Condensate
\item [CIC]			cloud-in-cell
\item [CPIC]		colored particle-in-cell
\item [EOM]			equation(s) of motion
\item [FDTD]		finite difference time domain
\item [FFT]			fast Fourier transformation
\item [GC]			Gauss constraint
\item [IR]			infrared
\item [JIMWLK]		Jalilian-Marian, Iancu, McLerran, Weigert, Leonidov, Kovner
\item [LC]			light cone
\item [LF]			leapfrog
\item [LHC]			Large Hadron Collider
\item [LHS]			left hand side
\item [LRF]			local rest frame
\item [MV]			McLerran, Venugopalan
\item [NCI]			numerical Cherenkov instability
\item [NGP]			nearest grid point
\item [PIC]			particle-in-cell
\item [QCD]			quantum chromodynamics
\item [QGP]			quark-gluon plasma
\item [RHIC]		Relativistic Heavy Ion Collider
\item [RHS]			right hand side
\item [SI]			semi-implicit
\item [SU]			special unitary
\item [UV]			ultraviolet
\item [w.r.t.]		with respect to
\item [YM]			Yang-Mills
\end{description}

\cleardoublepage
\phantomsection
\addcontentsline{toc}{chapter}{Bibliography}
\bibliographystyle{JHEP}
\bibliography{references}

\end{document}